\documentclass[10pt,a4paper]{article}

\usepackage[T1]{fontenc}
\usepackage[utf8]{inputenc}
\usepackage{amssymb,amsmath,amsthm,amsfonts}    %
\usepackage{stmaryrd}                           %
\usepackage{bm}                                 %
\usepackage{mathtools}                          %
\usepackage{breqn}                              %
\usepackage[textfont=footnotesize,labelfont={footnotesize,bf}]{caption}
\usepackage{subcaption}
\usepackage{booktabs}
\usepackage[inline]{enumitem}
\usepackage{graphicx}
\usepackage[dvipsnames]{xcolor}
\usepackage[top=22mm,right=22mm,left=22mm,bottom=22mm]{geometry}
\usepackage[colorlinks=true,linkcolor=blue,urlcolor=blue]{hyperref}
\usepackage[affil-it,auth-sc]{authblk}
\usepackage[colorinlistoftodos,textwidth=20mm,textsize=footnotesize]{todonotes}
\usepackage{lipsum}                             %
\usepackage[
    backend=biber,
    style=numeric,
    citestyle=numeric-comp,
    maxbibnames=99,
    minbibnames=99,
    maxcitenames=2,
    mincitenames=1,
    giveninits=true,
    sorting=none,
    sortlocale=auto,
    natbib=true,
    url=false,
    isbn=false,
    doi=true,
    eprint=true
]{biblatex}
\DeclareMathOperator{\tr}{tr}

\DeclareMathOperator{\Div}{Div}
\DeclareMathOperator{\grad}{grad}

\DeclarePairedDelimiter{\norm}{\lVert}{\rVert}

\newcommand{\scalp}{\bm \cdot}
\newcommand{\trans}[1]{#1^{\intercal}}
\newcommand{\conj}[1]{\overline{#1}}
\newcommand{\conjtrans}[1]{#1^{\mathsf{H}}}
\newcommand{\deriv}[2]{\frac{\partial #1}{\partial #2}}

\newcommand{\Reals}{\mathbb{R}}

\newcommand{\Integers}{\mathbb{Z}}

\newcommand{\bzero}{\bm 0}
\newcommand{\ba}{\bm a}
\newcommand{\bb}{\bm b}

\newcommand{\be}{\bm e}
\newcommand{\bef}{\bm f}
\newcommand{\bg}{\bm g}

\newcommand{\bk}{\bm k}

\newcommand{\bn}{\bm n}

\newcommand{\bp}{\bm p}
\newcommand{\bq}{\bm q}

\newcommand{\bt}{\bm t}
\newcommand{\bu}{\bm u}

\newcommand{\bx}{\bm x}
\newcommand{\by}{\bm y}

\newcommand{\bA}{\bm A}
\newcommand{\bB}{\bm B}

\newcommand{\bE}{\bm E}

\newcommand{\bI}{\bm I}
\newcommand{\bJ}{\bm J}
\newcommand{\bK}{\bm K}
\newcommand{\bL}{\bm L}
\newcommand{\bM}{\bm M}
\newcommand{\bN}{\bm N}

\newcommand{\bS}{\bm S}
\newcommand{\bT}{\bm T}

\newcommand{\bX}{\bm X}

\newcommand{\bZ}{\bm Z}
\newcommand{\balpha}{\bm \alpha}

\newcommand{\bvarphi}{\bm \varphi}

\newcommand{\bXi}{\bm \Xi}

\newcommand{\bGamma}{\bm \Gamma}

\newcommand{\fC}{\mathbb C}

\newcommand{\fE}{\mathbb E}

\newcommand{\mB}{\mathcal B}
\newcommand{\mC}{\mathcal C}

\newcommand{\mE}{\mathcal E}

\newcommand{\mL}{\mathcal L}

\newcommand{\mO}{\mathcal O}

\newcommand{\mS}{\mathcal S}
\newcommand{\mT}{\mathcal T}

\newcommand{\mV}{\mathcal V}

\graphicspath{{./}{./figures/}}                         %

\title{Incremental constitutive tensors and strain localization for
prestressed elastic lattices: Part II -- incremental dynamics}

\author[1]{G. Bordiga}
\author[2]{L. Cabras}
\author[1]{A. Piccolroaz}
\author[1]{D. Bigoni\footnote{Corresponding author: e-mail: \href{mailto:bigoni@ing.unitn.it}{bigoni@ing.unitn.it}; phone: +39\,0461\,282507.}}

\affil[1]{DICAM, University of Trento, Trento, Italy}
\affil[2]{DICATAM, University of Brescia, Brescia, Italy}

\date{}

\begin{document}

\maketitle

\begin{abstract}
\noindent
Floquet-Bloch wave asymptotics is used to homogenize the in-plane mechanical response of a periodic grillage of elastic Rayleigh rods, possessing a distributed mass density, together with rotational inertia.
The grid is subject to incremental time-harmonic dynamic motion, superimposed to a given state of axial loading of arbitrary magnitude. 
In contrast to the quasi-static energy match (addressed in Part I of the present study), the vibrational properties of the lattice are directly represented by the acoustic tensor of the equivalent solid (without passing through the constitutive tensor).
The acoustic tensor is shown to be independent of the rods' rotational inertia and allows the analysis of strong ellipticity of the equivalent continuum, evidencing coincidence with macro-bifurcation in the lattice.
On the other hand, micro-bifurcation corresponds to a vibration of vanishing frequency of the lowest dispersion branch of the lattice, occurring at finite wavelength.
Dynamic homogenization reveals the structure of the acoustic branches close to ellipticity loss and the analysis of forced vibrations (both in physical space and Fourier space) shows low-frequency wave localizations.
A comparison is presented between strain localization occurring near ellipticity loss and forced vibration of the lattice, both corresponding to the application of a concentrated pulsating force. 
The comparison shows that the homogenization technique allows an almost perfect representation of the lattice, with the exception of cases where micro-bifurcation occurs, which is shown to leave the equivalent solid unaffected.
Therefore, the presented results pave the way for the design of architected cellular materials to be used in applications where extreme deformations are involved.
\end{abstract}

\paragraph{Keywords} 
Dynamic homogenization \textperiodcentered\
Ellipticity loss \textperiodcentered\
Shear bands \textperiodcentered\
Bloch waves

\section{Introduction}
\label{sec:introduction}
\begin{figure}[htb!]
	\centering
	\includegraphics[width=0.55\linewidth]{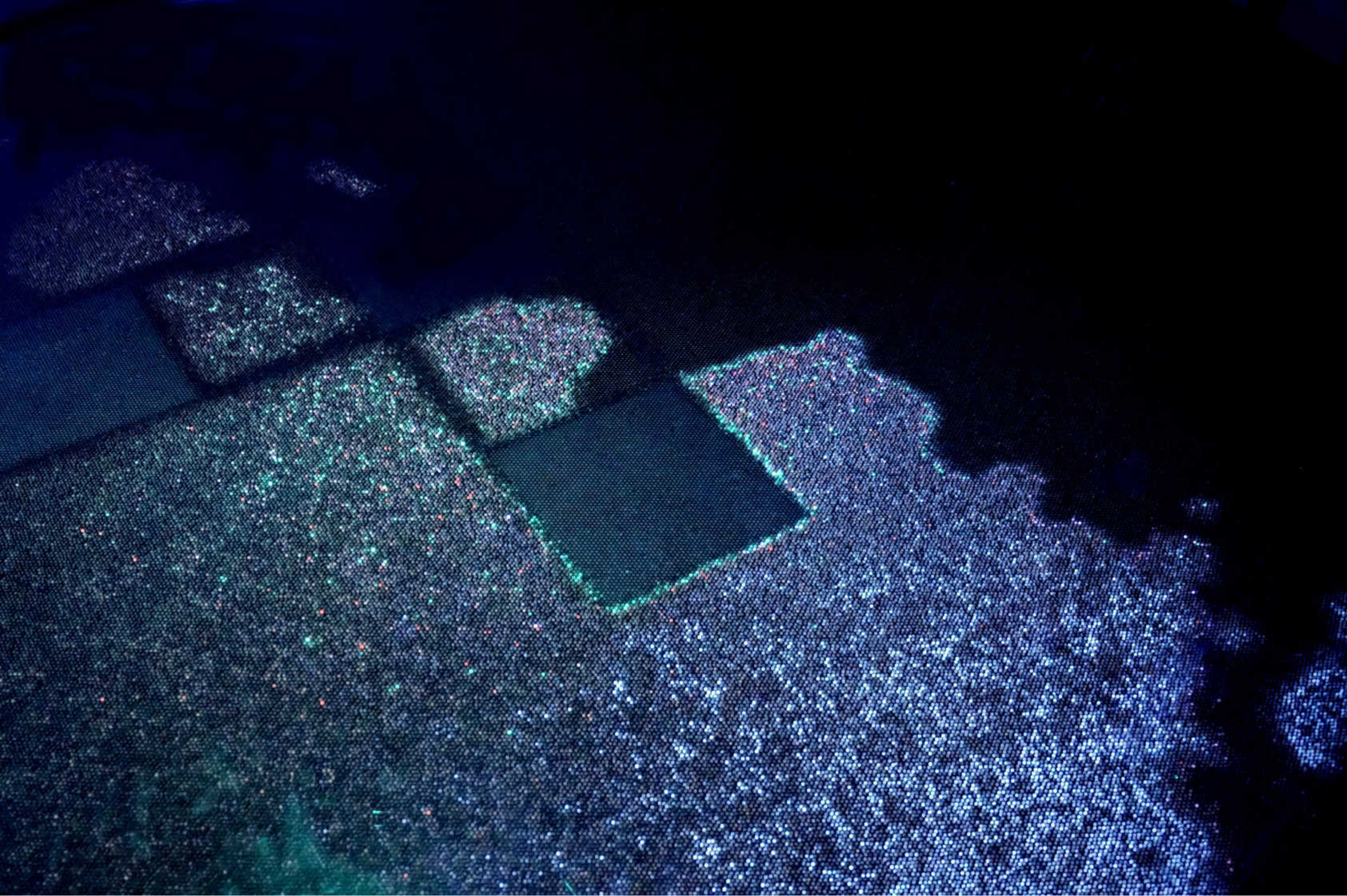}
	\caption{\label{fig:pinscreen}
		Discontinuity wavefronts (some rectilinear, other curvilinear) forming during the (out-of-plane) dynamics 
		of a periodic material (used as a toy, the so-called `pinscreen', invented by W. Fleming), which works on the boundary of ellipticity loss (photo taken at the Exploratorium, San Francisco).
	}
\end{figure}

A periodic grillage of axially-preloaded elastic rods, subject to in-plane incremental kinematics, has been shown (in Part I of this study, via quasi-static energy match with an equivalent continuum), to provide a model for the realization of materials evidencing macroscopic (strain localization) and microscopic (lattice buckling) instabilities within the elastic range. 
The structural grid represents an example of architected cellular material to be used for large strain applications where instabilities become a major issue. 
The latters are investigated in the present article within the dynamic time-harmonic regime, thus employing an asymptotic homogenization based on the Floquet-Bloch wave technique, exploited to a level of generality never achieved so far.
The vibrational properties of a cellular material are deeply affected by the emergence and development of localized signals, edge waves, and topologically protected modes~\cite{mishuris_2009a,wang_2015a,tallarico_2017,carta_2017,garau_2018,pal_2018,mazzotti_2019,bordiga_2019}, an example being that reported in Fig.~\ref{fig:pinscreen}. 
In the figure the dynamic emergence and propagation of discontinuity wavefronts (rectilinear and curvilinear) is shown in the so-called `pinscreen', a material (made up of a perforated plate having each hole filled with a movable pin) on the verge of ellipticity loss.

Homogenization techniques based on the asymptotic analysis of wave solutions dates back to Brillouin \cite{brillouin_1946} and Born \cite{born_1955}, and has received significant contributions in recent years when the case of random and periodic media has been considered \cite{parnell_2007,willis_2009,craster_2010,willis_2011,willis_2012,nemat-nasser_2011,nassar_2015,nassar_2017,kutsenko_2017} and extended to the high-frequency regime \cite{norris_2012,meng_2018,guzina_2019}.
With the exception of \cite{kutsenko_2017}, these developments have been so far produced for the analysis of wave propagation in continuous materials, not in structures, so that their practical implementation required the systematic use of numerical techniques (typically finite elements).
It is shown in the present article that low-frequency effective properties can be derived analytically for lattices composed of rods (incrementally loaded in-plane and subject to axial, shear and bending forces) through a direct computation of the wave asymptotics\footnote{The mathematical setting is two-dimensional for simplicity, but the three-dimensional extension is straightforward once the linearized dynamics of the rods is specified.}.
Recent results on beam grillages \cite{kutsenko_2017} are here extended to the case of elastic lattices, axially stressed up to an arbitrary amount, whose incremental dynamics is derived without restrictions on the rods' constitutive law and without neglecting the rotational inertia of the rods' cross section.
The low-frequency asymptotics of waves propagating in the lattice is shown to be governed by the spectral properties of the acoustic tensor associated to an effective continuum,  
equivalent to the discrete structure. 
This continuum is prestressed and, although coincident with that determined via energy match (Part I of this study), is obtained from the acoustic tensor, to which the Floquet-Bloch 
asymptotics directly leads. 
In this way, the correspondence between loss of ellipticity in the continuum and degeneracy of the acoustic properties in the lattice is \textit{directly} demonstrated.

The elastic rods are characterized by an axial mass density, equipped with rotational inertia (the Rayleigh model~\cite{piccolroaz_2017a,bordiga_2019a,nieves_2019}), so that it is possible to prove that the latter does not influence the vibrational properties, expressed by the acoustic tensor of the equivalent continuum. 
The homogenization, obtained analytically, is exploited to investigate the lattice response near macro-instability (coincident with the failure of strong ellipticity and thus of ellipticity in the effective continuum), therefore unveiling features of lattice dynamics loaded up to the verge of shear band formation.

The response to the application of a pulsating concentrated force is finally analyzed in the spirit of~\cite{bigoni_2005, piccolroaz_2006}. The force is applied both to the lattice (in the physical and Fourier spaces) and to the equivalent solid, with special detail on low-frequency wave localizations.

As shown in Part I of this study, in some cases micro-instabilities occur, which cannot be detected in the equivalent solid.
In these cases the lattice may be designed so that macro-instabilities \textit{prevail} over micro-instabilities.
The `explosive' nature of the latter has been addressed in Part I, while the former is here investigated under dynamic conditions. 
In this regard, the acoustic branches and the response to perturbation of the grid and its continuum approximation are shown to be in almost perfect agreement. 
This finding confirms that the lattice is an excellent candidate for the realization of architected cellular materials. These may be used to harness micro-buckling versus strain localization, two features which can control the dynamic response of a structure with tunable functionality~\cite{kochmann_2017,bertoldi_2017a}. 

This paper is organized as follows.
The mathematical setting for incremental wave propagation is developed in Section~\ref{sec:system_governing_equations} for an axially pre-loaded lattice of elastic rods, organized in an arbitrary periodic geometry.
The asymptotic analysis of lattice waves is derived in Section~\ref{sec:asymptotic_lattice_waves}, leading to the homogenization result that provides the acoustic tensor associated to the incremental effective elastic continuum, subject to a homogeneous state of prestress. 
The developed homogenization scheme is applied in Section~\ref{sec:ellipticity_loss_grid} to the same grid of elastic rods considered in Part I, so that failure of ellipticity and micro-buckling are investigated now through the analysis of the dispersion characteristics of the lattice.
In Section~\ref{sec:forced_response}, the emergence and characterization of dynamic strain localization is presented by means of a perturbative approach (applied to both lattice and effective continuum), which demonstrates how the \textit{macroscopic} localization direction, as well as the number of localization bands, can  effectively be designed through appropriate modifications of the \textit{underling} lattice structure.

\section{Incremental dynamics of preloaded lattices: governing equations}
\label{sec:system_governing_equations}
In this section, the governing equations for incremental wave propagation in an axially-preloaded lattice of elastic rods (connected to each other with joints capable of transmitting bending moment, shear, and axial forces) are formulated.
These are obtained (i) by solving for time-harmonic vibrations the incremental dynamics of a single rod (derived in Appendix~\ref{sec:linearized_elastica}), 
(ii) by using this solution to formulate the equations of motions for a single unit cell, and finally (iii) by applying the Bloch theorem to obtain the equations governing the incremental dynamics of the infinite lattice.
 
\begin{figure}[htb!]
    \centering
    \includegraphics[width=0.98\linewidth]{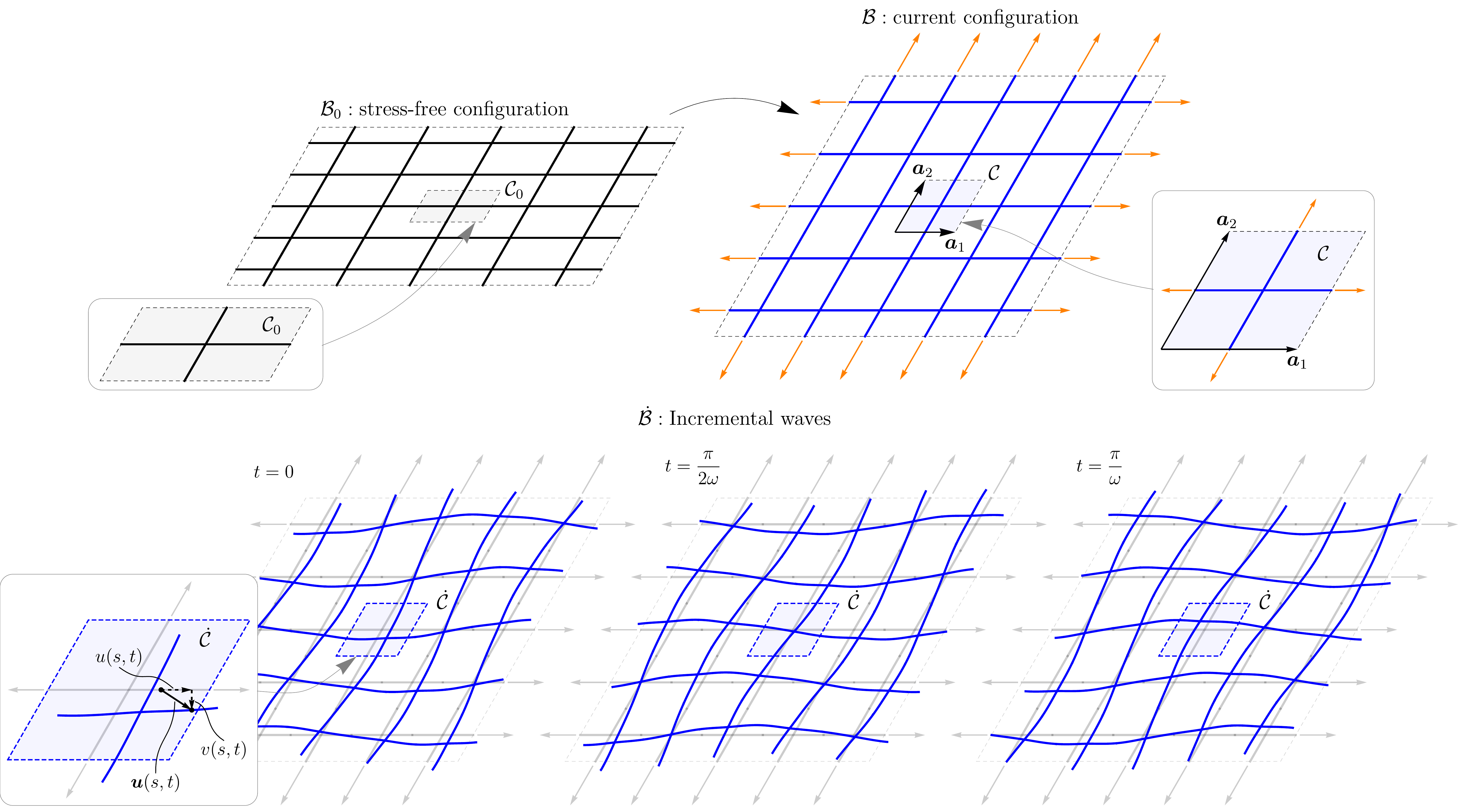}
    \caption{\label{fig:lattice_configurations}
    A periodic two-dimensional lattice of (axially and flexurally deformable) elastic rods is considered, preloaded from the stress-free configuration $\mB_0$ (upper part, on the left) by means of a pure axial loading state, transforming $\mB_0$ to the current, preloaded configuration $\mB$ (upper part, on the right). The latter configuration, used as reference in an updated Lagrangian description, can be represented as the tessellation of a single unit cell along the vectors of the direct basis $\{\ba_1,\ba_2\}$.
    The incremental dynamic response (lower part where an incremental deformation $\dot{\mB}$ is shown at three different instants of time) is defined on $\mB$ by the incremental displacement field of each rod $\bu(s,t)$, here decomposed in an axial and transverse component, $u(s,t)$ and $v(s,t)$.
    }
\end{figure}

An infinite two-dimensional lattice structure is considered, composed of nonlinear elastic rods which are axially preloaded (or prestretched) from an unloaded reference configuration $\mB_0$ to a preloaded configuration $\mB$ used as reference in an updated Lagrangian formulation of incremental dynamics (Fig.~\ref{fig:lattice_configurations}).

The configuration $\mB$ is assumed to be undistorted and to be described by the tessellation of a single unit cell along the vectors of the non-orthogonal direct basis 
$\{\ba_1,\ba_2\}$, as shown in Fig.~\ref{fig:lattice_configurations}. 
In the figure, in addition to the configurations $\mB_0$ and $\mB$, also incremental deformations $\dot{\mB}$ at 3 different instants of time are sketched, which highlight that the incremental analysis fully involves in-plane bending, stretching and shear.

By introducing a local coordinate $s_k$ for each rod of a given unit cell, the \textit{incremental} kinematics is described by means of the following fields
\begin{equation*}
    \bu_k(s_k,t)=\trans{\{ u_k(s_k,t), v_k(s_k,t) \}}, \qquad \forall k \in \{1,...,N_b\} \,,
\end{equation*}
where $N_b$ is the number of rods in the unit cell, and the two in-plane displacement components, respectively axial and transverse, are denoted by $u_k(s_k,t)$ and $v_k(s_k,t)$, while the rotation of the cross-section $\theta_k(s_k,t)$ is assumed to satisfy the unshearability condition $\theta_k(s_k,t)=v_k'(s_k,t)$\footnote{A dash will be used to denote differentiation with respect to the coordinate $s_k$.}. 
Time-harmonic solutions are sought, so that, by introducing the circular frequency $\omega$, the dependence on time $t$ can be represented as
\begin{equation}
    \label{eq:time_harmonic}
    \bu_k(s_k,t) = \hat{\bu}_k(s_k)\, e^{-i\,\omega\,t} \qquad \forall k \in \{1,...,N_b\} \,,
\end{equation}
where $\hat{\bu}_k(s_k)=\trans{\{ \hat{u}_k(s_k), \hat{v}_k(s_k) \}}$ are functions of the coordinate $s_k$ only and $i=\sqrt{-1}$.

In the following the superscript symbol $\hat{~}$ will be omitted, so it will be assumed that all quantities depend on time as prescribed by equation \eqref{eq:time_harmonic}.

\subsection{Time-harmonic solution for a preloaded elastic rod}
\label{sec:time-harmonic_solution_beam}
The analytic representation for time-harmonic vibrations of a Rayleigh rod is briefly introduced.
In the framework of a linear theory, the equations of motion governing the incremental time-harmonic dynamics of an axially pre-stretched Rayleigh rod are the following
\begin{subequations}
\label{eq:governing_beam_EB}
\begin{gather}
    \label{eq:governing_beam_EB_u}
    -\gamma(\lambda_0)\,\omega^2 u(s) - A(\lambda_0)\, u''(s) = 0 \,, \\
    \label{eq:governing_beam_EB_v}
    -\gamma(\lambda_0)\,\omega^2 v(s) + \gamma_r(\lambda_0)\,\omega^2 v''(s) + B(\lambda_0)\, v''''(s) - P(\lambda_0)\,v''(s) = 0 \,,
\end{gather}
\end{subequations}
where $\gamma(\lambda_0)$ is the \textit{current} linear mass density, $\gamma_r(\lambda_0)$ is the \textit{current} rotational inertia, $\lambda_0$ is the axial pre-stretch and $P(\lambda_0)$ the corresponding axial preload (assumed positive in tension), while $A(\lambda_0)$ and $B(\lambda_0)$ are, respectively, the \textit{current} axial and bending stiffness. 
The analytic derivation of  equations~\eqref{eq:governing_beam_EB} is reported in Appendix~\ref{sec:linearized_elastica}, where the formulation provided in Part I of this work is extended to the dynamic case.
Moreover, the derivation of the current stiffnesses $A(\lambda_0)$ and $B(\lambda_0)$ from strain-energy functions, as well as the their identification for rods made up of an incompressible nonlinear elastic material such as Mooney-Rivlin, can be found in Appendix A of Part I.
In the following, the parameters $\gamma(\lambda_0)$, $\gamma_r(\lambda_0)$, $A(\lambda_0)$, and $B(\lambda_0)$ will simply be denoted as, $\gamma$, $\gamma_r$, $A$, and $B$, and treated as independent quantities for generality. 

The substitution of Eq.~\eqref{eq:time_harmonic} into Eq.~\eqref{eq:governing_beam_EB} leads to a system of linear ODEs for the functions $\hat{u}(s)$ and $\hat{v}(s)$.
As the system is fully decoupled, the solution is easily obtained in the form
\begin{equation}
    \label{eq:u_v_sol}
    u(s) = \sum_{j=1}^2 C_{j}^u\, e^{i\,\beta_j^u\,s} \,, \qquad v(s) = \sum_{j=1}^4 C_{j}^v\, e^{i\,\beta_j^v\,s} \,,
\end{equation}
where $\{C_{1}^u,C_{2}^u,C_{1}^v,...,C_{4}^v\}$ are 6 arbitrary complex constants and the characteristic roots $\beta_j^{u}$ and $\beta_j^{v}$ are given by
\begin{equation*}
    \beta_{1,2}^u = \pm \frac{\tilde{\omega}}{l} \,, \qquad
    \beta_{1,2,3,4}^v= \pm \frac{1}{l\sqrt{2}}\sqrt{-p +r\,\tilde{\omega}^2 \pm \sqrt{p^2 + (4\Lambda^2 - 2\,p\,r)\,\tilde{\omega}^2 + r^2\,\tilde{\omega}^4}} \,,
\end{equation*}
with $l$ being the current length of the rod, $\tilde{\omega} = \omega\, l \sqrt{\gamma/A}$ the non-dimensional angular frequency, $p = P l^2/B$ the non-dimensional preload, $\Lambda=l/\sqrt{B/A}$ the slenderness of the rod, and $r = \gamma_r A/(\gamma B)$ is the dimensionless rotational inertia.

\subsection{Exact time-harmonic shape functions, mass and stiffness matrices}
\label{sec:shape_functions}
To facilitate the asymptotic expansion, it is instrumental to identify the 6 constants $\trans{\{C_{1}^u,C_{2}^u,C_{1}^v,...,C_{4}^v\}}$, with the \textit{degrees of freedom} at the rod's ends (i.e. its nodal displacements).
This allows a dimensional reduction through a direct application of the compatibility conditions at the joints.

For any given rod of length $l$ the following notation for the nodal parameters is introduced 
\begin{equation}
    \label{eq:nodaldisp}
    u(0) = u_1 \,, \quad v(0) = v_1 \,, \quad \theta(0) = \theta_1 \,, \quad u(l) = u_2 \,, \quad v(l) = v_2 \,, \quad \theta(l) = \theta_2 \,. 
\end{equation}
Collecting the degrees of freedom at the two ends of the rod in the vector $\bq=\trans{\{u_1,v_1,\theta_1,u_2,v_2,\theta_2\}}$ yields 
the solution of system \eqref{eq:nodaldisp} in the form
\begin{equation}
    \label{eq:time_harmonic_sf}
    \bu(s) = \bN(s;\omega, P)\, \bq \,.
\end{equation}
The 2-by-6 matrix $\bN(s;\omega, P)$ acts as a matrix of frequency-dependent and preload-dependent `shape functions' which is the \textit{exact} functional basis in which the time-harmonic response of the rod can be represented.
Equation~\eqref{eq:time_harmonic_sf} can also be considered as the definition of a `finite element' endowed with shape functions built from the exact solution.
Moreover, these time-harmonic shape functions reduce to the quasi-static solution when $\omega\to0$ (so that for vanishing preload, in the limit $\lim_{\omega\to0}\bN(s;\omega, 0)$ the usual shape functions for beam elements, employed for instance in \cite{phani_2006}, are recovered). 
In the following $\bN(s;\omega, P)$ will be denoted as $\bN(s;\omega)$, to simplify notation.

By employing Eq.~\eqref{eq:time_harmonic_sf}, the exact mass and stiffness matrices for a rod subject to time-harmonic vibration can be computed.
For the $k$-th rod the kinetic energy and the elastic strain energy at second order are given by\footnote{A superimposed dot is used to denote time differentiation unless explicitly stated otherwise.}
\begin{subequations}
\label{eq:kinetic_strain_energy}
\begin{align}
    &
    \label{eq:kinetic_energy}
    \begin{aligned}
        \mT_k &= \frac{1}{2} \int_0^{l_k} \gamma_k \left(\dot{u}_k(s_k,t)^2+\dot{v}_k(s_k,t)^2\right) ds_k + \frac{1}{2} \int_0^{l_k} \gamma_{r,k} \dot{v}'_k(s_k,t)^2 ds_k \\
        &
        =\frac{1}{2} \,\trans{\dot{\bq}_{k}(t)} \left(\int_0^{l_k}  \trans{\bN_{k}(s_k;\omega)}\bJ_{k}\, \bN_{k}(s_k;\omega) ds_k\right) \dot{\bq}_{k}(t)
        + \frac{1}{2} \,\trans{\dot{\bq}_{k}(t)} \left(\int_0^{l_k} \gamma_{r,k}\, \trans{\bb_{k}(s_k;\omega)} \bb_{k}(s_k;\omega) \,ds_k \right) \dot{\bq}_{k}(t) \,, 
    \end{aligned} \\
    &
    \label{eq:strain_energy}
        \mE_k = \frac{1}{2} \int_0^{l_k} \left(A_k\,u'_k(s_k,t)^2 + B_k\,v''_k(s_k,t)^2\right) ds_k
	    = \frac{1}{2} \,\trans{\bq_{k}(t)} \left(\int_0^{l_k} \trans{\bB_{k}(s_k;\omega)}\bE_k\,\bB_{k}(s_k;\omega) ds_k \right) \bq_{k}(t) \,,
\end{align}
\end{subequations}
where $\bJ_{k}$ and $\bE_{k}$ are matrices collecting the inertia and stiffness terms, respectively, while $\bB_{k}(s_k;\omega)$ is the strain-displacement matrix, which are defined as follows
\begin{equation*}
    \bJ_{k}=\begin{bmatrix}
        \gamma_k 	& 0 \\
        0  		& \gamma_k
    \end{bmatrix} \,,
    \qquad
    \bE_{k}=\begin{bmatrix}
        A_k 	& 0 \\
        0  		& B_k
    \end{bmatrix} \,,
    \qquad
    \bB_{k}(s_k;\omega) = \begin{bmatrix}
        \deriv{}{s_k} 	& 0 \\[2mm]
        0  	 	        & \deriv{^2}{s_k^2}
    \end{bmatrix} \bN_{k}(s_k;\omega) \,,
\end{equation*}
and $\bb_{k}(s_k;\omega)=\begin{bmatrix}0 & \deriv{}{s_k}\end{bmatrix}\bN_{k}(s_k;\omega)$ is a row vector containing the derivative of the shape functions corresponding to the transverse displacement $v$.
Note that the kinetic energy~\eqref{eq:kinetic_energy} accounts for translational as well as rotational inertia of the rod.
In addition, the contribution of the axial preload has to be included in the second-order potential energy as (details are provided in Appendix~\ref{sec:linearized_elastica})
\begin{equation}
\label{eq:prestress_energy}
    \mV_k^g = \frac{1}{2} P_k \int_0^{l_k} v'_k(s_k,t)^2 \,ds_k
            = \frac{1}{2} \,\trans{\bq_{k}(t)} \left(P_k \int_0^{l_k} \trans{\bb_{k}(s_k;\omega)} \bb_{k}(s_k;\omega) \,ds_k \right) \bq_{k}(t) \,.
\end{equation}
By combining Eqs.~\eqref{eq:strain_energy} and~\eqref{eq:prestress_energy}, the second-order potential energy of the $k$-th rod is denoted as
\begin{equation}
    \label{eq:potential_energy}
    \mV_k = \mE_k + \mV_k^g \,.
\end{equation}

From Eqs.~\eqref{eq:kinetic_strain_energy},~\eqref{eq:prestress_energy} and~\eqref{eq:potential_energy} the frequency-dependent mass and stiffness matrices are naturally defined as
\begin{subequations}
\label{eq:M_K_beam}
\begin{align}
    \label{eq:M_beam}
    \bM_{k}(\omega) &= \int_0^{l_k} \trans{\bN_{k}(s_k;\omega)}\bJ_{k}\,\bN_{k}(s_k;\omega)\,ds_k + \int_0^{l_k} \gamma_{r,k}\, \trans{\bb_{k}(s_k;\omega)} \bb_{k}(s_k;\omega)\,ds_k \,, \\
    \label{eq:K_beam}
    \bK_{k}(\omega) &= \int_0^{l_k} \trans{\bB_{k}(s_k;\omega)}\bE_{k}\,\bB_{k}(s_k;\omega)\,ds_k +
    P_k \int_0^{l_k} \trans{\bb_{k}(s_k;\omega)} \bb_{k}(s_k;\omega) \,ds_k \,.
\end{align}
\end{subequations}
The matrices for the quasi-static case can be obtained by evaluating the limit $\omega\to0$.
In particular, the quasi-static stiffness matrix $\lim_{\omega\to0}\bK_{k}(\omega)$ is found to be identical to that reported in Part I, where it is used to formulate the incremental equilibrium and analyze the bifurcation of preloaded lattices.

\subsection{Equations of motion for the unit cell}
\label{sec:equations_motion_cell}
As expressions~\eqref{eq:kinetic_strain_energy}--\eqref{eq:potential_energy} govern the \textit{incremental} dynamics of a single rod, the kinetic and potential energies of a single unit cell can be obtained through a summation of the contributions from each rod
\begin{equation*}
    \mT = \sum_{k=1}^{N_b} \mT_k \,, \qquad \mV = \sum_{k=1}^{N_b} \mV_k \,,
\end{equation*}
and the \textit{incremental} equations of motion for the unit cell can be derived from the Lagrangian
\begin{equation}
    \label{eq:lagrangian}
    \mL(\bq,\dot{\bq}) = \mT(\dot{\bq}) - \mV(\bq) \,,
\end{equation}
where $\bq$ is the vector collecting all the degrees of freedom of the unit cell.

The Lagrangian~\eqref{eq:lagrangian} leads to the following Euler-Lagrange equations
\begin{equation}
\label{eq:equations_of_motion_cell}
    \bM(\omega)\,\ddot{\bq}(t) + \bK(\omega)\,\bq(t) = \bef(t) \,,
\end{equation}
where $\bM(\omega)$ and $\bK(\omega)$ are, respectively, the mass and stiffness matrices of the unit cell\footnote{The matrices $\bM(\omega)$ and $\bK(\omega)$ can be easily obtained by assembling the matrices expressed by Eqs.~\eqref{eq:M_K_beam} for all the rods in the unit cell.} and the force vector $\bef$ collects the \textit{incremental} forces acting on the boundary nodes of the unit cell. 
As in Part I, no external forces are assumed on the nodes inside the unit cell due to the assumption of null body forces acting on the lattice.

By recalling the time-harmonic assumption, Eq.~\eqref{eq:equations_of_motion_cell} is rewritten as
\begin{equation}
    \label{eq:system_isolated}
    \bA(\omega)\, \bq = \bef \,,
\end{equation}
with the definition $\bA(\omega)=-\omega^2\bM(\omega)+\bK(\omega)$.
Note that the dimension of the linear system~\eqref{eq:system_isolated} is $3N_j$, where $N_j$ is the number of nodes within the unit cell.

\subsection{Bloch's theorem}
\label{sec:Bloch_theorem}
Wave propagation in an infinite periodic elastic medium can effectively be analyzed through the application of Bloch's theorem.
Essentially, the theorem states that the time-harmonic solutions of the equations of motion possess a modulation in space having the same periodicity of the medium, a condition expressed by the following requirement\footnote{Note that the representation~\eqref{eq:bloch_theorem} holds also when the displacement is replaced by fields defining the generalized forces internal to the rods.}
\begin{subequations}
\label{eq:bloch_theorem}
\begin{equation}
    \label{eq:bloch_theorem_1}
    \bu(\bx) = \bvarphi(\bx) \,e^{i\,\bk \scalp \bx} \,,
\end{equation}
where $\bk$ is the Bloch vector and the modulation $\bvarphi(\bx)$ is periodic with respect to the direct basis, so that it satisfies
\begin{equation*}
    \bvarphi(\bx+n_j\ba_j)=\bvarphi(\bx) \qquad \forall \{n_1,n_2\}\in\Integers^2 \quad\forall\bx\in\Reals^2 \,.
\end{equation*}
Eq.~\eqref{eq:bloch_theorem_1} can be equivalently expressed as 
\begin{equation}
    \label{eq:bloch_theorem_2}
    \bu(\bx+n_j\ba_j) = \bu(\bx) \,e^{i\,\bk \scalp (n_j\ba_j)} \,.
\end{equation}
\end{subequations}

Note that, in the case of a lattice made up of rods, the waveform $\bvarphi(\bx)$ (as well as the field $\bu(\bx,t)$) is described by the displacement of the rods forming the unit cell, and thus it is defined only for $\bx$ corresponding to the location of the structural elements.

The importance of the Bloch's theorem lies in the fact that it allows the response of an infinite and periodic structure to be described through the equations of motion for a unit cell, the latter complemented  by suitable boundary conditions.
These so-called `Floquet-Bloch' conditions enforce the periodicity of the lattice, by relating nodal displacements and forces on the boundary of the unit cell, according to the property expressed by Eq.~\eqref{eq:bloch_theorem_2}.
The application of these conditions to periodic beam lattices is well-known~\cite{langley_1993, phani_2006} and is briefly summarized in the following.
\begin{figure}[htb!]
    \centering
    \begin{subfigure}{0.5\textwidth}
    \centering
    \caption{\label{fig:FB_joints_q}}
    \includegraphics[height=40mm]{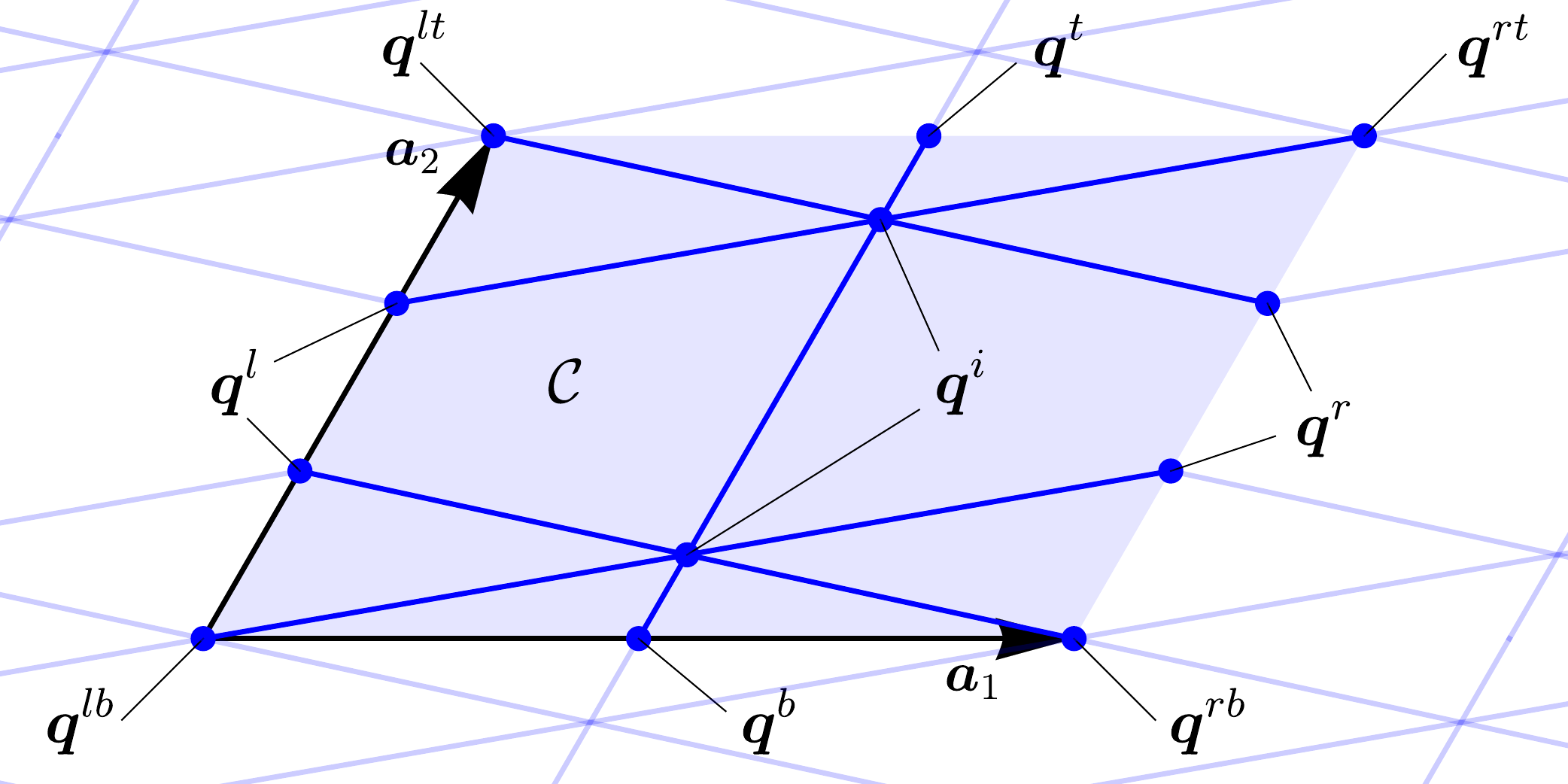}
    \end{subfigure}%
    \begin{subfigure}{0.4\textwidth}
    \centering
    \caption{\label{fig:FB_joints_f}}
    \includegraphics[height=40mm]{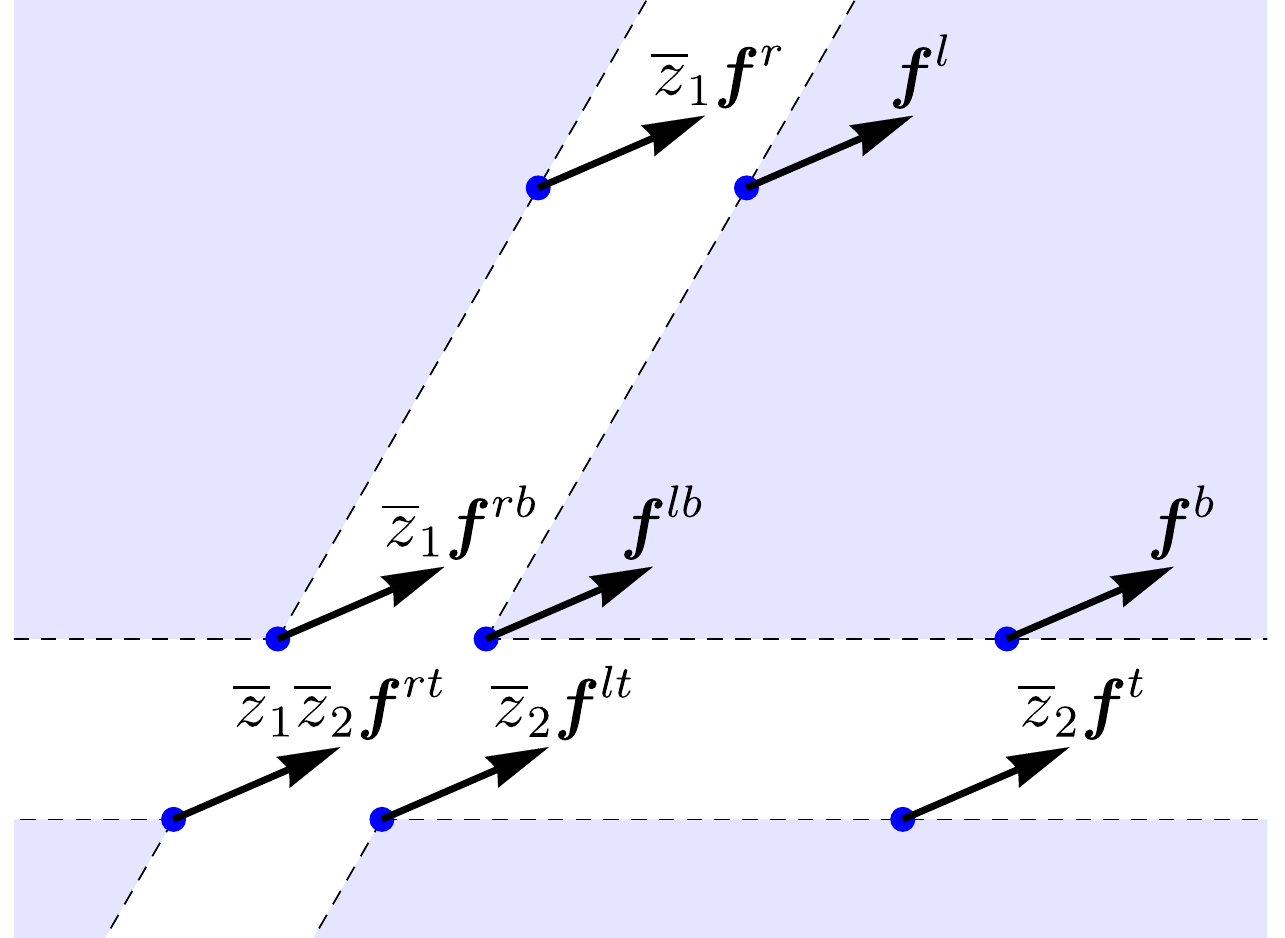}
    \end{subfigure}
    \caption{\label{fig:FB_joints}
		The imposition of the Floquet-Bloch conditions required by Eq.~\eqref{eq:bloch_theorem_2} is obtained using vector $\bq$, which collects the degrees of freedom of the unit cell.
		$\bq$ is partitioned by distinguishing the sets of internal nodes $\bq^i$ from the nodes at the boundary, located at corners $\{\bq^{lb}, \bq^{lt}, \bq^{rb}, \bq^{rt}\}$ and on the edges $\{\bq^l, \bq^b, \bq^r, \bq^t\}$~(\subref{fig:FB_joints_q}).
		The corresponding force vector $\bef$ is partitioned in the same way~(\subref{fig:FB_joints_f}) and the forces acting on the boundary nodes have to satisfy equilibrium as well as the Floquet-Bloch conditions (the `shift factors' $\bar{z}_j$, $j=1,2$ are the complex conjugate of $z_j = e^{i\,\bk\scalp\ba_j}$ used in the Bloch representation, Eq.~\eqref{eq:bloch_theorem_2}).
    }
\end{figure}

Equation~\eqref{eq:bloch_theorem_2} requires
\begin{equation}
    \label{eq:FB_shifts}
    \bq_q = \bq_p \, e^{i\,\bk\scalp(\bx_q-\bx_p)} \,,
\end{equation}
for all pairs of nodes $\{p,q\}$ such that $\bx_q-\bx_p$ is an integer linear combination of the lattice vectors $\{\ba_1,\ba_2\}$.
Therefore, the relations to be imposed on the boundary of the unit cell derive directly from Eq.~\eqref{eq:bloch_theorem_2}, evaluated at $\bx\in\partial\mC$ and for $n_j\in\{0,1\}$, hence obtaining
\begin{subequations}
\label{eq:FB_conditions}
\begin{equation}
    \label{eq:FB_conditions_long}
    \bq =
    \begin{Bmatrix}
        \bq^i \\
        \bq^l \\
        \bq^b \\
        \bq^{lb} \\
        \bq^r \\
        \bq^t \\
        \bq^{rb} \\
        \bq^{lt} \\
        \bq^{rt}
    \end{Bmatrix}
    =
    \begin{bmatrix}
        \bI & \bzero & \bzero & \bzero \\
        \bzero & \bI & \bzero & \bzero \\
        \bzero & \bzero & \bI & \bzero \\
        \bzero & \bzero & \bzero & \bI \\
        \bzero & z_1\bI & \bzero & \bzero \\
        \bzero & \bzero & z_2\bI & \bzero \\
        \bzero & \bzero & \bzero & z_1\bI \\
        \bzero & \bzero & \bzero & z_2\bI \\
        \bzero & \bzero & \bzero & z_1 z_2 \bI
    \end{bmatrix}
    \begin{Bmatrix}
        \bq^i \\
        \bq^l \\
        \bq^b \\
        \bq^{lb}
    \end{Bmatrix} \,,
\end{equation}
succinctly written as
\begin{equation}
    \label{eq:FB_conditions_short}
    \bq = \bZ(\bk)\, \bq^* \,,
\end{equation}
\end{subequations}
where $\bZ(\bk)$ and $\bq^*$ are defined according to Eq.~\eqref{eq:FB_conditions_long}, and $z_j = e^{i\,\bk\scalp\ba_j}$, with $j=1,2$.
In Eq.~\eqref{eq:FB_conditions_long} the vector $\bq$ has been partitioned to denote the inner and boundary nodes according to the notation sketched in Fig.~\ref{fig:FB_joints}.
The same partitioning is also introduced for the force vector $\bef$.

A substitution of Eqs.~\eqref{eq:FB_conditions} into Eq.~\eqref{eq:system_isolated} provides
\begin{equation*}
    \bA(\omega)\bZ(\bk)\, \bq^* = \bef \,,
\end{equation*}
so that a left multiplication by 
$\conjtrans{\bZ(\bk)}$, where the superscript $\conjtrans{\,}$ denotes the complex conjugate transpose operation\footnote{
The transpose of the conjugate of a matrix $\bM$ is defined~ as $\conjtrans{M_{ij}}=\conj{M}_{ji}$, where the bar denotes the complex conjugate.
}, leads to the reduced system
\begin{equation}
    \label{eq:system_f}
    \conjtrans{\bZ(\bk)}\bA(\omega)\bZ(\bk)\, \bq^* = \bef^* \,,
\end{equation}
where the following definition is introduced
\begin{equation*}
    \bef^* = \conjtrans{\bZ(\bk)} \bef =
    \begin{Bmatrix}
        \bef^i \\
        \bef^l + \conj{z}_1 \bef^r \\
        \bef^b + \conj{z}_2 \bef^t \\
        \bef^{lb} + \conj{z}_1 \bef^{rb} + \conj{z}_2 \bef^{lt} + \conj{z}_1 \conj{z}_2 \bef^{rt}
    \end{Bmatrix} \,.
\end{equation*}
Note that the dimension of system~\eqref{eq:system_f} is smaller than the dimension of system \eqref{eq:system_isolated}. 
In fact, the imposition of the Floquet-Bloch conditions allows to express the equations of motion for the lattice only in terms of the reduced variables $\bq^*$ and $\bef^*$.

If external loads are not present in the infinite lattice (so that the boundary forces shown in Fig.~\ref{fig:FB_joints_f} are purely internal), Eq.~\eqref{eq:bloch_theorem_2} implies $\bef^*=\bzero$ and therefore the following system of equations is obtained, governing Floquet-Bloch wave propagation within the lattice
\begin{equation}
    \label{eq:system}
    \bA^*(\omega,\bk) \, \bq^* = \bzero \,,
\end{equation}
where the matrix of the reduced system is
\begin{equation}
    \bA^*(\omega,\bk) = \conjtrans{\bZ(\bk)}\bA(\omega)\bZ(\bk).
\end{equation}
Note that this matrix is non-symmetric, but for a conservative system is always Hermitian, so that $\bA^*(\omega,\bk)=\conjtrans{\bA^*(\omega,\bk)}$ and $\omega^2\in\Reals$.

\subsection{Generalized eigenvalue problem for the dynamics of an infinite lattice}
Eq.~\eqref{eq:system} defines a homogeneous linear system for the unknown vector $\bq^*$, in which the angular frequency $\omega$ and the wave vector $\bk$ are for the moment undetermined.

It is important to note that the matrix-valued complex function $\bA^*(\omega,\bk)$ depends on all mechanical and geometrical parameters of the unit cell.
Eq.~\eqref{eq:system} is homogeneous, so that all non-trivial solutions are obtained by imposing the condition
\begin{equation}
    \label{eq:dispersion}
    \det\,\bA^*(\omega,\bk) = 0 \,,
\end{equation}
that defines, although implicitly, the \textit{dispersion relation}, providing $\omega$ as a function of $\bk$ for an infinite and periodic grillage of preloaded elastic rods.

For every given wave vector $\bk$ and for each of the corresponding roots $\omega(\bk)$ obtained from Eq.~\eqref{eq:dispersion}, the non-trivial solutions of Eq.~\eqref{eq:system} provide the modes $\bq^*$ of the Floquet-Bloch waves propagating through the lattice at each frequency.
This means that the vector $\bq^*$ is an implicit function of $\omega$ and therefore of $\bk$, so that the dependence on $\bk$ can be made explicit and Eq.~\eqref{eq:system} is rewritten as 
\begin{equation}
    \label{eq:system_dispersion}
    \bA^*(\omega(\bk),\bk)\,\bq^*(\omega(\bk),\bk)=\bzero \, ,
\end{equation}
an expression that will help clarifying the asymptotic expansion that will be performed in the next section.

\section{Dynamic homogenization of the lattice}
\label{sec:asymptotic_lattice_waves}
The scope of this section is the analysis of the low-frequency/long-wavelength asymptotics of the lattice dynamics, which will enable the identification of the effective continuum material. 
Identification requires first appropriate reference to the equivalent continuum.

\subsection{Wave propagation in a prestressed elastic continuum}
\label{sec:wave_propagation_Cauchy}
Before introducing the homogenization technique, it is worth recalling the fundamental equations governing incremental wave propagation in a prestressed hyperelastic continuum.
An appropriate form of the equations for the prestressed continuum has to be selected, to result compatible with the formulation of the lattice dynamics introduced in Section~\ref{sec:system_governing_equations}.
Specifically, it is observed that the equations of motion for the lattice are
\begin{enumerate*}[label=(\roman*)]
    \item obtained in the context of a linearized theory, and
    \item referred to a preloaded reference configuration.
\end{enumerate*}
Therefore, the dynamics of the unknown `equivalent' continuum has to be formulated in the context of the incremental theory of nonlinear elasticity by means of a relative Lagrangian description~\cite{bigoni_2012}. This is based on \textit{incremental constitutive laws} relating the increment of the first Piola-Kirchhoff stress $\dot{\bS}$ to the incremental deformation gradient  $\bL=\grad{\bu}$
\begin{equation}
    \label{eq:incremental_constitutive_law}
    \dot{\bS} = \fC[\bL] \,,
\end{equation}
through the elasticity tensor $\fC$
\begin{equation}
    \label{eq:constitutive_operator}
    \fC = \fE + \bI \boxtimes \bT \qquad \mbox{in components} \qquad \fC_{ijkl} = \fE_{ijkl} + \delta_{ik} T_{jl} \,,
\end{equation}
where $\bT$ is the Cauchy stress, defining here the \textit{prestress}, and $\fE$ is a fourth-order tensor endowed with the left and right minor symmetries and the major symmetry.
Moreover, Eq. \eqref{eq:constitutive_operator} implies that the number of unknown components of $\fC$ is at most 9 for a 2D material and 27 in the 3D case.

In the absence of body forces, the incremental equations of motion for the continuum can be written in the usual form
\begin{equation}
    \label{eq:equations_of_motion_continuum}
    \Div{\dot{\bS}} = \rho_h \ddot{\bu} \,,
\end{equation}
where $\bu$ is the incremental displacement field and $\rho_h>0$ the mass density.

Assuming the usual plane wave representation for incremental displacement, Eq.~\eqref{eq:equations_of_motion_continuum} leads to the following eigenvalue problem
\begin{equation}
    \label{eq:eigenvalue_problem_cauchy}
    \left(\bA^{(\fC)}(\bk) -\rho_h \, \omega_h^2\bI \right) \ba = \bzero \,,
\end{equation}
governing the wave propagation in a homogeneous elastic material whose acoustic tensor $\bA^{(\fC)}(\bk)$ is defined with reference to the unit vector $\bk$ as
\begin{equation*}
    A^{(\fC)}_{pr}(\bk) = k_q\,\fC_{pqrs}\,k_s \,.
\end{equation*}
The eigenvectors $\ba$ represent the wave amplitudes, while the eigenvalues $\omega_h^2$ are the roots of the characteristic equation
\begin{equation}
    \label{eq:characteristic_equation_cauchy}
     \det(\bA^{(\fC)}(\bk)-\rho_h \, \omega_h^2\bI) = 0 \,.
\end{equation}

Note that the subscript $h$ differentiates $\omega_h(\bk)$ from $\omega(\bk)$, so that the former defines the angular frequency of a wave propagating through the equivalent homogeneous elastic  continuum, while the latter the dispersion relation for the lattice.

\subsection{Asymptotic expansion of Floquet-Bloch waves}
\label{sec:wave_expansion}
A perturbation method is now developed for the equations governing wave propagation in a lattice made of prestressed elastic rods, through a generalization of the technique proposed by Born \cite{born_1955} for lattices involving only point-like mass interactions. 
A rigorous link is established between the low-frequency solutions of Eq.~\eqref{eq:system_dispersion} and spectral characteristics of the equivalent continuum governed by Eq.~\eqref{eq:eigenvalue_problem_cauchy}.

Suppose for a moment that the grillage of rods would satisfy the following conditions:
\begin{itemize}
\item a \textit{linear dispersion relation}, obtained by writing $\bk = \epsilon\,\bn$ (with $\bn$ being a unit vector), 
\begin{equation}
    \label{eq:omega_linear}
    \omega(\epsilon\,\bn) = \omega_{\bn}^{(1)}\,\epsilon \,,
\end{equation}
where $\omega_{\bn}^{(1)}$ depends only on the direction $\bn$;
\item  a \textit{uniform spatial modulation}
\begin{equation}
    \label{eq:modulation_constant}
    \bvarphi_{\bk}(\bx)=\ba(\bk) \, ,
\end{equation}
\end{itemize}
in this case wave propagation in the grillage would be non-dispersive and with uniform amplitude, a situation which cannot occur for every $\bk$, but is instead typical of a homogeneous medium. 
However, conditions~\eqref{eq:omega_linear}--\eqref{eq:modulation_constant} can be met in the limit $\bk \rightarrow \bzero$, so that in this case an equivalent material can be defined and vector $\ba$ in Eq.~\eqref{eq:modulation_constant} becomes the eigenmode defined by Eq.~\eqref{eq:eigenvalue_problem_cauchy} and $\omega_h = \omega_{\bn}^{(1)}\,\epsilon$.
This is the way in which homogenization will performed. 

The linear relation~\eqref{eq:omega_linear} can be considered as the first-order term in the asymptotic expansion of the dispersion relation $\omega(\epsilon\,\bn)$ centered at $\epsilon=0$, where  $\omega=0$ and $\bk=\bzero$, and along the direction $\bn$ in the $\bk$-space.
This asymptotic expansion, truncated at the $N$-th term is 
\begin{equation}
    \label{eq:omega_expansion}
    \omega(\epsilon\,\bn) \sim \omega_{\bn}^{(1)}\,\epsilon + \omega_{\bn}^{(2)}\,\epsilon^2 +...+\omega_{\bn}^{(N)}\,\epsilon^N = \mS_{\bn}^N (\omega)(\epsilon) \,,
\end{equation}
where the $\mO(\epsilon^0)$ term vanishes in the above expansion because the point $\{\omega,\bk\}=\bzero$ satisfies the dispersion equation~\eqref{eq:dispersion}.
This follows from the fact that, setting $\{\omega,\bk\}=\bzero$, Eq.~\eqref{eq:system_dispersion} becomes 
\begin{equation}
    \label{eq:A0}
    \bA^*(\bzero)\,\bq^*(\bzero) = \conjtrans{\bZ(\bzero)}\bK(0)\bZ(\bzero)\,\bq^*(\bzero) = \bzero \,,
\end{equation}
so that the only possible solutions $\bq^*(\bzero)$ correspond to 2 rigid-body translations, because the matrix $\bZ(\bzero)$ prescribes \textit{equal} displacements on corresponding sides of the unit cell thus preventing rigid-body rotations. 
The two rigid-body translations, plus the rigid-body rotation, represent the nullspace of $\bK(0)$.
The translations represent the limit of the eigenmodes $\bvarphi_{\bk}(\bx)$ for $\norm{\bk}\to0$. 
Their derivation requires first the construction of an asymptotic expansion for $\bvarphi_\bk(\bx)$, in complete analogy to Eq.~\eqref{eq:omega_expansion},
\begin{equation}
    \label{eq:phi_expansion}
    \bvarphi_{\epsilon\,\bn}(\bx) \sim \bvarphi^{(0)}_{\bn}(\bx) + \bvarphi^{(1)}_{\bn}(\bx)\,\epsilon + \bvarphi^{(2)}_{\bn}(\bx)\,\epsilon^2 +...+ \bvarphi^{(N)}_{\bn}(\bx)\,\epsilon^N
    =\mS_{\bn}^N (\bvarphi)(\epsilon) \,,
\end{equation}
and then the computation of the limit $\epsilon\to0$.
As a result, the zeroth-order term of the waveform $\bvarphi^{(0)}_{\bn}$ is indeed \textit{uniform} in space (independent of $\bx$), and therefore the acoustic properties of the equivalent elastic continuum have to satisfy
\begin{subequations}
\label{eq:omega_phi_match}
\begin{align}
    \label{eq:omega_match}
    \omega_h(\epsilon\,\bn) &= \omega_{\bn}^{(1)}\,\epsilon , \\
    \label{eq:phi_match}
    \ba(\epsilon\,\bn) &= \bvarphi^{(0)}_{\bn} , \qquad \forall \bn\in\Reals^2 \, ,
\end{align}
\end{subequations}
conditions which define an `acoustic equivalence' between the lattice and the continuum. 

An effective method to obtain the series expansions~\eqref{eq:omega_expansion} and~\eqref{eq:phi_expansion} is outlined in the following.
As the waveform $\bvarphi_{\bk}(\bx)$ for a lattice made up of rods is governed by the vector of degrees of freedom $\bq^*$, solution of the eigenvalue problem~\eqref{eq:system_dispersion}, the expansion of $\bq^*(\omega(\bk),\bk)$ is performed along an arbitrary direction $\bn$ in the $\bk$-space
\begin{equation}
    \label{eq:q_expansion}
    \bq^*(\omega(\epsilon\,\bn),\epsilon\,\bn) \sim \bq^{*(0)}_{\bn} + \bq^{*(1)}_{\bn}\,\epsilon + \bq^{*(2)}_{\bn}\,\epsilon^2 + ... \,,
\end{equation}
so that the first term $\bq^{*(0)}_{\bn}$ can be used to identify the left-hand side of Eq.~\eqref{eq:phi_match}.
To this end, the matrix $\bA^*(\omega(\bk),\bk)$ is expanded as
\begin{equation}
    \label{eq:A_expansion}
    \bA^*(\omega(\epsilon\,\bn),\epsilon\,\bn) \sim \bA^{*(0)}_{\bn} + \bA^{*(1)}_{\bn}\,\epsilon + \bA^{*(2)}_{\bn}\,\epsilon^2 + ... \,,
\end{equation}
so that the eigenvalue problem~\eqref{eq:system_dispersion} is rewritten through a substitution of the series representations~\eqref{eq:q_expansion} and~\eqref{eq:A_expansion} as
\begin{equation}
    \label{eq:system_expansion}
    (\bA^{*(0)}_{\bn} + \bA^{*(1)}_{\bn}\,\epsilon + \bA^{*(2)}_{\bn}\,\epsilon^2 + ...)(\bq^{*(0)}_{\bn} + \bq^{*(1)}_{\bn}\,\epsilon + \bq^{*(2)}_{\bn}\,\epsilon^2 + ...) = \bzero \,.
\end{equation}

Since the dispersion relation $\omega(\bk)$ is formally inserted in the above expansions,  Eq.~\eqref{eq:system_expansion} has to be satisfied for every value of $\epsilon$, which means that the left-hand side has to vanish at every order in $\epsilon$.
Thus the following sequence of linear systems is obtained
\begin{equation}
    \label{eq:system_sequence}
    \begin{aligned}
        &\mO(\epsilon^0): \quad \bA^{*(0)}_{\bn}\,\bq^{*(0)}_{\bn} = \bzero \,, \\
        &\mO(\epsilon^1): \quad \bA^{*(0)}_{\bn}\,\bq^{*(1)}_{\bn} + \bA^{*(1)}_{\bn}\,\bq^{*(0)}_{\bn} = \bzero \,, \\
        &\mO(\epsilon^2): \quad \bA^{*(0)}_{\bn}\,\bq^{*(2)}_{\bn} + \bA^{*(1)}_{\bn}\,\bq^{*(1)}_{\bn} + \bA^{*(2)}_{\bn}\,\bq^{*(0)}_{\bn} = \bzero \,, \\
        &\quad\vdots \qquad \qquad\qquad \qquad\qquad\qquad \vdots \\
        &\mO(\epsilon^j): \quad \bA^{*(0)}_{\bn}\,\bq^{*(j)}_{\bn} + \sum_{h=1}^{j} \bA^{*(h)}_{\bn}\,\bq^{*(j-h)}_{\bn} = \bzero, \qquad \forall j>0  \,, \\
    \end{aligned}
\end{equation}
which has to be solved for the unknown vectors $\bq^{*(j)}_{\bn}$.
It is clear that the computation of these vectors starts from the solution of the zeroth-order equation and then, sequentially, the higher-order terms are to be obtained. At the $j$-th order, the matrix of the linear system is $\bA^{*(0)}_{\bn}$ and multiplies the unknown vector $\bq^{*(j)}_{\bn}$, so that the constant term (not involving the unknown $\bq^{*(j)}_{\bn}$) contains all the previously determined vectors $\{\bq^{*(0)}_{\bn},...,\bq^{*(j-1)}_{\bn}\}$.
Moreover, it is important to observe that the terms $\bA^{*(j)}_{\bn}$ in the expansion~\eqref{eq:A_expansion} can be computed explicitly once the series $\mS_{\bn}^N (\omega)(\epsilon)$ has been determined.

It is recalled that, as shown by Eq.~\eqref{eq:A0}, the matrix of each linear system $\bA^{*(0)}_{\bn}$ is singular and it has a two-dimensional nullspace spanned by two linearly independent vectors, $\bt_1$ and $\bt_2$, which represent the two in-plane rigid-body translations.\footnote{Since $\bt_1$ and $\bt_2$ describe two arbitrary rigid translations, $\bt_1$ and $\bt_2$ can be conveniently chosen as the rigid translations aligned parallel to $\be_1$ and $\be_2$, respectively.}
Thus, every linear combination in the form
\begin{equation}
    \label{eq:q0}
    \bq^{*(0)}_{\bn} = \alpha_1\bt_1 + \alpha_2\bt_2 , \qquad \forall \{\alpha_1,\alpha_2\}\in\Reals^2 \,,
\end{equation}
is a solution of the zeroth-order equation in~\eqref{eq:system_sequence}.
This implies that the matrix $\bA^{*(0)}_{\bn}$ is not invertible, so that the solvability of the $j$-th linear system depends on the form of its right-hand side, which has to satisfy the following condition, known as the Fredholm alternative theorem
\begin{equation}
    \label{eq:solvability}
    \left(\sum_{h=1}^{j} \bA^{*(h)}_{\bn}\,\bq^{*(j-h)}_{\bn}\right) \scalp \by = 0 , \qquad \forall \by\in\ker(\bA^{*(0)}_{\bn}), \qquad \forall j>0 \,,
\end{equation}
or, equivalently, using Eq.~\eqref{eq:q0}, the condition
\begin{equation*}
    \left(\sum_{h=1}^{j} \bA^{*(h)}_{\bn}\,\bq^{*(j-h)}_{\bn}\right) \scalp \bt_1 = 0 \,, \qquad \left(\sum_{h=1}^{j} \bA^{*(h)}_{\bn}\,\bq^{*(j-h)}_{\bn}\right) \scalp \bt_2 = 0, \qquad \forall j>0 \,.
\end{equation*}

In principle, Eqs.~\eqref{eq:system_sequence} and \eqref{eq:solvability} are sufficient to compute the series representations~\eqref{eq:q_expansion} and~\eqref{eq:omega_expansion}, thus making conditions~\eqref{eq:omega_phi_match} explicit.

\subsection{The acoustic tensor for a lattice of elastic rods}
\label{sec:acoustic_tensor_lattice}
The perturbation method outlined in Section~\ref{sec:wave_expansion} is general enough to provide, up to the desired order, the series representation of the acoustic properties of a preloaded lattice subject to incremental dynamics.

It will be proved in the following that it is always possible to employ the above-described perturbation technique to construct an eigenvalue problem governing the propagation of waves in a lattice (where elements are subject to both axial and flexural deformation) in the low-frequency and long-wavelength regime.
In particular, this eigenvalue problem will possess the following properties:
\begin{enumerate}[label=(\roman*)]
    \item the eigenvalues identify the first-order term $\omega_{\bn}^{(1)}$ of both acoustic branches of the dispersion relation;
    \item the eigenvectors govern the zeroth-order term of the Floquet-Bloch waveform for both the two acoustic waves, through coefficients $\{\alpha_1,\alpha_2\}$ in the linear combination~\eqref{eq:q0};
    \item the algebraic structure of the problem is \textit{exactly equivalent} to that governing wave propagation in a prestressed elastic material, Eq.~\eqref{eq:eigenvalue_problem_cauchy}.
\end{enumerate}

\textit{The construction of the above eigenvalue problem allows the rigorous definition of the `acoustic tensor for a lattice of elastic rods' and from the latter the identification of the elasticity tensor representing a material equivalent to the lattice}.
In fact, this eigenvalue problem defines eigenvalues and eigenvectors satisfying the conditions of acoustic equivalence, Eq.~\eqref{eq:omega_phi_match}.

In order to construct the eigenvalue problem, the solution of the sequence of the linear systems~\eqref{eq:system_sequence} is obtained  up to the order $\mO(\epsilon^2)$.
The equations involve the following terms of the series~\eqref{eq:A_expansion}
\begin{equation}
    \label{eq:A0_A1_A2}
    \begin{aligned}
        \bA^{*(0)} &= \conjtrans{{\bZ^{(0)}}}\bK^{(0)}\bZ^{(0)} \,, \\
        \bA^{*(1)}_{\bn} &= \conjtrans{{\bZ^{(0)}}}\bK^{(0)}\bZ^{(1)}_{\bn} + \conjtrans{{\bZ^{(1)}_{\bn}}}\bK^{(0)}\bZ^{(0)} \,, \\
        \bA^{*(2)}_{\bn} &= \conjtrans{{\bZ^{(1)}_{\bn}}}\bK^{(0)}\bZ^{(1)}_{\bn} + \conjtrans{{\bZ^{(0)}}}\bK^{(0)}\bZ^{(2)}_{\bn} + \\
        &\quad+\conjtrans{{\bZ^{(2)}_{\bn}}}\bK^{(0)}\bZ^{(0)} - \conjtrans{{\bZ^{(0)}}}\bM^{(0)}\bZ^{(0)} \left(\omega_{\bn}^{(1)}\right)^2 \,,
    \end{aligned}
\end{equation}
where a series expansion has been introduced for the matrices $\bK(\omega(\epsilon\,\bn))$, $\bM(\omega(\epsilon\,\bn))$ and $\bZ(\epsilon\,\bn)$ as $\epsilon\to0$.
It is important to note that: 
\begin{enumerate}[label=(\roman*)]
    \item up to the order $\mO(\epsilon^2)$, only the zeroth-order terms of the matrices $\bK$ and $\bM$ (which correspond to the quasi-static limit, $\bK^{(0)} = \lim_{\omega\to0}\bK(\omega)$, $\bM^{(0)}=\lim_{\omega\to0}\bM(\omega)$) are present;
    \item the zeroth-order matrix $\bZ^{(0)}$, and consequently $\bA^{*(0)}$, is independent of the direction $\bn$ (owing to continuity of $\bZ(\bk)$); 
    while $\bA^{*(1)}_{\bn}$ is linear in $\bn$, $\bA^{*(2)}_{\bn}$ is quadratic in $\bn$;
    \item the linear term $\omega_{\bn}^{(1)}$ starts to appear at order $\mO(\epsilon^2)$. 
\end{enumerate}

In the following, the first and second-order equations in the sequence of equations~\eqref{eq:system_sequence} are considered and their solvability conditions derived, Eq.~\eqref{eq:solvability}.
By means of Eq.~\eqref{eq:q0}, the first-order equation in the sequence~\eqref{eq:system_sequence} reads as
\begin{equation}
    \label{eq:system_first_order}
    \mO(\epsilon^1): \quad \bA^{*(0)}\,\bq^{*(1)}_{\bn} + \bA^{*(1)}_{\bn}(\alpha_1\bt_1+\alpha_2\bt_2) = \bzero \,,
\end{equation}
and its solvabilty condition requires
\begin{equation*}
    \bt_1 \scalp \bA^{*(1)}_{\bn}(\alpha_1\bt_1+\alpha_2\bt_2) = 0 \,, \qquad 
    \bt_2 \scalp \bA^{*(1)}_{\bn}(\alpha_1\bt_1+\alpha_2\bt_2) = 0 \,,
\end{equation*}
two conditions which are always satisfied.
In fact, a use of Eq.~\eqref{eq:A0_A1_A2}$_2$ yields
\begin{equation*}
    \bt_j\scalp\bA^{*(1)}_{\bn}\bt_i = \bt_j \scalp \left(\conjtrans{{\bZ^{(0)}}}\bK^{(0)}\bZ^{(1)}_{\bn} + \conjtrans{{\bZ^{(1)}_{\bn}}}\bK^{(0)}\bZ^{(0)}\right)\bt_i \,,
\end{equation*}
a scalar product which vanishes because $\bZ^{(0)}\bt_i$ is a rigid-body translation, so that it cannot produce any stress, hence $\bK^{(0)}\bZ^{(0)}\bt_i=\bzero$.
Since Eq.~\eqref{eq:system_first_order} is always solvable, all its solutions can be expressed in the form
\begin{equation}
    \label{eq:q1}
    \bq^{*(1)}_{\bn} = \alpha_1\bt_1'(\bn) + \alpha_2\bt_2'(\bn) \qquad \forall \{\alpha_1,\alpha_2\}\in\Reals^2 \,,
\end{equation}
where $\bt_1'$ and $\bt_2'$ are the solutions of the following two linear systems
\begin{equation*}
    \bA^{*(0)}\,\bt_1'(\bn) + \bA^{*(1)}_{\bn}\bt_1 = \bzero \,, \qquad \bA^{*(0)}\,\bt_2'(\bn) + \bA^{*(1)}_{\bn}\bt_2 = \bzero \,.
\end{equation*}
Note that $\bt_1'$ and $\bt_2'$ are defined up to an arbitrary rigid-body translation.

By employing Eqs.~\eqref{eq:q0} and~\eqref{eq:q1}, the linear system of order $\mO(\epsilon^2)$ reads as
\begin{equation}
    \label{eq:system_second_order}
    \mO(\epsilon^2): \quad \bA^{*(0)}\,\bq^{*(2)}_{\bn} = -\alpha_1 (\bA^{*(1)}_{\bn}\bt_1'(\bn) + \bA^{*(2)}_{\bn}\bt_1)-\alpha_2 (\bA^{*(1)}_{\bn}\bt_2'(\bn) + \bA^{*(2)}_{\bn}\bt_2) \,,
\end{equation}
which (because $\bA^{*(0)}$ is singular) admits a solution if and only if the right-hand side is orthogonal to both $\bt_1$ and $\bt_2$, namely
\begin{equation*}
    \begin{aligned}
        \alpha_1 (\bA^{*(1)}_{\bn}\bt_1'(\bn) + \bA^{*(2)}_{\bn}\bt_1)\scalp\bt_1+\alpha_2 (\bA^{*(1)}_{\bn}\bt_2'(\bn) + \bA^{*(2)}_{\bn}\bt_2)\scalp\bt_1 = 0 \,, \\
        \alpha_1 (\bA^{*(1)}_{\bn}\bt_1'(\bn) + \bA^{*(2)}_{\bn}\bt_1)\scalp\bt_2+\alpha_2 (\bA^{*(1)}_{\bn}\bt_2'(\bn) + \bA^{*(2)}_{\bn}\bt_2)\scalp\bt_2 = 0 \,,
    \end{aligned}
\end{equation*}
that in matrix form can be written as\footnote{
Note that vectors $\bt_j'(\bn)$ may contain an arbitrary rigid-body translation. This would apparently lead to a non-uniqueness in the form of Eq.~\eqref{eq:system_solvability_second_order}, because the terms $\bt_j\scalp\bA^{*(1)}_{\bn}\bt_i'(\bn)$ are present. This lack of uniqueness is only apparent, because $\bt_j\scalp\bA^{*(1)}_{\bn}\bt_i=0$.
}
\begin{equation}
    \label{eq:system_solvability_second_order}
    \begin{bmatrix}
        (\bA^{*(1)}_{\bn}\bt_1'(\bn) + \bA^{*(2)}_{\bn}\bt_1)\scalp\bt_1 & (\bA^{*(1)}_{\bn}\bt_2'(\bn) + \bA^{*(2)}_{\bn}\bt_2)\scalp\bt_1 \\
        (\bA^{*(1)}_{\bn}\bt_1'(\bn) + \bA^{*(2)}_{\bn}\bt_1)\scalp\bt_2 & (\bA^{*(1)}_{\bn}\bt_2'(\bn) + \bA^{*(2)}_{\bn}\bt_2)\scalp\bt_2
    \end{bmatrix}
    \begin{Bmatrix}
        \alpha_1\\
        \alpha_2
    \end{Bmatrix}=
    \begin{Bmatrix}
        0\\
        0
    \end{Bmatrix} \,.
\end{equation}

Up to order $\mO(\epsilon^1)$ the coefficients $\{\alpha_1,\alpha_2\}$ and the linear term $\omega_{\bn}^{(1)}$ remain \textit{completely arbitrary}, but now they have to satisfy  system~\eqref{eq:system_solvability_second_order} in order to make Eq.~\eqref{eq:system_second_order} solvable.
In fact, the homogeneous system~\eqref{eq:system_solvability_second_order} represents an eigenvalue problem with eigenvectors $\{\alpha_1,\alpha_2\}$ and eigenvalues $\left(\omega_{\bn}^{(1)}\right)^2$.
To see this point more explicitly, expressions~\eqref{eq:A0_A1_A2} can be substituted into Eq.~\eqref{eq:system_solvability_second_order} to obtain\footnote{The resulting expression has been simplified using again the property $\bK^{(0)}\bZ^{(0)}\bt_i=\bzero$.}
\begin{equation}
    \label{eq:eigenvalue_problem_omega1}
    \begin{aligned}
        \overbrace{
        \begin{bmatrix}
            \bt_1\scalp\tilde{\bK}^{(1)}_{\bn}\bt_1 + \bt_1'(\bn)\scalp\tilde{\bK}^{(0)}\bt_1'(\bn) & \bt_2\scalp\tilde{\bK}^{(1)}_{\bn}\bt_1 + \bt_2'(\bn)\scalp\tilde{\bK}^{(0)}\bt_1'(\bn) \\
            \bt_1\scalp\tilde{\bK}^{(1)}_{\bn}\bt_2 + \bt_1'(\bn)\scalp\tilde{\bK}^{(0)}\bt_2'(\bn) & \bt_2\scalp\tilde{\bK}^{(1)}_{\bn}\bt_2 + \bt_2'(\bn)\scalp\tilde{\bK}^{(0)}\bt_2'(\bn)
        \end{bmatrix}}^{\bXi}
        \begin{Bmatrix}
            \alpha_1\\
            \alpha_2
        \end{Bmatrix} + \\[1mm]
        -\left(\omega_{\bn}^{(1)}\right)^2
        \underbrace{
        \begin{bmatrix}
            \bt_1\scalp\tilde{\bM}^{(0)}\bt_1 & \bt_2\scalp\tilde{\bM}^{(0)}\bt_1 \\
            \bt_1\scalp\tilde{\bM}^{(0)}\bt_2 & \bt_2\scalp\tilde{\bM}^{(0)}\bt_2
        \end{bmatrix}}_{\bGamma}
        \begin{Bmatrix}
            \alpha_1\\
            \alpha_2
        \end{Bmatrix} =
        \begin{Bmatrix}
            0\\
            0
        \end{Bmatrix} \,,
    \end{aligned}
\end{equation}
with the following definitions
\begin{equation*}
    \tilde{\bK}^{(0)} = \conjtrans{{\bZ^{(0)}}}\bK^{(0)}\bZ^{(0)} \,, \quad
    \tilde{\bK}^{(1)}_{\bn} = \conjtrans{{\bZ^{(1)}_{\bn}}}\bK^{(0)}\bZ^{(1)}_{\bn} \,, \quad
    \tilde{\bM}^{(0)} = \conjtrans{{\bZ^{(0)}}}\bM^{(0)}\bZ^{(0)} \,.
\end{equation*}

Eq.~\eqref{eq:eigenvalue_problem_omega1} is an eigenvalue problem, and the following properties can be deduced:
\begin{enumerate}[label=(\roman*)]
    \item As the matrices $\tilde{\bK}^{(0)}$, $\tilde{\bK}^{(1)}_{\bn}$ and $\tilde{\bM}^{(0)}$ are real and symmetric, also the matrices $\bXi$ and $\bGamma$ are real and symmetric, hence the eigenvalues $\left(\omega_{\bn}^{(1)}\right)^2$ are real;
    \item Since $\bZ^{(0)}$, $\bt_1$ and $\bt_2$ are independent of $\bn$ and $\bZ^{(1)}_{\bn}$, $\bt_1'(\bn)$ and $\bt_2'(\bn)$ are all linear in $\bn$, each component of the 2-by-2 matrix $\bXi$ is a quadratic form in $\bn$;
    \item The components of the 2-by-2 matrix $\bGamma$ admit the following simplifications
    \begin{equation*}
        \bt_2\scalp\tilde{\bM}^{(0)}\bt_1 = \bt_1\scalp\tilde{\bM}^{(0)} \bt_2 = 0 \,,
        \quad \bt_1\scalp\tilde{\bM}^{(0)}\bt_1 = \bt_2\scalp\tilde{\bM}^{(0)}\bt_2= \rho_h\, \lvert\mC\lvert \,,
    \end{equation*}
	where $\rho_h$ is the average mass density 
	\begin{equation}
		\rho_h = \frac{1}{\lvert\mC\lvert}\sum_{k=1}^{N_b} \gamma_k\,l_k ,
	\end{equation}
	and $\lvert\mC\lvert$ the area of the unit cell.
    Note that the matrix $\bGamma$ contains only terms in the form $\bt_i\scalp\tilde{\bM}^{(0)}\bt_j$, where $\bt_i$ are rigid translations, and therefore \textit{the rotational inertia of the rods plays no role} (recall the definition~\eqref{eq:M_beam}).
\end{enumerate}

Finally the eigenvalue problem~\eqref{eq:eigenvalue_problem_omega1} can be written in the standard form
\begin{equation}
    \label{eq:eigenvalue_problem_omega1_compact}
    \left[\bA^{\text{(L)}}(\bn) - \rho_h \left(\omega_{\bn}^{(1)}\right)^2 \bI \right] \balpha = \bzero \,,
\end{equation}
where $\balpha = \trans{\{\alpha_1,\alpha_2\}}$ and tensor $\bA^{\text{(L)}}(\bn)$ reads as
\begin{equation}
    \label{eq:acoustic_tensor_lattice}
    \bA^{\text{(L)}}(\bn) = \frac{1}{\lvert\mC\lvert}
    \begin{bmatrix}
        \bt_1\scalp\tilde{\bK}^{(1)}_{\bn}\bt_1 + \bt_1'(\bn)\scalp\tilde{\bK}^{(0)}\bt_1'(\bn) & \bt_2\scalp\tilde{\bK}^{(1)}_{\bn}\bt_1 + \bt_2'(\bn)\scalp\tilde{\bK}^{(0)}\bt_1'(\bn) \\
        \bt_1\scalp\tilde{\bK}^{(1)}_{\bn}\bt_2 + \bt_1'(\bn)\scalp\tilde{\bK}^{(0)}\bt_2'(\bn) & \bt_2\scalp\tilde{\bK}^{(1)}_{\bn}\bt_2 + \bt_2'(\bn)\scalp\tilde{\bK}^{(0)}\bt_2'(\bn)
    \end{bmatrix} .
\end{equation}

It is important to note at this stage, that the eigenvalue problem~\eqref{eq:eigenvalue_problem_omega1_compact} has exactly the same structure of Eq.~\eqref{eq:eigenvalue_problem_cauchy}.
Furthermore, tensor $\bA^{\text{(L)}}(\bn)$, the `acoustic tensor of the lattice', uniquely defines  the eigenvalues and eigenvectors appearing on the right-hand side of the equivalence conditions~\eqref{eq:omega_phi_match}, so that $\omega_h = \omega_{\bn}^{(1)} \epsilon$ and $\ba = \bvarphi_{\bn}^{(0)} = \balpha$.
\textit{This implies that the acoustic equivalence holds if and only if the `acoustic tensor of the lattice' coincides with the acoustic tensor of the elastic material.}
Therefore, the equivalent elastic continuum has to satisfy the acoustic equivalence condition (valid for every unit vector $\bn$)
\begin{equation}
    \label{eq:equivalence_acoustic_tensors}
    \bA^{(\fC)}(\bn) = \bA^{\text{(L)}}(\bn) .
\end{equation}

It is important to note that the equivalence condition has been obtained without introducing restrictive assumptions on the lattice structure, so that the homogenization method is completely general and \textit{includes a generic state of axial preload acting on the lattice}.
Moreover, the presented technique can easily be extended to three-dimensional lattices.

\subsection{Identification of the continuum equivalent to a preloaded lattice}
\label{sec:identification_constitutive_tensor}
The perturbation method developed in Section~\ref{sec:asymptotic_lattice_waves} leads to the determination of the acoustic tensor of an effective prestressed elastic continuum, equivalent to the low-frequency response of a preloaded grillage of rods. 
As the method is entirely based on the dynamics of the periodic medium, the acoustic tensor is obtained directly, without any prior computation of the effective constitutive tensor, which is instead traditional in standard energy-based homogenization techniques~\cite{willis_2002,hutchinson_2006,elsayed_2010,bacca_2013,abdoul-anziz_2018}, see also Part I.

In this section the steps for retrieving the incremental (or `tangent') constitutive tensor $\fC$ are outlined from the acoustic tensor given by Eq.~\eqref{eq:equivalence_acoustic_tensors}.

As the condition~\eqref{eq:equivalence_acoustic_tensors} has to hold for an arbitrary direction of propagation, it can equivalently be  expressed by applying the Hessian with respect to $\bn$ on both sides of Eq.~\eqref{eq:equivalence_acoustic_tensors} to obtain
\begin{equation}
    \label{eq:hessian_acoustic_tensor}
    \fC_{ikjl} + \fC_{iljk} = \frac{\partial^2 A^{\text{(L)}}_{ij}(\bn)}{\partial n_k \partial n_l},
\end{equation}
where the right-hand side can be regarded as a data defined by the lattice structure, namely, the Hessian of tensor~\eqref{eq:acoustic_tensor_lattice}.
By considering the symmetry with respect to the $\{k,l\}$ indices, Eq.~\eqref{eq:hessian_acoustic_tensor} provides a linear system of 54 equations in a three-dimensional setting or 12 equations in a two-dimensional setting, while the rank of the system is found to be 26 or 8, respectively.
By recalling that the unknown tensor $\fC$ has the form~\eqref{eq:constitutive_operator}, it is clear that, if the system is solvable, all but one of the unknown components of $\fC$ can be determined as these are 27 for a three-dimensional lattice and 9 for a two-dimensional.

In order to solve for the identification, results obtained by Max Born~\cite{born_1955} can now be generalized to prove that:
\begin{enumerate}[label=(\roman*)]
    \item the system is solvable when the equations of motions of the lattice satisfy the rotational invariance and
    \item the solution is unique, except for the spherical part of the prestress (i.e. $\tr{\bT}/3$ in 3D and $\tr{\bT}/2$ in 2D) which remains undetermined for the system~\eqref{eq:hessian_acoustic_tensor}, but can be determined by matching the first-order incremental work introduced in Part I.
\end{enumerate}

\section{Bifurcation and loss of ellipticity in a lattice of preloaded elastic rods}
\label{sec:ellipticity_loss_grid}
In order to demonstrate the effectiveness of the homogenization method developed in Section~\ref{sec:asymptotic_lattice_waves}, bifurcation and loss of ellipticity are investigated in a preloaded two-dimensional grid lattice of elastic rods. 
The vibrations of the grillage will be directly analyzed with the Floquet-Bloch technique reviewed in Section \ref{sec:system_governing_equations} and results will be compared to those obtained on the equivalent elastic material, Section \ref{sec:asymptotic_lattice_waves}.

\begin{figure}[htb!]
    \centering
    \includegraphics[width=0.55\linewidth]{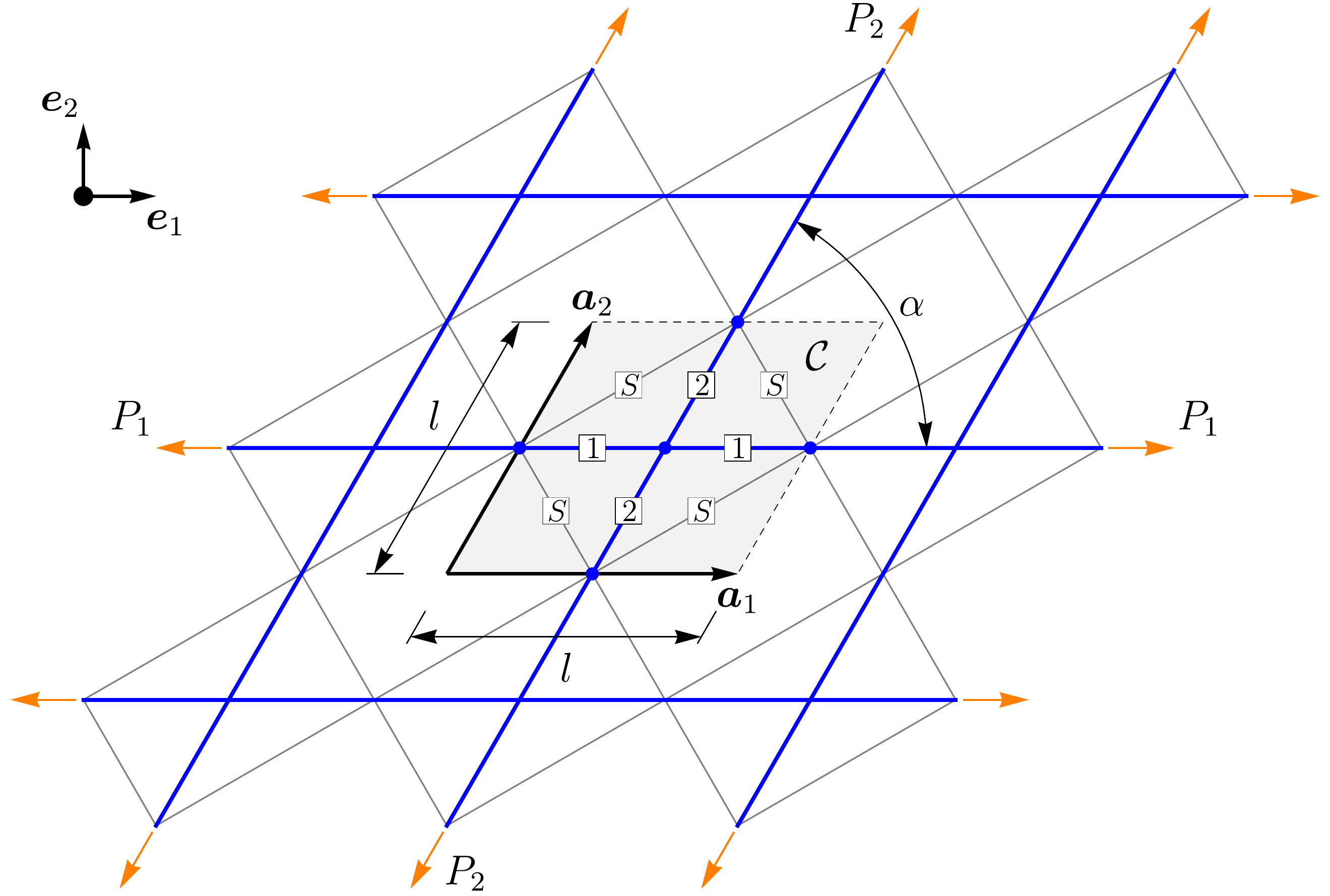}
    \caption{\label{fig:geometry_grid_and_unit_cell}
    Current configuration of a rhombic lattice of preloaded elastic rods, with the associated unit cell $\mC$ (highlighted in gray).
    The direct basis for the lattice is denoted by the pair of vectors $\{\ba_1,\ba_2\}$.
    Labels 1, 2, and $S$ denote the horizontal rods, the inclined rods, and the diagonal springs, respectively.
    The stiffness of inclined springs, the axial and flexural rigidity of the rods, the preloads $P_1$ and $P_2$, as well as the grid angle $\alpha$ can all be varied to investigate different incremental responses.
    }
\end{figure}

The geometry of the reference configuration is sketched in Fig.~\ref{fig:geometry_grid_and_unit_cell} and is composed of two sets of rods, inclined at an angle $\alpha$ and characterized by an axial stiffness $A_1=A_2=A$ and slenderness ratios $\Lambda_1 = l \sqrt{A/B_1}$, $\Lambda_2 = l \sqrt{A/B_2}$, where the subscripts $1$ and $2$ are relative to the horizontal and inclined rods, respectively.
For simplicity, the linear mass density is assumed to be the same for both rods $\gamma_1=\gamma_2=\gamma$, while the rotational inertia has been shown in Section~\ref{sec:acoustic_tensor_lattice} to be negligible for the homogenization.
The resulting grid of rods is also considered stiffened by diagonal linear springs (Fig.~\ref{fig:geometry_grid_and_unit_cell}), whose stiffness $k_{\text{S}}$ is conveniently made dimensionless by introducing the parameter $\kappa=k_{\text{S}}l/A$.
In this configuration, the lattice is subject to a preload state defined by the axial forces $P_1$ and $P_2$, made dimensionless respectively as 
$p_1 = P_1 l^2/B_1$ and $p_2 = P_2 l^2/B_2$. 
Hence, a lattice configuration is completely defined by the parameter set $\{p_1,p_2,\Lambda_1,\Lambda_2,\kappa,\alpha\}$.
Note also that the lattice structure considered includes the simpler case of a rectangular grid that has been analyzed in~\cite{triantafyllidis_1993}.

\subsection{Macro and micro bifurcations as degeneracies of the dispersion relation}
The effect of the diagonal reinforcement (springs labeled with $S$ in Fig.~\ref{fig:geometry_grid_and_unit_cell}) on the bifurcation of the lattice has been systematically investigated in Part I of this study, where it has been shown to play a fundamental role in determining the wavelength critical for bifurcation. 
Specifically, it has been demonstrated that an increase in the spring stiffness induces a transition of the critical bifurcation from macroscopic to microscopic, and in particular the bifurcation is characterized by an infinite wavelength when $\kappa=0$.
\begin{figure}[htb!]
    \centering
    \begin{subfigure}{0.245\textwidth}
        \centering
        \includegraphics[width=0.95\linewidth]{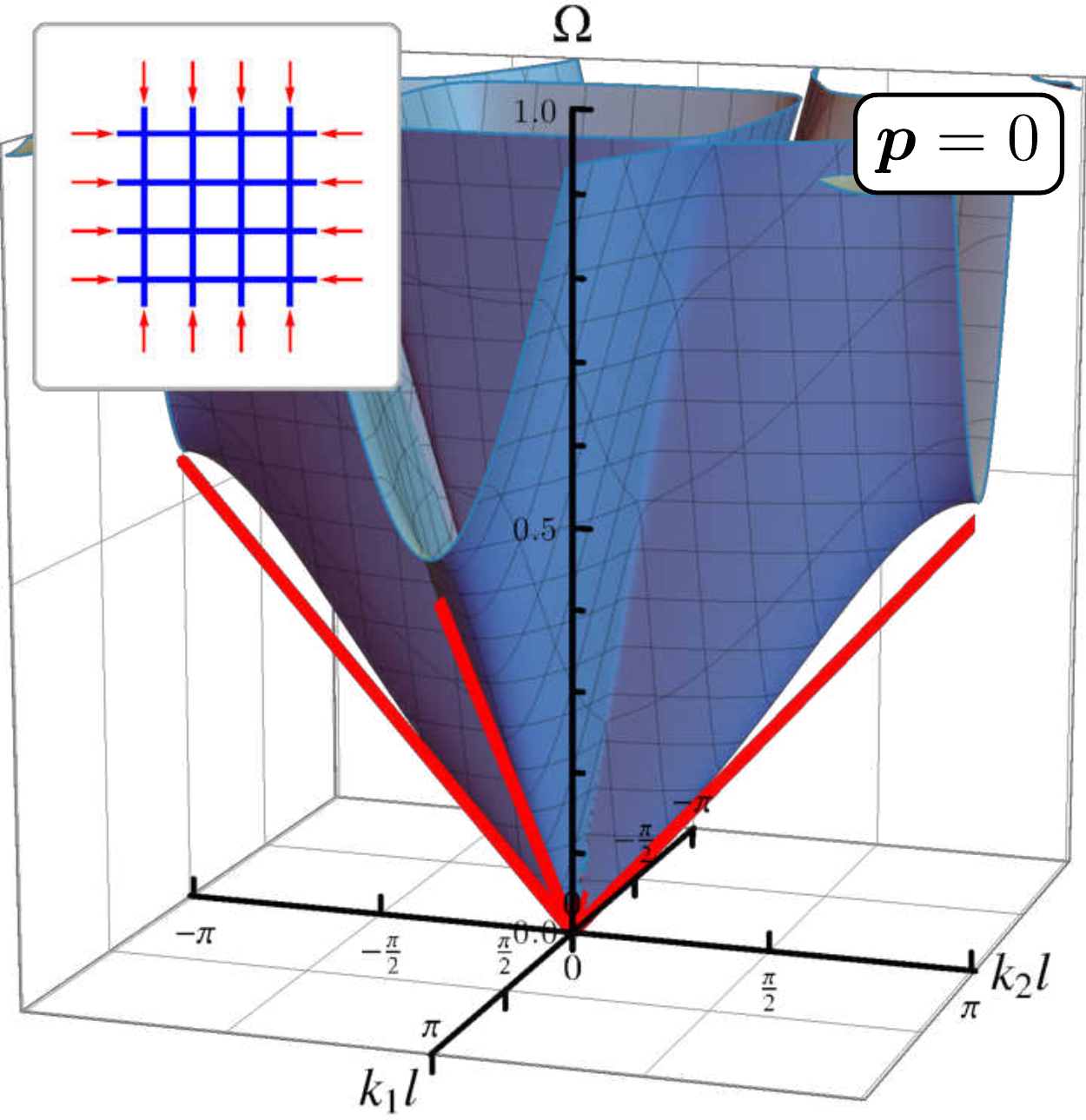}
    \end{subfigure}%
    \begin{subfigure}{0.245\textwidth}
        \centering
        \includegraphics[width=0.95\linewidth]{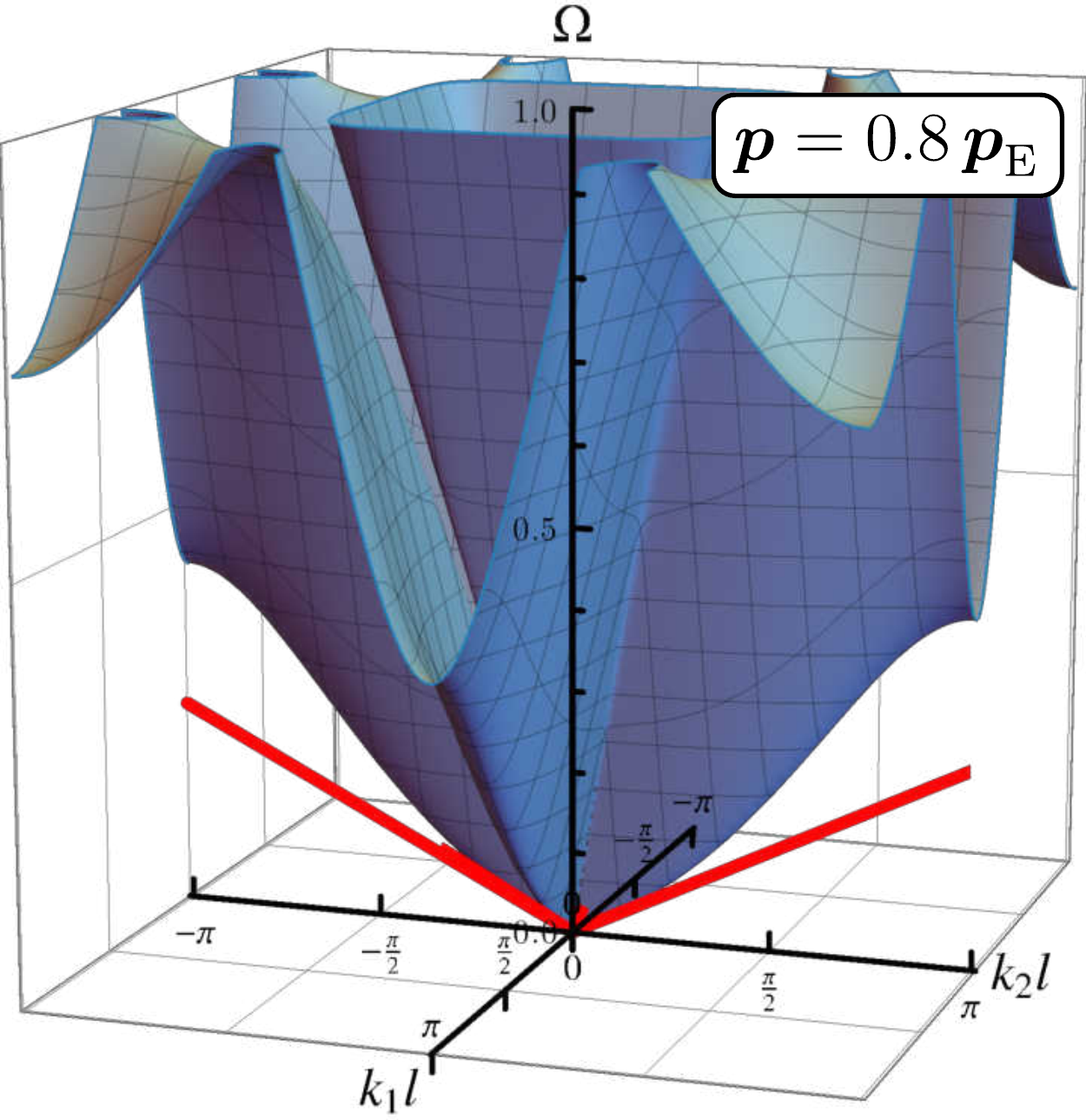}
    \end{subfigure}%
    \begin{subfigure}{0.245\textwidth}
        \centering
        \includegraphics[width=0.95\linewidth]{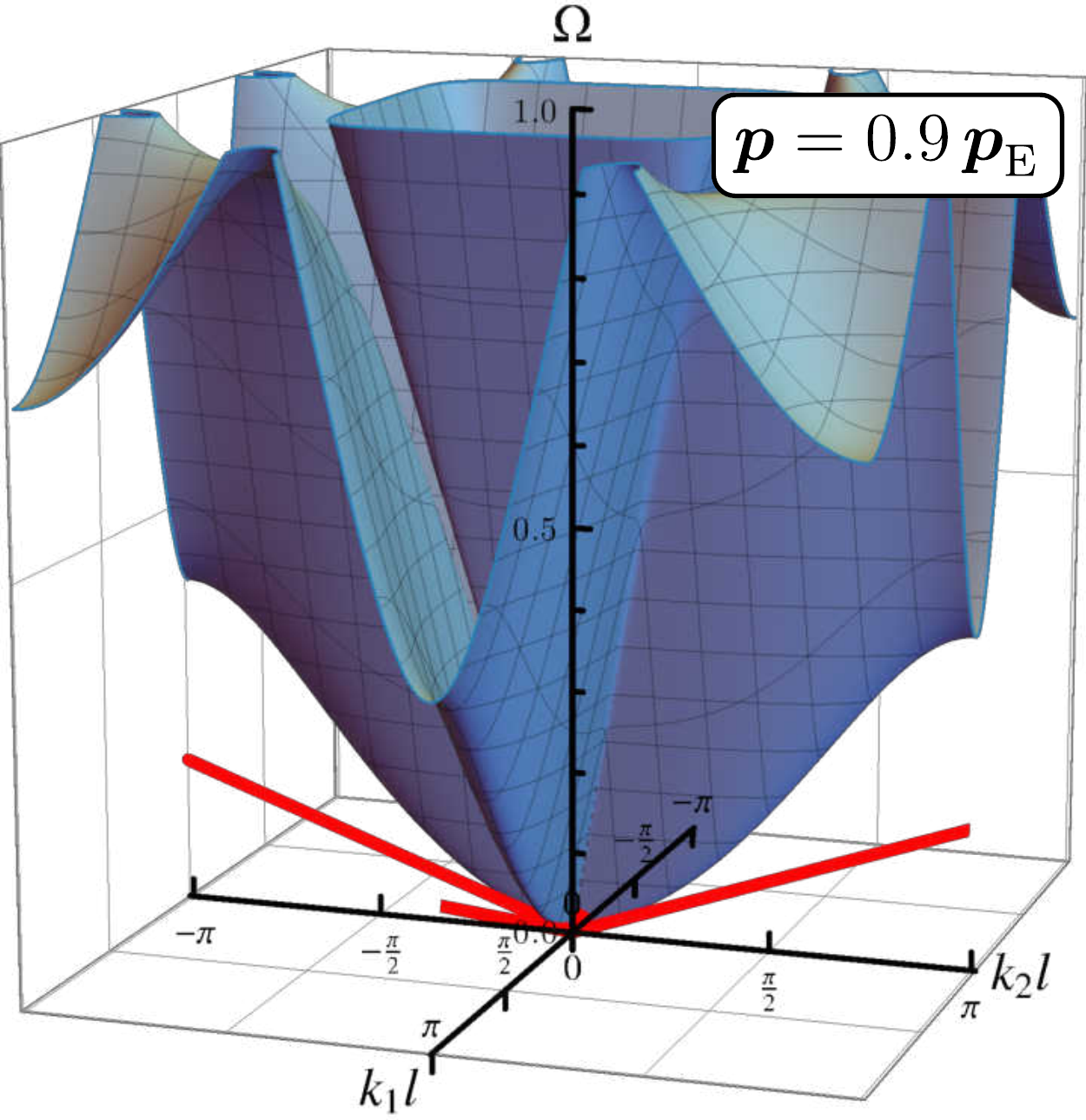}
    \end{subfigure}%
    \begin{subfigure}{0.245\textwidth}
        \centering
        \includegraphics[width=0.95\linewidth]{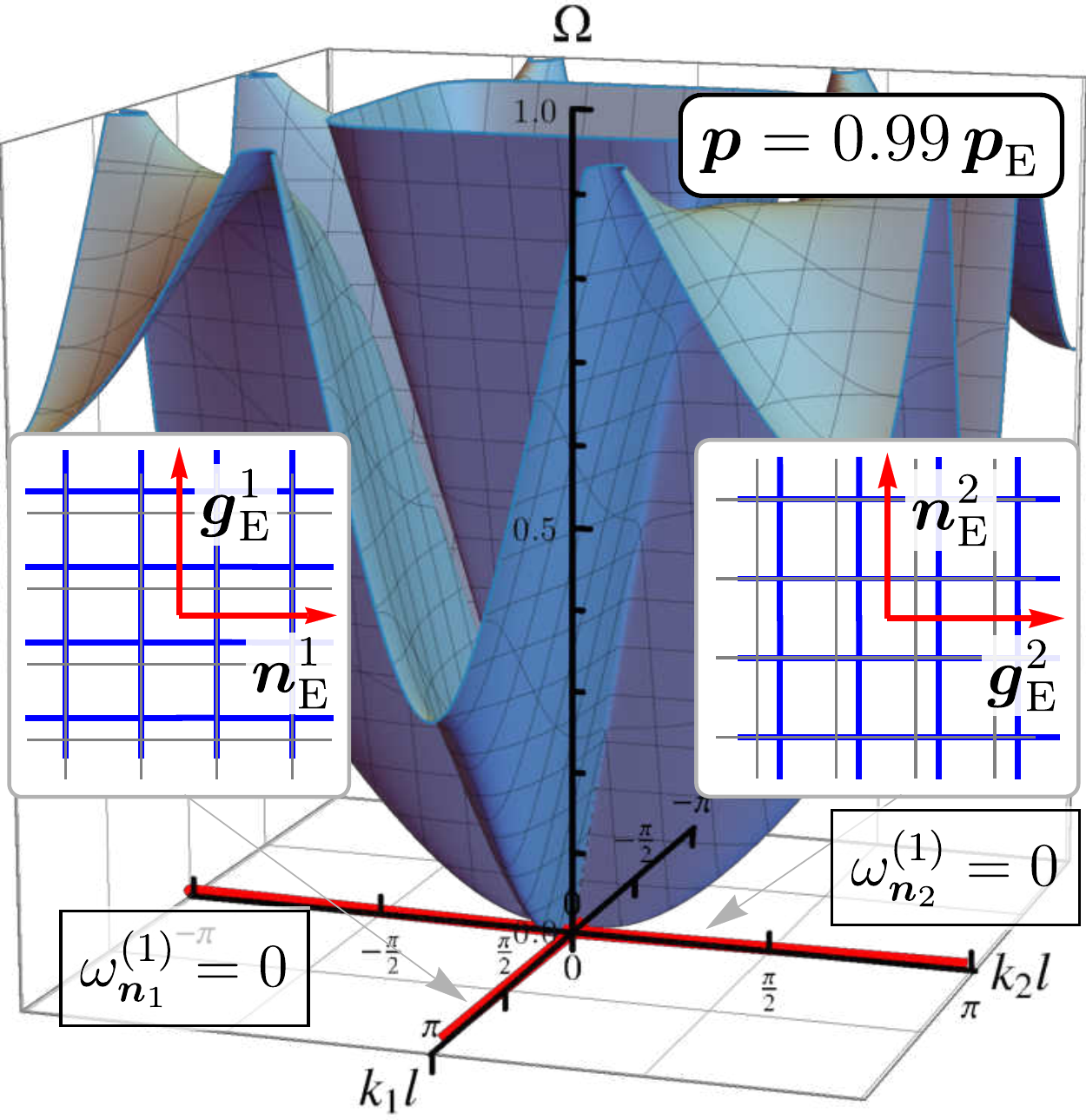}
    \end{subfigure}\\ \vspace{2mm}
    \begin{subfigure}{0.245\textwidth}
        \centering
        \includegraphics[width=0.95\linewidth]{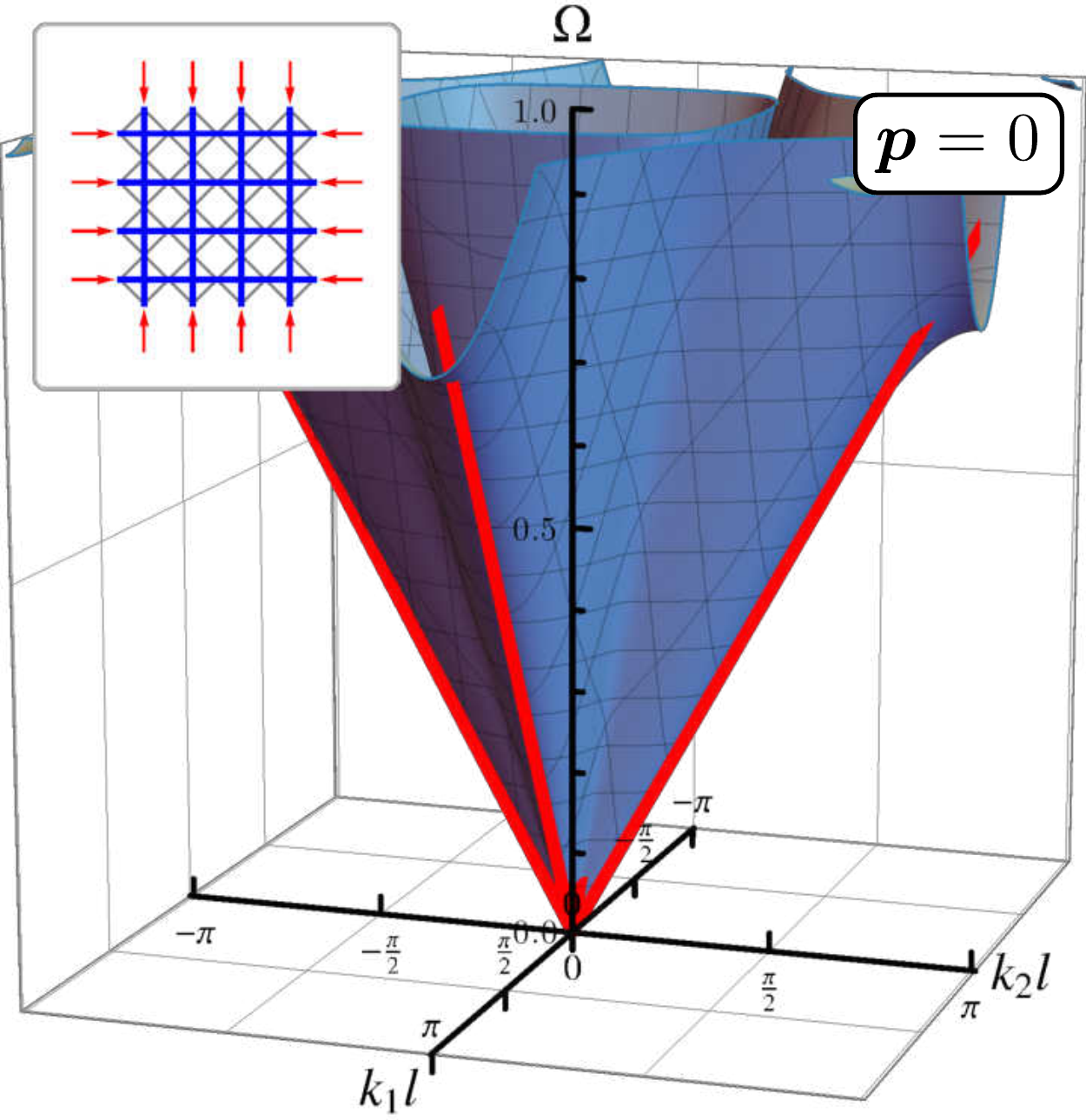}
    \end{subfigure}%
    \begin{subfigure}{0.245\textwidth}
        \centering
        \includegraphics[width=0.95\linewidth]{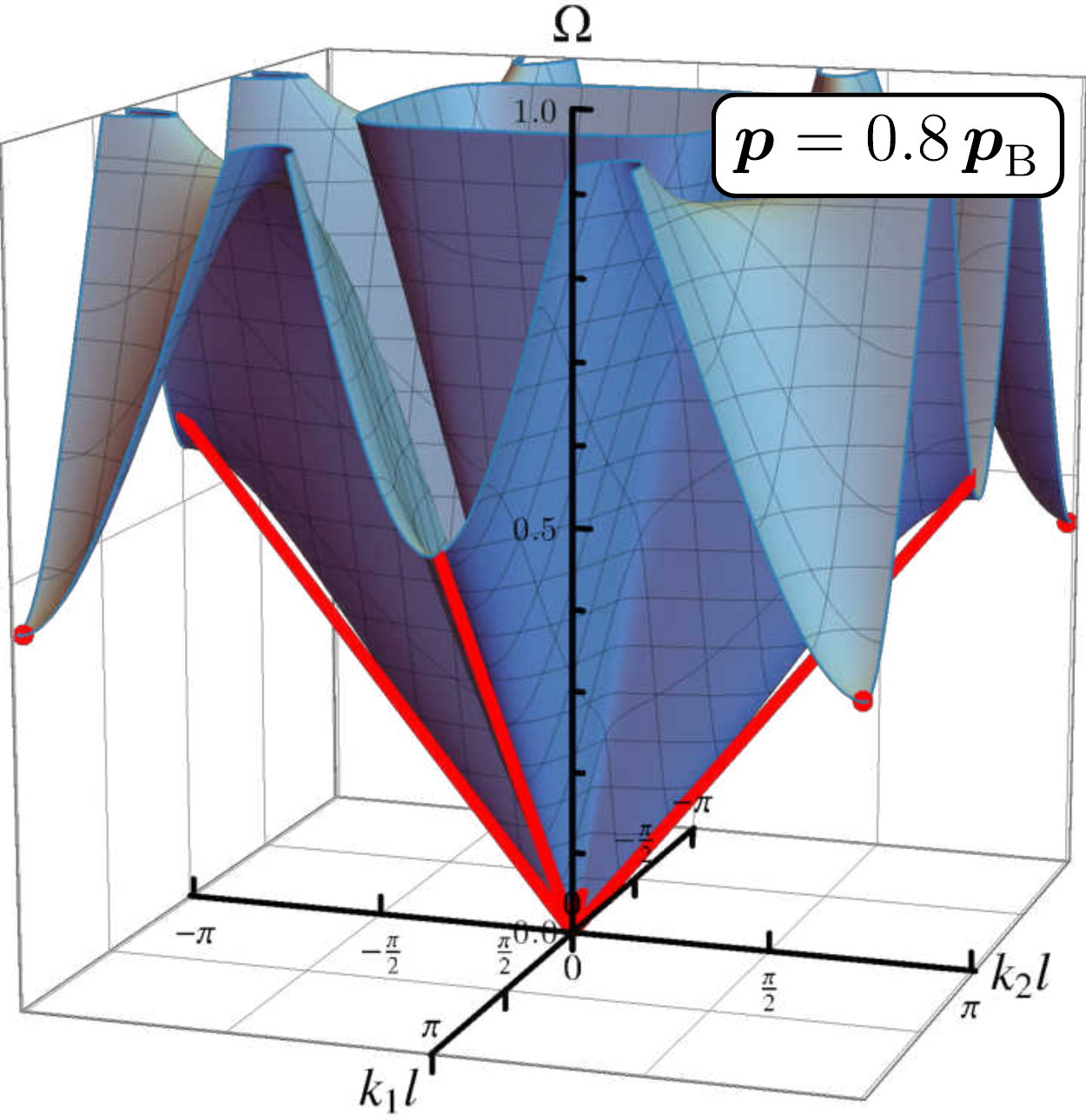}
    \end{subfigure}%
    \begin{subfigure}{0.245\textwidth}
        \centering
        \includegraphics[width=0.95\linewidth]{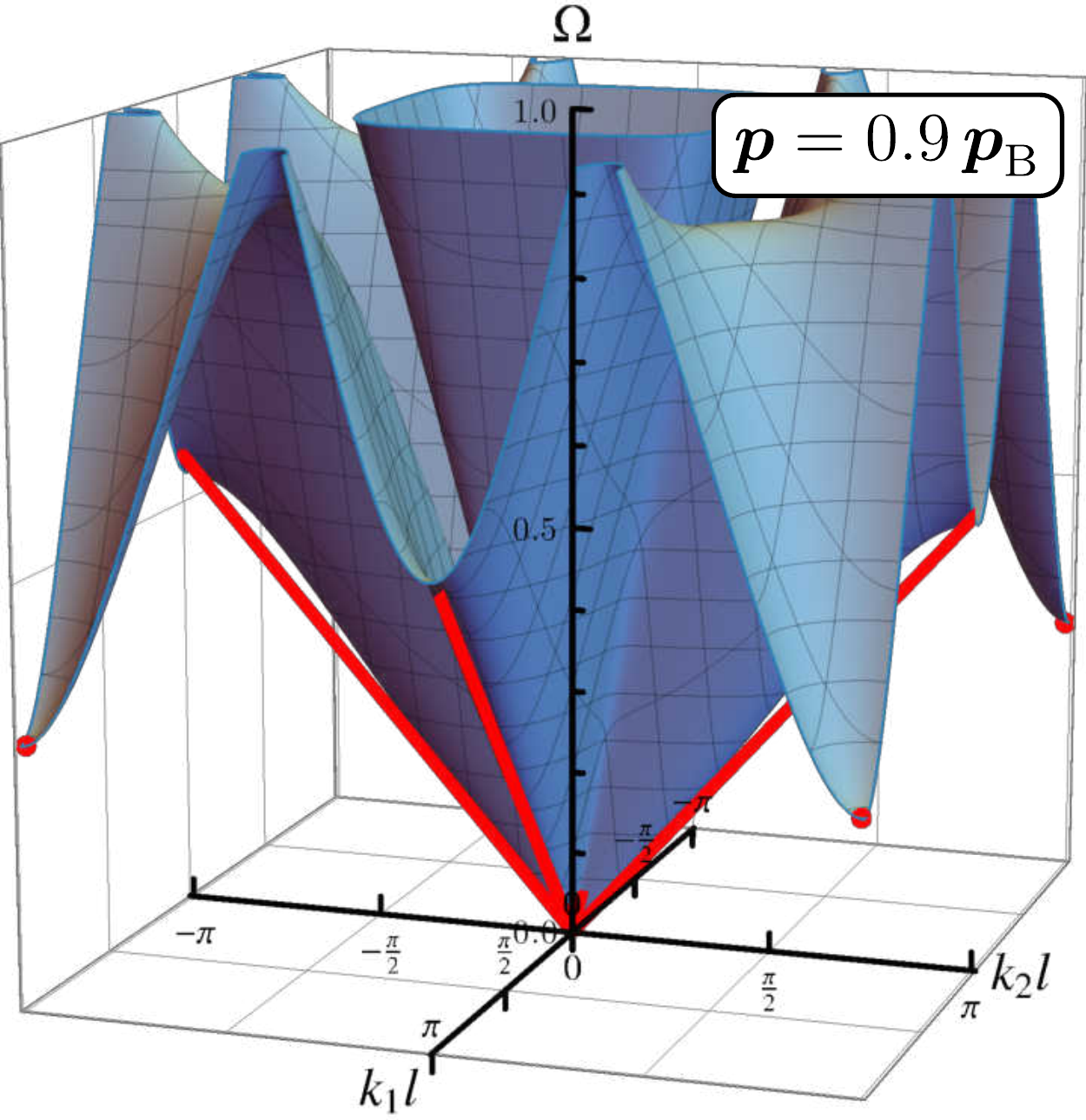}
    \end{subfigure}%
    \begin{subfigure}{0.245\textwidth}
        \centering
        \includegraphics[width=0.95\linewidth]{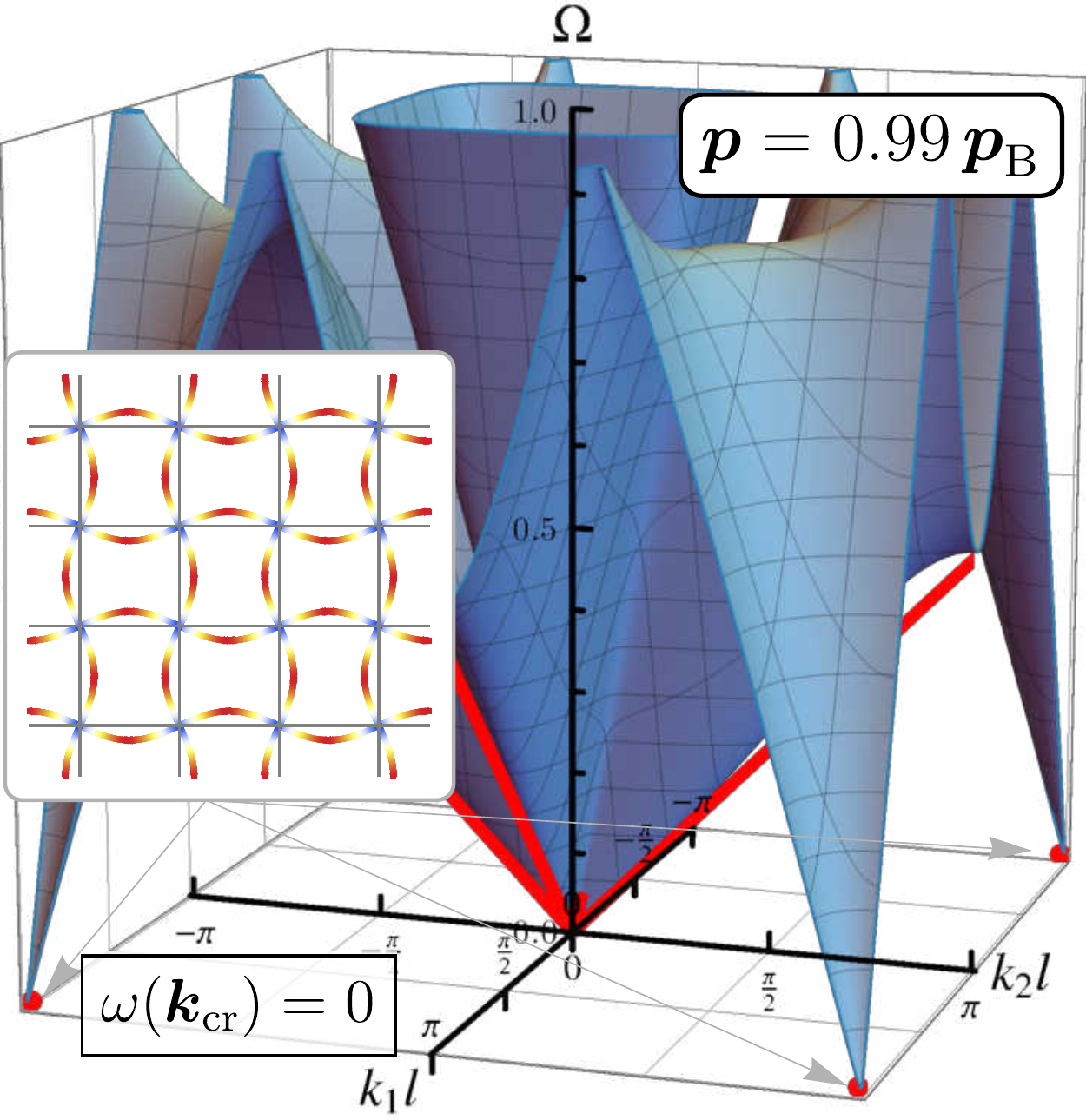}
    \end{subfigure}
    \caption{\label{fig:macro_micro_bifurcation}
    The dispersion surfaces, computed for states of preload of increasing magnitude (from left to right), demonstrate the difference between macroscopic (upper part) and microscopic (lower part) bifurcations occurring in a square grid of elastic rods.
    The stiffness of the diagonal springs can be tuned to cause a switching of the critical bifurcation mode from macroscopic (low spring stiffness) to microscopic (high spring stiffness).
    The four surfaces reported in the upper part refer to equibiaxial compression of a square grid with $\Lambda_1=\Lambda_2=10$ and not reinforced with springs. In this case an infinite-wavelength bifurcation occurs at vanishing slope of the acoustic branches in the origin (when the preload reaches a critical value $\bp_{\text{E}}=\{-5.434,-5.434\}$).
    The four surfaces reported in the lower part refer to the same grid but reinforced with springs ($\kappa=0.4$), which induces a `stiffening' of the acoustic branches at the origin. This triggers a critical bifurcation at a preload $\bp_{\text{B}}=\{-\pi^2,-\pi^2\}$, when the lowering of the dispersion surface causes the generation of a zero-frequency wave with non-null wave vector (corresponding to a finite wavelength buckling).
    }
\end{figure}
On the other hand, the time-harmonic dynamics analyzed here provides a mechanical interpretation to instabilities different from that obtained through a quasi-static approach. 
The difference between macroscopic and microscopic bifurcation will be specially focused, as the latter is lost in the homogenization approach.
The homogenization scheme introduced in Section~\ref{sec:asymptotic_lattice_waves} proves that the long-wavelength asymptotics for waves propagating in the lattice is governed by the acoustic tensor of the effective medium, Eq.~\eqref{eq:acoustic_tensor_lattice}. Therefore, it becomes now clear that a macro-bifurcation in the lattice has to be  equivalent to failure of ellipticity in its equivalent continuum. 
Hence, a macro-bifurcation occurs when the velocity of the acoustic long waves of the lattice vanishes along some directions.
Moreover, a clear interpretation of short-wavelength bifurcations (micro-bifurcations) is also provided by the analysis of the dispersion relation of the lattice~\eqref{eq:dispersion}, interpreted now as a function of the preload state. The latter can be used to identify the condition of buckling in the lattice as the `propagation' of a Bloch wave at vanishing frequency.
In fact, regardless of the critical wavelength, macro and micro bifurcations can be visualized by plotting the evolution of the dispersion surfaces along a loading path up to loss of stability.

The essential difference between the two kinds of bifurcation is exemplified in Fig.~\ref{fig:macro_micro_bifurcation} for two square grids ($\alpha=\pi/2$ and $\Lambda_1=\Lambda_2=10$), one without diagonal springs (upper row in the figure) and the other with $\kappa=0.4$ (lower row in the figure), subject to equibiaxial compression ($p_1=p_2$) of increasing magnitude (from left to right in the figure).

The dispersion surfaces (plotted in the non-dimensional space $\{k_1 l,k_2 l,\Omega\}$ with $\Omega=\omega\,l\sqrt{\gamma/A}$ and $\gamma_r=0$) show that \textit{the macro-bifurcation in the grid without springs occurs with the progressive lowering, and eventually vanishing, of the slope of the acoustic branches at the origin, while the dispersion surface attains non-null frequency for every other wave vector}.
On the contrary, the micro-bifurcation occurring in the grid reinforced with springs is characterized by non-vanishing slope of the acoustic branches at the origin. Moreover, instead of the preload-induced lowering of the dispersion surface observed for macro-bifurcations, a zero-frequency wave is generated with non-null wave vector, which corresponds to a finite wavelength bifurcation.
These dispersion surfaces can be considered the dynamic counterpart of the bifurcation surfaces presented in Part I of this study. 

Failure of ellipticity and the consequent emergence of strain localization in the lattice is now investigated following the following steps:
\begin{enumerate}[label=(\roman*)]
    \item the acoustic tensor $\bA^{(\fC)}(\bn)$ of the homogenized continuum is analytically calculated as an explicit function of the parameter set $\{p_1,p_2,\Lambda_1,\Lambda_2,\kappa,\alpha\}$;
    \item loss of ellipticity is analyzed for cubic, orthotropic, and fully anisotropic lattices by identifying the prestress states leading to a vanishing eigenvalue of the acoustic tensor, while the corresponding eigenvector identifies the localization mode;
    \item the ellipticity domain in the $\{p_1,p_2\}$-space and its dependence on lattice parameters $\{\Lambda_1,\Lambda_2,\alpha\}$ is determined;
    \item the predicted directions and modes of strain localization are compared with the maps of displacements resulting in the equivalent elastic material 
    from the application of a concentrated pulsating force (the time-harmonic Green's function), thus following the so-called `perturbative
    approach to localization' introduced in~\cite{bigoni_2002a}; 
    \item the behaviour of the equivalent elastic solid is validated through numerical simulations (with finite elements) of the forced low-frequency response of the corresponding preloaded lattice.
\end{enumerate}

\subsection{Acoustic tensor, eigenvalues, eigenvectors, and ellipticity domain}
\label{sec:acoustic_tensor_grid}
With reference to the orthonormal basis defined by the two unit vectors $\{\be_1,\be_2\}$, the acoustic tensor $\bA^{(\fC)}(\bn)$ for the continuum equivalent, in a homogenized sense, 
to the lattice shown in Fig.~\ref{fig:geometry_grid_and_unit_cell} is represented as 
\begin{equation}
    \label{eq:acoustic_tensor_grid}
    \bA^{(\fC)}(\bn) = A^{(\fC)}_{11}(\bn)\,\be_1\otimes\be_1 + A^{(\fC)}_{12}(\bn)\,\be_1\otimes\be_2 + A^{(\fC)}_{21}(\bn)\,\be_2\otimes\be_1 + A^{(\fC)}_{22}(\bn)\,\be_2\otimes\be_2 \,,
\end{equation}
where the components, computed via Eq.~\eqref{eq:acoustic_tensor_lattice}, are expressed as follows
\begin{equation*}
\begin{aligned}
    A^{(\fC)}_{11}(\bn) &= \left(h_{1111}\,n_1^2 + h_{1112}\,n_1 n_2 + h_{1122}\,n_2^2\right) A/l \,, \\
    A^{(\fC)}_{12}(\bn) = A^{(\fC)}_{21}(\bn) &= \left(h_{1211}\,n_1^2 + h_{1212}\,n_1 n_2 + h_{1222}\,n_2^2\right) A/l \,, \\
    A^{(\fC)}_{22}(\bn) &= \left(h_{2211}\,n_1^2 + h_{2212}\,n_1 n_2 + h_{2222}\,n_2^2\right) A/l \,,
\end{aligned}
\end{equation*}
with the coefficients $h_{ijkl}$ being function of the parameter set $\{p_1,p_2,\Lambda_1,\Lambda_2,\kappa,\alpha\}$.
Using the same notation as in Part I, the contribution of the rods' grid and the springs are denoted as $h_{ijkl}^{\text{G}}$ and $h_{ijkl}^{\text{S}}$, respectively, so that
\begin{equation}
\label{eq:acoustic_tensor_grid_contributions}
    h_{ijkl}(p_1,p_2,\Lambda_1,\Lambda_2,\kappa,\alpha) = h_{ijkl}^{\text{G}} (p_1,p_2,\Lambda_1,\Lambda_2,\alpha) + h_{ijkl}^{\text{S}}(\kappa,\alpha) \,.
\end{equation}
The non-linear dependence on the prestress $\bp$ causes the full expression for the functions $h_{ijkl}^{\text{G}}$ to be quite lengthy and therefore the complete result is deferred to Appendix~\ref{sec:homogenized_acoustic_tensor}, while the first-order expansion with respect to $\bp$ is given in the following (components that have to be equal by symmetry are not reported)
\begin{dgroup*}[style={\footnotesize},breakdepth={20}]
    \begin{dmath*}
        h_{1111}^{\text{G}} \sim \frac{12 \sin\alpha \cos^2\alpha}{\Lambda_1^2+\Lambda_2^2}+\csc\alpha+\cos^3\alpha \cot\alpha
        +p_1 \frac{\Lambda_1^2 \sin\alpha \cos^2\alpha}{5 \left(\Lambda_1^2+\Lambda_2^2\right)^2}
        +p_2 \frac{\sin\alpha \cos^2\alpha \left(5 \Lambda_1^4+10 \Lambda_1^2 \Lambda_2^2+6 \Lambda_2^4\right)}{5 \Lambda_2^2 \left(\Lambda_1^2+\Lambda_2^2\right)^2} 
        ,
    \end{dmath*}
    \begin{dmath*}
        h_{1112}^{\text{G}} \sim \frac{2 \cos\alpha \left(\cos^2\alpha \left(\Lambda_1^2+\Lambda_2^2-12\right)+12\right)}{\Lambda_1^2+\Lambda_2^2}
        +p_1 \frac{2 \Lambda_1^2 \sin^2\alpha \cos\alpha}{5 \left(\Lambda_1^2+\Lambda_2^2\right)^2}
        +p_2 \frac{2 \sin^2\alpha \cos\alpha \left(5 \Lambda_1^4+10 \Lambda_1^2 \Lambda_2^2+6 \Lambda_2^4\right)}{5 \Lambda_2^2 \left(\Lambda_1^2+\Lambda_2^2\right)^2}
        ,
    \end{dmath*}
    \begin{dmath*}
        h_{1122}^{\text{G}} \sim \frac{12 \sin^3\alpha}{\Lambda_1^2+\Lambda_2^2}+\sin\alpha \cos^2\alpha
        +p_1 \frac{\Lambda_1^2 \sin^3\alpha}{5 \left(\Lambda_1^2+\Lambda_2^2\right)^2}
        +p_2 \frac{\sin^3\alpha \left(5 \Lambda_1^4+10 \Lambda_1^2 \Lambda_2^2+6 \Lambda_2^4\right)}{5 \Lambda_2^2 \left(\Lambda_1^2+\Lambda_2^2\right)^2}
        ,
    \end{dmath*}
    \begin{dmath*}
        h_{1211}^{\text{G}} \sim \frac{\cos\alpha \left(\cos^2\alpha \left(\Lambda_1^2+\Lambda_2^2-12\right)+12\right)}{\Lambda_1^2+\Lambda_2^2}
        +p_1 \frac{\Lambda_1^2 \sin^2\alpha \cos\alpha}{5 \left(\Lambda_1^2+\Lambda_2^2\right)^2}
        +p_2 \frac{\Lambda_2^4 \cos\alpha-\cos^3\alpha \left(5 \Lambda_1^4+10 \Lambda_1^2 \Lambda_2^2+6 \Lambda_2^4\right)}{5 \Lambda_2^2 \left(\Lambda_1^2+\Lambda_2^2\right)^2}
        ,
    \end{dmath*}
    \begin{dmath*}
        h_{1212}^{\text{G}} \sim \frac{\sin\alpha \left(\cos (2 \alpha ) \left(\Lambda_1^2+\Lambda_2^2-12\right)+\Lambda_1^2+\Lambda_2^2\right)}{\Lambda_1^2+\Lambda_2^2}
        -p_1 \frac{\Lambda_1^2 \sin\alpha \cos (2 \alpha )}{5 \left(\Lambda_1^2+\Lambda_2^2\right)^2}
        +p_2 \frac{\sin\alpha \left(\Lambda_2^4-2 \cos^2\alpha \left(5 \Lambda_1^4+10 \Lambda_1^2 \Lambda_2^2+6 \Lambda_2^4\right)\right)}{5 \Lambda_2^2 \left(\Lambda_1^2+\Lambda_2^2\right)^2}
        ,
    \end{dmath*}
    \begin{dmath*}
        h_{1222}^{\text{G}} \sim \frac{\sin^2\alpha \cos\alpha \left(\Lambda_1^2+\Lambda_2^2-12\right)}{\Lambda_1^2+\Lambda_2^2}
        -p_1 \frac{\Lambda_1^2 \sin^2\alpha \cos\alpha}{5 \left(\Lambda_1^2+\Lambda_2^2\right)^2}
        +p_2 -\frac{\sin^2\alpha \cos\alpha \left(5 \Lambda_1^4+10 \Lambda_1^2 \Lambda_2^2+6 \Lambda_2^4\right)}{5 \Lambda_2^2 \left(\Lambda_1^2+\Lambda_2^2\right)^2}
        ,
    \end{dmath*}
    \begin{dmath*}
        h_{2211}^{\text{G}} \sim \frac{12 \sin^3\alpha}{\Lambda_1^2+\Lambda_2^2}+\sin\alpha \cos^2\alpha
        +p_1 \left( \frac{\Lambda_1^2 \sin^3\alpha}{5 \left(\Lambda_1^2+\Lambda_2^2\right)^2}+\frac{\csc\alpha}{\Lambda_1^2} \right)
        +p_2 \frac{4 \cos (2 \alpha ) \left(5 \Lambda_1^4+10 \Lambda_1^2 \Lambda_2^2+4 \Lambda_2^4\right)+(\cos (4 \alpha )+3) \left(5 \Lambda_1^4+10 \Lambda_1^2 \Lambda_2^2+6 \Lambda_2^4\right)}{40 \Lambda_2^2 \left(\Lambda_1^2+\Lambda_2^2\right)^2 \sin\alpha}
        ,
    \end{dmath*}
    \begin{dmath*}
        h_{2212}^{\text{G}} \sim \frac{\sin\alpha \sin (2 \alpha ) \left(\Lambda_1^2+\Lambda_2^2-12\right)}{\Lambda_1^2+\Lambda_2^2}
        -p_1 \frac{2 \Lambda_1^2 \sin^2\alpha \cos\alpha}{5 \left(\Lambda_1^2+\Lambda_2^2\right)^2}
        +p_2 \frac{\cos\alpha \left(\cos (2 \alpha ) \left(5 \Lambda_1^4+10 \Lambda_1^2 \Lambda_2^2+6 \Lambda_2^4\right)+5 \Lambda_1^4+10 \Lambda_1^2 \Lambda_2^2+4 \Lambda_2^4\right)}{5 \Lambda_2^2 \left(\Lambda_1^2+\Lambda_2^2\right)^2}
        ,
    \end{dmath*}
    \begin{dmath*}
        h_{2222}^{\text{G}} \sim \frac{12 \sin\alpha \cos^2\alpha}{\Lambda_1^2+\Lambda_2^2}+\sin^3\alpha
        +p_1 \frac{\Lambda_1^2 \sin\alpha \cos^2\alpha}{5 \left(\Lambda_1^2+\Lambda_2^2\right)^2}
        +p_2 \frac{\sin\alpha \cos^2\alpha \left(5 \Lambda_1^4+10 \Lambda_1^2 \Lambda_2^2+6 \Lambda_2^4\right)}{5 \Lambda_2^2 \left(\Lambda_1^2+\Lambda_2^2\right)^2}
        .
    \end{dmath*}
\end{dgroup*}
The components $h_{ijkl}^{\text{S}}$, ruling the effect of diagonal springs, can be written as 
\begin{align*}
    h^{\text{S}}_{1111} &= \kappa \frac{5+3 \cos (2 \alpha)}{4\sin\alpha}  \,, &
    h^{\text{S}}_{1112} &= 2 \kappa  \cos \alpha  \,, &
    h^{\text{S}}_{1211} &= \kappa  \cos \alpha \,,\\
    h^{\text{S}}_{1212} &= \kappa  \sin \alpha \,, &
    h^{\text{S}}_{1122} &= h^{\text{S}}_{2211} = h^{\text{S}}_{2222} = \frac{1}{2} \kappa  \sin \alpha  \,, &
    h^{\text{S}}_{1222} &= h^{\text{S}}_{2212} = 0 \,.
\end{align*}

For the special case of a square grid $\alpha=\pi/2$ and in the absence of prestress ($\bp=\bzero$), the acoustic tensor can be further simplified to 
\begin{align*}
    \lim_{\bp\to\bzero} \left.\bA^{(\fC)}(\bn)\right\rvert_{\alpha=\pi/2} &= 
    \frac{A}{l}\left(n_1^2+\frac{12 n_2^2}{\Lambda_1^2+\Lambda_2^2}\right)\,\be_1\otimes\be_1 + \\ 
    & + \frac{A}{l}\frac{12 n_1 n_2}{\Lambda_1^2+\Lambda_2^2}\,(\be_1\otimes\be_2 + \be_2\otimes\be_1)
    + \frac{A}{l}\left(\frac{12 n_1^2}{\Lambda_1^2+\Lambda_2^2}+n_2^2\right)\,\be_2\otimes\be_2 \,.
\end{align*}

It is recalled that the homogenized elastic material is defined to be \textit{strongly elliptic} (SE) when the acoustic tensor is positive definite, 
\begin{equation}
    \label{eq:strong_ellipticity}
    \bg \scalp \bA^{(\fC)}(\bn)\, \bg > 0, \quad \forall \bg\neq\bzero, \quad \forall \bn \neq\bzero \qquad \Longleftrightarrow \qquad \textrm{(SE)} \,,
\end{equation}
while \textit{ellipticity} is defined as the non-singularity of the acoustic tensor, 
\begin{equation}
    \label{eq:ellipticity}
    \det\bA^{(\fC)}(\bn) \neq 0, \quad \forall \bn\neq\bzero \qquad \Longleftrightarrow \qquad \textrm{(E)} \,.
\end{equation}

As shown in Section~\ref{sec:acoustic_tensor_lattice}, the acoustic tensor resulting from homogenization is symmetric, which implies that its eigenvalues are always real.
This means that, letting $\{c_1^2, c_2^2\}$ be the eigenvalues of a symmetric $\bA^{(\fC)}(\bn)$, (SE) is equivalent to the strict positiveness of the eigenvalues,  $c_1^2>0,\, c_2^2>0$, and (E) is equivalent to the condition of non-vanishing eigenvalues, $c_1^2\neq0,\, c_2^2\neq0$ (for all unit vectors $\bn$).
\begin{table}[htb!]
    \centering
    \begin{tabular}{lllll}
        \toprule
        \textit{Geometry}       & \textit{Slenderness}          & \textit{Symmetry}   & $\bp_{\text{E}}$               & $\theta_{\text{cr}}$\\ \midrule
        Square $\alpha=\pi/2$   & $\Lambda_1=\Lambda_2=10$      & Cubic               & $\{-5.434,-5.434\}$   & $0^\circ,90^\circ$\\
                                & $\Lambda_1=7,\,\Lambda_2=15$  & Orthotropic         & $\{-2.071,-2.071\}$   & $0^\circ$\\
        Rhombus $\alpha=\pi/3$  & $\Lambda_1=\Lambda_2=10$      & Orthotropic         & $\{-5.345,-5.345\}$   & $88.2^\circ,151.8^\circ$\\
                                & $\Lambda_1=7,\,\Lambda_2=15$  & Anisotropic         & $\{-2.043,-2.043\}$   & $151.4^\circ$\\
        \bottomrule
    \end{tabular}
    \caption{\label{tab:cases_analyzed}
    Loss of ellipticity for different geometric configurations of the preloaded grid-like lattice (Fig.~\ref{fig:geometry_grid_and_unit_cell}) in the absence of diagonal springs ($\kappa=0$).
    The symmetry class is referred to the unloaded configuration.
    The preload $\bp_{\text{E}}$ and the localization direction $\bn_{\text{E}}=\cos(\theta_{\text{cr}})\be_1+\sin(\theta_{\text{cr}})\be_2$ are obtained by solving the loss of ellipticity condition~\eqref{eq:ellipticity_loss} (assuming a radial path $\bp=\{p_1,p_1\}$).
    }
\end{table}
\begin{figure}[htb!]
    \centering
    \begin{subfigure}{0.45\textwidth}
        \centering
        \caption{\label{fig:eigenvalue_square_10_10}$p_1=p_2=-5.434,\,\Lambda_1=\Lambda_2=10,\,\alpha=\pi/2$}
        \includegraphics[width=0.98\linewidth]{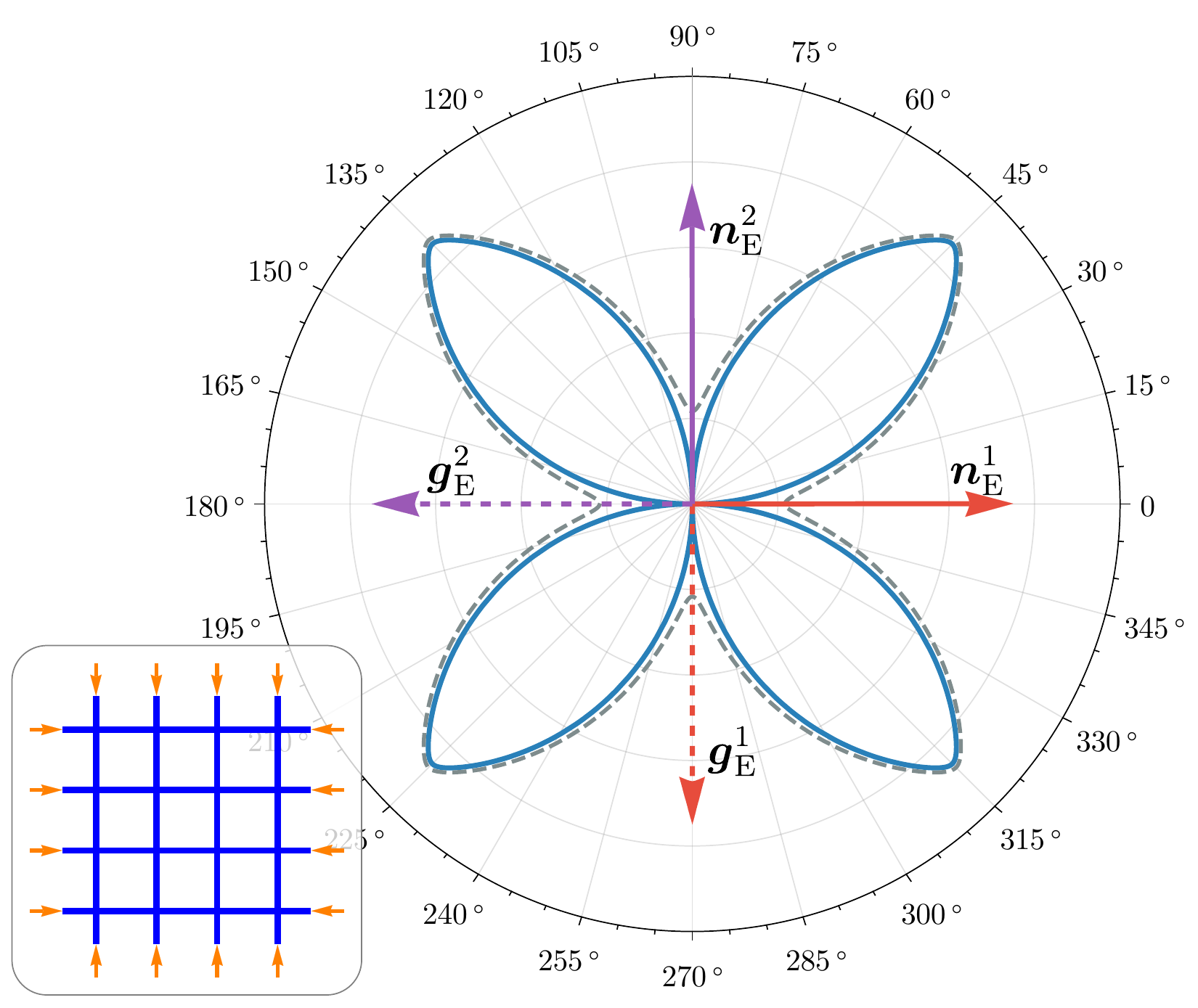}
    \end{subfigure}%
    \begin{subfigure}{0.45\textwidth}
        \centering
        \caption{\label{fig:eigenvalue_square_7_15}$p_1=p_2=-2.071,\,\Lambda_1=7,\, \Lambda_2=15,\,\alpha=\pi/2$}
        \includegraphics[width=0.98\linewidth]{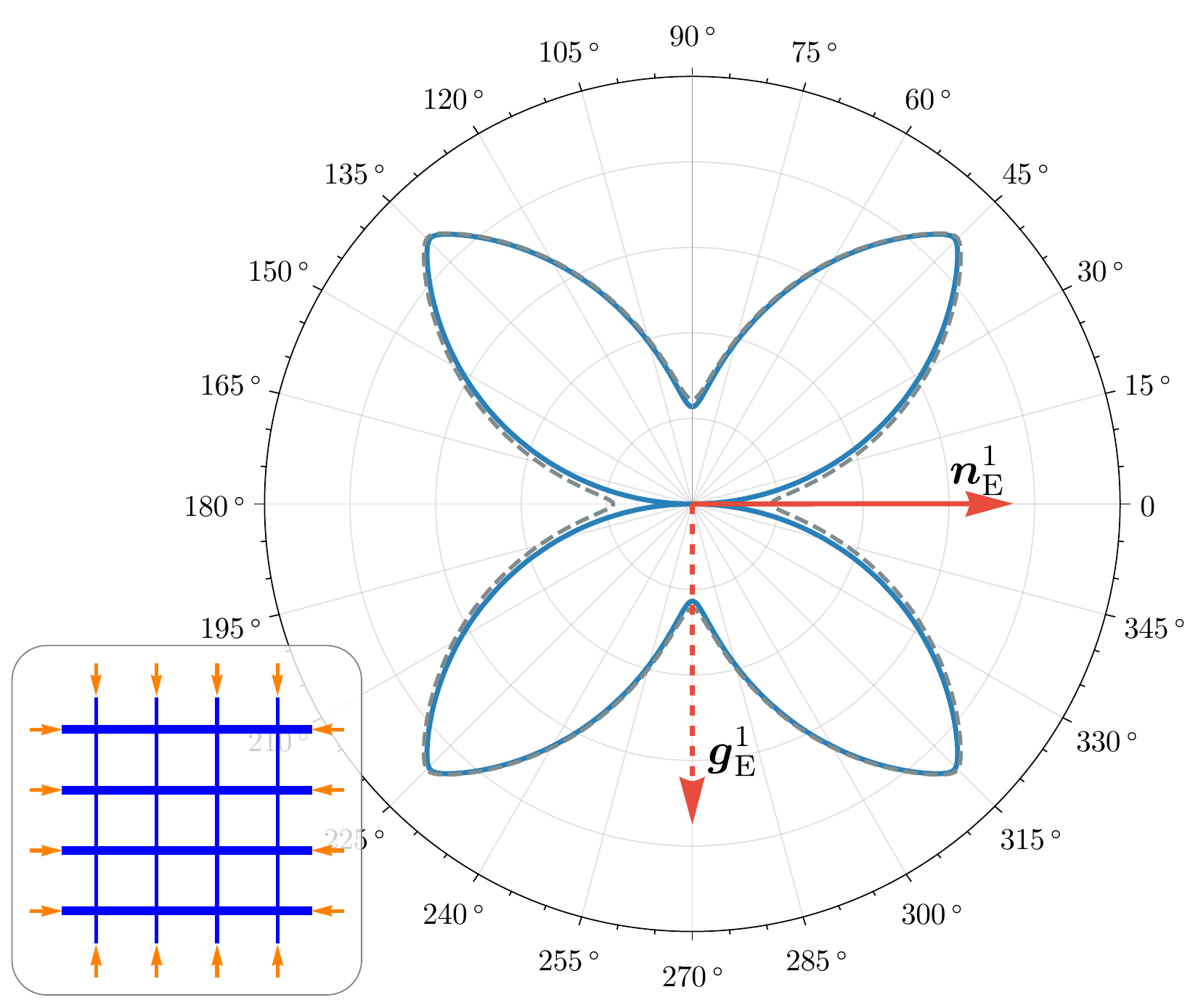}
    \end{subfigure}\\ \vspace{4mm}
    \begin{subfigure}{0.45\textwidth}
        \centering
        \caption{\label{fig:eigenvalue_rhombus_10_10}$p_1=p_2=-5.345,\,\Lambda_1=\Lambda_2=10,\,\alpha=\pi/3$}
        \includegraphics[width=0.98\linewidth]{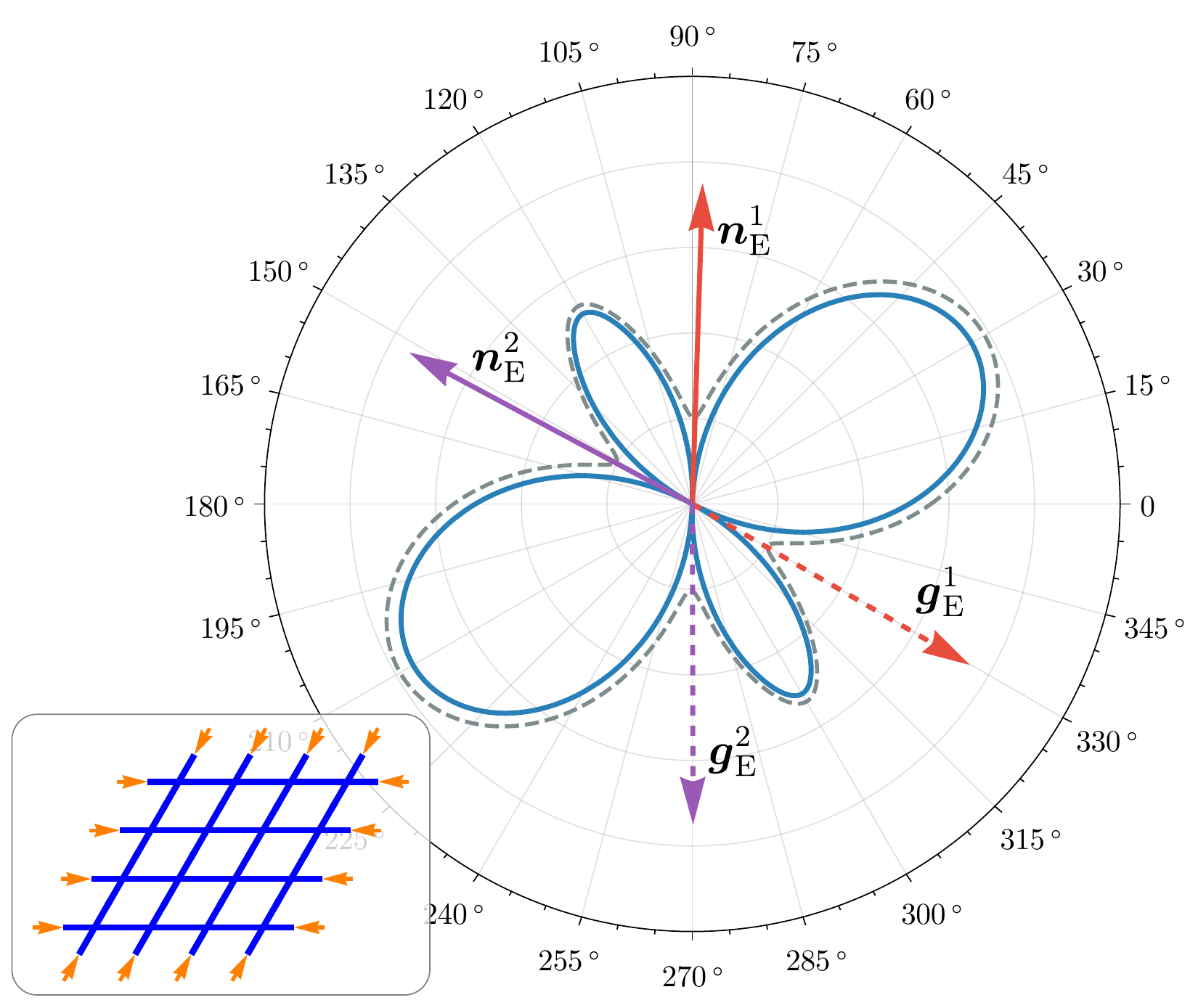}
    \end{subfigure}%
    \begin{subfigure}{0.45\textwidth}
        \centering
        \caption{\label{fig:eigenvalue_rhombus_7_15}$p_1=p_2=-2.043,\,\Lambda_1=7,\, \Lambda_2=15,\,\alpha=\pi/3$}
        \includegraphics[width=0.98\linewidth]{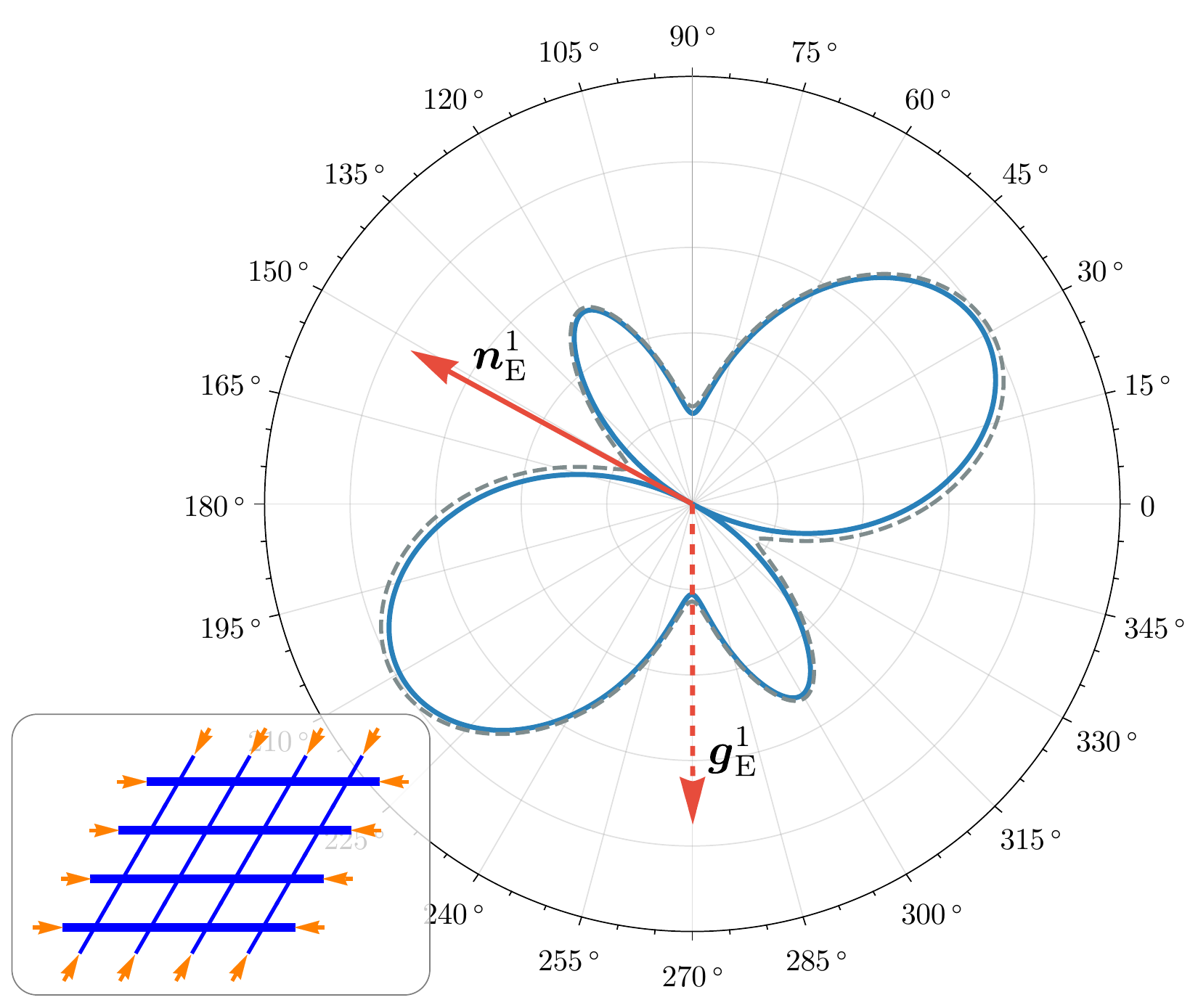}
    \end{subfigure}%
    \caption{\label{fig:eigenvalue_square_rhombus}
    Polar plots of the lowest eigenvalue 
		of the acoustic tensor~\eqref{eq:acoustic_tensor_grid}, as a function of $\bn$, 		
		for a prestress state at $50\%$ (dashed gray line) and $100\%$ (continuous blue line) of the limit value for ellipticity  loss ($\bp_{\text{E}}$ reported in Table~\ref{tab:cases_analyzed}). 
    A square lattice and a rhombic lattice are considered with $\Lambda_1=\Lambda_2=10$ and $\Lambda_1=7$, $\Lambda_2=15$. 
    In the orthotropic cases, both for squared and rhombic lattices, the loss of ellipticity is characterized by the simultaneous vanishing of the eigenvalue of the acoustic tensor along two directions $\bn^1_{\text{E}}$ and $\bn^2_{\text{E}}$ (the associated wave amplitudes are reported as $\bg^1_{\text{E}}$ and $\bg^2_{\text{E}}$). 
    The fully anisotropic cases, displays the vanishing of the eigenvalue along only one direction $\bn^1_{\text{E}}$.
    Note that for the rhombic lattice, the relative orientation of $\bn_{\text{E}}$ with respect to $\bg_{\text{E}}$ shows that the mode of localization is neither a pure shear nor a pure expansion wave, but a mixing of the two.
    Conversely, the square lattice always reaches loss of ellipticity through the formation of bands of pure shear strain.
    }
\end{figure}

The objective is now to characterize \textit{failure of (E)} by studying the eigenvalues of~\eqref{eq:acoustic_tensor_grid} as functions of the preload state applied to the grillage. 
To this end, solutions are sought for the following loss of ellipticity condition
\begin{equation}
    \label{eq:ellipticity_loss}
    c_1^2(\bn,\bp)\, c_2^2(\bn,\bp) = 0 \,, 
\end{equation}
where $\bn$ is the usual unit vector defining the direction of propagation and $\bp=\{p_1,p_2\}$ is a vector simply collecting the preload parameters. 
In Eq.~\eqref{eq:ellipticity_loss} the dependence on the geometric parameters $\{\alpha,\Lambda_1,\Lambda_2\}$ is omitted for brevity and moreover, without loss of generality, it is assumed that $c_1^2\leq c_2^2$.
For every solution $\{\bn_{\text{E}},\bp_{\text{E}}\}$ of Eq.~\eqref{eq:ellipticity_loss}, the eigenvector $\bg_{\text{E}}=\bg(\bn_{\text{E}})$ associated to the vanishing eigenvalue can be computed.
Vectors $\bn_{\text{E}}$ and $\bg_{\text{E}}$ will be respectively referred as the \textit{direction} (more precisely, the normal to) and deformation \textit{mode} of the strain localization band.

It can be directly verified that in the absence of preload, $\bp\to\bzero$, the considered grid has $c_1^2>0$ and $c_2^2>0$, so that (SE) holds, except in the case of an extreme material, where the stiffness of the the rods becomes vanishing small, as analyzed for instance in~\cite{gourgiotis_2016}.

Due to the symmetry of $\bA^{(\fC)}(\bn)$, it follows that, starting from the unloaded state $\bp=\bzero$ with rods of finite stiffness and continuously varying the prestress, the material remains both (SE) and (E) until both conditions simultaneously fail. 
Therefore, solutions of Eq.~\eqref{eq:ellipticity_loss} are sought as pairs $\{\bn_{\text{E}},\bp_{\text{E}}\}$ such that $\bp_{\text{E}}$ represents the terminal point of a path starting at $\bp=\bzero$ and entirely contained in the (SE) domain;
in other words, $\bp_{\text{E}}$ is on both boundaries of (SE) and (E).
The set of these points $\bp_{\text{E}}$ is referred as the \textit{elliptic boundary}.

In order to explore loss of ellipticity for lattice configurations characterized by different symmetry classes, a square and a rhombic grid are considered, respectively with $\alpha=\pi/2$ and $\alpha=\pi/3$.
For both examples, the slenderness $\Lambda_1=\Lambda_2=10$, and $\Lambda_1=7$, $\Lambda_2=15$ are selected. 

In Table~\ref{tab:cases_analyzed}, for each geometry considered, the first solution to Eq.~\eqref{eq:ellipticity_loss} for equal prestress components $p_1=p_2$ is reported, together with the associated directions of localization, denoted as $\bn_{\text{E}}=\cos(\theta_{\text{cr}})\be_1+\sin(\theta_{\text{cr}})\be_2$.
Note that the symmetry class is referred here to the unloaded configuration, so that the symmetry of incremental response may change as an effect of loading.
With the assumed values for grid angle $\alpha$ and slenderness, the cubic, orthotropic, and fully anisotropic cases (10 constants for planar elasticity) can be investigated. 

In order to better visualize the direction $\bn_{\text{E}}$ and the associated mode $\bg_{\text{E}}$, 
a polar plot of the square root of the lowest eigenvalue $c_1(\bn,\bp)$ is reported 
 in Fig.~\ref{fig:eigenvalue_square_rhombus}, 
for the cases listed in Table~\ref{tab:cases_analyzed}, at two levels of preload, namely $0.5\,\bp_{\text{E}}$ (dashed gray line) and $\bp_{\text{E}}$ (continuous blue line).
In Fig.~\ref{fig:eigenvalue_square_10_10} the square lattice with $\Lambda_1=\Lambda_2=10$ is subject to an isotropic prestress in the two directions, $p_1=p_2$, and therefore the cubic symmetry is maintained in the prestressed state.
Owing to this symmetry, ellipticity is lost along two orthogonal directions $\bn^1_{\text{E}}$ and $\bn^2_{\text{E}}$. 
Moreover, the associated wave amplitudes $\bg^1_{\text{E}}$ and $\bg^2_{\text{E}}$ are perpendicular to the vectors $\bn^1_{\text{E}}$ and $\bn^2_{\text{E}}$ respectively, hence indicating that the modes of localization are pure shear waves, the so-called \textit{shear bands}. 

For the orthotropic square lattice ($\Lambda_1=7$, $\Lambda_2=15$), the polar plot is given in Fig.~\ref{fig:eigenvalue_square_7_15}. 
In this case, owing to the orthotropy, waves propagating along the horizontal and vertical direction possess different velocities and therefore ellipticity is lost when the smallest of these velocities vanishes, leading to a single shear band (in this case with a normal $\bn^1_{\text{E}}$ aligned parallel to the horizontal direction).

Quite remarkably, \textit{the shear wave responsible for the ellipticity loss is the one propagating along the direction of the `stiffer' elastic link} (having the  lowest slenderness), \textit{while intuitively a `shear mechanism' would be expected in the direction of the `soft' elastic link}.
This effect will be confirmed and explained further with the computation of the forced response in Section~\ref{sec:forced_response}.
Moreover, it is worth noting that the shear band directions for the square lattice, both cubic and orthotropic, are aligned parallel to the directions of the rods forming the lattice.

For the rhombic lattice with $\Lambda_1=\Lambda_2=10$ and isotropic preload, shown in Fig.~\ref{fig:eigenvalue_rhombus_10_10}, the mechanical behaviour is orthotropic and therefore two directions of localization are obtained.
The associated wave amplitudes $\bg^1_{\text{E}}$ and $\bg^2_{\text{E}}$ both have respectively a component orthogonal and parallel  to the vector $\bn^1_{\text{E}}$ and $\bn^2_{\text{E}}$, so that a `mixture' of shear and compression waves is involved. 

The fully anisotropic version for the rhombic lattice (Fig.~\ref{fig:eigenvalue_rhombus_7_15}) can be obtained by changing the slenderness values ($\Lambda_1=7$, $\Lambda_2=15$), so that one of the two localizations is suppressed, while the other is preserved. 
It is also worth noting that, in contrast to the square case, the directions of localization for the rhombic lattice are not perfectly aligned parallel to the rods' normal, instead, they result slightly inclined, as will be confirmed by the computations reported in Section~\ref{sec:forced_response}.

\subsection{The effect of the prestress directionality on the ellipticity loss and localization directions}
\label{sec:ellipticity_domains}
In the previous section, the loss of ellipticity has been determined for the four lattice's configurations reported Table~\ref{tab:cases_analyzed}, under the assumption of isotropic preload $p_1=p_2$.
\begin{figure}[htb!]
    \centering
    \begin{subfigure}{0.45\textwidth}
        \centering
        \caption{\label{fig:ellipticity_domains_10_10}$\Lambda_1=\Lambda_2=10$}
        \includegraphics[width=\linewidth]{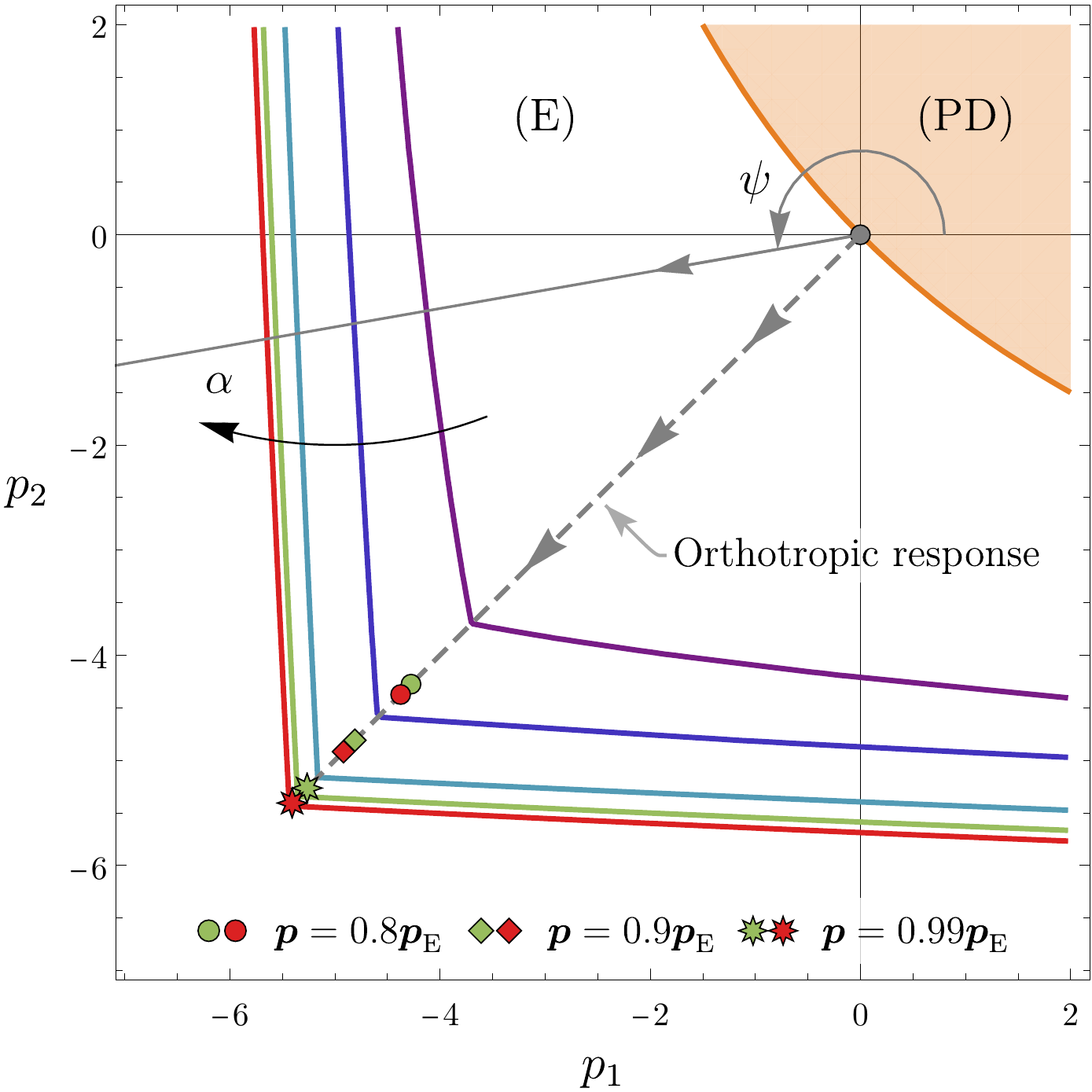}
    \end{subfigure} \hspace{2mm}
    \begin{subfigure}{0.45\textwidth}
        \centering
        \caption{\label{fig:ellipticity_domains_7_15}$\Lambda_1=7,\, \Lambda_2=15$}
        \includegraphics[width=\linewidth]{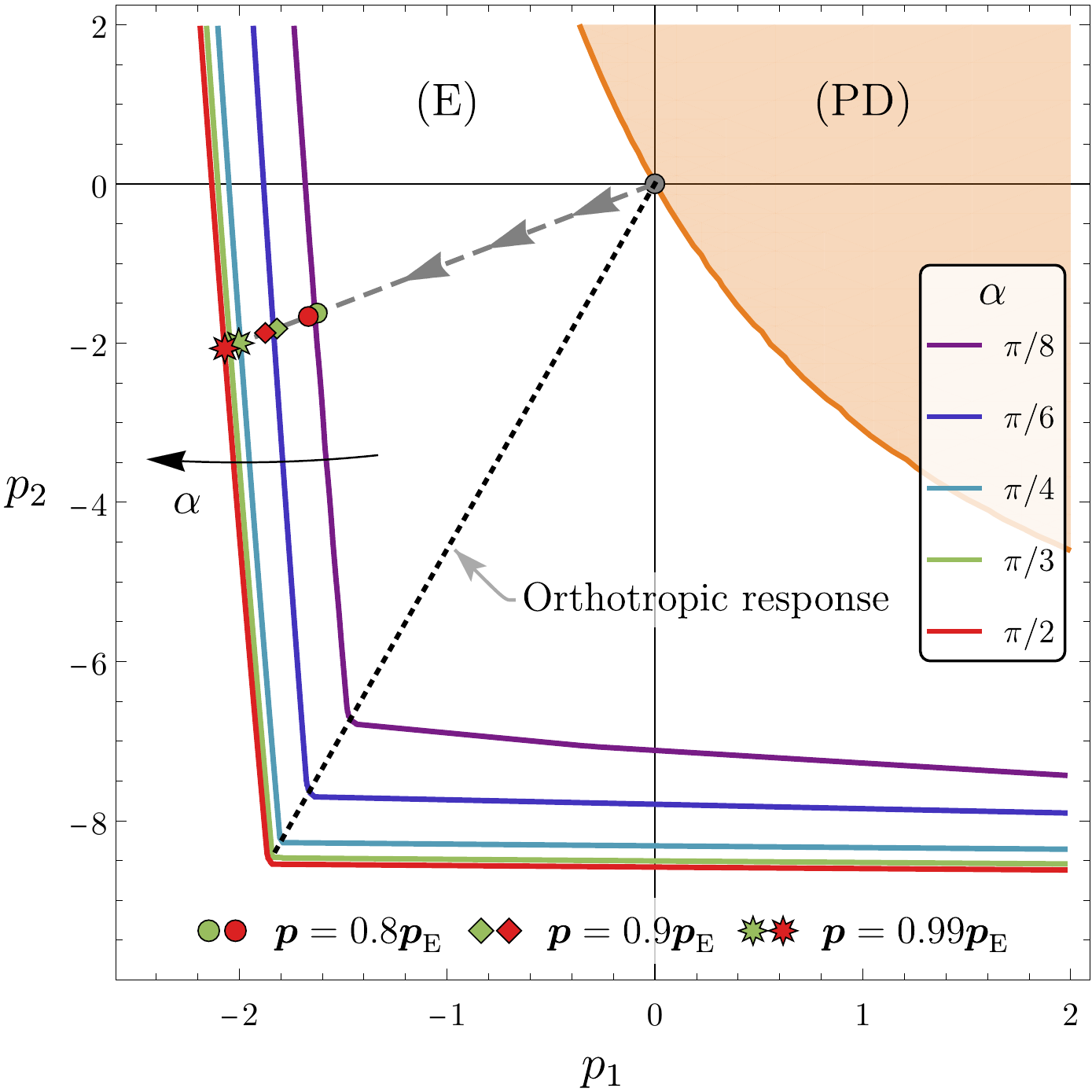}
    \end{subfigure}\\ \vspace{2mm}
    \begin{subfigure}{0.47\textwidth}
        \centering
        \caption{\label{fig:angle_n_10_10}$\Lambda_1=\Lambda_2=10$}
        \includegraphics[width=\linewidth]{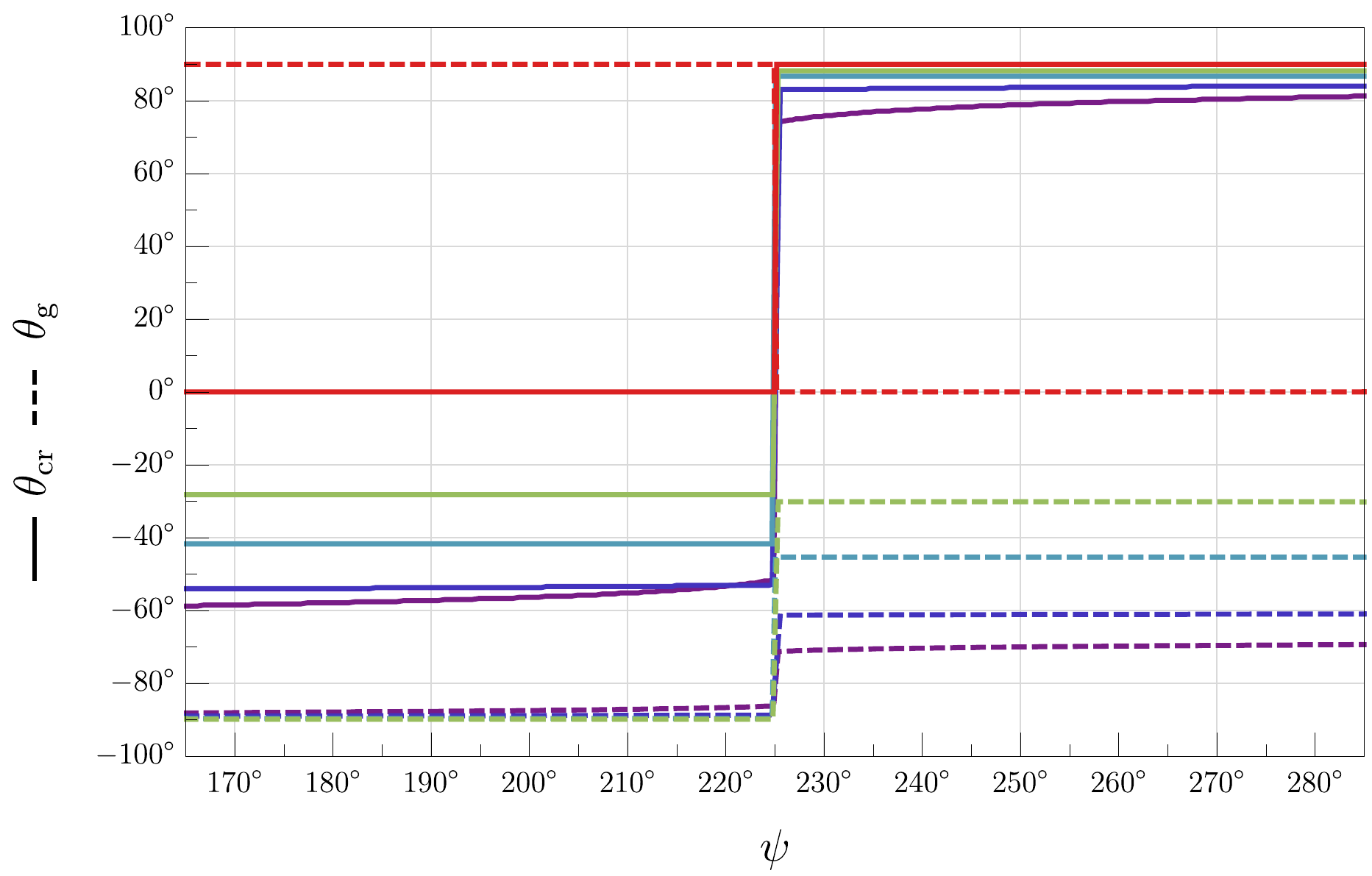}
    \end{subfigure} %
    \begin{subfigure}{0.47\textwidth}
        \centering
        \caption{\label{fig:angle_n_7_15}$\Lambda_1=7,\, \Lambda_2=15$}
        \includegraphics[width=\linewidth]{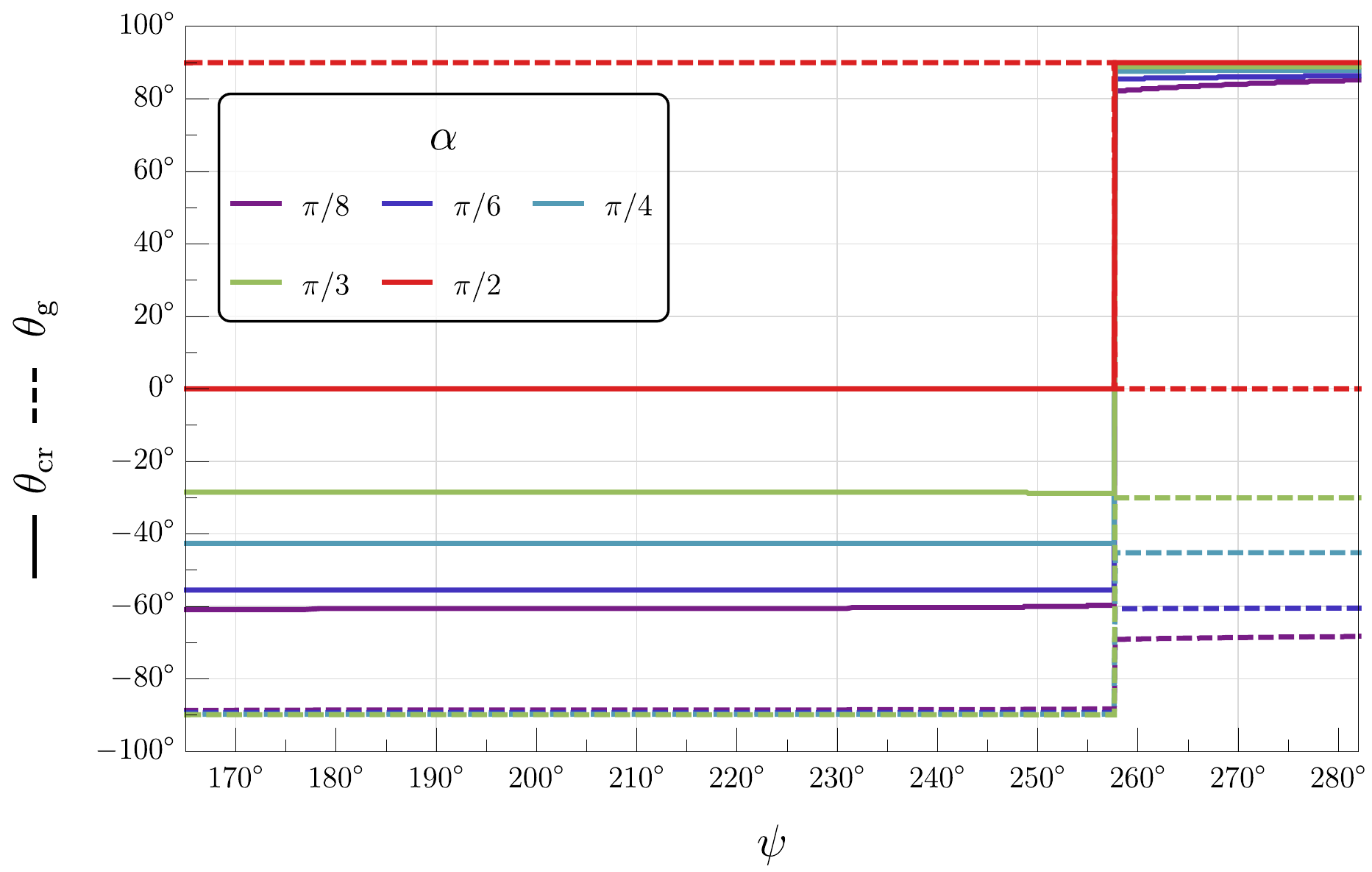}
    \end{subfigure}%
    \caption{\label{fig:ellipticity_domains}
    Ellipticity domains for a grillage of preloaded elastic rods, for different values of rods' angle $\alpha$ and assuming ~(\subref{fig:ellipticity_domains_10_10}) orthotropy ($\Lambda_1=\Lambda_2=10$) or~(\subref{fig:ellipticity_domains_7_15}) full anisotropy ($\Lambda_1=7,\,\Lambda_2=15$).
    The region of positive definiteness of the elasticity tensor $\fC$ is also reported (orange region).
    The loading path (gray dashed arrow) refers to condition of isotropic preload $p_1=p_2$ on which the markers (in red for the case $\alpha = \pi/2$ and green for $\alpha=\pi/3$) identify the preload levels used in the computation of the forced response (Section~\ref{sec:forced_response}).
    (\subref{fig:angle_n_10_10})~and~(\subref{fig:angle_n_7_15}) report the relationship between the direction in the preload space (defined by the angle $\psi=\arctan(p_1,p_2)$, see~(\subref{fig:ellipticity_domains_10_10})) and the dyad $\bn_{\text{E}}\otimes\bg_{\text{E}}$ defined by the angles $\theta_{\text{cr}}$ and $\theta_{\text{g}}$ singling out, respectively, the direction of ellipticity loss and of the localization mode (i.e. $\bn_{\text{E}}=\cos\theta_{\text{cr}}\be_1+\sin\theta_{\text{cr}}\be_2$ and$\bg_{\text{E}}=\cos\theta_{\text{g}}\be_1+\sin\theta_{\text{g}}\be_2$).
    Notice that the case of localization along two directions (corresponding to the jumps visible in the graphs of part~(\subref{fig:angle_n_10_10})~and~(\subref{fig:angle_n_7_15})) only occurs for a prestress state at the corner point of the elliptic boundary, which corresponds to an incrementally orthotropic material behaviour.
    }
\end{figure}
This hypothesis is now removed to explore the influence of the `directionality' of the preload state on the ellipticity loss, in terms of number and direction of strain localizations.
The analysis is performed by computing the full set of solutions $\{\bn_{\text{E}},\bp_{\text{E}}\}$ of Eq.~\eqref{eq:ellipticity_loss}, to determine the elliptic boundary in the $\{p_1,p_2\}$-space, for several values of rods' inclination $\alpha$ and slenderness $\lambda_{1,2}$.

Fig.~\ref{fig:ellipticity_domains_10_10} and~\ref{fig:ellipticity_domains_7_15} show the ellipticity domains computed, respectively, for the orthotropic ($\Lambda_1=\Lambda_2=10$) and the anisotropic grid ($\Lambda_1=7,\,\Lambda_2=15$).
The following observations can be drawn.
\begin{enumerate}[label=(\roman*)]
    \item The elliptic region is unbounded for tensile (positive) preload and bounded when at least one of the preloads is compressive (negative); this is an expected feature, as the contribution of a tensile preload to the potential energy, Eq.~\eqref{eq:prestress_energy}, is positive definite; 
    \item any deviation of the angle $\alpha$ from $\pi/2$ reduces the size of the elliptic region;
    \item in the orthotropic case, the domain is symmetric with respect to the bisector $p_1=p_2$ (Fig.~\ref{fig:ellipticity_domains_10_10});
    \item the elliptic boundary appears to be smooth everywhere except at a corner point;
    \item the anisotropy induced by different values of slenderness causes the corner to move; the elliptic region is reduced in size along the direction of the smallest  slenderness.
\end{enumerate}

In the same figure, the region of \textit{positive definiteness} (PD) of the elastic tensor, $\bX\scalp\fC[\bX]>0\,\,\forall\bX\neq\bzero$, is also reported.
Surprisingly, the (PD) region is found to be independent of the rods' angle $\alpha$.

The directions of localization in terms of critical angle $\theta_{\text{cr}}$ are reported in Figs.~\ref{fig:angle_n_10_10} and~\ref{fig:angle_n_7_15} for different values of the lattice inclination $\alpha$, with the purpose of quantifying the influence of the loading path (parameterized by means of the angle $\psi = \arctan(p_1,p_2)$).
These plots show that only one localization occurs, except for a loading direction corresponding to a corner point on the elliptic boundary, where two simultaneous localizations occur.
Even though this is expected for the orthotropic case (due to symmetry), the anisotropic grid can also exhibit a sort of `orthotropic' response (displaying two localizations) when a proper value of preload $p_1\neq p_2$ is applied (along the dashed black line of Fig.~\ref{fig:ellipticity_domains_7_15}).
Furthermore, the angle between the vectors $\bn^{1,2}_{\text{E}}$ and the horizontal axis (Fig.~\ref{fig:angle_n_10_10} and~\ref{fig:angle_n_7_15}), which identifies the direction of the localization bands, depends weakly on the prestress (except at the corner point of the elliptic boundary), namely, the directions of $\bn^{1,2}_{\text{E}}$ follow the orientations of the grid's ligaments, even if they result slightly misaligned.

\section{Time-harmonic forced response near the elliptic boundary}
\label{sec:forced_response}
The analysis of the homogenized continuum, equivalent to a preloaded grid of elastic rods (presented in Section~\ref{sec:ellipticity_loss_grid}) predicts that the incremental response can display strain localizations due to prestress-induced loss of ellipticity.
In this section, the homogenization is validated through comparisons between the actual low-frequency forced response of the lattice (simulated numerically with a finite element technique) and the time-harmonic Green's function (which can be found in~\cite{piccolroaz_2006}) for the equivalent continuum, both computed at increasing levels of the preload, so that  the elliptic boundary is approached.
The comparison is performed for the four geometric configurations reported in Table~\ref{tab:cases_analyzed} and for four prestress levels, namely $\{0,0.8,0.9,0.99\}\,\bp_{\text{E}}$ (with $\bp_{\text{E}}$ being the prestress state leading to ellipticity loss), which are marked on the loading path of Fig.~\ref{fig:ellipticity_domains_10_10} and~\ref{fig:ellipticity_domains_7_15}.

The lattice response is numerically analyzed using the COMSOL Multiphysics\textsuperscript{\circledR} finite elements program in the frequency response mode.
A square finite-size computational window with a width of 350 unit cells (of dimension $350 l_1$, with $l_1$ denoting the cell edge) is considered, with a perfectly matched layer (PML) along the boundaries, so that here waves are not reflected, rather absorbed, and the response of an infinite body is simulated. 
\begin{figure}[htb!]
\centering
\begin{subfigure}{0.265\textwidth}
\centering
\includegraphics[height=40mm]{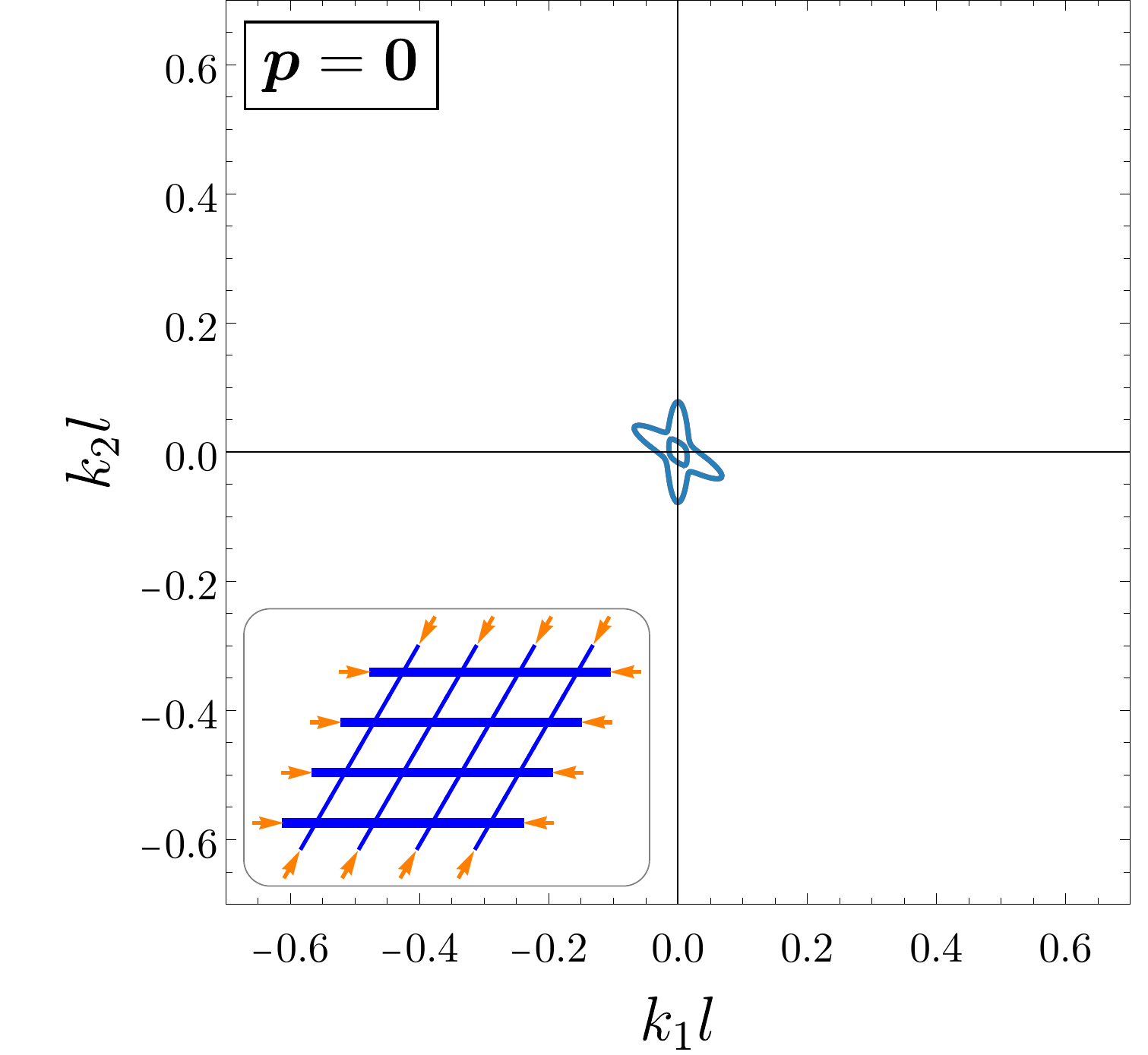}
\end{subfigure}%
\begin{subfigure}{0.245\textwidth}
\centering
\includegraphics[height=40mm]{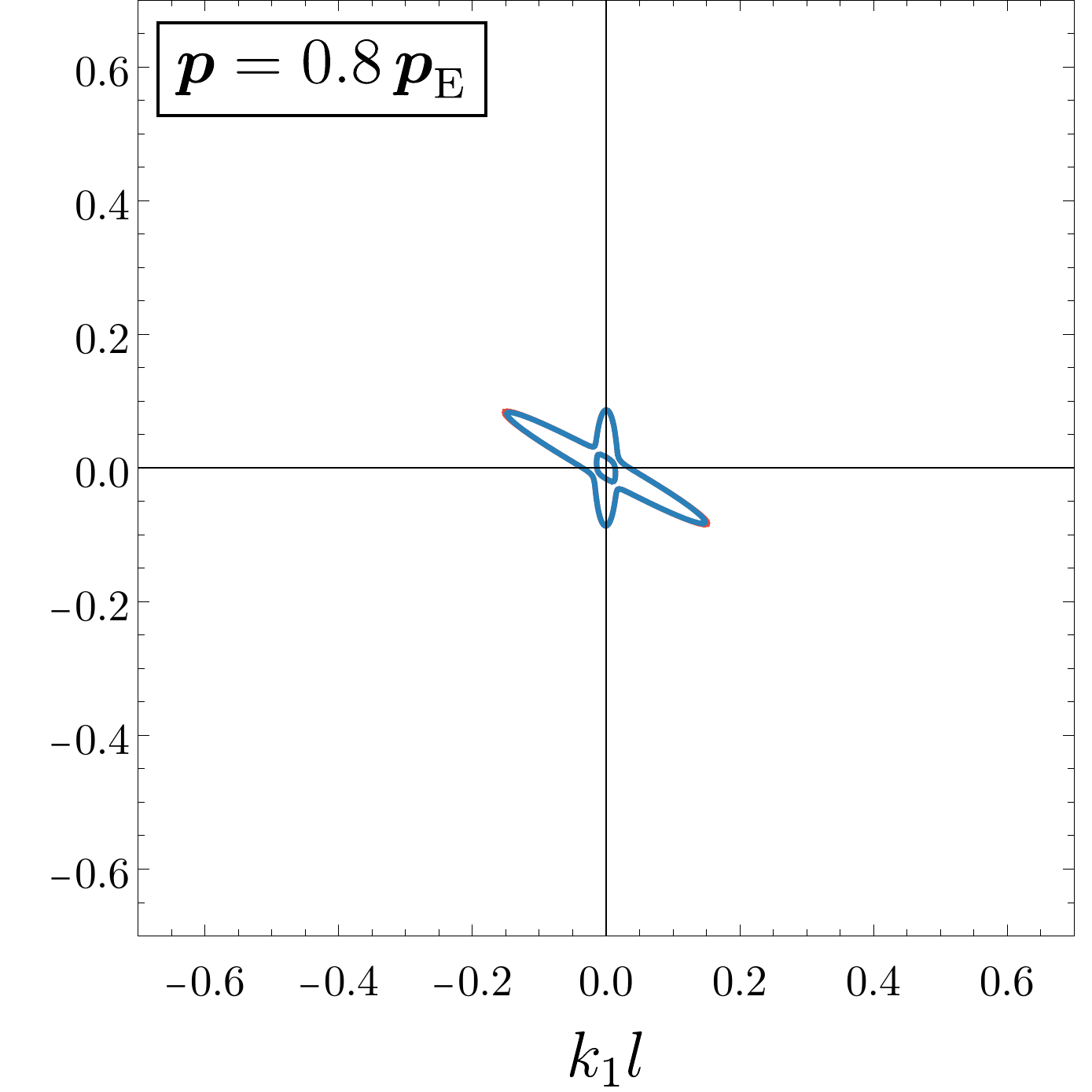}
\end{subfigure}%
\begin{subfigure}{0.245\textwidth}
\centering
\includegraphics[height=40mm]{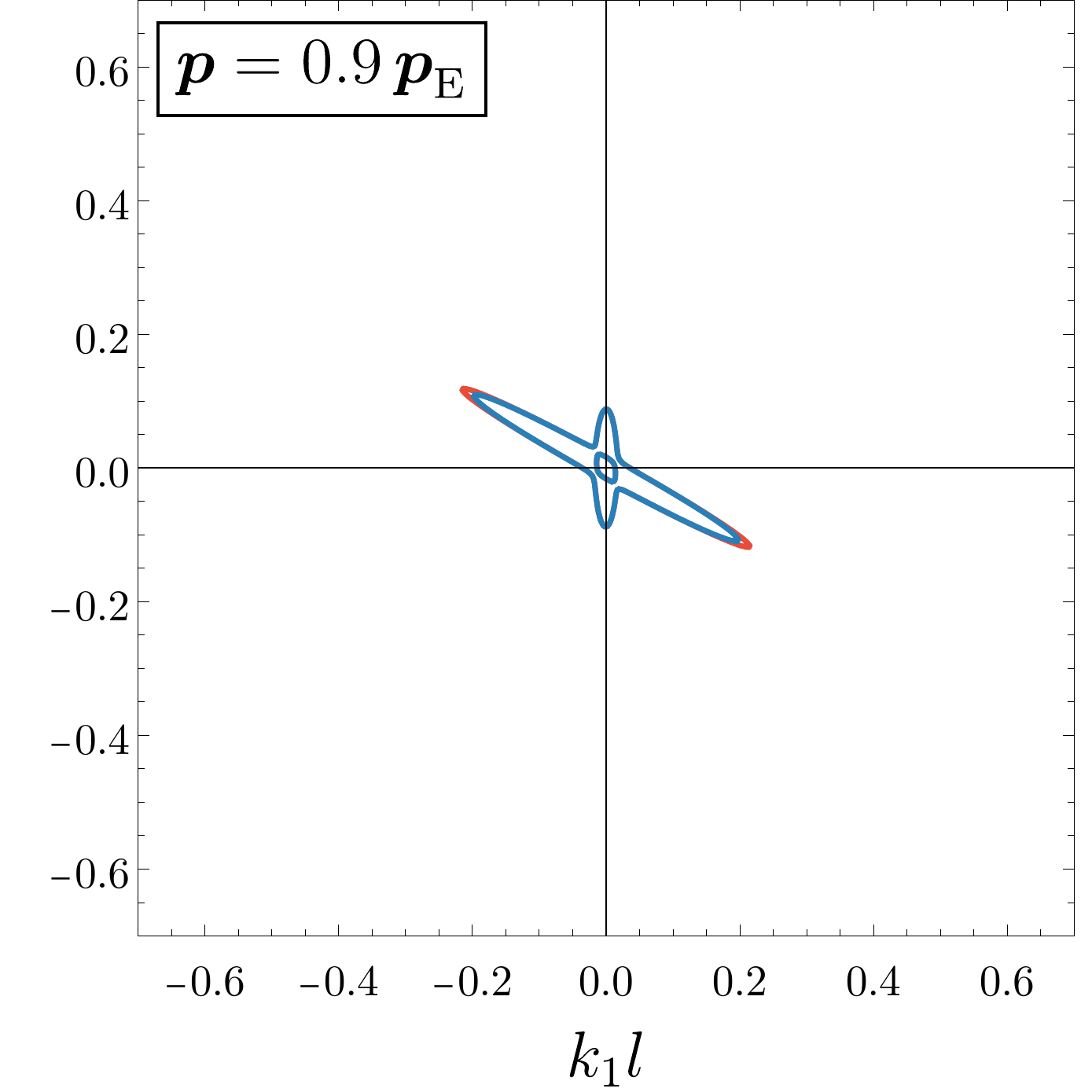}
\end{subfigure}%
\begin{subfigure}{0.245\textwidth}
\centering
\includegraphics[height=40mm]{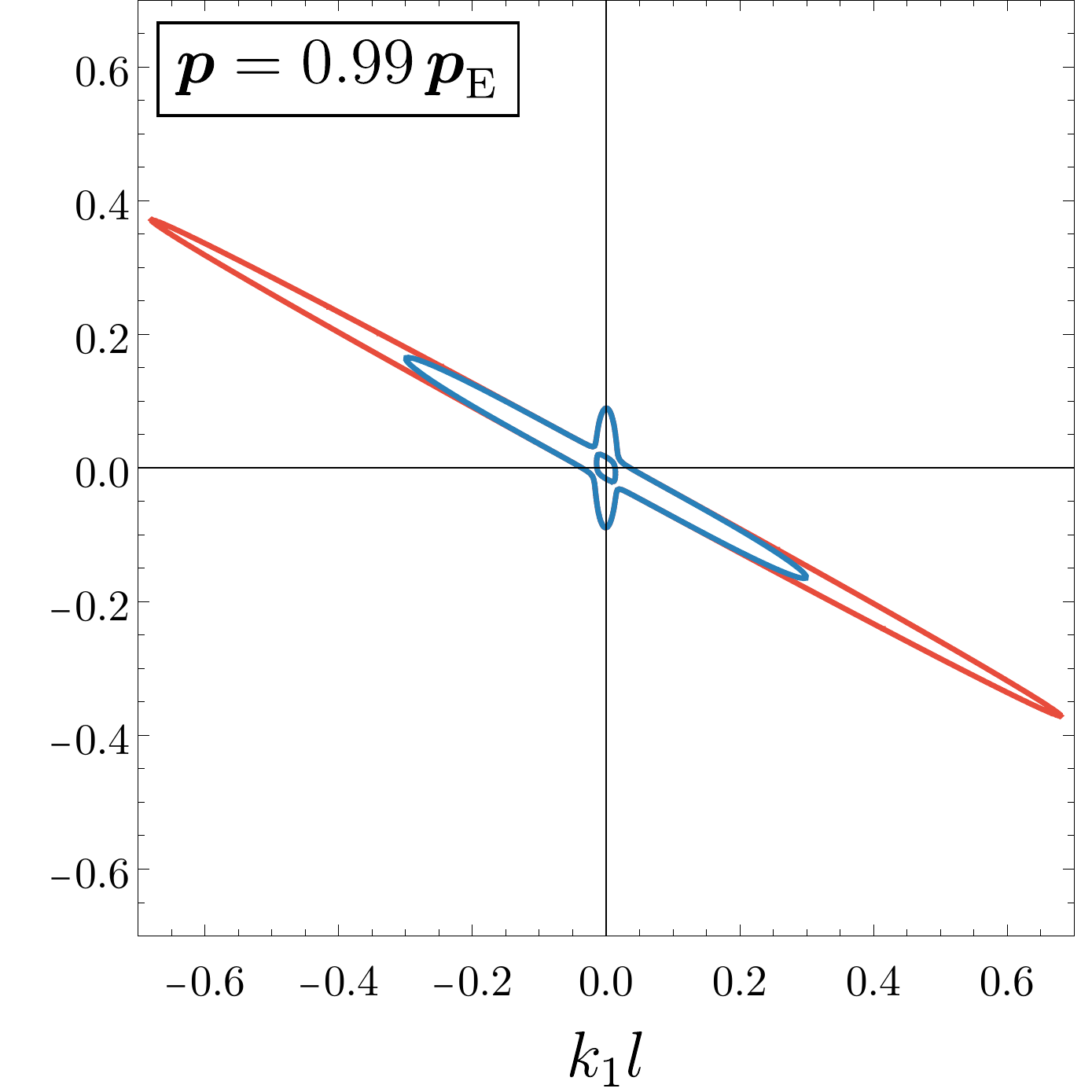}
\end{subfigure}\\ \vspace{2mm}
\begin{subfigure}{0.265\textwidth}
\centering
\includegraphics[height=38mm]{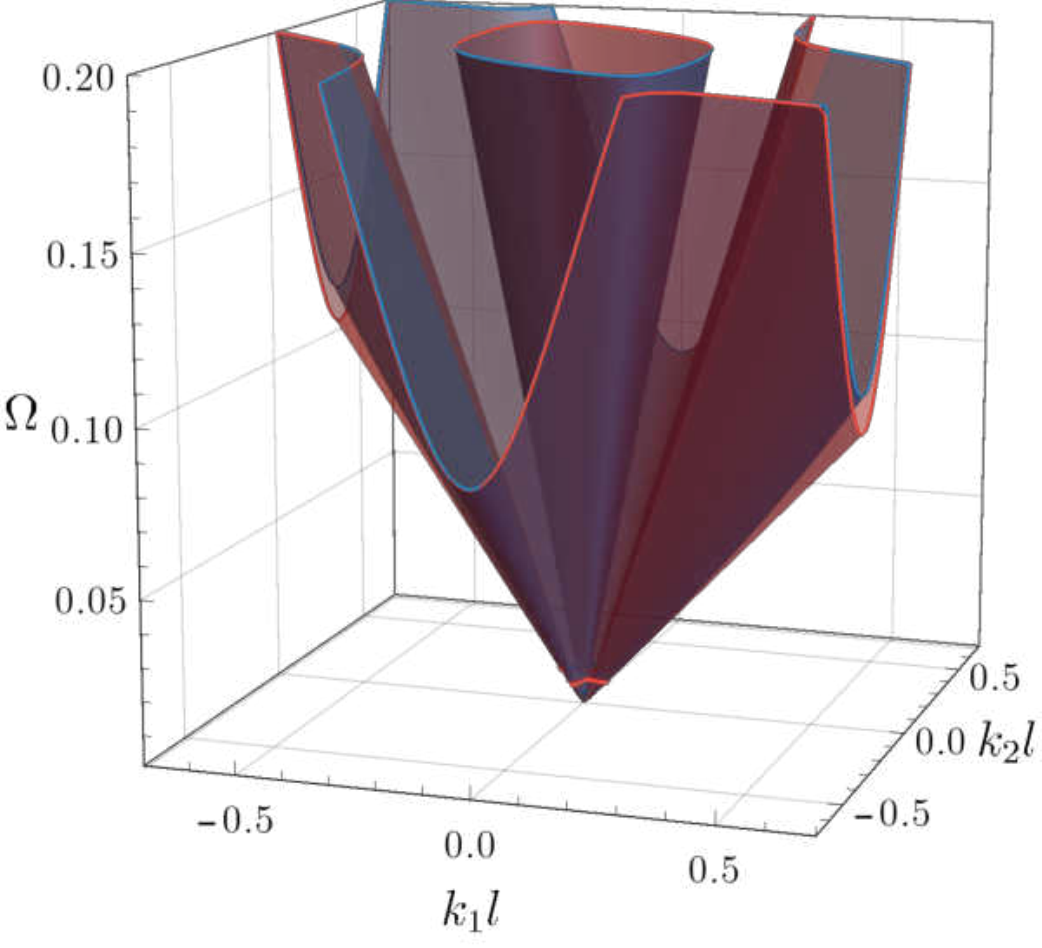}
\end{subfigure}%
\begin{subfigure}{0.245\textwidth}
\centering
\includegraphics[height=38mm]{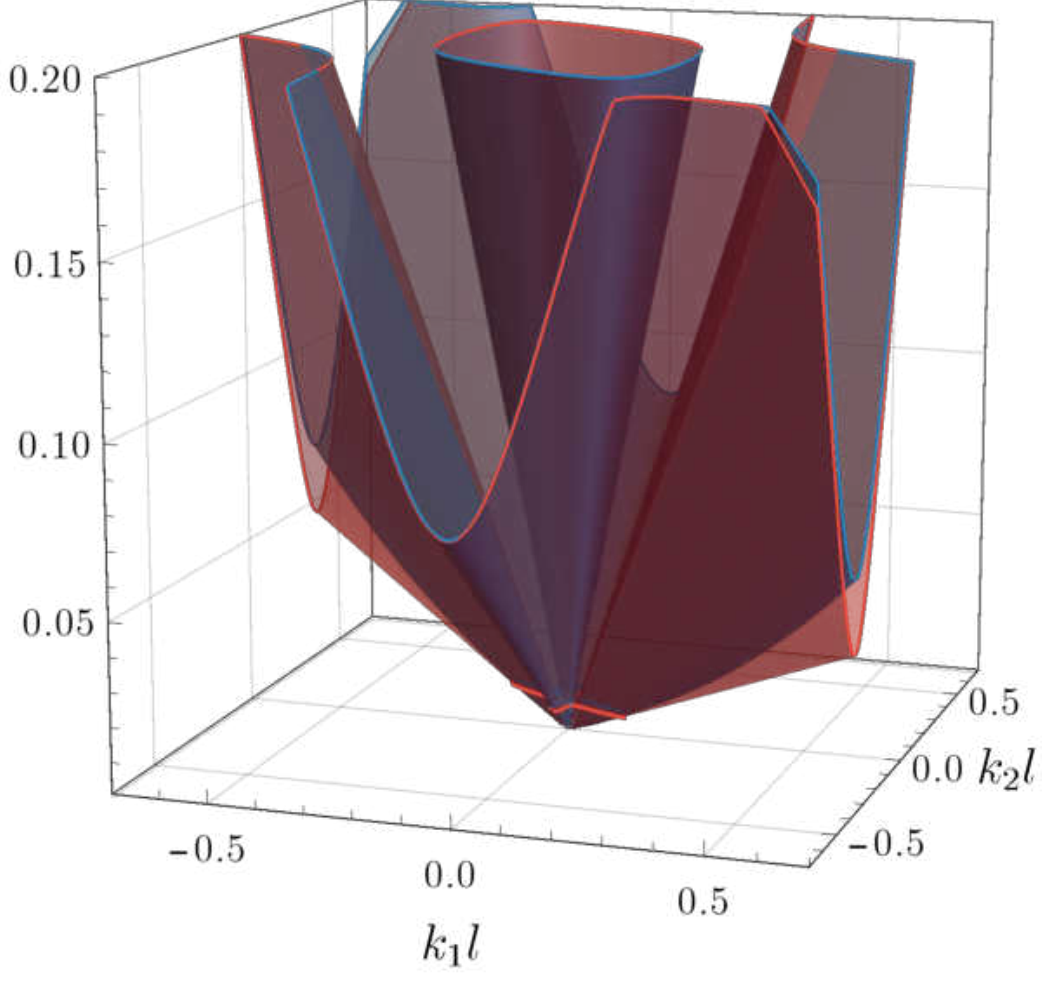}
\end{subfigure}%
\begin{subfigure}{0.245\textwidth}
\centering
\includegraphics[height=38mm]{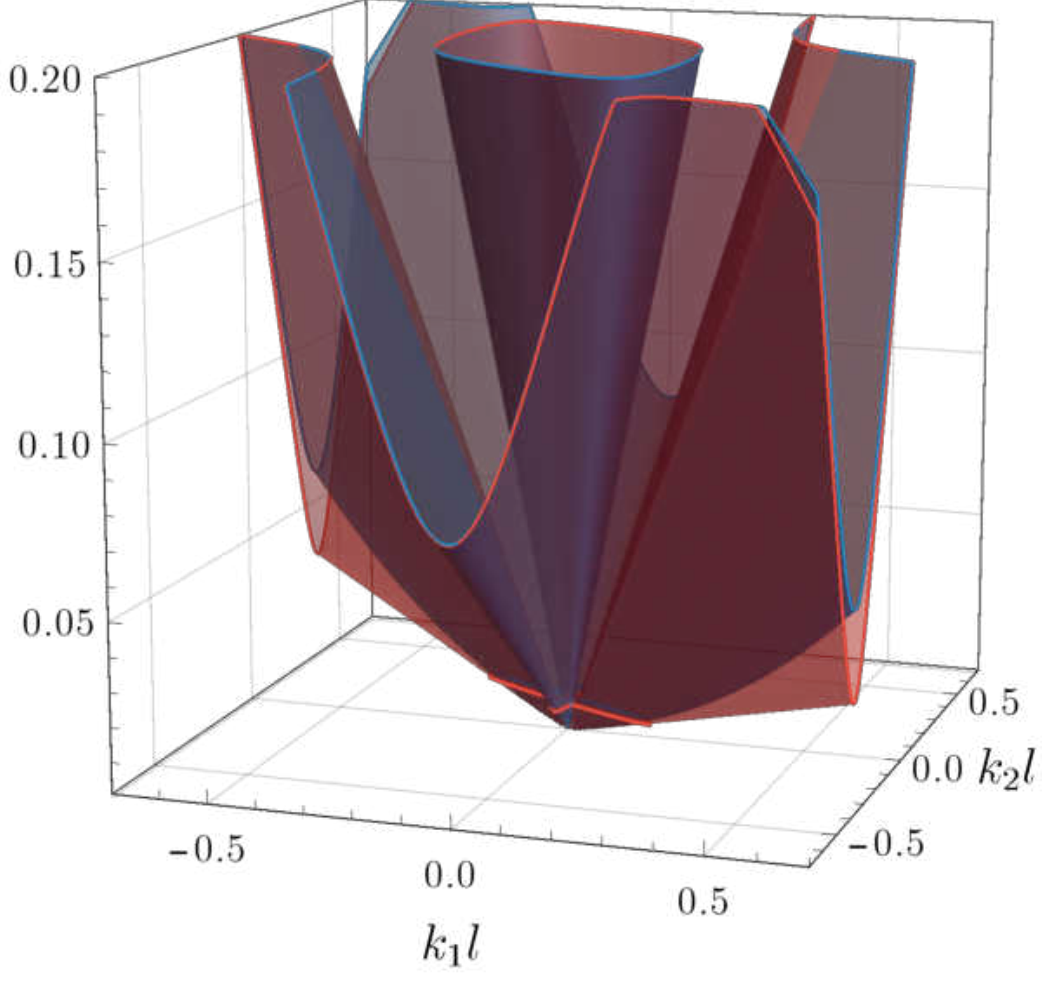}
\end{subfigure}%
\begin{subfigure}{0.245\textwidth}
\centering
\includegraphics[height=38mm]{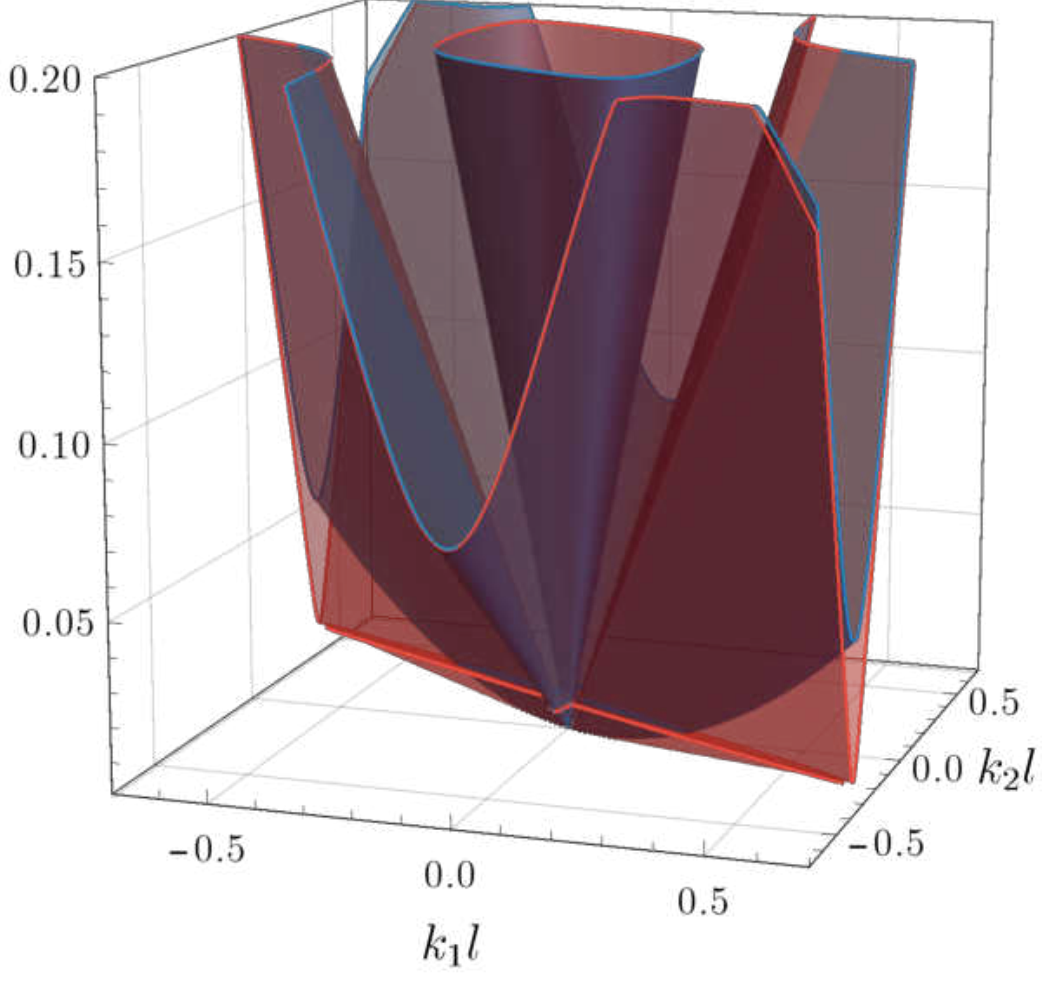}
\end{subfigure}%
\caption{\label{fig:contours_rhombus_7_15}
    Slowness contours and dispersion surfaces for a rhombic anisotropic lattice with $\Lambda_1=7,\,\Lambda_2=15$ (in blue) and for the effective elastic continuum 
    (in red) at frequency $\Omega=0.01$.
    The evolution of the contours and dispersion surfaces induced by a compressive preload of equal components $p_1=p_2$ along two inclined directions (the considered loading path is depicted in Fig.~\ref{fig:ellipticity_domains_7_15}) demonstrates that the nonlinear dispersion of the lattice is negligible, except when the material is very close to the elliptic boundary, namely, for a prestress above $0.9\,\bp_{\text{E}}$. The comparison between the behaviours of the lattice and its equivalent continuum shows a great agreement. 
    }
\end{figure}

A pulsating concentrated force, in-plane applied, is considered acting at the center of the computational domain.
For a given load, the complex displacement field $\bu(\bx)$, with horizontal and vertical components $u_1(x_1,x_2)$ and $u_2(x_1,x_2)$, is computed and the results are plotted in terms of the modulus of the displacement associated to its real part only, $\delta_R(x_1,x_2) = \sqrt{(\Re u_1)^2+(\Re u_2)^2}$ (the plots of the imaginary part of the displacement is omitted for brevity).

In all the following analyses the frequency of the pulsating force is set to be $\Omega=\omega\,l\sqrt{\gamma/A}=0.01$, a low value providing a reasonable match, in term of acoustic properties, between the effective continuum and the lattice. 
In fact, the mismatch between the two is different from zero for any non-vanishing frequency, although becomes zero in the limit $\Omega\to0$. 
When the elliptic boundary is approached, this mismatch is expected to become wider for those waves which propagate parallel to the direction of ellipticity loss.
This is easily explained by the fact that, as the \textit{linear} term in the dispersion relation tends to vanish (in a direction $\bn_{\text{E}}$), the \textit{nonlinear} dispersion of the lattice becomes non-negligible at any \textit{non-vanishing} frequency.

By considering for instance the rhombic anisotropic grid ($\Lambda_1=7,\,\Lambda_2=15$), the deviation between the responses of the lattice (reported in blue in Fig.~\ref{fig:contours_rhombus_7_15}) and its equivalent continuum (reported in red in Fig.~\ref{fig:contours_rhombus_7_15}) can be visualized in terms of slowness contours computed at the frequency $\Omega=0.01$.
By comparing the contours for the four preload states, it can be appreciated that these are superimposed up $0.9\,\bp_{\text{E}}$, so that the nonlinear dispersion of the lattice becomes non-negligible only when the material is very close to the elliptic boundary, namely, at a preload $0.99\,\bp_{\text{E}}$, and only for waves close to the direction of ellipticity loss.
It is also worth noting that when $\bp\approx\bp_{\text{E}}$, the slowness contour pertinent to the lattice (reported in blue) is always contained inside the contour relative to the continuum (reported in red), so that the nonlinear dispersion implies that waves speeds are slightly \textit{higher} for the lattice than for the effective elastic medium.

\subsection{Square lattice}
\label{sec:square_lattice}
Cubic and orthotropic square grids are considered, subject to a pulsating diagonal force (inclined at $45^\circ$ with respect to the rods' axes), with the purpose of revealing  the emergence of strain localizations aligned parallel and orthogonal to the rod axes.
\begin{figure}[htb!]
\centering
\begin{subfigure}{0.24\textwidth}
\centering
\phantomsubcaption{\label{fig:square_10_10_force_d_0}}
\includegraphics[width=0.98\linewidth]{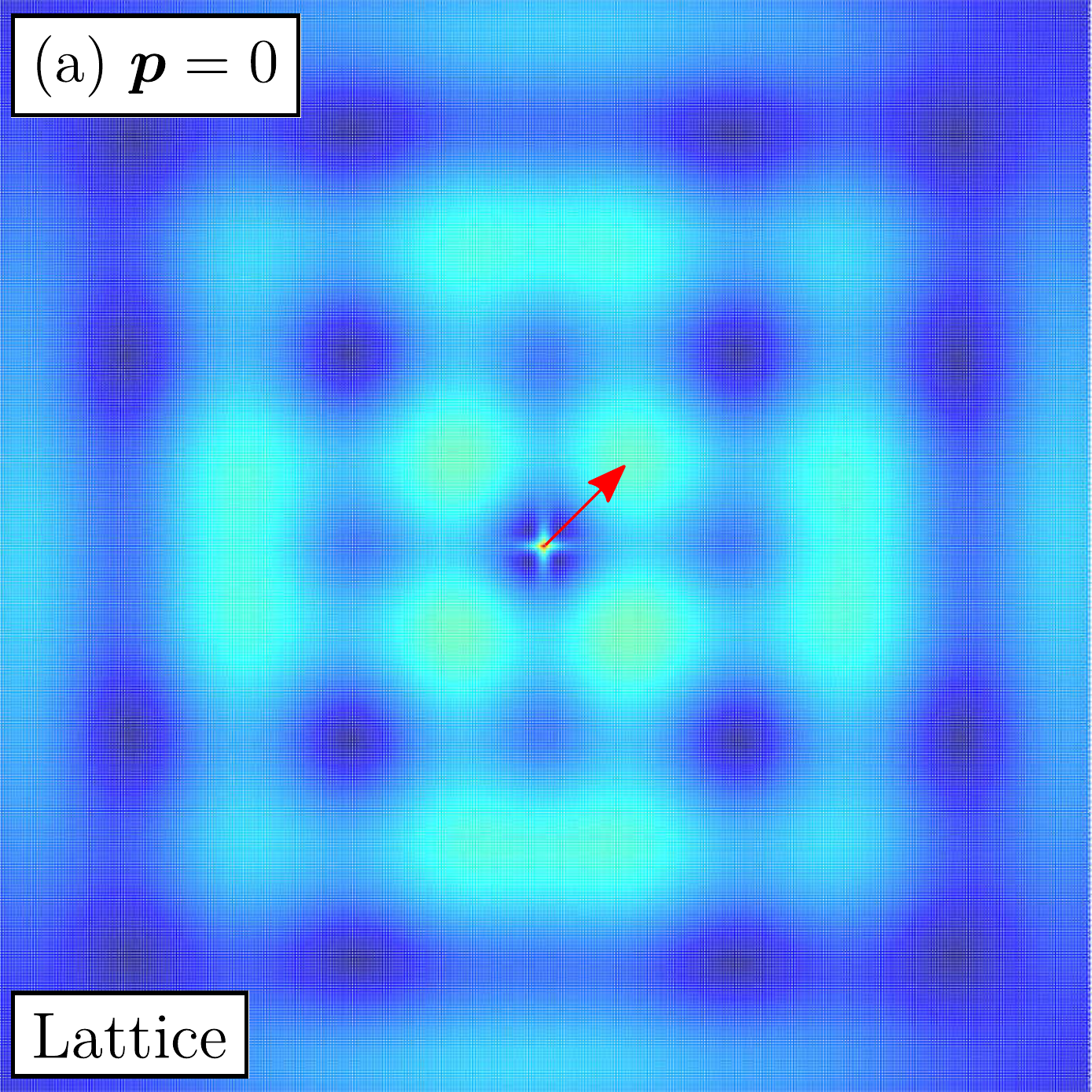}
\end{subfigure}
\begin{subfigure}{0.24\textwidth}
\centering
\phantomsubcaption{\label{fig:square_10_10_force_d_80}}
\includegraphics[width=0.98\linewidth]{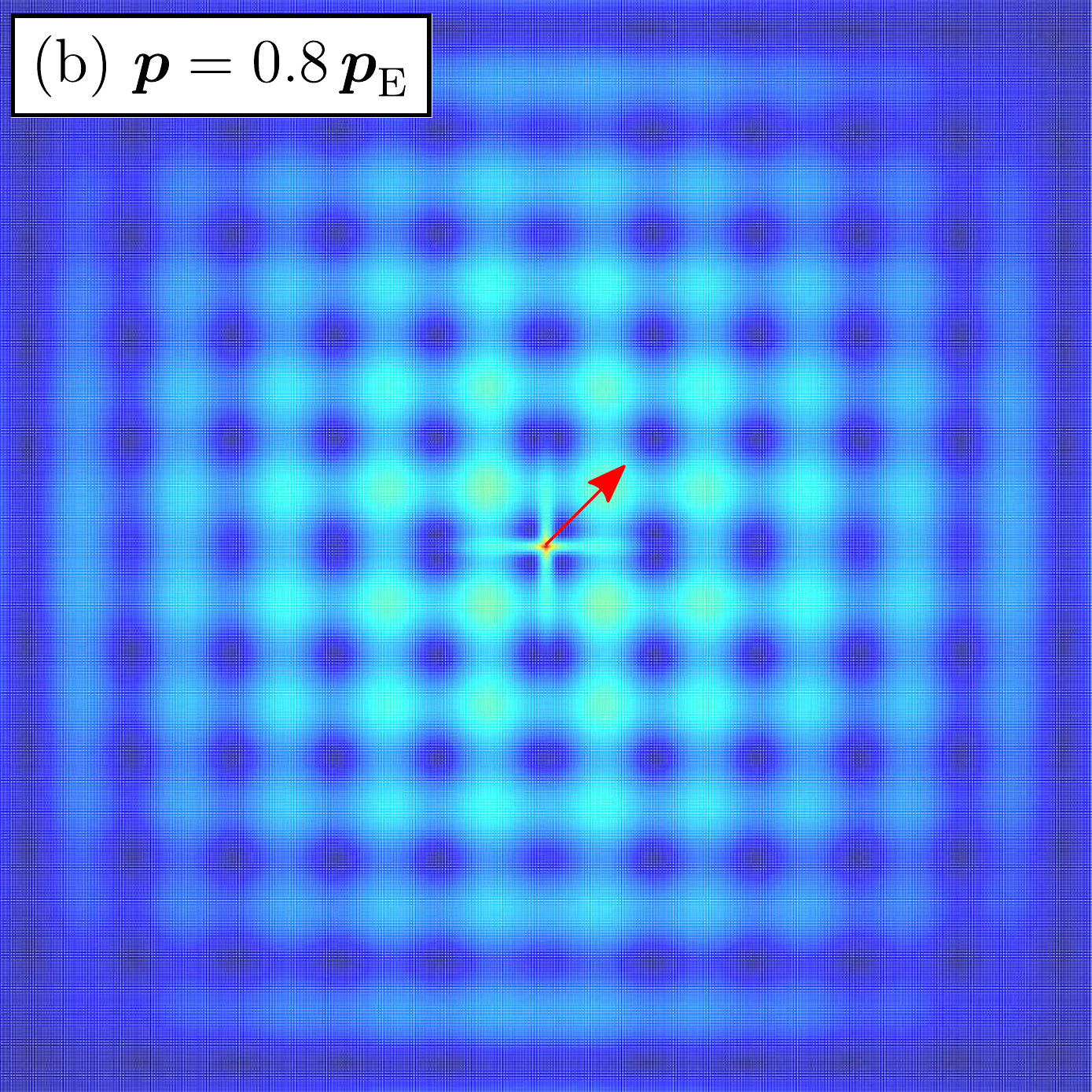}
\end{subfigure}
\begin{subfigure}{0.24\textwidth}
\centering
\phantomsubcaption{\label{fig:square_10_10_force_d_90}}
\includegraphics[width=0.98\linewidth]{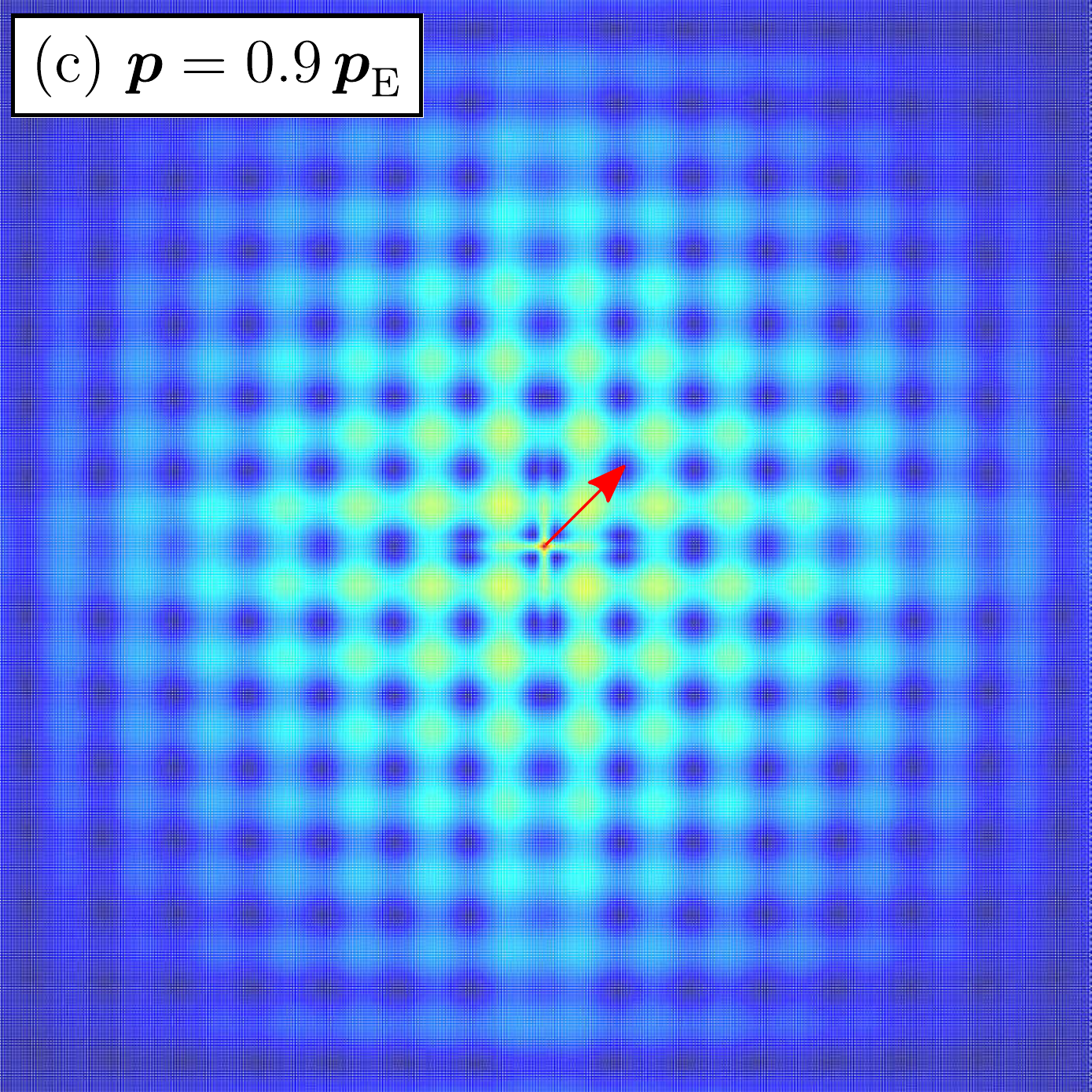}
\end{subfigure}
\begin{subfigure}{0.24\textwidth}
\centering
\phantomsubcaption{\label{fig:square_10_10_force_d_99}}
\includegraphics[width=0.98\linewidth]{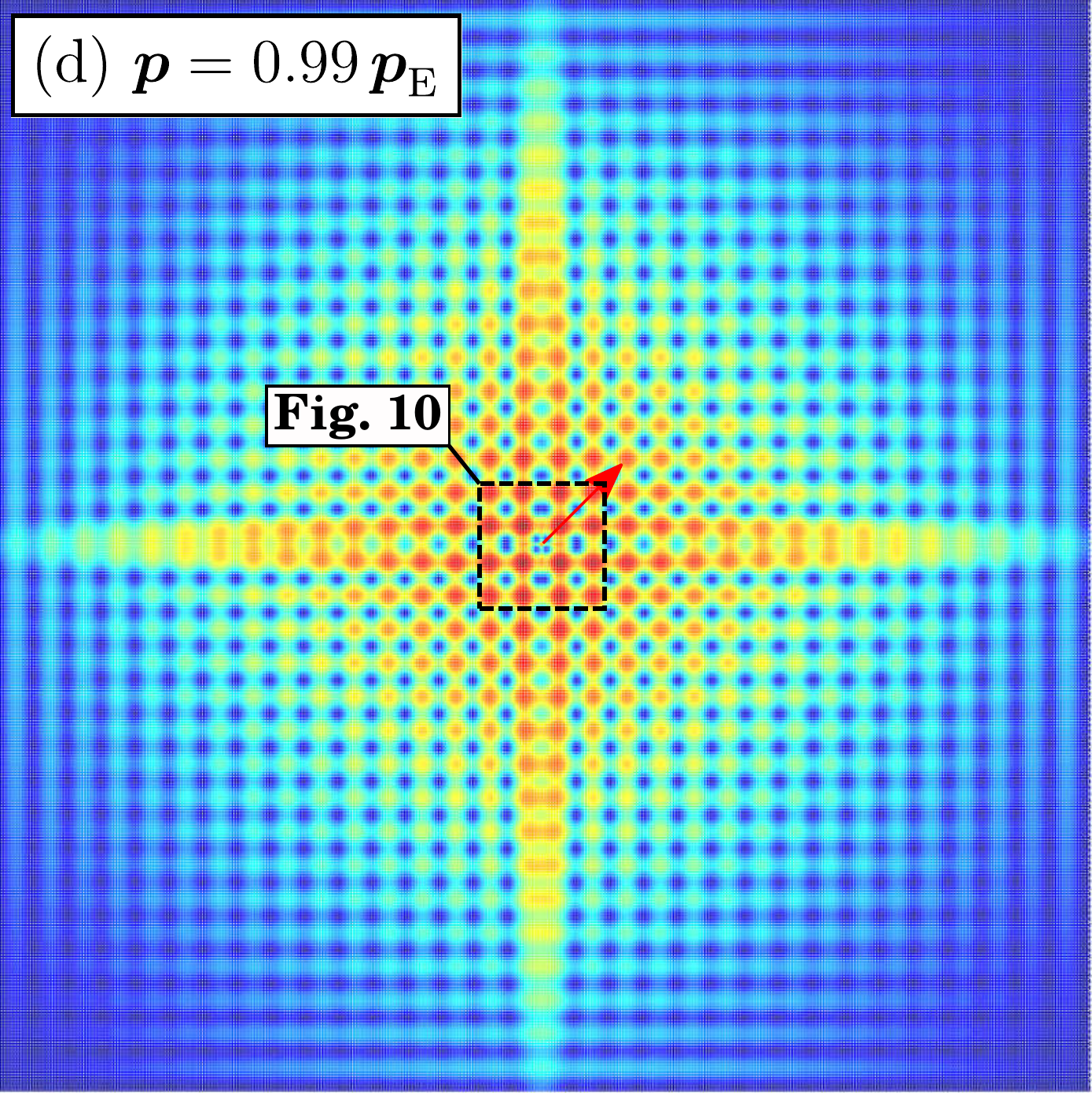}
\end{subfigure}\\
\vspace{0.01\linewidth}
\centering
\begin{subfigure}{0.24\textwidth}
\centering
\phantomsubcaption{\label{fig:square_10_10_force_d_0_gf}}
\includegraphics[width=0.98\linewidth]{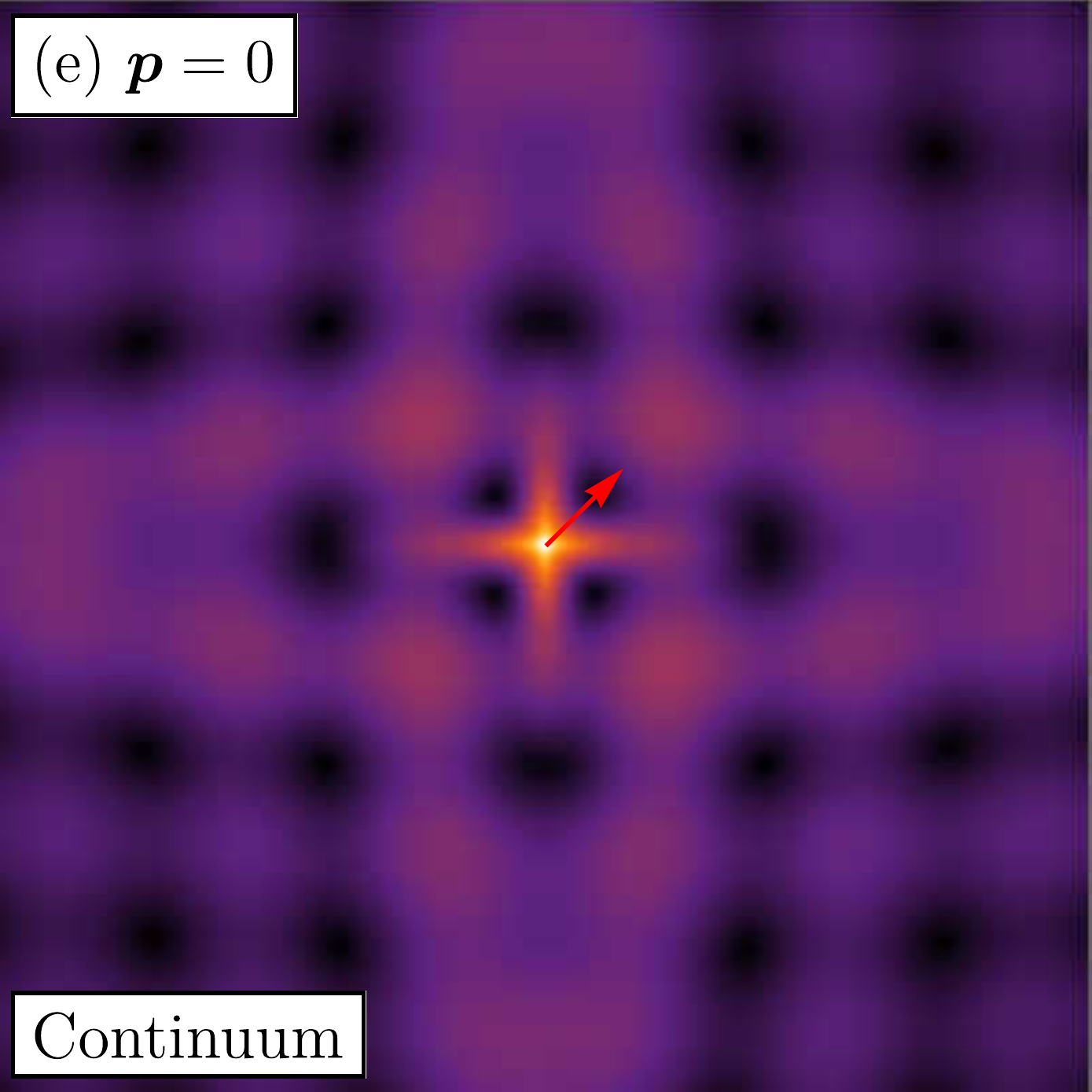}
\end{subfigure}
\begin{subfigure}{0.24\textwidth}
\centering
\phantomsubcaption{\label{fig:square_10_10_force_d_80_gf}}
\includegraphics[width=0.98\linewidth]{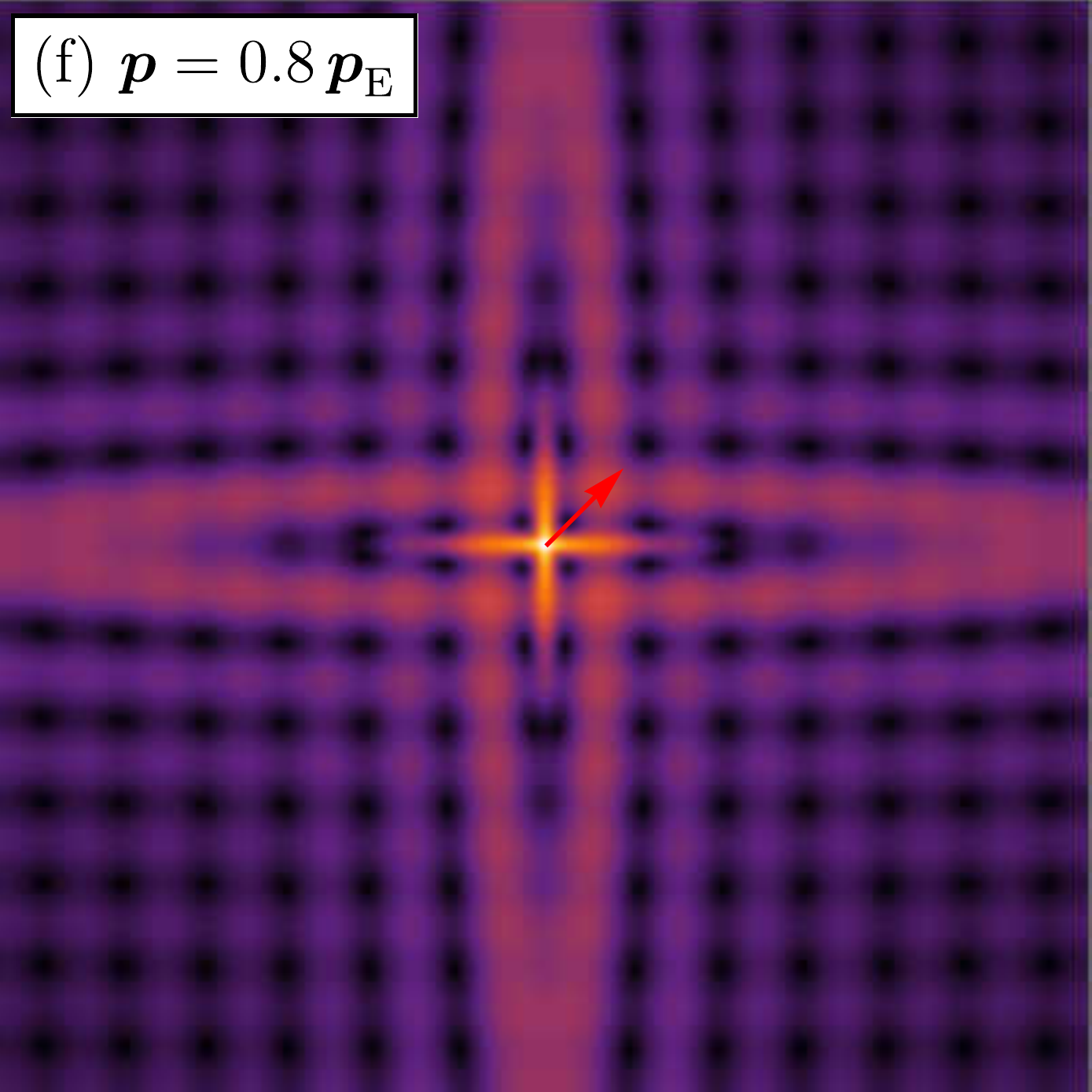}
\end{subfigure}
\begin{subfigure}{0.24\textwidth}
\centering
\phantomsubcaption{\label{fig:square_10_10_force_d_90_gf}}
\includegraphics[width=0.98\linewidth]{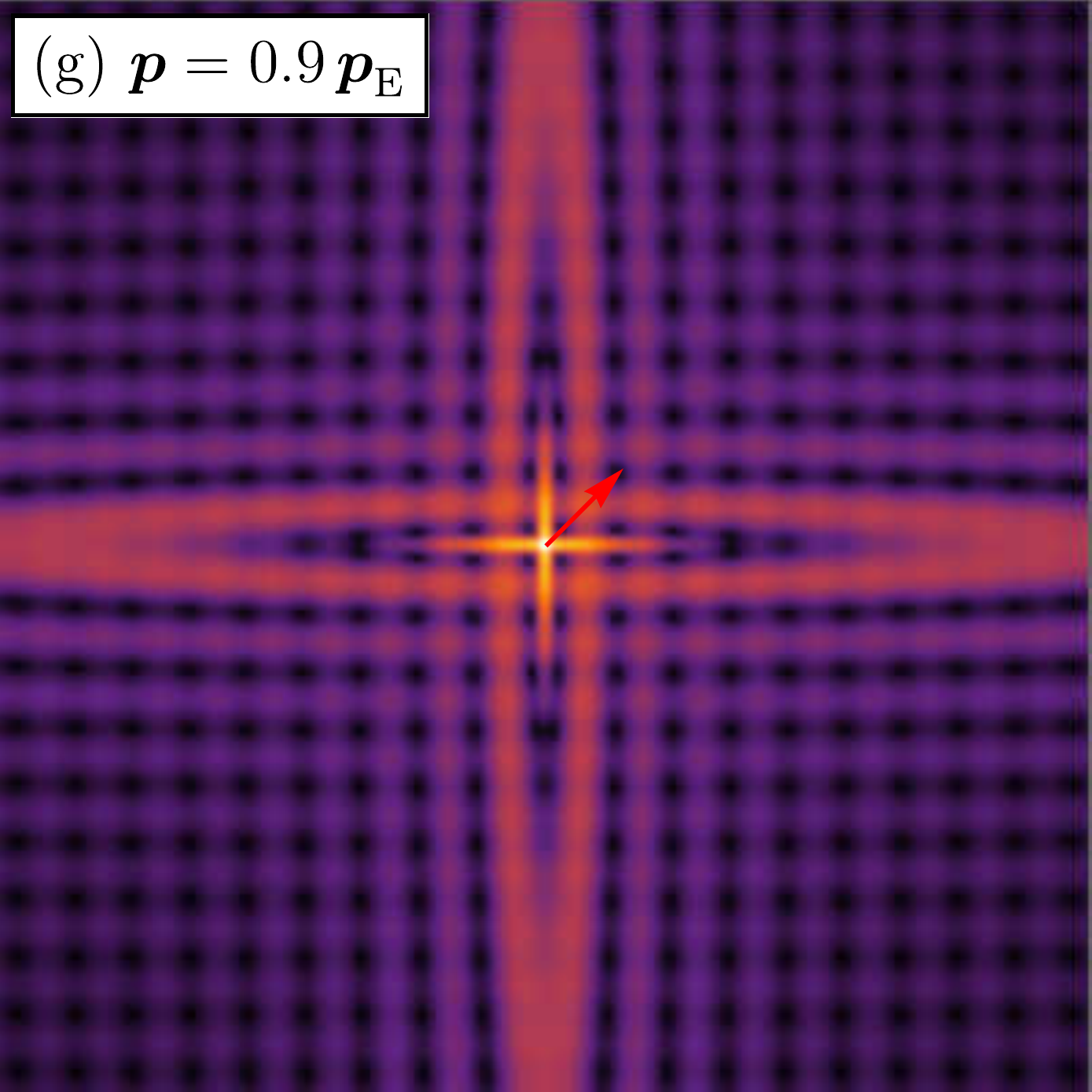}
\end{subfigure}
\begin{subfigure}{0.24\textwidth}
\centering
\phantomsubcaption{\label{fig:square_10_10_force_d_99_gf}}
\includegraphics[width=0.98\linewidth]{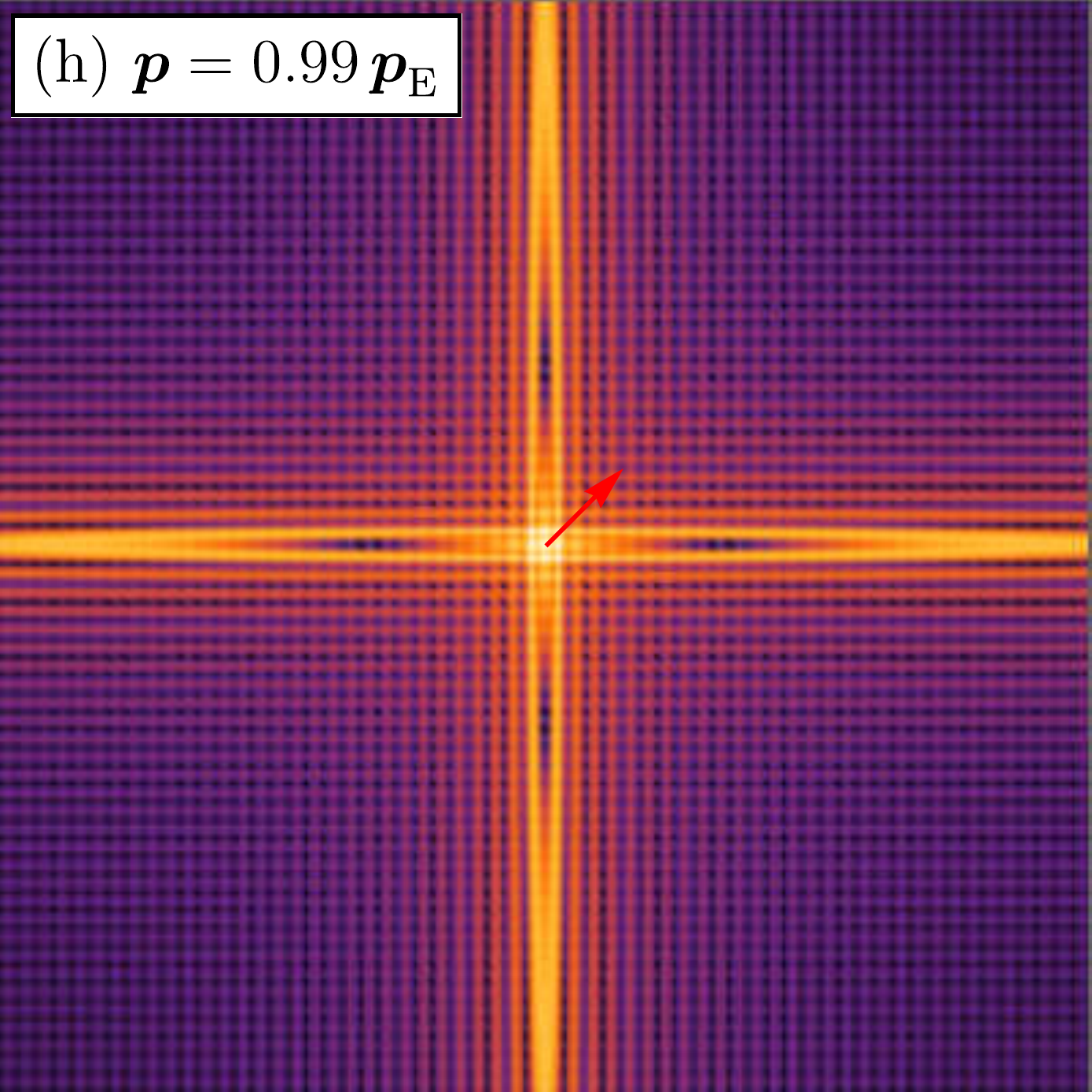}
\end{subfigure}
\caption{\label{fig:square_10_10_force}
    The displacement field generated by a pulsating diagonal force (denoted with a red arrow and applied to the square lattice with cubic symmetry, $\Lambda_1=\Lambda_2=10$) is simulated via f.e.m., see (\subref{fig:square_10_10_force_d_0})--(\subref{fig:square_10_10_force_d_99}), and compared to the response of the homogenized continuum, see (\subref{fig:square_10_10_force_d_0_gf})--(\subref{fig:square_10_10_force_d_99_gf}), at different levels of prestress $\bp$ (preload path shown in Fig.~\ref{fig:ellipticity_domains_10_10}).
    Note the emergence of two orthogonal shear bands, aligned parallel to the directions predicted for failure of ellipticity, see Fig.~\ref{fig:eigenvalue_square_10_10}.
}
\end{figure}

The case of cubic symmetry ($\Lambda_1=\Lambda_2=10$) is analyzed in Fig.~\ref{fig:square_10_10_force}, where the displacement field, numerically computed for the square grid (subject to a pulsating concentrated force, upper row), is compared to the response of the homogenized continuum (subject to the same concentrated force, solved via Green's function, lower row), for four values of prestress (increasing from left to right, $p_1=p_2=\{0,-3.347,-4.891,-5.380\}$).
As the elliptic boundary is approached, the emergence of two strain localizations becomes evident and confirms the predicted vanishing of an eigenvalue (wave speed) reported in Fig.~\ref{fig:eigenvalue_square_10_10}.
\begin{figure}[htb!]
\centering
\begin{subfigure}{0.24\textwidth}
\centering
\caption*{$t=0$}
\includegraphics[width=0.98\linewidth]{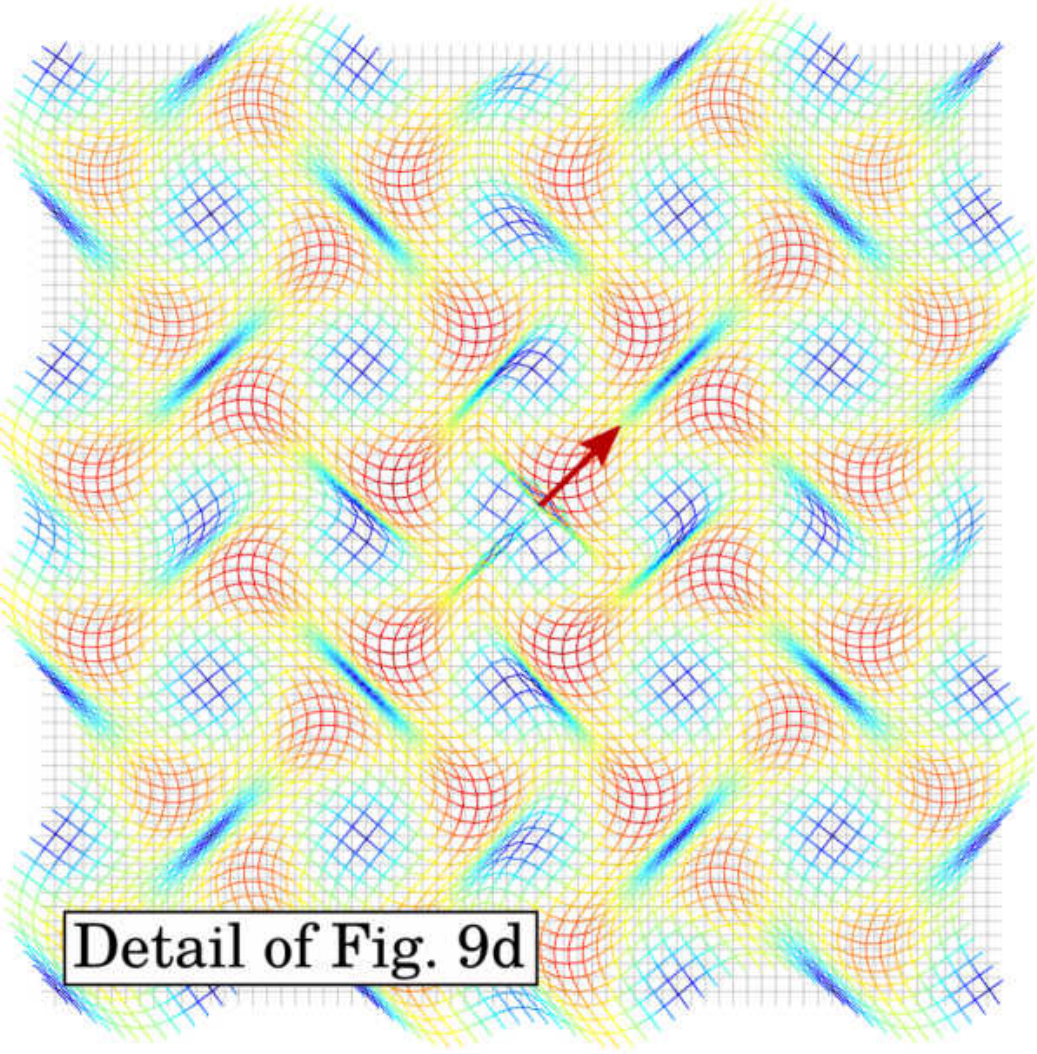}
\end{subfigure}
\begin{subfigure}{0.24\textwidth}
\centering
\caption*{$t=\frac{\pi}{2\omega}$}
\includegraphics[width=0.98\linewidth]{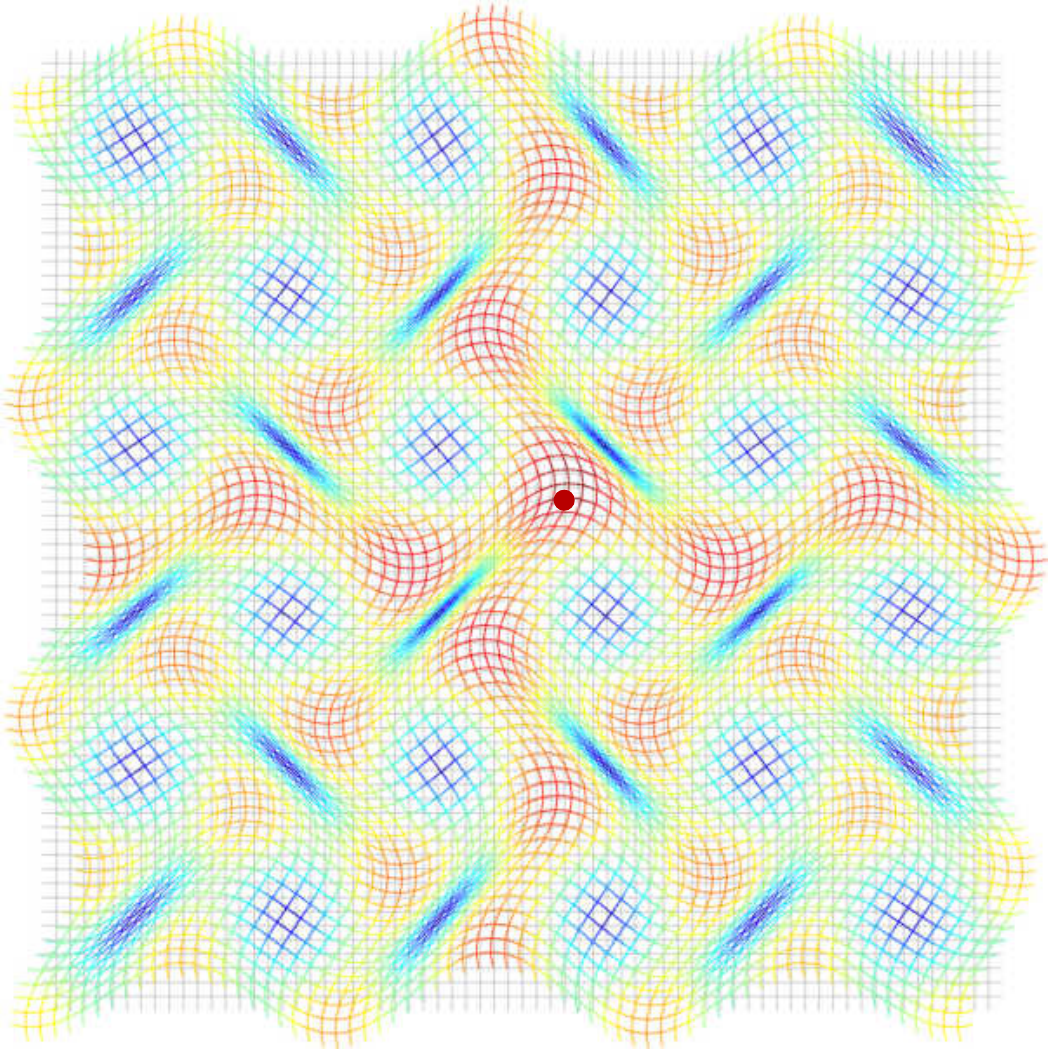}
\end{subfigure}
\begin{subfigure}{0.24\textwidth}
\centering
\caption*{$t=\frac{\pi}{\omega}$}
\includegraphics[width=0.98\linewidth]{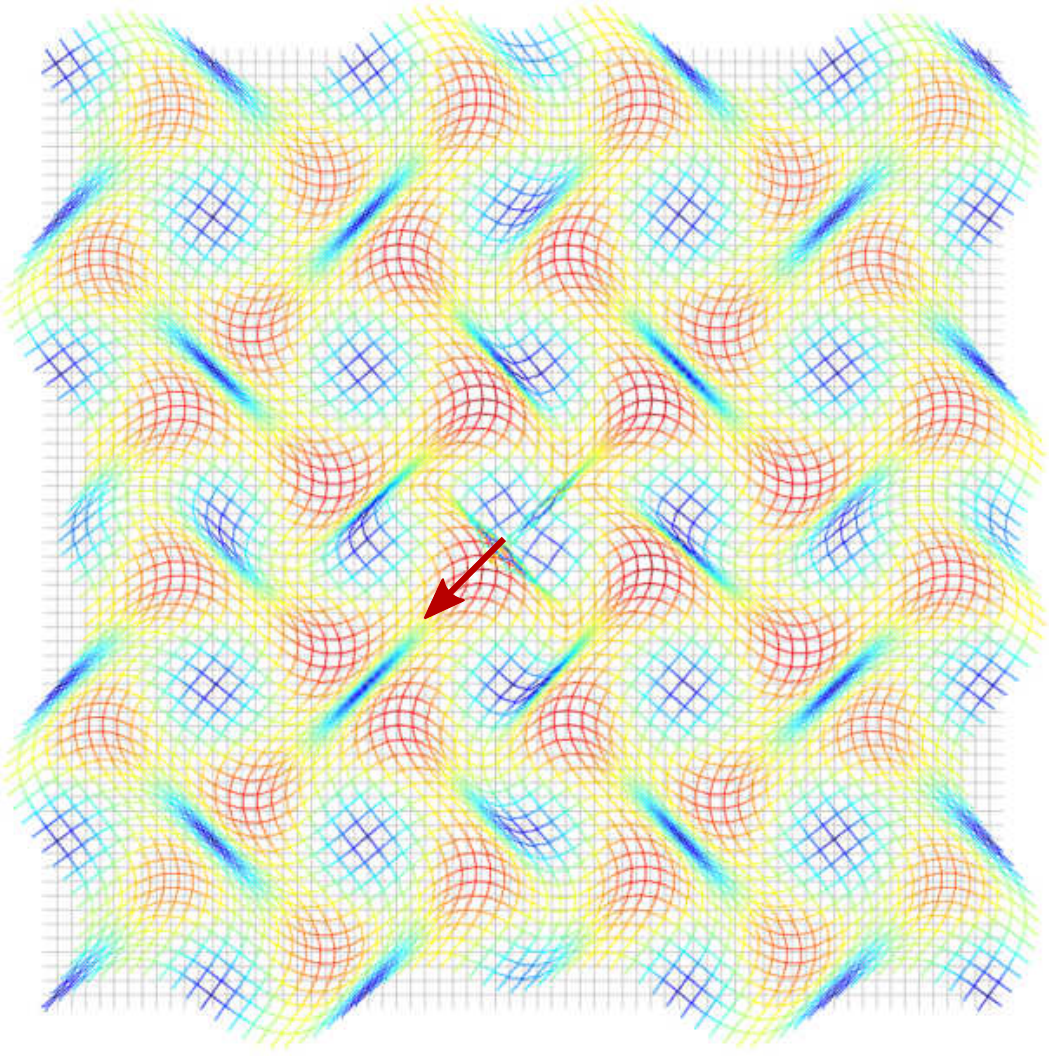}
\end{subfigure}
\begin{subfigure}{0.24\textwidth}
\centering
\caption*{$t=\frac{3\pi}{2\omega}$}
\includegraphics[width=0.98\linewidth]{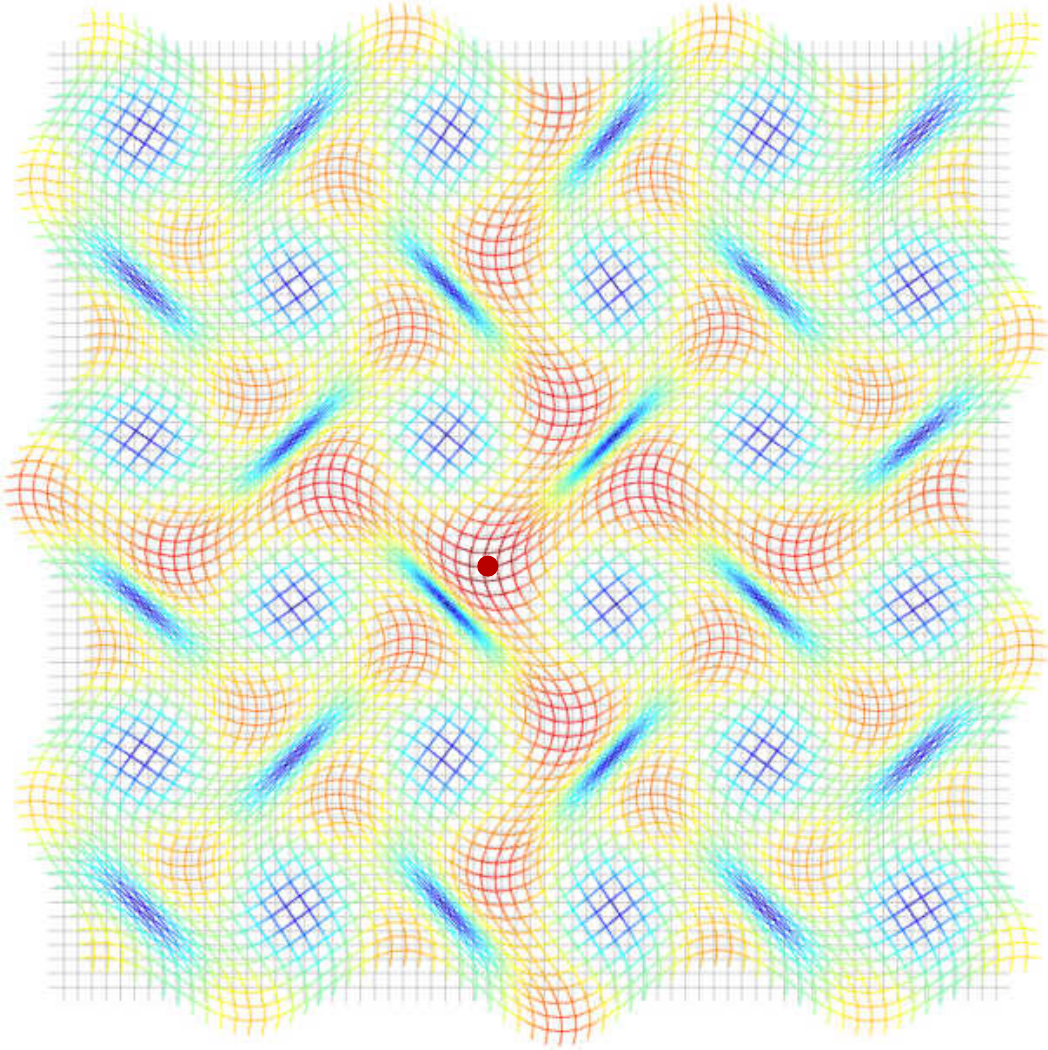}
\end{subfigure}
\caption{\label{fig:square_10_10_deformed}
    Deformed configurations of a square lattice ($\Lambda_1=\Lambda_2=10$) near (the zone is indicated in Fig. \ref{fig:square_10_10_force_d_99}) the point of application of a pulsating force (denoted with a red arrow) at a level of prestress close to the elliptic boundary ($\bp=0.99\,\bp_{\text{E}}$).
    The pattern shows a motion resulting from the superposition of two shear localizations induced by a pulsating diagonal load. 
}
\end{figure}

Snapshots of the displacement map at different instants of time, during the dynamic response of the grid and near the point of application of the pulsating force, reveal the actual localization mode activated by the applied diagonal force.
In fact, deformed configurations calculated in the grid (the zone is indicated in Fig. \ref{fig:square_10_10_force_d_99}) through a finite element simulation and plotted in Fig.~\ref{fig:square_10_10_deformed} display a characteristic motion resulting from the superposition of two shear  localizations emanating from the loading point. 
\begin{figure}[htb!]
\centering
\begin{subfigure}{0.24\textwidth}
\centering
\phantomsubcaption{\label{fig:square_7_15_force_d_0}}
\includegraphics[width=0.98\linewidth]{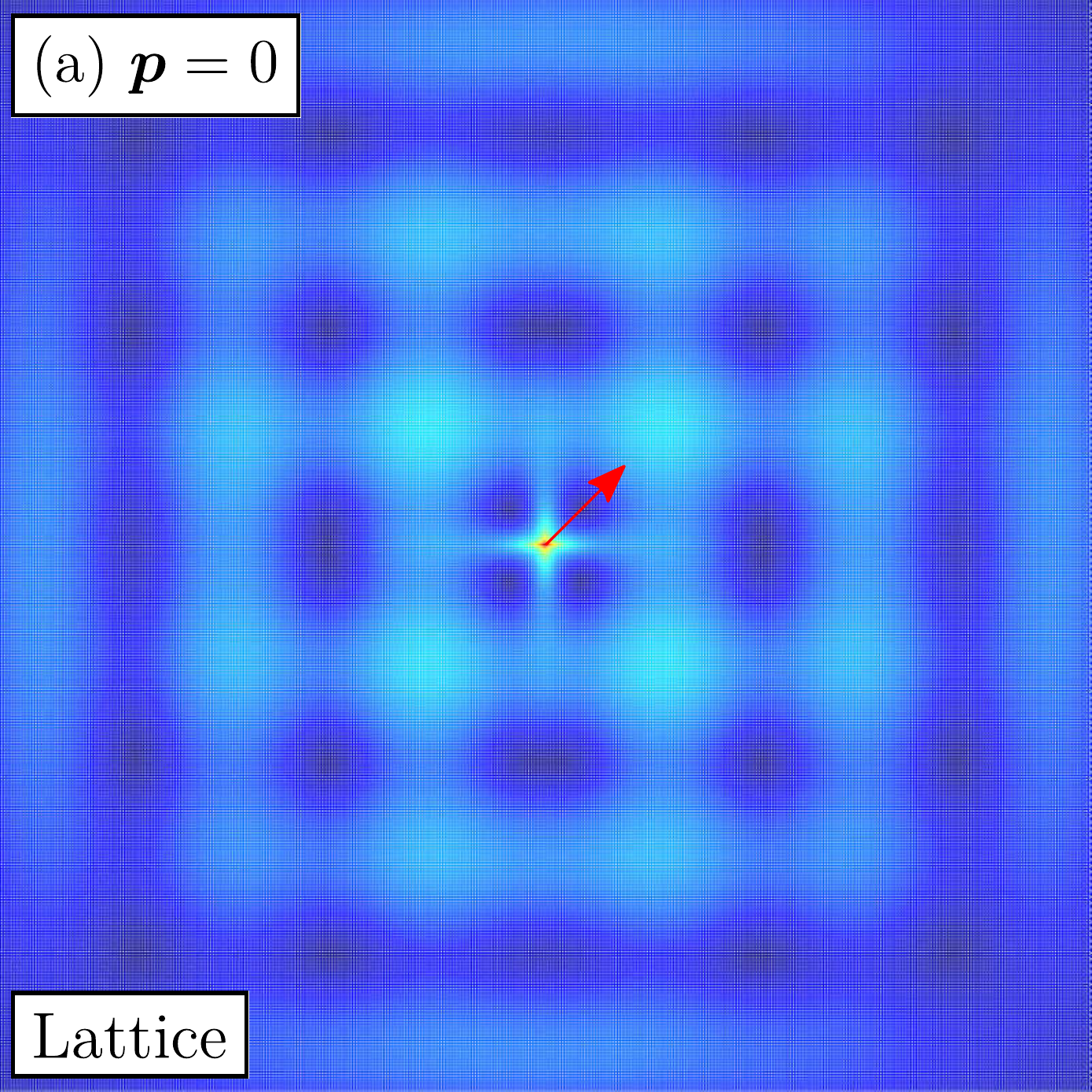}
\end{subfigure}
\begin{subfigure}{0.24\textwidth}
\centering
\phantomsubcaption{\label{fig:square_7_15_force_d_80}}
\includegraphics[width=0.98\linewidth]{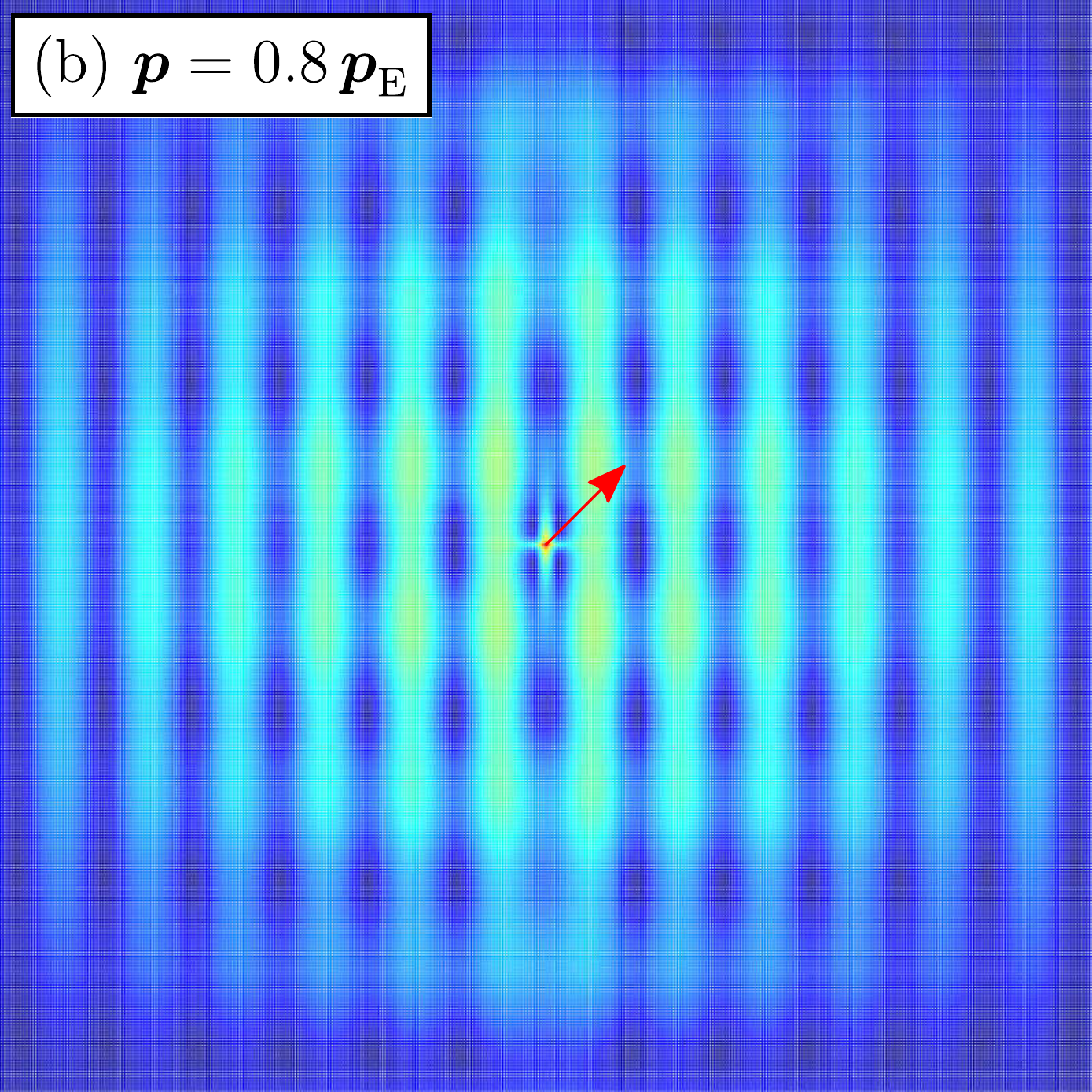}
\end{subfigure}
\begin{subfigure}{0.24\textwidth}
\centering
\phantomsubcaption{\label{fig:square_7_15_force_d_90}}
\includegraphics[width=0.98\linewidth]{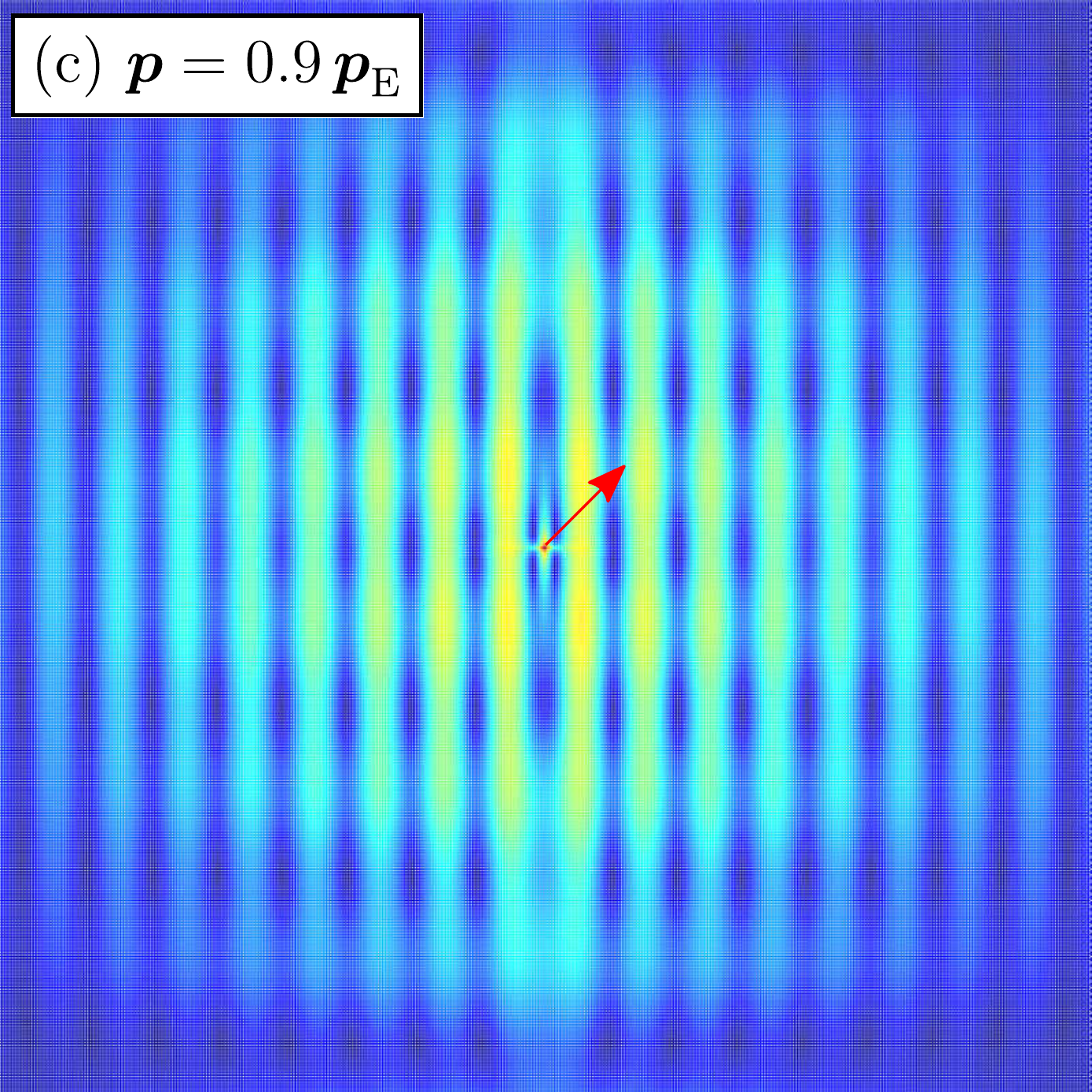}
\end{subfigure}
\begin{subfigure}{0.24\textwidth}
\centering
\phantomsubcaption{\label{fig:square_7_15_force_d_99}}
\includegraphics[width=0.98\linewidth]{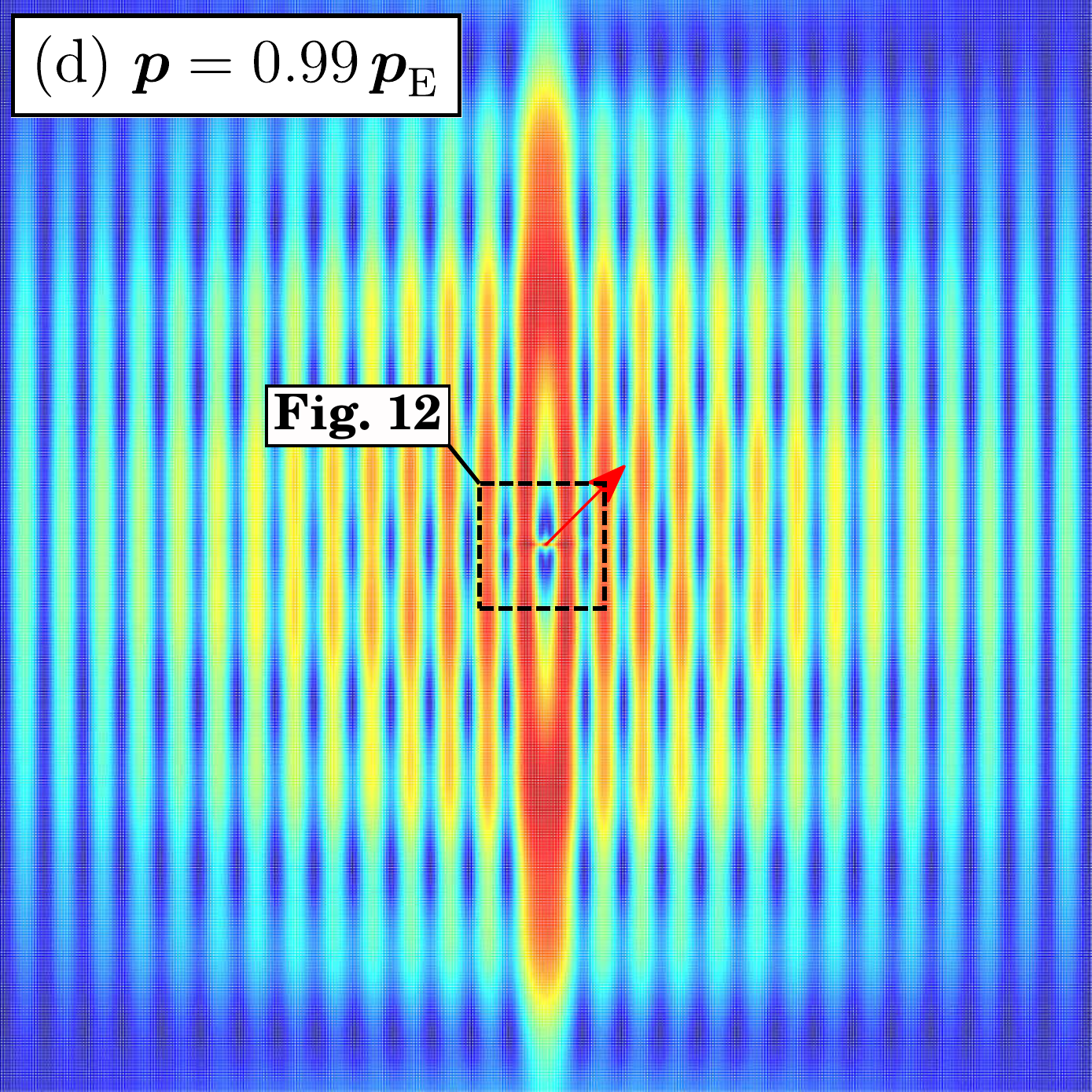}
\end{subfigure}\\
\vspace{0.01\linewidth}
\begin{subfigure}{0.24\textwidth}
\centering
\phantomsubcaption{\label{fig:square_7_15_force_d_0_gf}}
\includegraphics[width=0.98\linewidth]{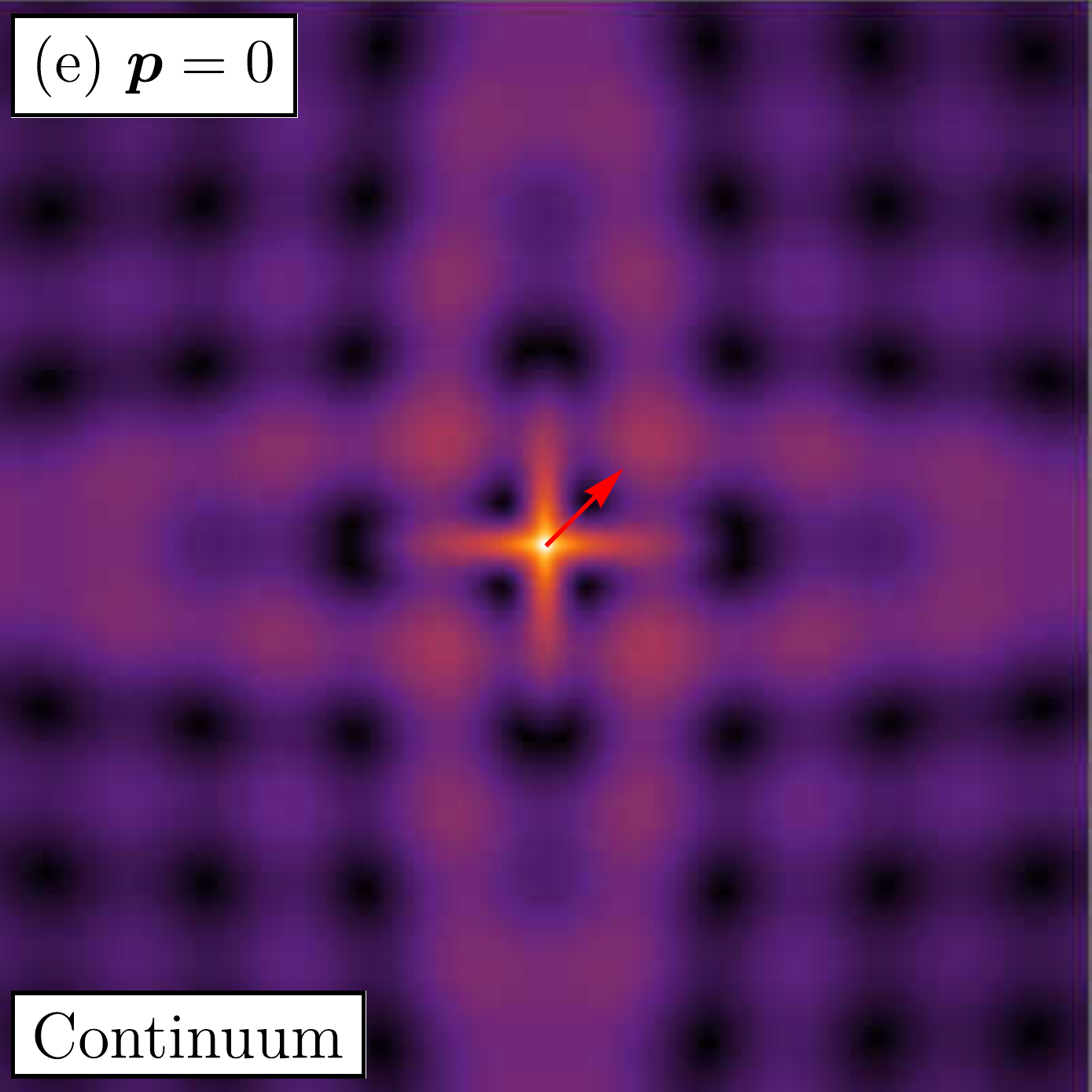}
\end{subfigure}
\begin{subfigure}{0.24\textwidth}
\centering
\phantomsubcaption{\label{fig:square_7_15_force_d_80_gf}}
\includegraphics[width=0.98\linewidth]{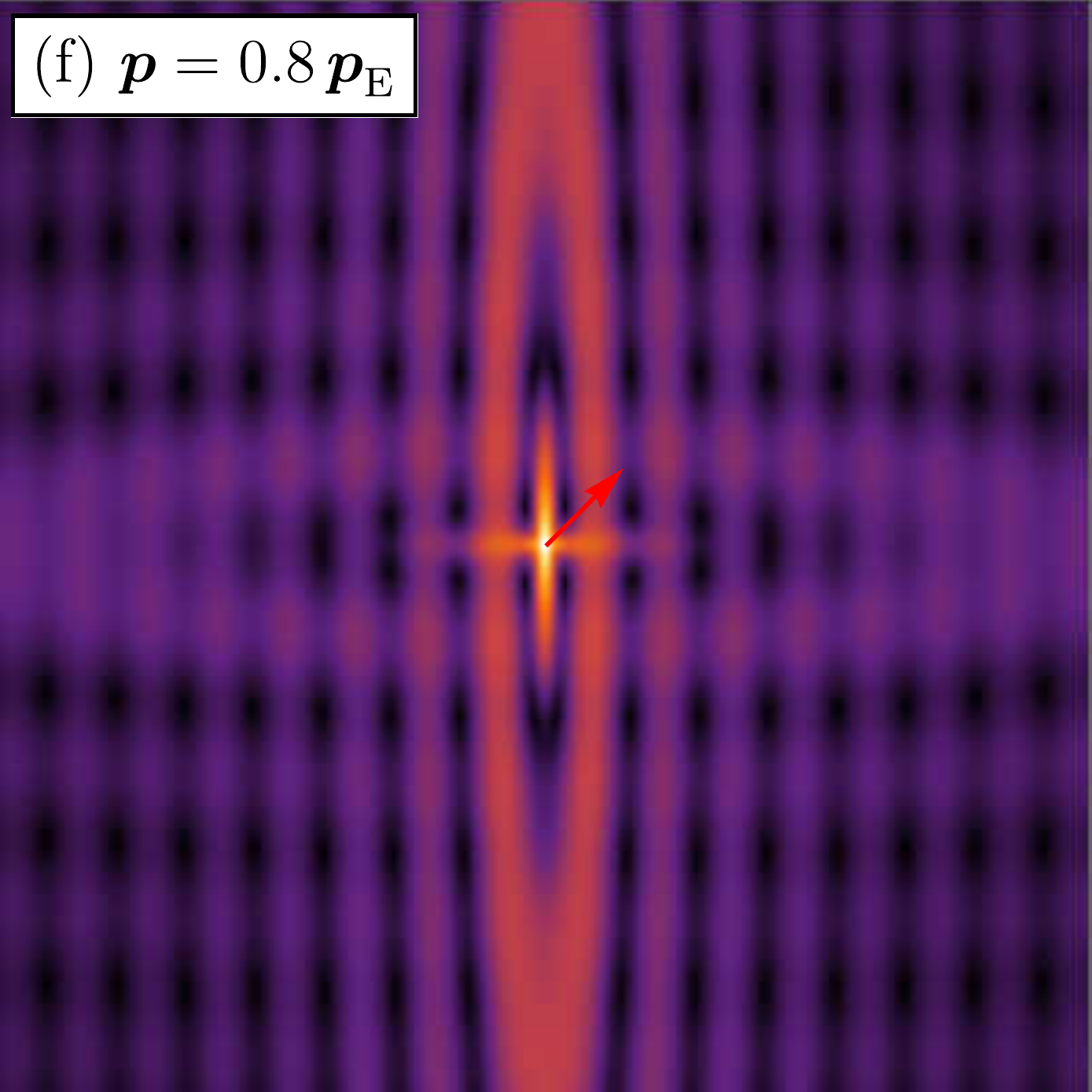}
\end{subfigure}
\begin{subfigure}{0.24\textwidth}
\centering
\phantomsubcaption{\label{fig:square_7_15_force_d_90_gf}}
\includegraphics[width=0.98\linewidth]{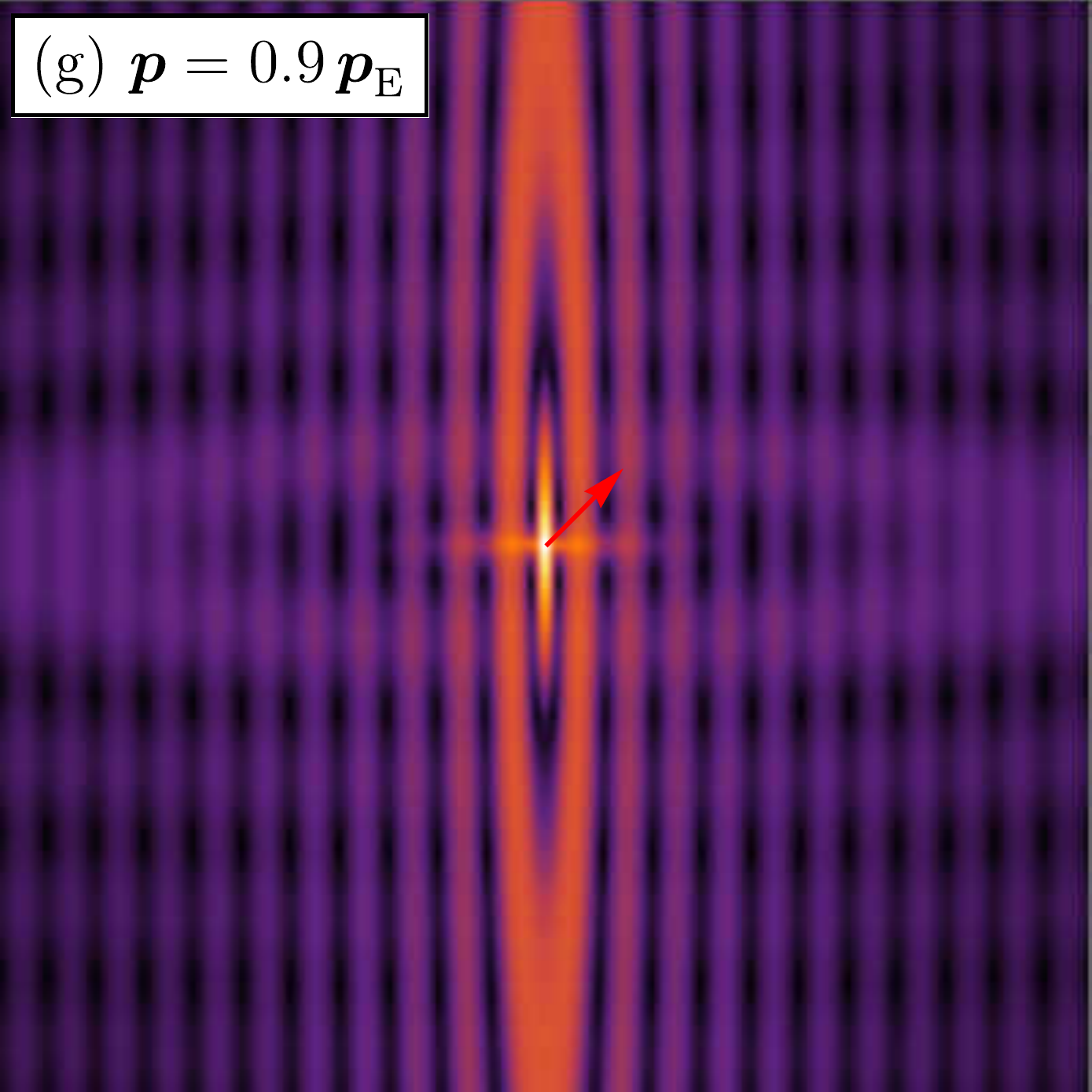}
\end{subfigure}
\begin{subfigure}{0.24\textwidth}
\centering
\phantomsubcaption{\label{fig:square_7_15_force_d_99_gf}}
\includegraphics[width=0.98\linewidth]{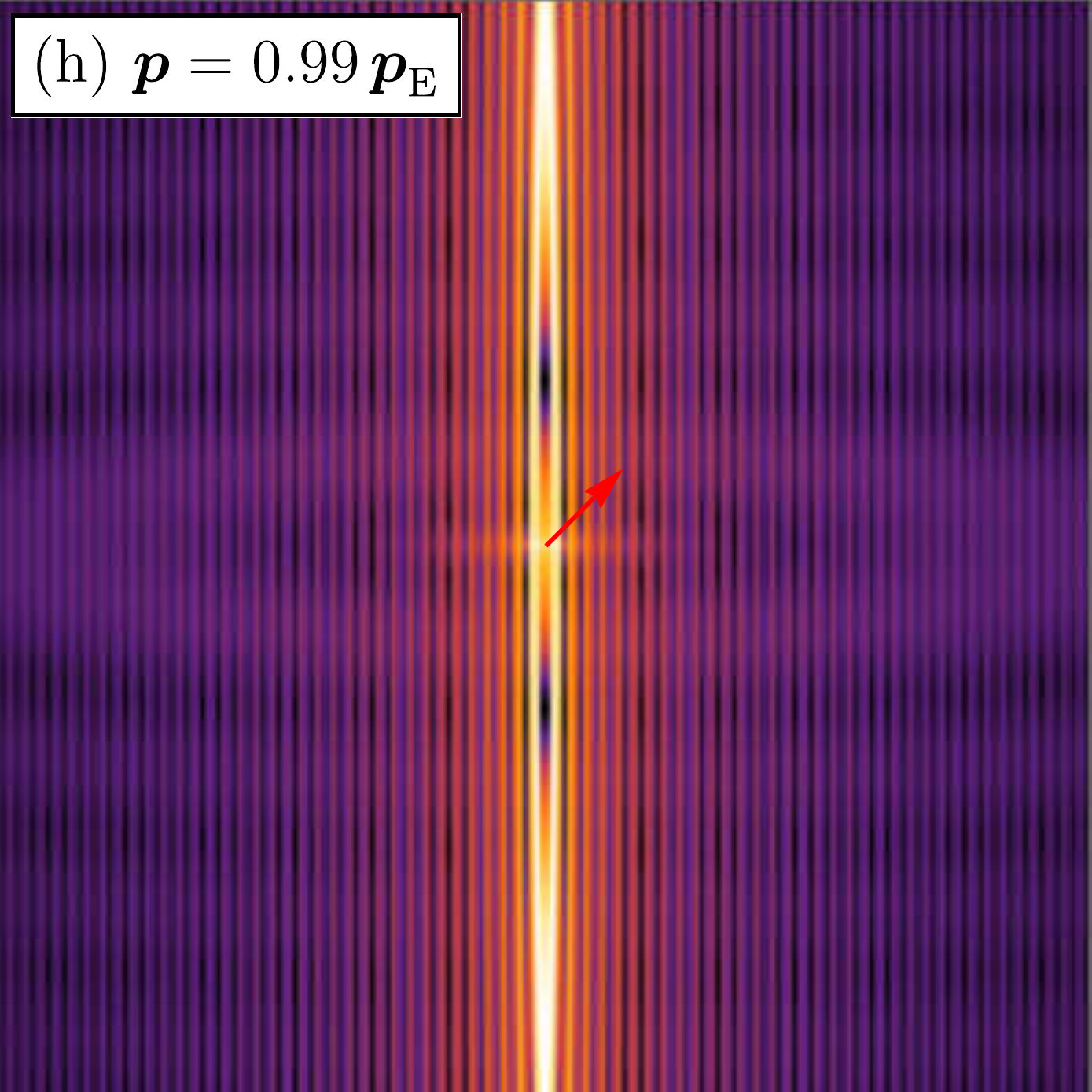}
\end{subfigure}
\caption{\label{fig:square_7_15_force}
    The displacement field generated by a pulsating diagonal force (denoted with a red arrow and applied to the orthotropic square lattice, $\Lambda_1=7,\,\Lambda_2=15$) is simulated via f.e.m., see (\subref{fig:square_7_15_force_d_0})--(\subref{fig:square_7_15_force_d_99}), and compared to the response of the homogenized continuum, see(\subref{fig:square_7_15_force_d_0_gf})--(\subref{fig:square_7_15_force_d_99_gf}), at different levels of prestress $\bp$ (preload path shown in Fig.~\ref{fig:ellipticity_domains_7_15}).
    Note the emergence of a vertical shear band, as predicted in Fig.~\ref{fig:eigenvalue_square_7_15}.
}
\end{figure}

\begin{figure}[htb!]
\centering
\begin{subfigure}{0.24\textwidth}
\centering
\caption*{$t=0$}
\includegraphics[width=0.98\linewidth]{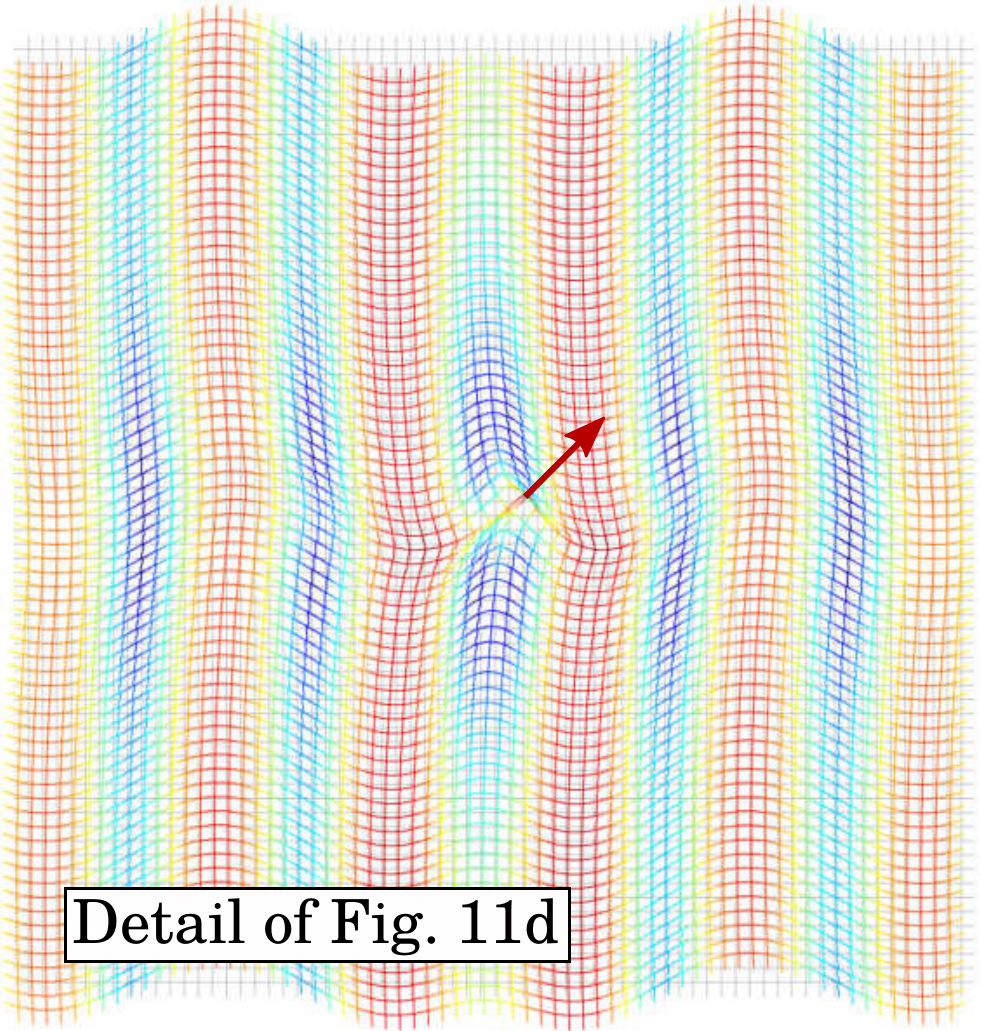}
\end{subfigure}
\begin{subfigure}{0.24\textwidth}
\centering
\caption*{$t=\frac{\pi}{2\omega}$}
\includegraphics[width=0.98\linewidth]{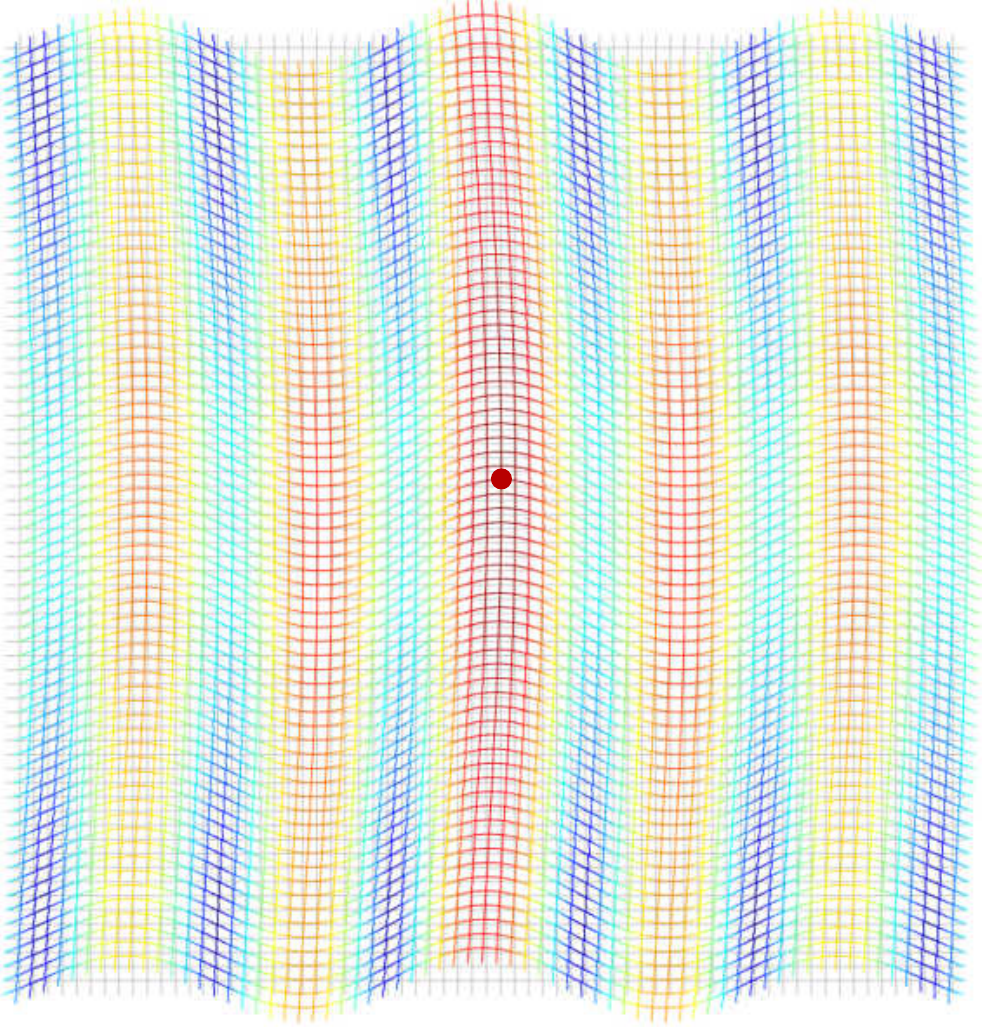}
\end{subfigure}
\begin{subfigure}{0.24\textwidth}
\centering
\caption*{$t=\frac{\pi}{\omega}$}
\includegraphics[width=0.98\linewidth]{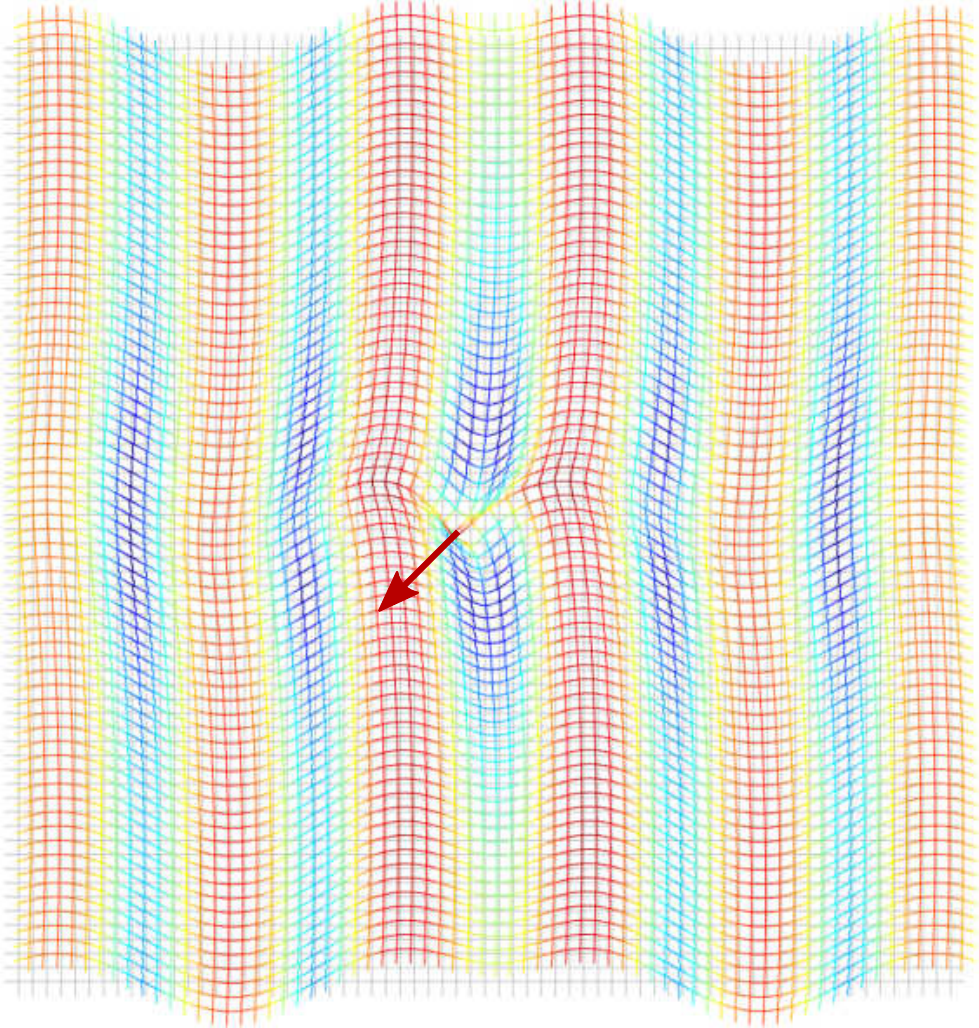}
\end{subfigure}
\begin{subfigure}{0.24\textwidth}
\centering
\caption*{$t=\frac{3\pi}{2\omega}$}
\includegraphics[width=0.98\linewidth]{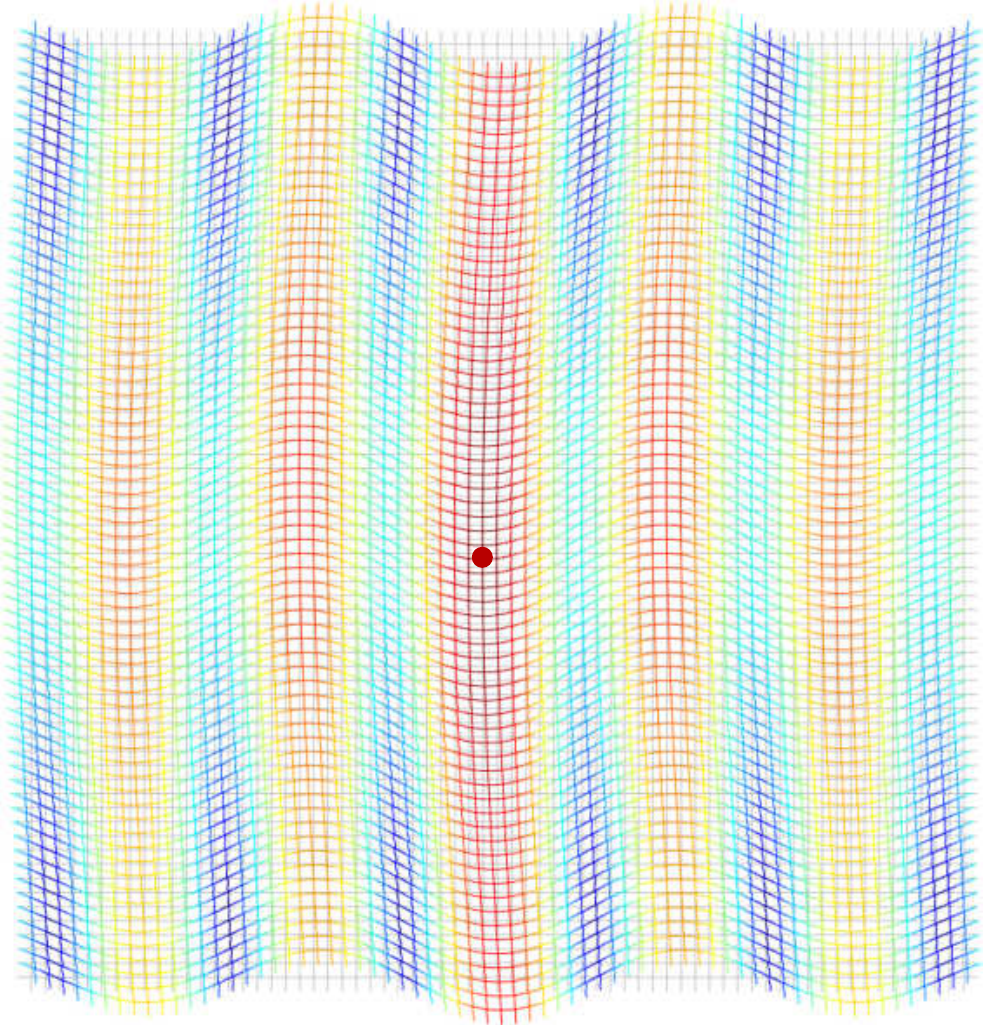}
\end{subfigure}
\caption{\label{fig:square_7_15_deformed}
    Deformed configurations of the square lattice ($\Lambda_1=7,\,\Lambda_2=15$) near (the zone is indicated in Fig. \ref{fig:square_7_15_force_d_99}) the point of application of a pulsating concentrated force (denoted with a red arrow) at a level of prestress close to the elliptic boundary ($\bp=0.99\,\bp_{\text{E}}$).
    The pattern shows the characteristic motion of a vertical shear band, induced by the pulsating load.
    The localization emerges from a prevalent bending deformation of the `soft' vertical links and an approximately rigid rotation of the `stiff' horizontal rods allowed by a significant rotation of the grid nodes.
}
\end{figure}

The square grid displays a strain localization into a \textit{single} shear band  when the slenderness values of the two orthogonal elastic links are set to be different, thus breaking the cubic symmetry, but preserving orthotropy.
The response of the grid with $\Lambda_1=7$ and $\Lambda_2=15$ is reported in Fig.~\ref{fig:square_7_15_force} for four preload states corresponding to $p_1=p_2=\{0,-1.657,-1.864,-2.050\}$.
As already revealed by Fig.~\ref{fig:eigenvalue_square_7_15}, a single vertical shear band emerges, thus confirming the counter-intuitive result obtained in the previous section, namely that \textit{the shear wave responsible for the ellipticity loss is the one propagating along the direction of the `stiffest' elastic link} (which possesses the lowest slenderness).
The mechanism underlying this effect is displayed by analyzing the actual deformed configuration of the grid reported in Fig.~\ref{fig:square_7_15_deformed} (plotted at different instants of time and obtained via f.e.m. simulations). The figure, which refers to the zone indicated in Fig. \ref{fig:square_7_15_force_d_99},  reveals that the vertical strain localization emerges from a prevalent bending deformation of the `soft' vertical links accompanied by an approximately rigid rotation of the `stiff' horizontal rods, which is allowed by a large rotation of the nodes.
\begin{figure}[htb!]
\centering
\begin{subfigure}{0.4\textwidth}
\centering
\caption{\label{fig:square_10_10_force_d_FFT}Cubic square lattice ($\Lambda_1=\Lambda_2=10$)}
\includegraphics[width=0.98\linewidth]{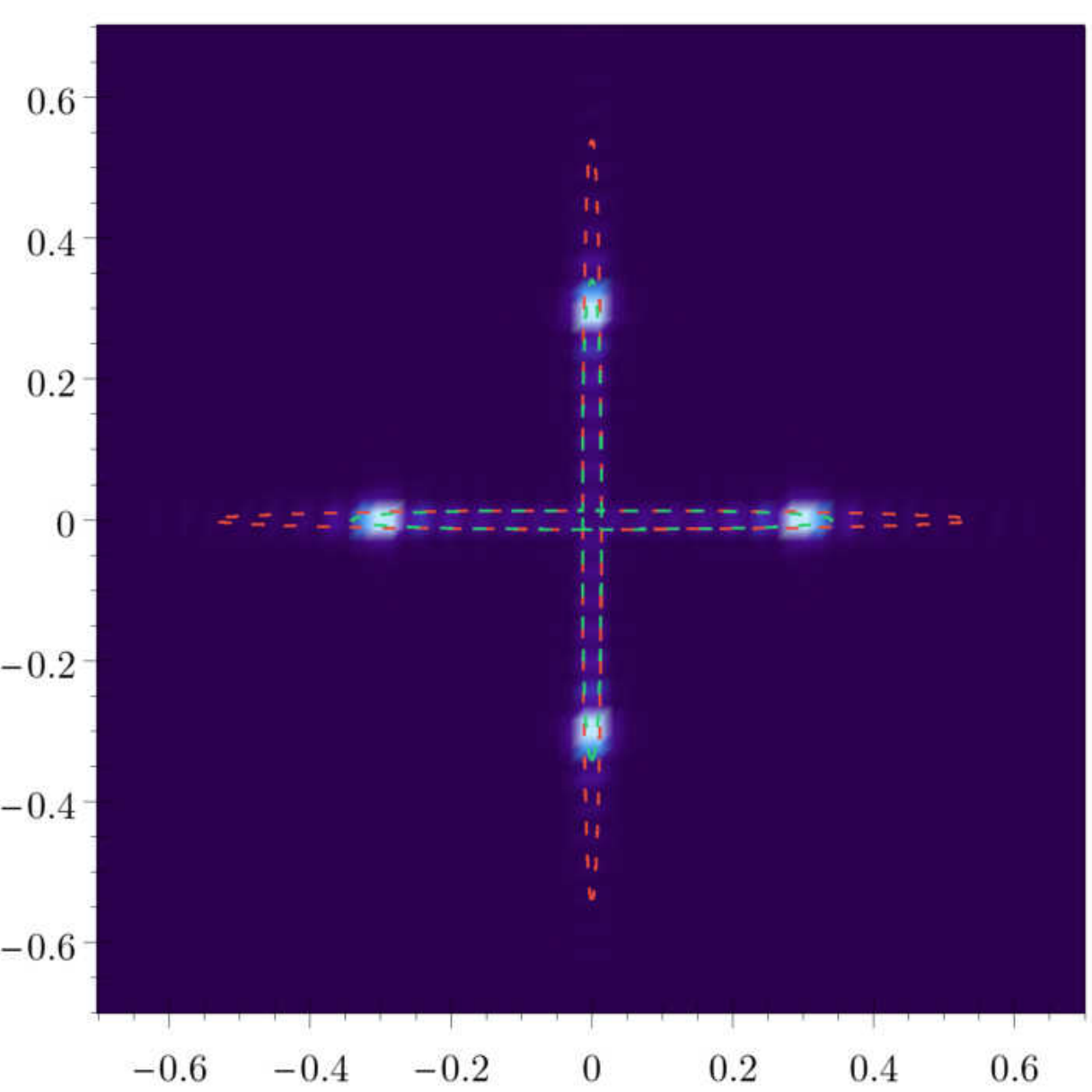}
\end{subfigure} \hspace{2mm}
\begin{subfigure}{0.4\textwidth}
\centering
\caption{\label{fig:square_7_15_force_d_FFT}Orthotropic square lattice ($\Lambda_1=7,\, \Lambda_2=15$)}
\includegraphics[width=0.98\linewidth]{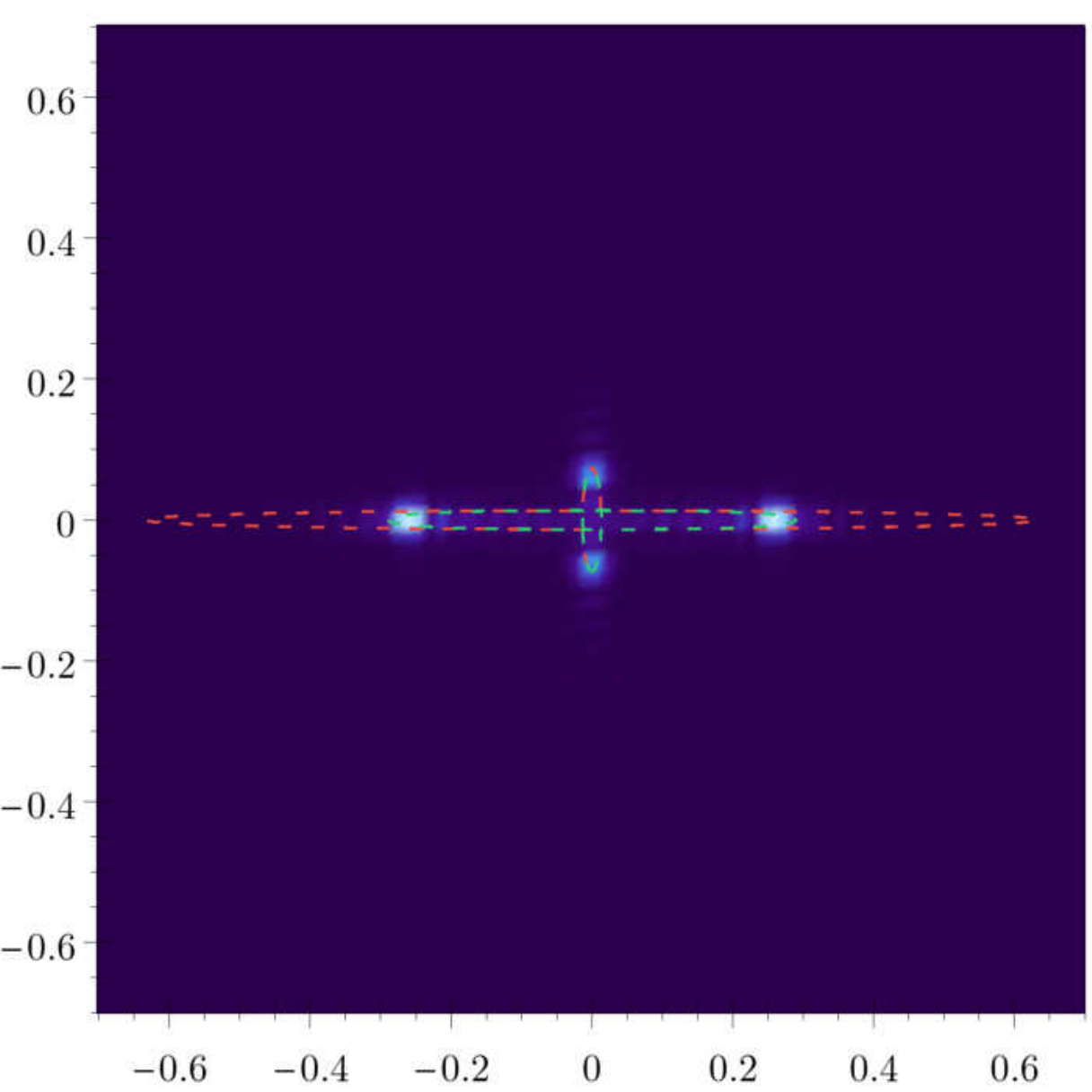}
\end{subfigure}
\caption{\label{fig:square_force_d_FFT}
    Fourier transform of the complex displacement fields for the cubic~(\subref{fig:square_10_10_force_d_FFT}) and orthotropic~(\subref{fig:square_7_15_force_d_FFT}) square lattice subject to a diagonal force applied at the center. A prestress state of $\bp=0.99\,\bp_{\text{E}}$, close to the elliptic boundary, has been assumed. 
    The slowness contours of the lattice (dashed green) and the effective continuum (dashed red) are superimposed to highlight the Bloch spectrum of waves excited by the forcing source.
    Note that the strong focus of the spectrum indicates that few plane waves, namely those `slow' waves that are close to cause the ellipticity loss, prevail on the response. This is expected for  
    the orthotropic grid~(\subref{fig:square_7_15_force_d_FFT}) near the elliptic boundary, where the waves propagating vertically remain almost inactive, when compared to those propagating horizontally.
}
\end{figure}

The comparison between the responses of the grid and of the homogenized continuum that has been presented in Figs.~\ref{fig:square_10_10_force} and~\ref{fig:square_7_15_force} shows an almost perfect agreement from low to high prestress levels, up to values close to the elliptic boundary. 
However, the agreement can be further tested by considering the lattice's complex displacement field (reported in the last column of Figs.~\ref{fig:square_10_10_force} and~\ref{fig:square_7_15_force}, prestressed at $\bp=0.99\,\bp_{\text{E}}$), computing its Fourier transform and superimposing this to the corresponding slowness contour. 
This is reported in Fig.~\ref{fig:square_force_d_FFT}, where the Fourier transform shows that, for both considered square grids, the Bloch spectrum of waves excited by the diagonal load matches the slowness contour of the lattice (reported in green) and is also highly focused around the directions of ellipticity loss where it is at the maximum distance from the contour of the continuum (reported in red).
It is worth noting that the strong focus of the spectrum confirms the fact that few plane waves, namely those `slow' waves that are close to cause the ellipticity loss, prevail on the response, as it is expected for a material near the elliptic boundary.

\subsection{Rhombic lattice}
\label{sec:rhombic_lattice}
In the previous section, the square lattice was shown to display only localizations in the form of `pure' shear bands, i.e. in which shear strain prevails, perfectly aligned parallel to the elastic ligaments.
However, on the basis of the analysis performed in Section~\ref{sec:ellipticity_loss_grid}, the formation of localizations is expected along different directions and with different deformation modes, when a rhombic grid is considered, $\alpha\neq\pi/2$.
\begin{figure}[htb!]
\centering
\begin{subfigure}{0.24\textwidth}
\centering
\phantomsubcaption{\label{fig:rhombus_10_10_force_h_0}}
\includegraphics[width=0.98\linewidth]{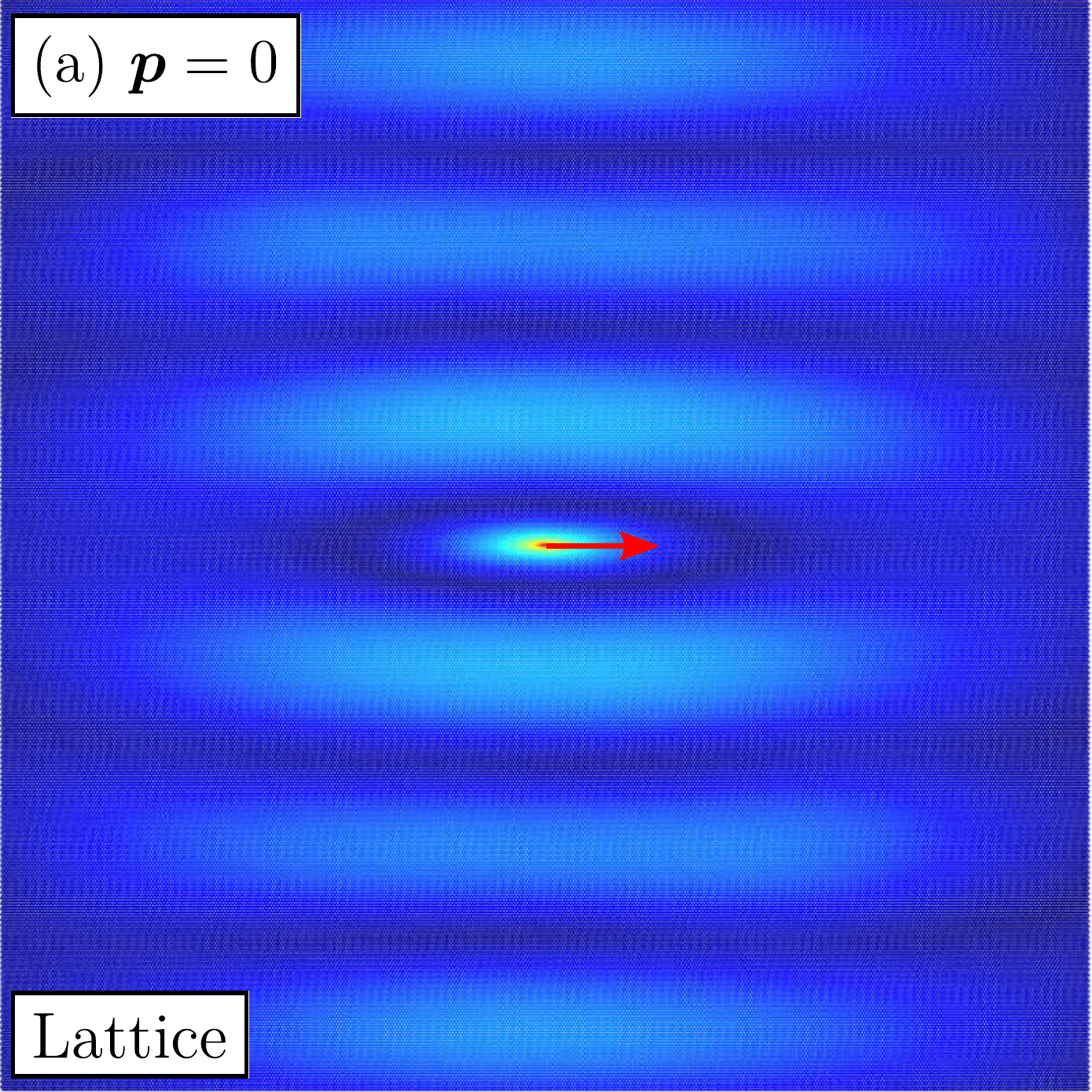}
\end{subfigure}
\begin{subfigure}{0.24\textwidth}
\centering
\phantomsubcaption{\label{fig:rhombus_10_10_force_h_80}}
\includegraphics[width=0.98\linewidth]{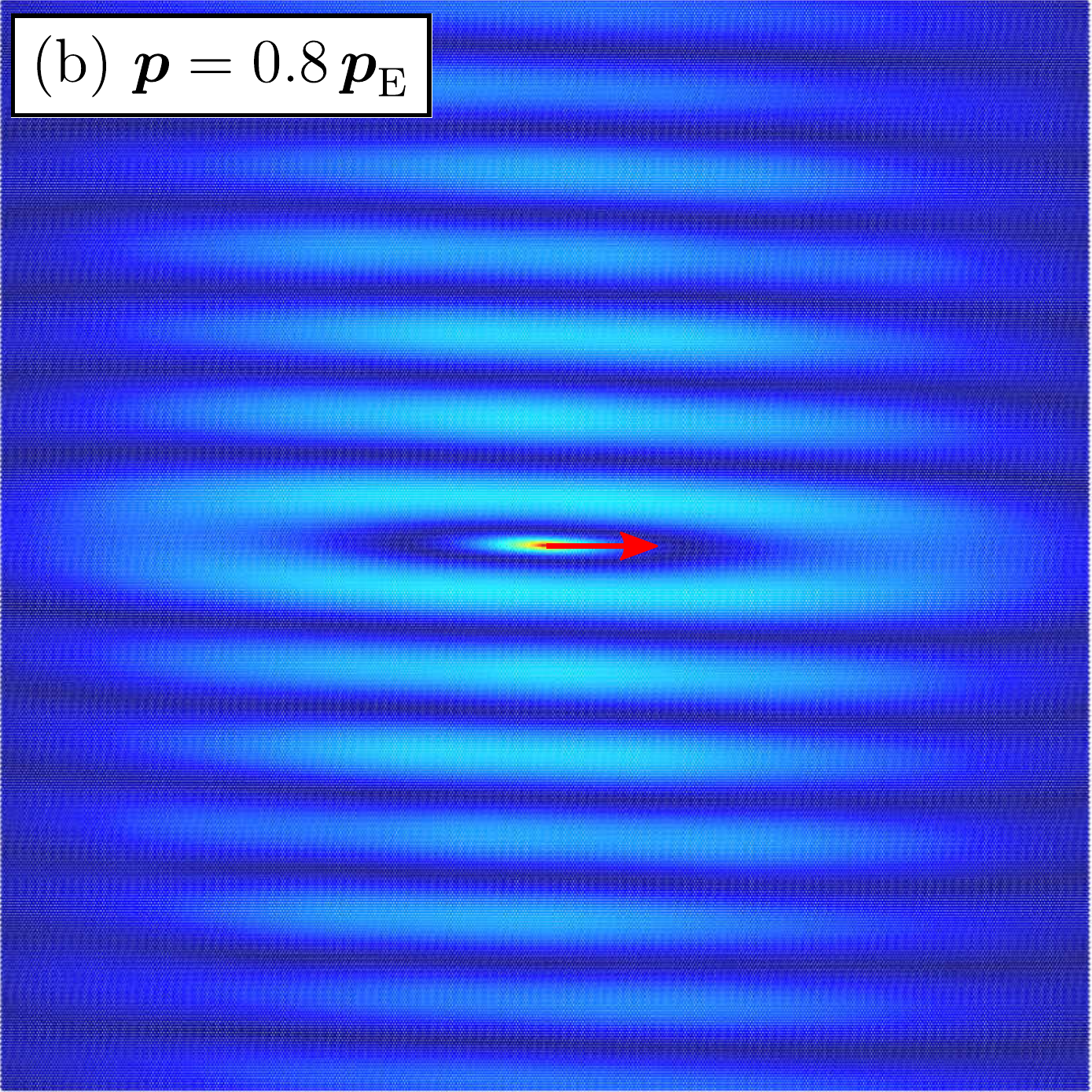}
\end{subfigure}
\begin{subfigure}{0.24\textwidth}
\centering
\phantomsubcaption{\label{fig:rhombus_10_10_force_h_90}}
\includegraphics[width=0.98\linewidth]{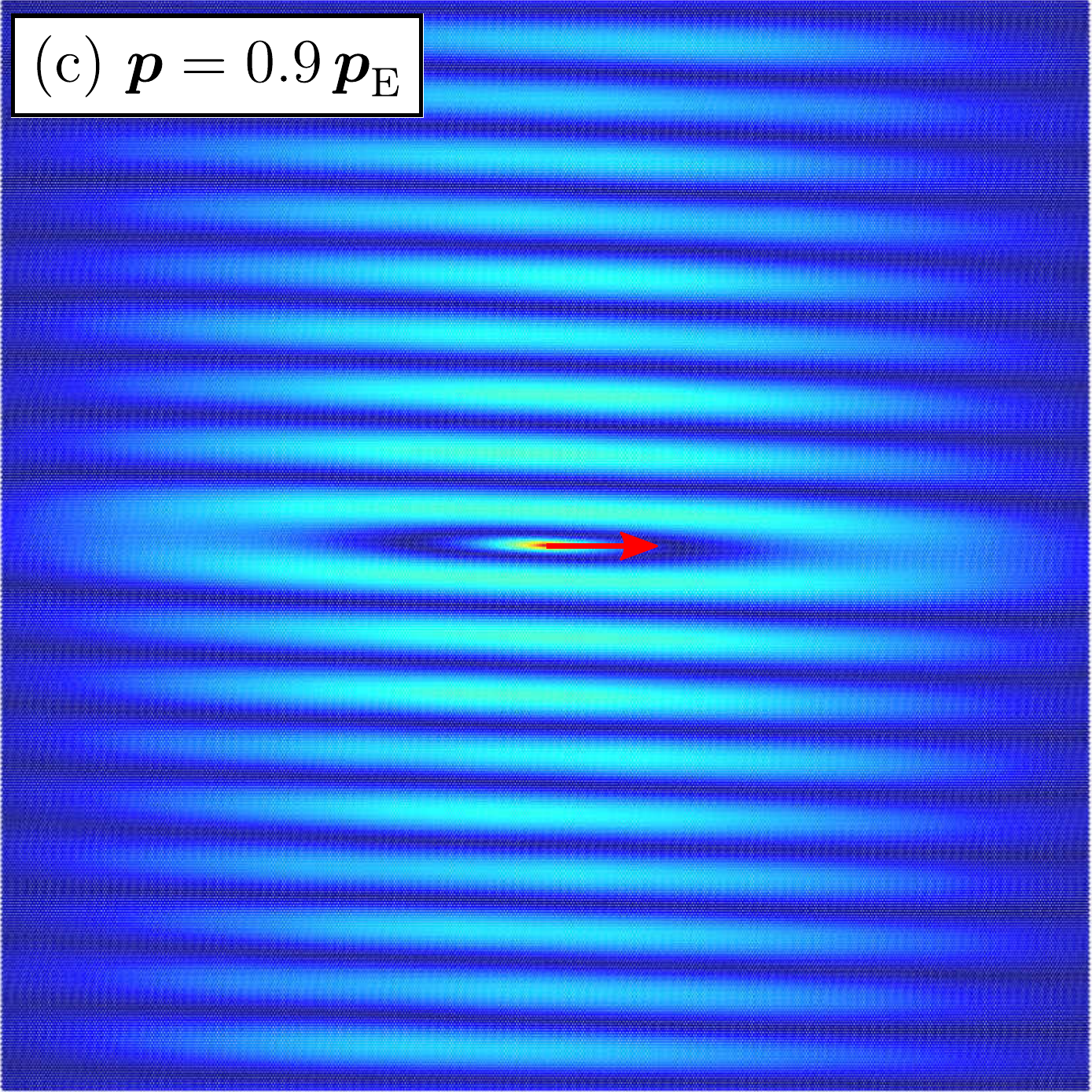}
\end{subfigure}
\begin{subfigure}{0.24\textwidth}
\centering
\phantomsubcaption{\label{fig:rhombus_10_10_force_h_99}}
\includegraphics[width=0.98\linewidth]{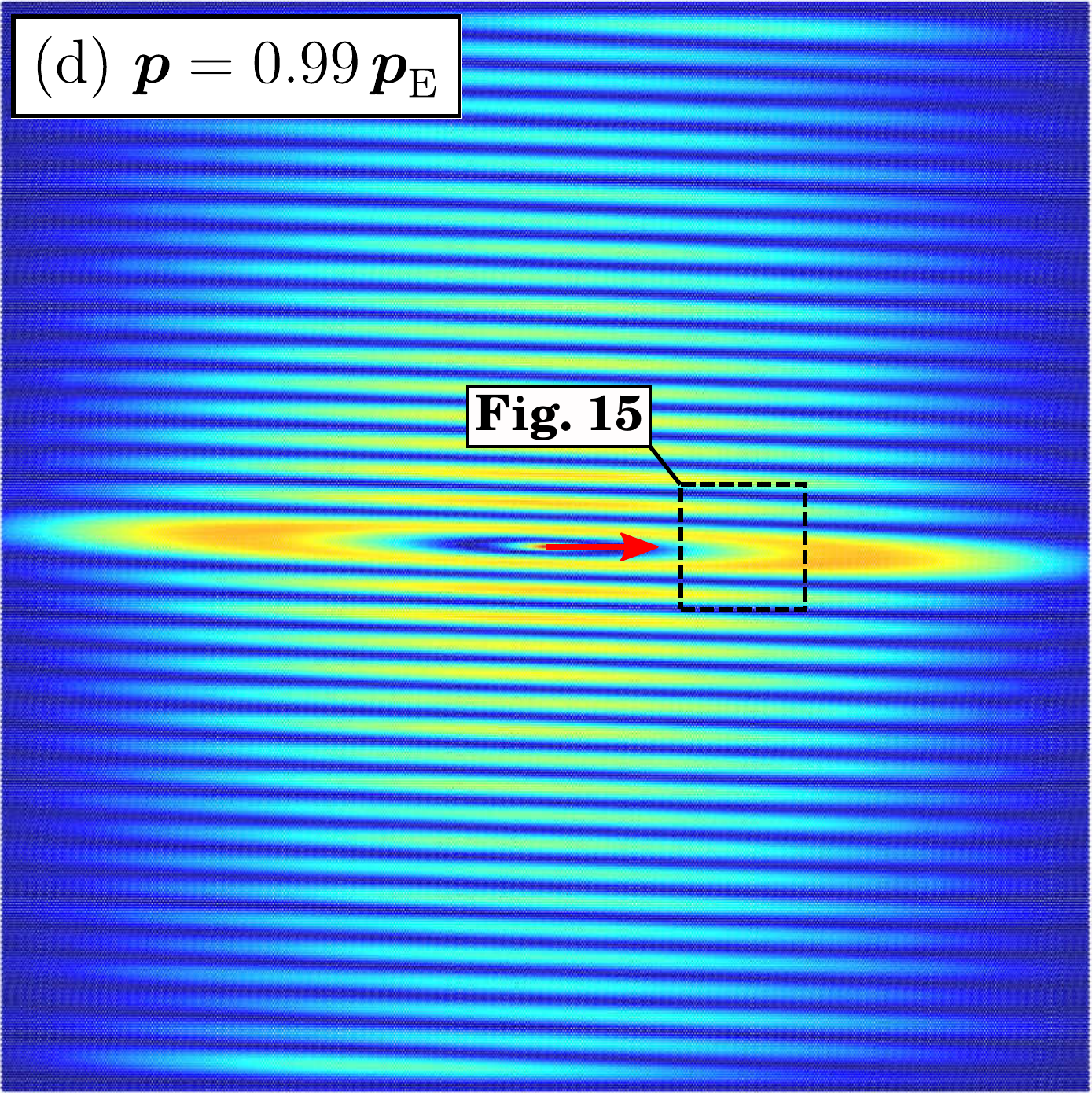}
\end{subfigure}\\
\vspace{0.01\linewidth}
\begin{subfigure}{0.24\textwidth}
\centering
\phantomsubcaption{\label{fig:rhombus_10_10_force_h_0_gf}}
\includegraphics[width=0.98\linewidth]{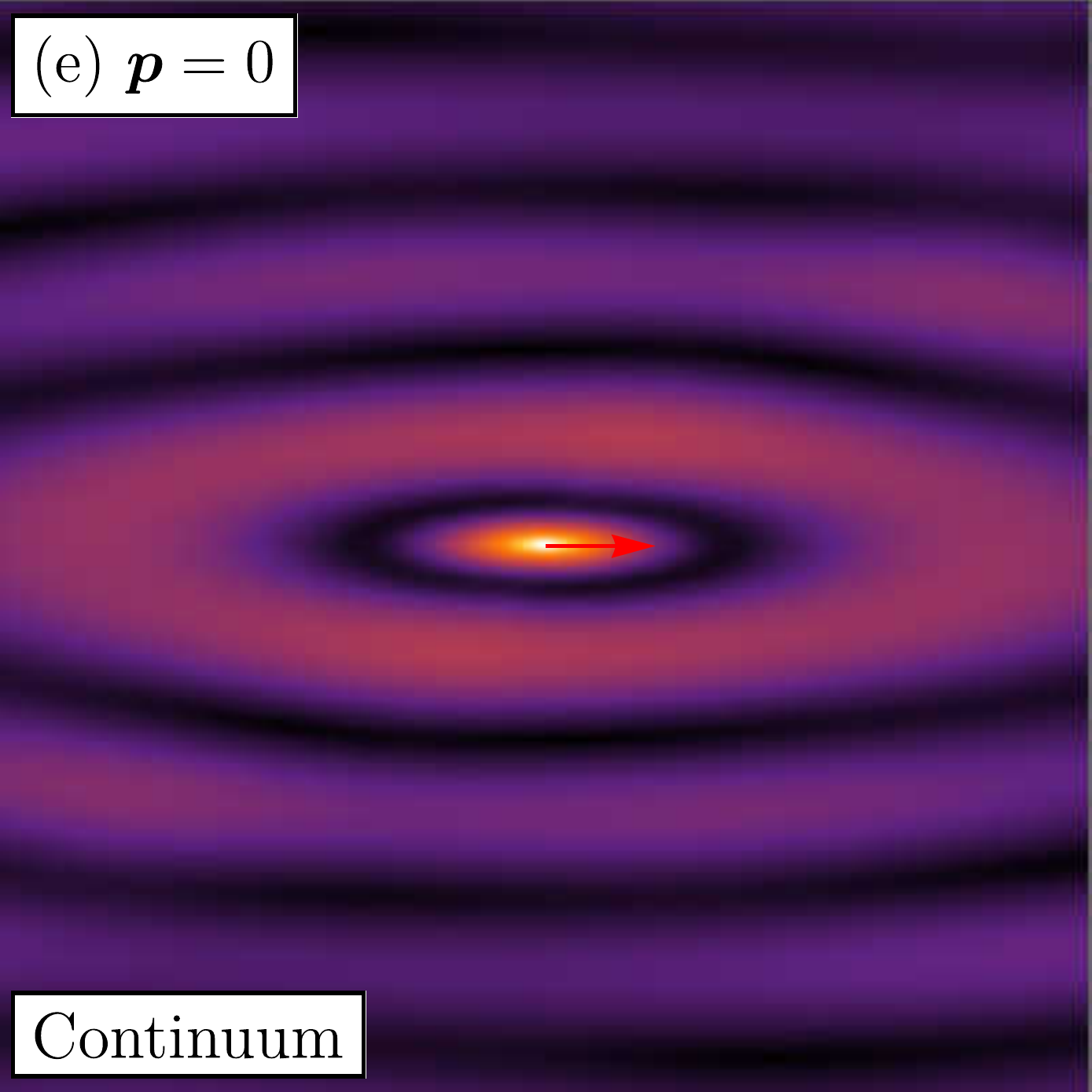}
\end{subfigure}
\begin{subfigure}{0.24\textwidth}
\centering
\phantomsubcaption{\label{fig:rhombus_10_10_force_h_80_gf}}
\includegraphics[width=0.98\linewidth]{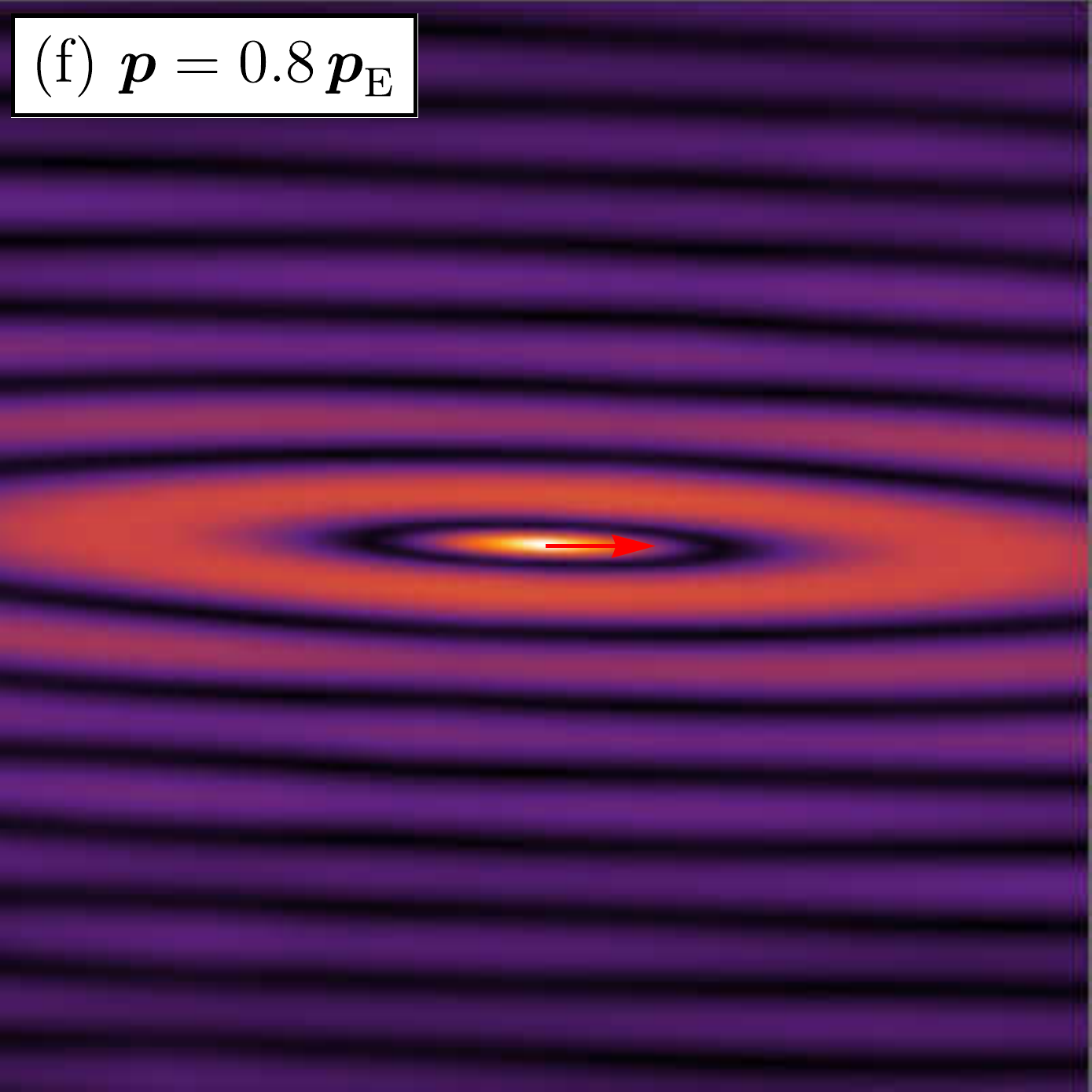}
\end{subfigure}
\begin{subfigure}{0.24\textwidth}
\centering
\phantomsubcaption{\label{fig:rhombus_10_10_force_h_90_gf}}
\includegraphics[width=0.98\linewidth]{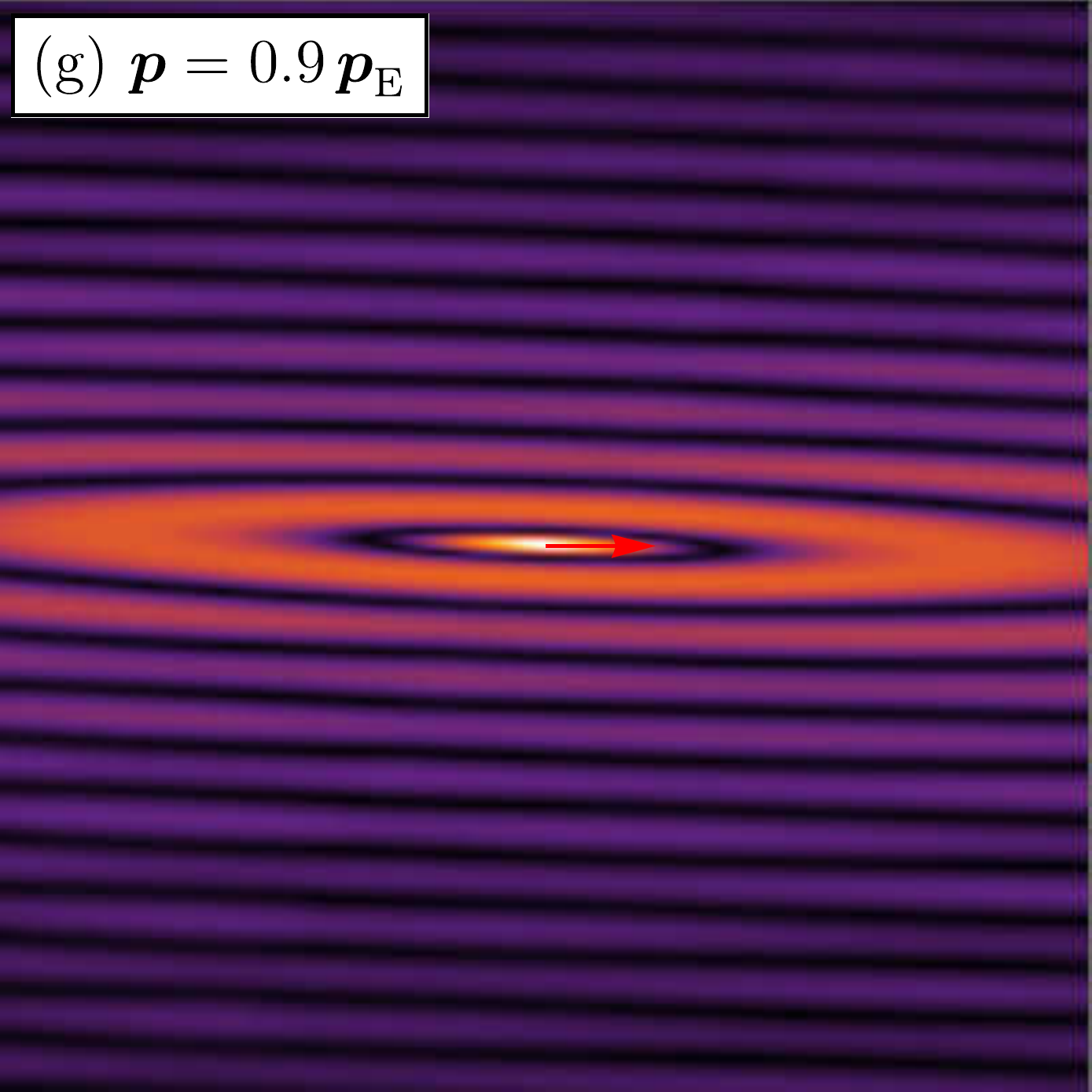}
\end{subfigure}
\begin{subfigure}{0.24\textwidth}
\centering
\phantomsubcaption{\label{fig:rhombus_10_10_force_h_99_gf}}
\includegraphics[width=0.98\linewidth]{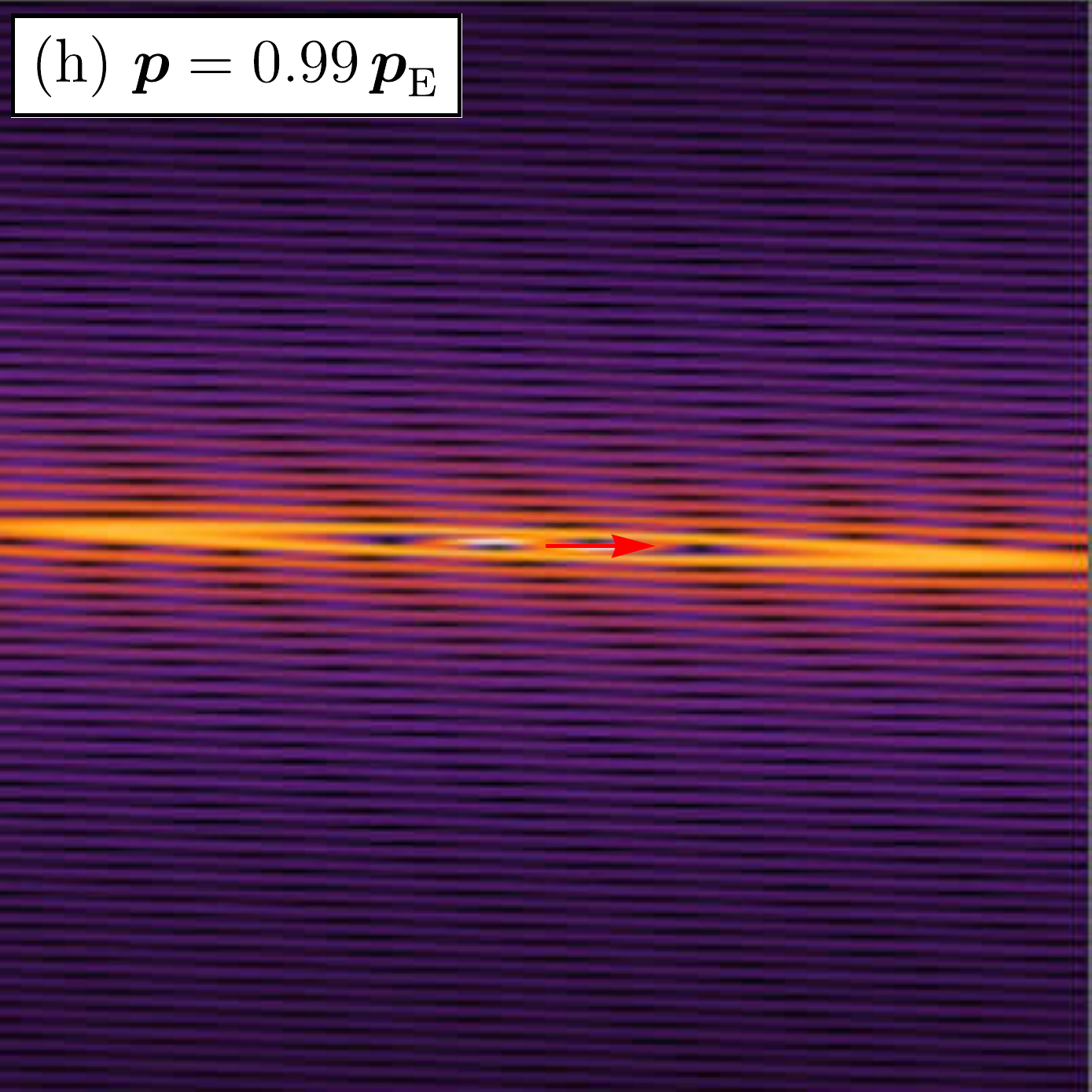}
\end{subfigure}\\
\vspace{0.015\linewidth}
\begin{subfigure}{0.24\textwidth}
\centering
\phantomsubcaption{\label{fig:rhombus_10_10_force_v_0}}
\includegraphics[width=0.98\linewidth]{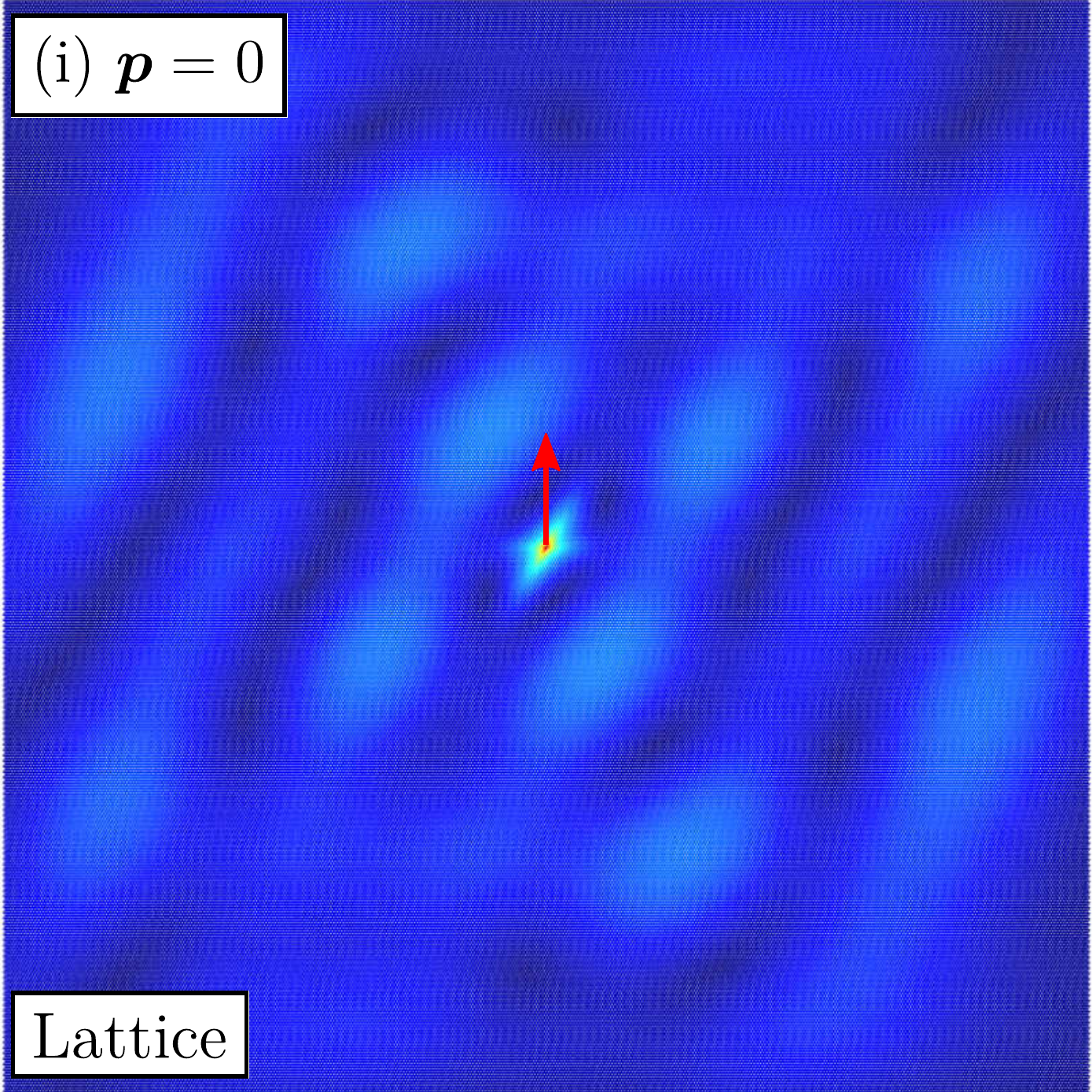}
\end{subfigure}
\begin{subfigure}{0.24\textwidth}
\centering
\phantomsubcaption{\label{fig:rhombus_10_10_force_v_80}}
\includegraphics[width=0.98\linewidth]{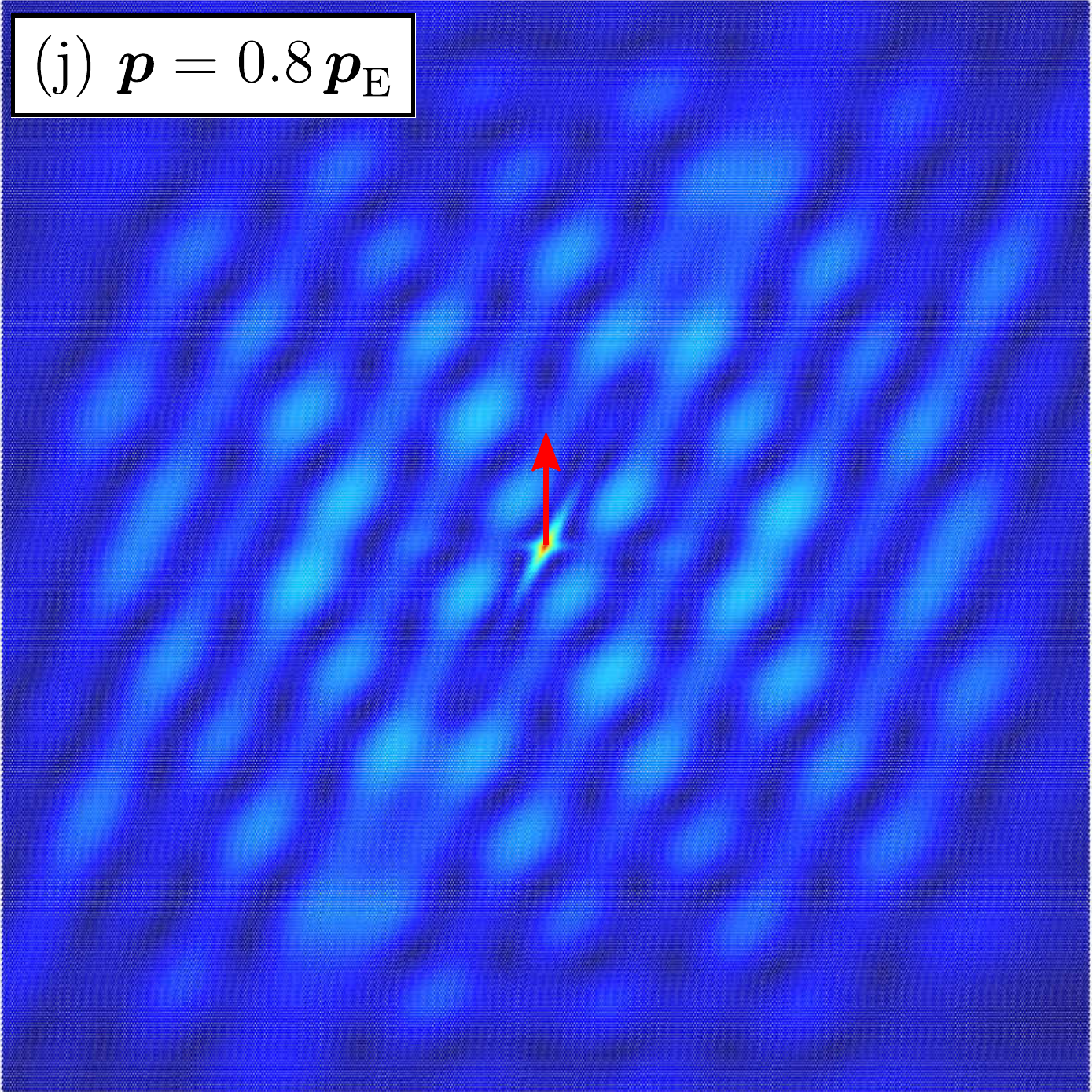}
\end{subfigure}
\begin{subfigure}{0.24\textwidth}
\centering
\phantomsubcaption{\label{fig:rhombus_10_10_force_v_90}}
\includegraphics[width=0.98\linewidth]{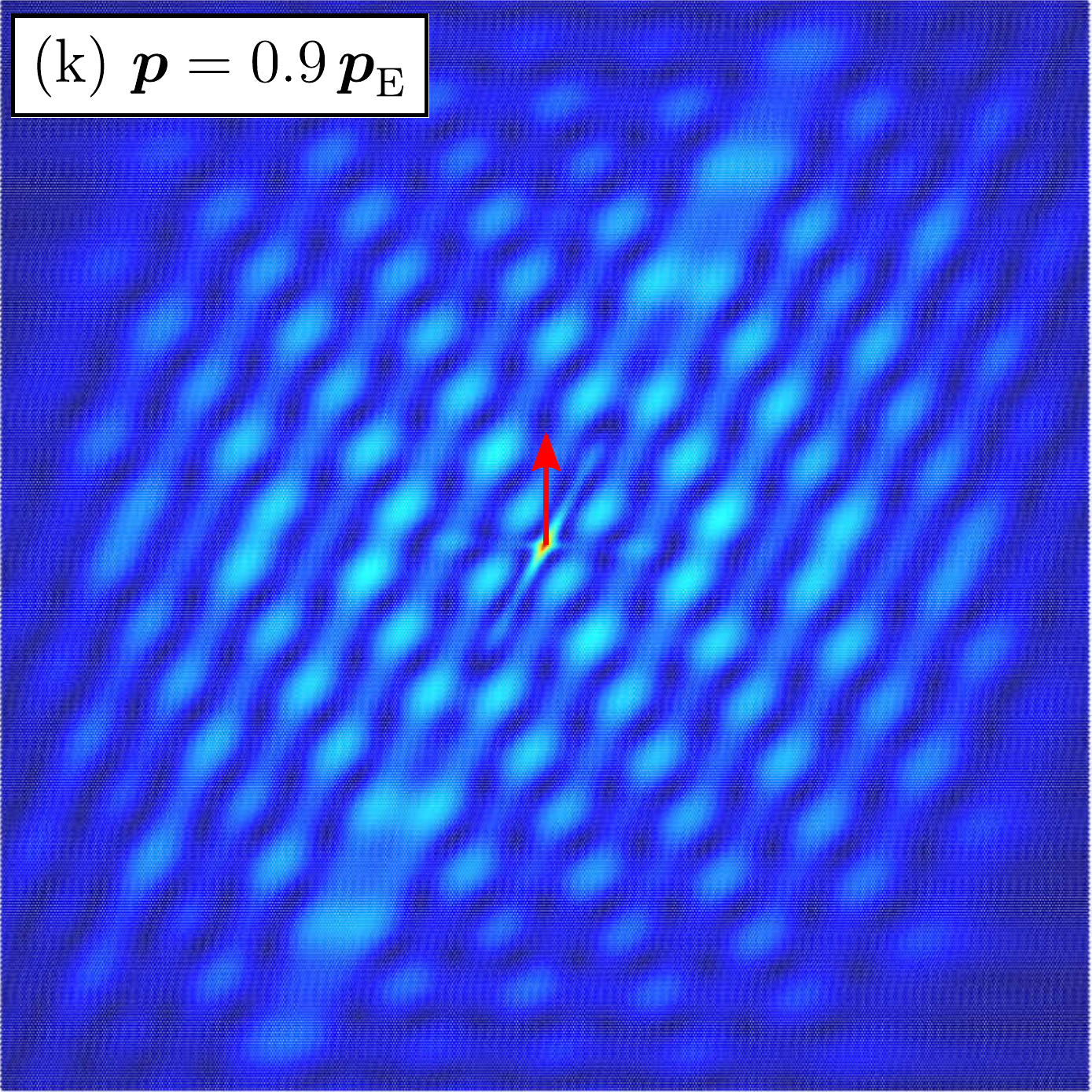}
\end{subfigure}
\begin{subfigure}{0.24\textwidth}
\centering
\phantomsubcaption{\label{fig:rhombus_10_10_force_v_99}}
\includegraphics[width=0.98\linewidth]{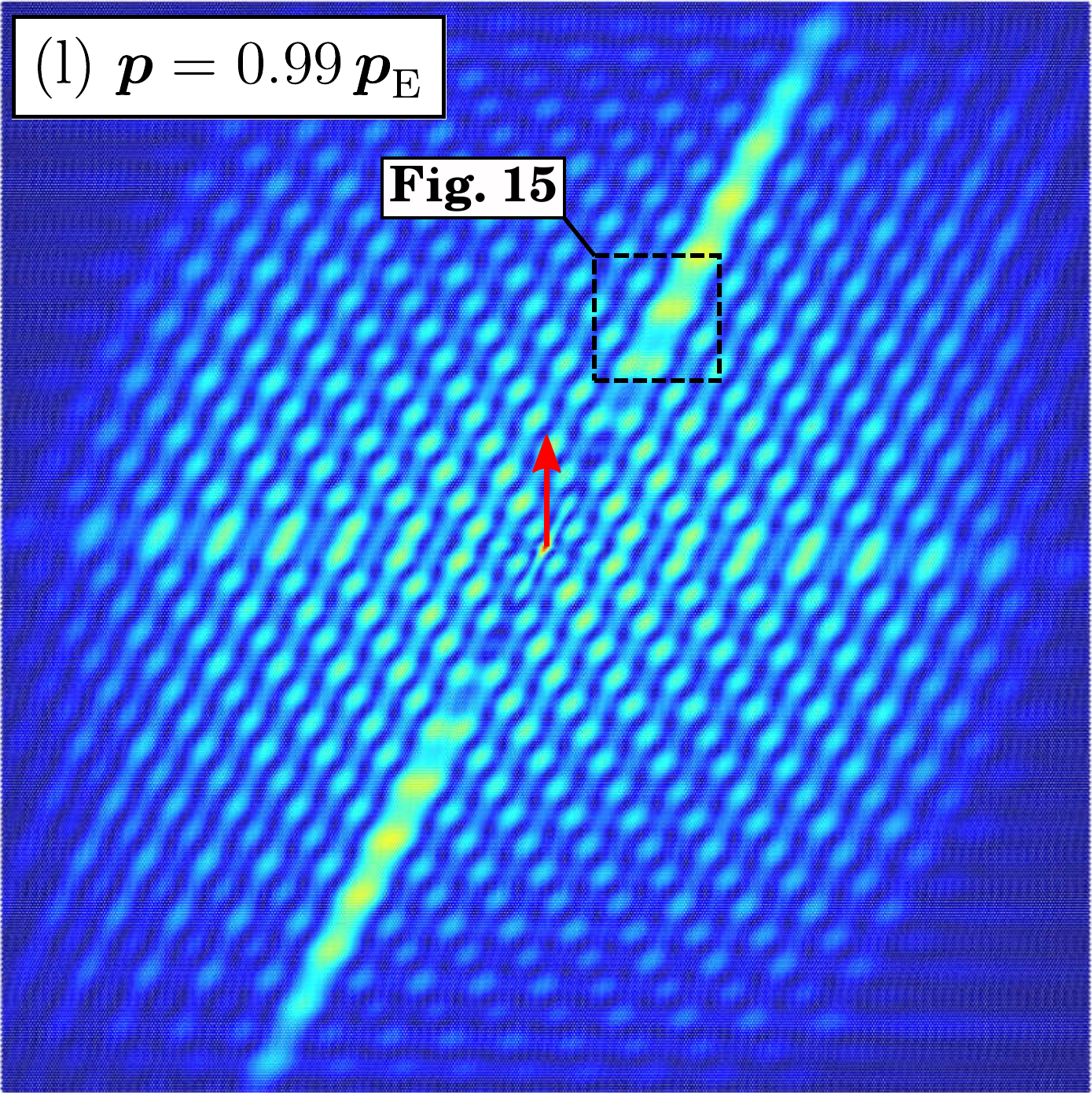}
\end{subfigure}\\
\vspace{0.01\linewidth}
\begin{subfigure}{0.24\textwidth}
\centering
\phantomsubcaption{\label{fig:rhombus_10_10_force_v_0_gf}}
\includegraphics[width=0.98\linewidth]{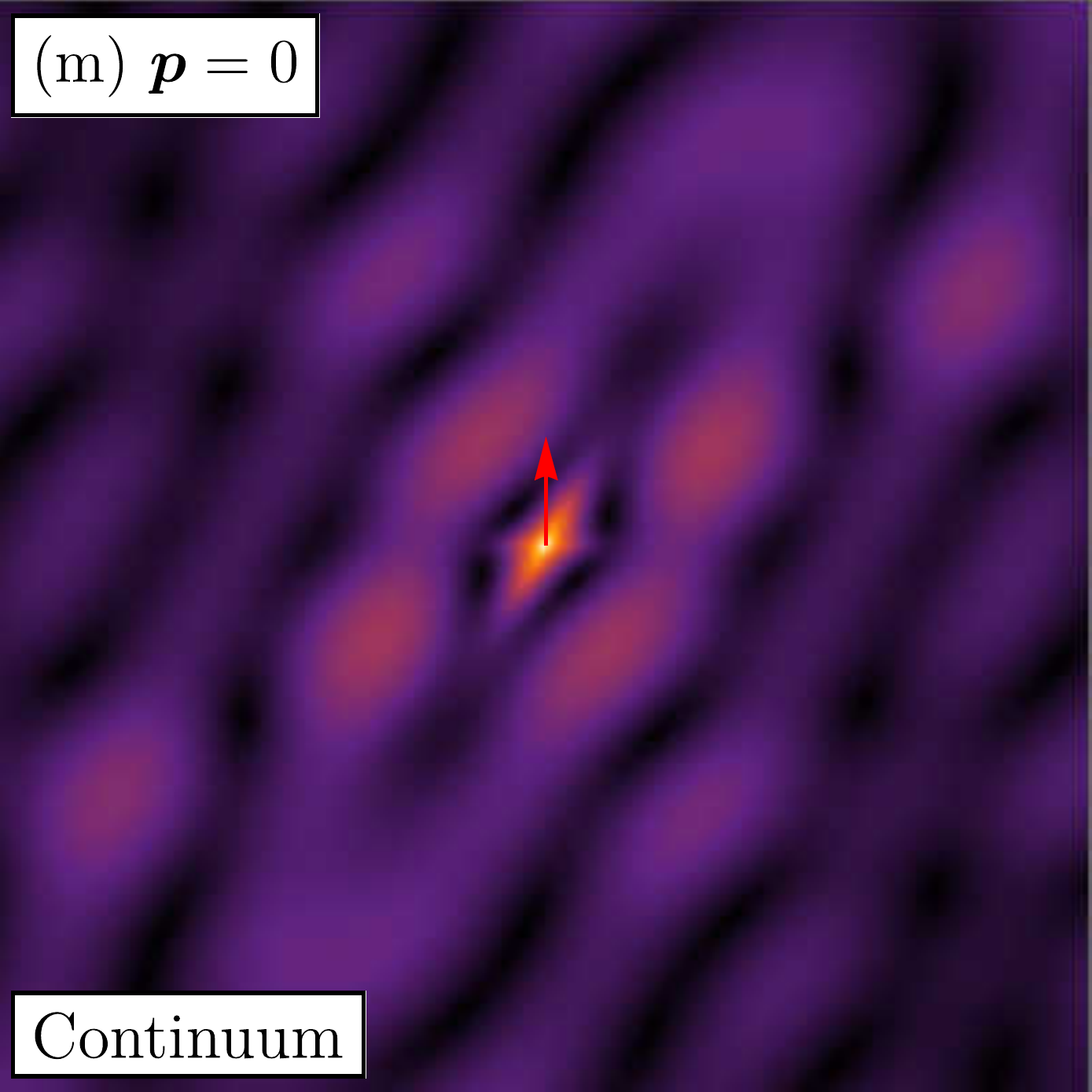}
\end{subfigure}
\begin{subfigure}{0.24\textwidth}
\centering
\phantomsubcaption{\label{fig:rhombus_10_10_force_v_80_gf}}
\includegraphics[width=0.98\linewidth]{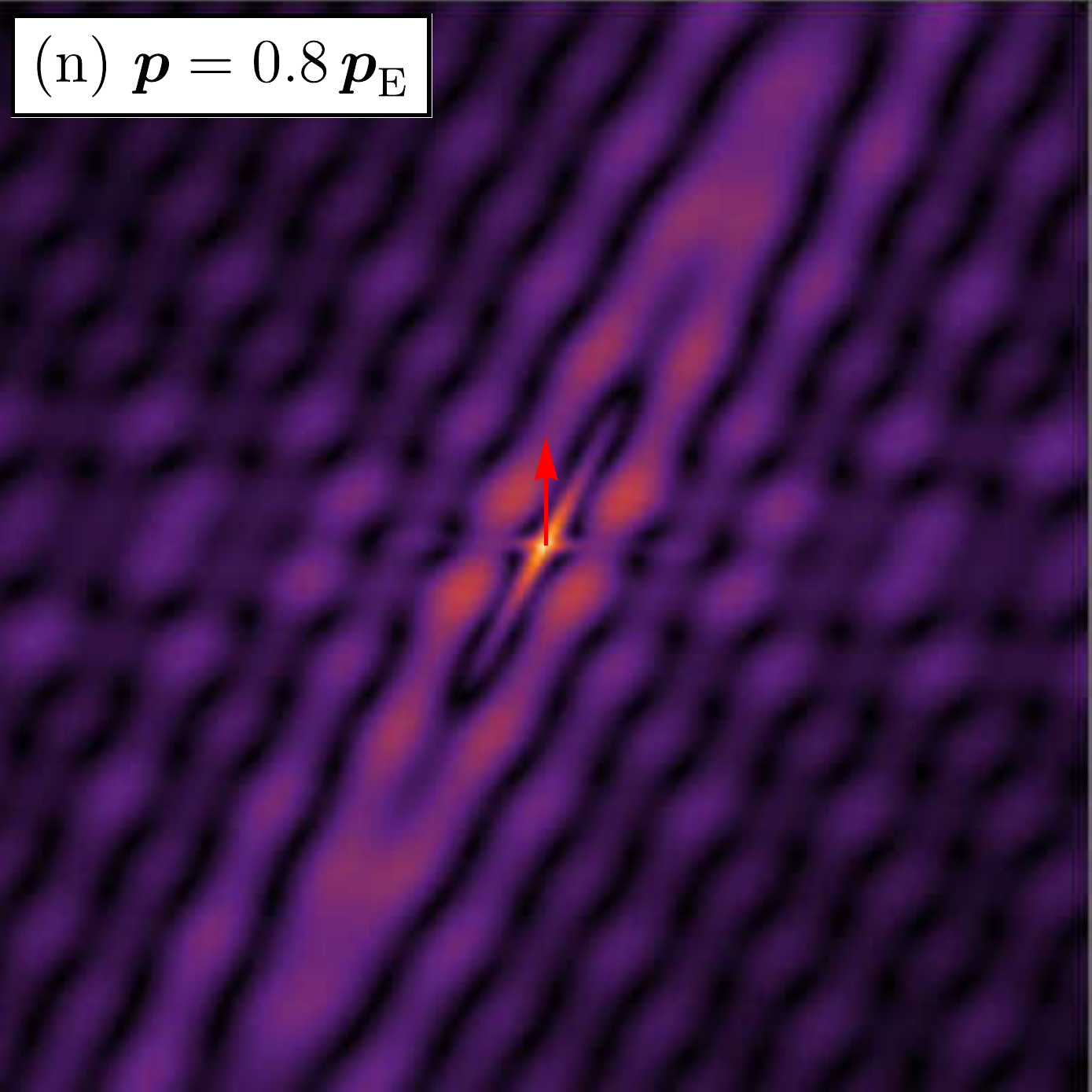}
\end{subfigure}
\begin{subfigure}{0.24\textwidth}
\centering
\phantomsubcaption{\label{fig:rhombus_10_10_force_v_90_gf}}
\includegraphics[width=0.98\linewidth]{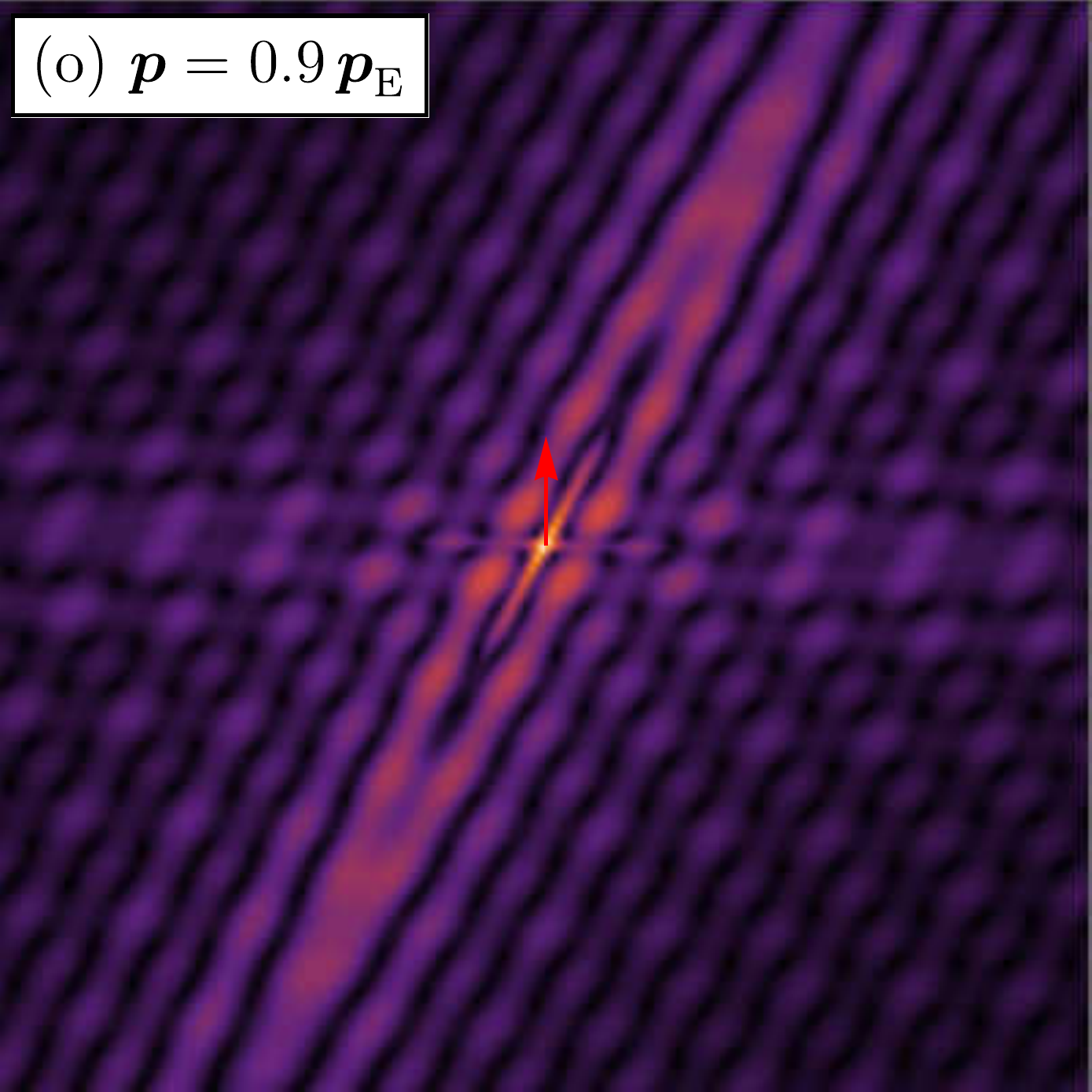}
\end{subfigure}
\begin{subfigure}{0.24\textwidth}
\centering
\phantomsubcaption{\label{fig:rhombus_10_10_force_v_99_gf}}
\includegraphics[width=0.98\linewidth]{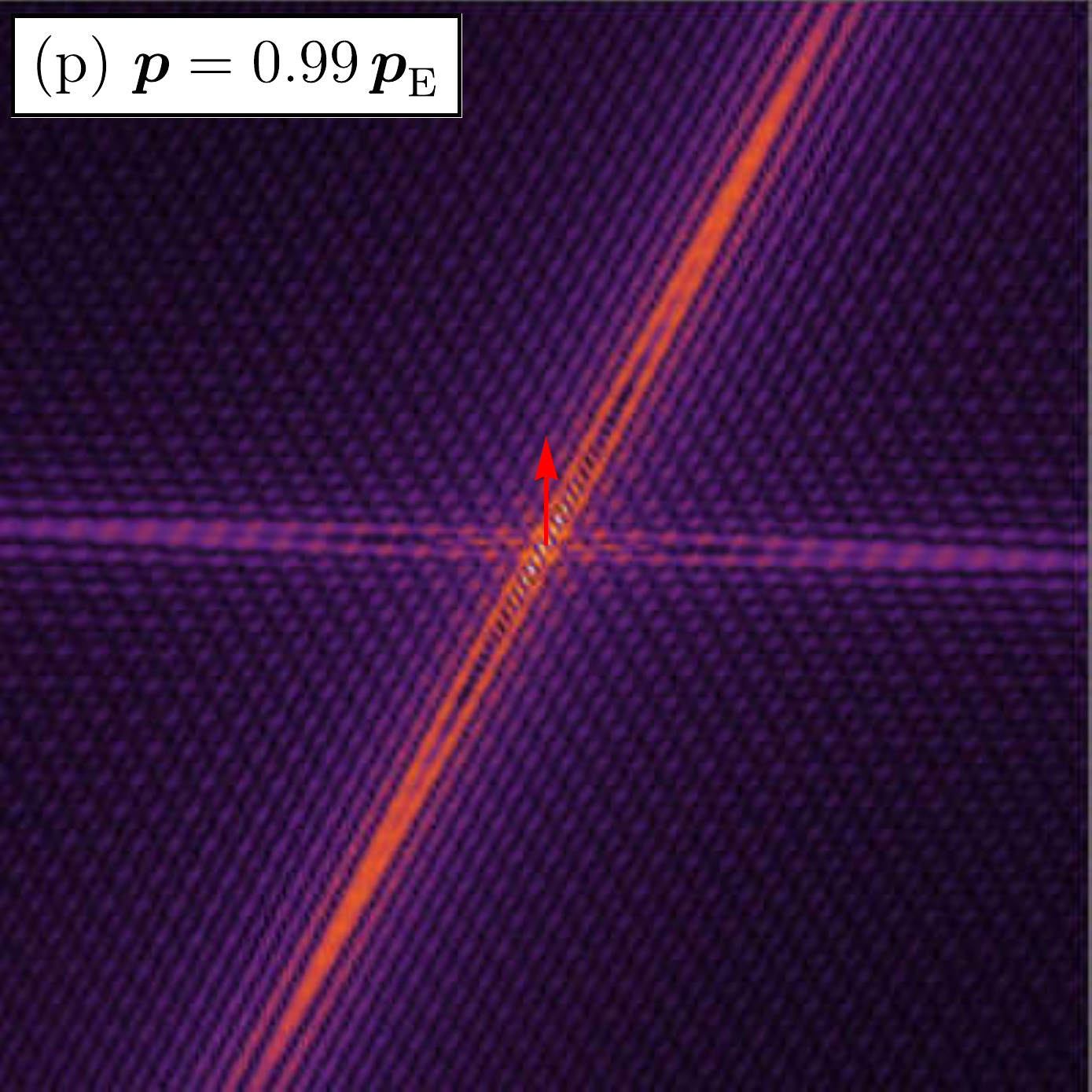}
\end{subfigure}
\caption{\label{fig:rhombus_10_10_force}
    The displacement field generated by a pulsating \textit{horizontal} force 
    (denoted with a red arrow and 
    applied to the orthotropic rhombic lattice, $\Lambda_1=\Lambda_2=10$) is simulated via f.e.m., see (\subref{fig:rhombus_10_10_force_h_0})--(\subref{fig:rhombus_10_10_force_h_99}), and compared to the response of the homogenized continuum, see (\subref{fig:rhombus_10_10_force_h_0_gf})--(\subref{fig:rhombus_10_10_force_h_99_gf}), at different levels of prestress $\bp$ (preload path shown in Fig.~\ref{fig:ellipticity_domains_10_10}).
    The same comparison is reported in (\subref{fig:rhombus_10_10_force_v_0})--(\subref{fig:rhombus_10_10_force_v_99}) and (\subref{fig:rhombus_10_10_force_v_0_gf})--(\subref{fig:rhombus_10_10_force_v_99_gf}) for a pulsating \textit{vertical} force.
    Even though ellipticity is lost along the two directions predicted in Fig.~\ref{fig:eigenvalue_rhombus_10_10}, the activation of strain localization depends on the orientation of the load, so that two bands are activated by the vertical force, while only one is generated by the horizontal load.
    Furthermore, note that the directions of the localization bands (with angles of the normal equal to $\theta_{\text{cr}}=88.2^\circ,151.8^\circ$) is slightly misaligned with respect to the rod's inclination.
}
\end{figure}

In order to investigate the response of the orthotropic ($\Lambda_1=\Lambda_2=10$) and anisotropic ($\Lambda_1=7,\,\Lambda_2=15$) rhombic lattices ($\alpha=\pi/3$), both horizontal and vertical concentrated forces will be considered, so to observe a dependence of the number of strain localizations on the loading orientation.
Furthermore, in contrast to what happens in the case of the square grid, the directions of localization are expected to occur with a slight misalignment with respect to the directions of the rods, as predicted in Figs.~\ref{fig:eigenvalue_rhombus_10_10} and~\ref{fig:eigenvalue_rhombus_7_15}.

In Fig.~\ref{fig:rhombus_10_10_force} the displacement field computed via f.e.m. for the orthotropic rhombic lattice (horizontally and vertically loaded with a pulsating force and reported on first and third row from the top of the figure) is compared to the response of the homogenized continuum (reported in the second and fourth row) at four values of preload (increasing from left to right) $p_1=p_2=\{0,-4.276,-4.811,-5.292\}$.

A comparison between Figs.~\ref{fig:rhombus_10_10_force_h_0}--\ref{fig:rhombus_10_10_force_h_99} and Figs.~\ref{fig:rhombus_10_10_force_h_0_gf}--\ref{fig:rhombus_10_10_force_h_99_gf} and a comparison between Figs.~\ref{fig:rhombus_10_10_force_v_0}--\ref{fig:rhombus_10_10_force_v_99} and Figs.~\ref{fig:rhombus_10_10_force_v_0_gf}--\ref{fig:rhombus_10_10_force_v_99_gf}, shows an excellent agreement between the lattice response and its homogenized continuum counterpart, for each state of lattice's preload.
With reference to a prestress state $\bp=0.99\,\bp_{\text{E}}$ (last column on the right of the figure), while two localization bands are activated by the vertical force, only one is generated by the horizontal force.

Note also that a slight misalignment between the localization direction and the rod angle remains hardly visible \textit{until the material is close to elliptic boundary} (compare for example the case of vanishing prestress, Fig.~\ref{fig:rhombus_10_10_force_h_0}, to the case $\bp=0.99\,\bp_{\text{E}}$, Fig.~\ref{fig:rhombus_10_10_force_h_99}).
\begin{figure}[htb!]
\centering
\begin{subfigure}{0.24\textwidth}
\centering
\caption*{$t=0$}
\includegraphics[width=0.98\linewidth]{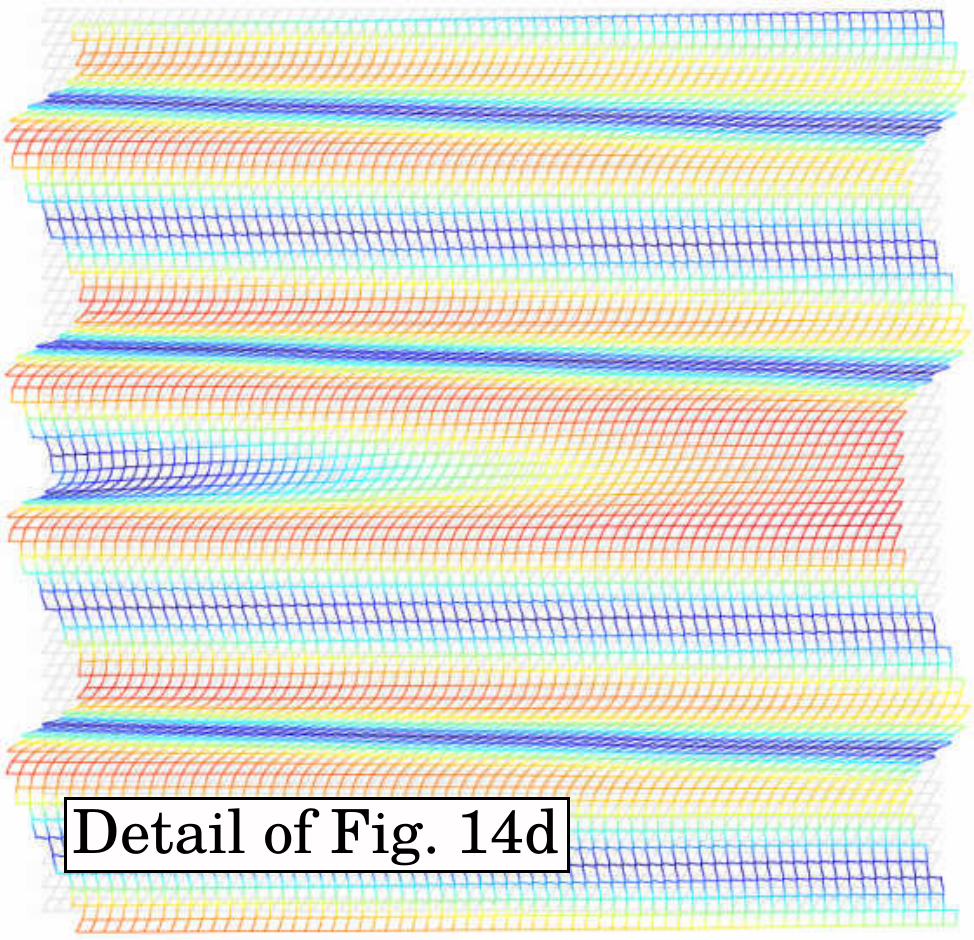}
\end{subfigure}
\begin{subfigure}{0.24\textwidth}
\centering
\caption*{$t=\frac{\pi}{2\omega}$}
\includegraphics[width=0.98\linewidth]{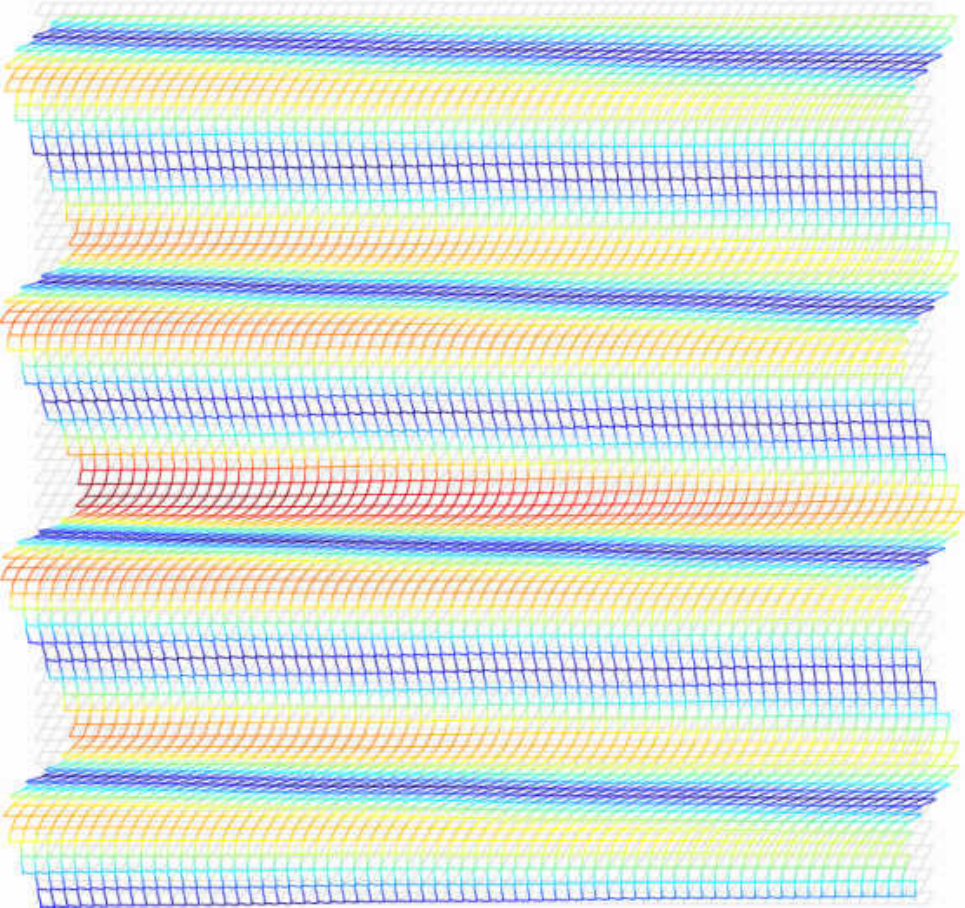}
\end{subfigure}
\begin{subfigure}{0.24\textwidth}
\centering
\caption*{$t=\frac{\pi}{\omega}$}
\includegraphics[width=0.98\linewidth]{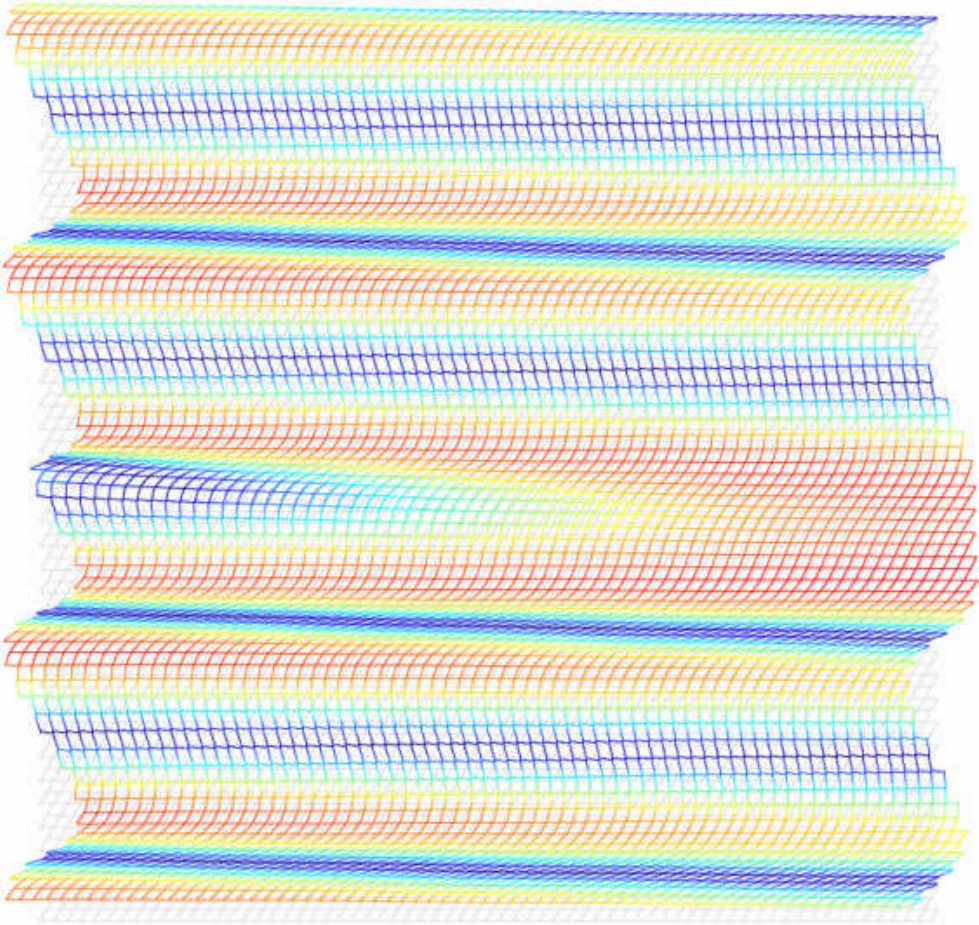}
\end{subfigure}
\begin{subfigure}{0.24\textwidth}
\centering
\caption*{$t=\frac{3\pi}{2\omega}$}
\includegraphics[width=0.98\linewidth]{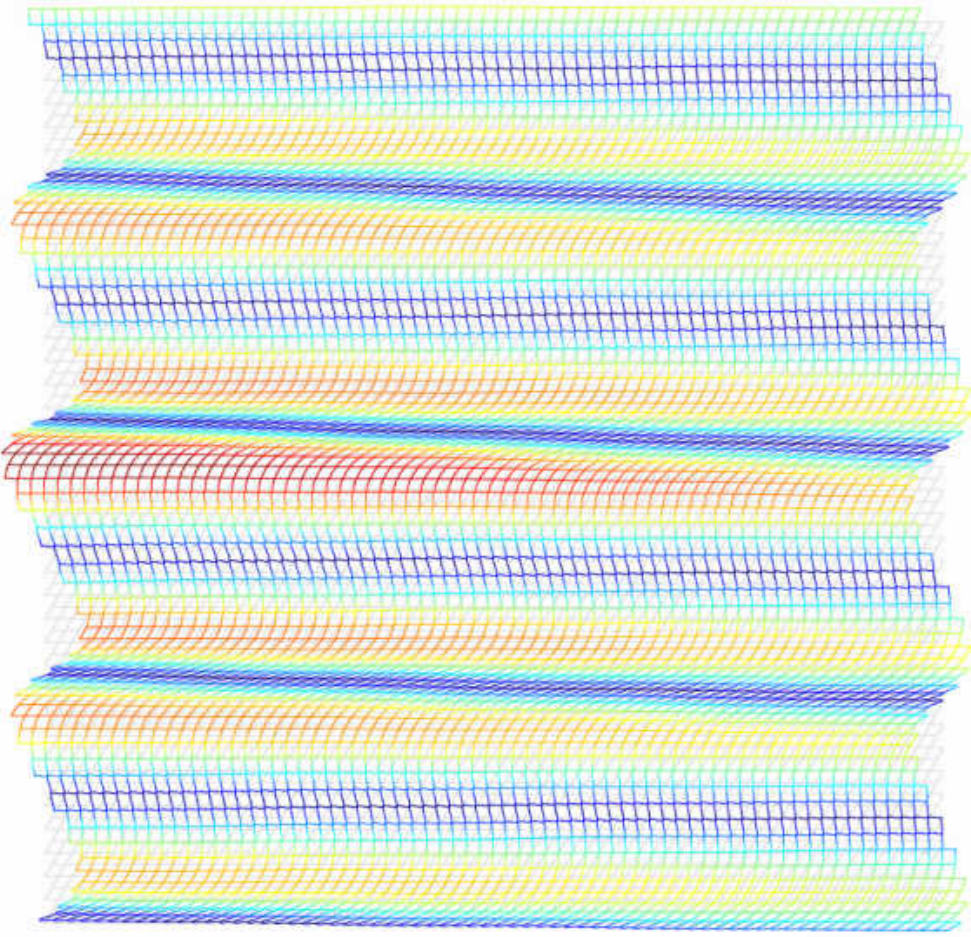}
\end{subfigure}\\
\vspace{1mm}
\begin{subfigure}{0.24\textwidth}
\centering
\includegraphics[width=0.98\linewidth]{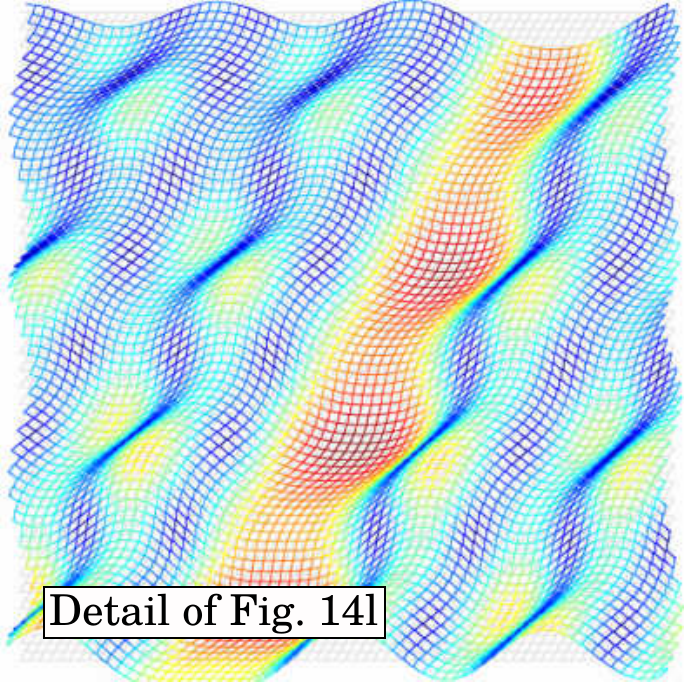}
\end{subfigure}
\begin{subfigure}{0.24\textwidth}
\centering
\includegraphics[width=0.98\linewidth]{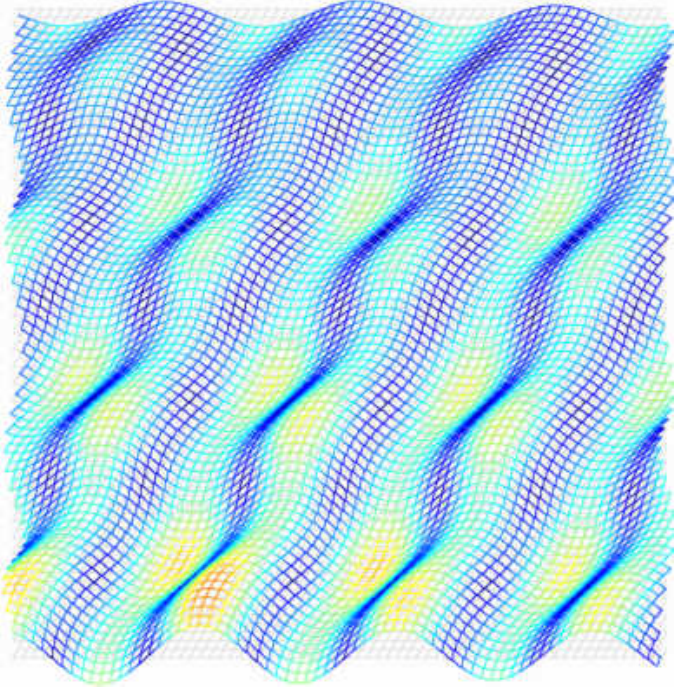}
\end{subfigure}
\begin{subfigure}{0.24\textwidth}
\centering
\includegraphics[width=0.98\linewidth]{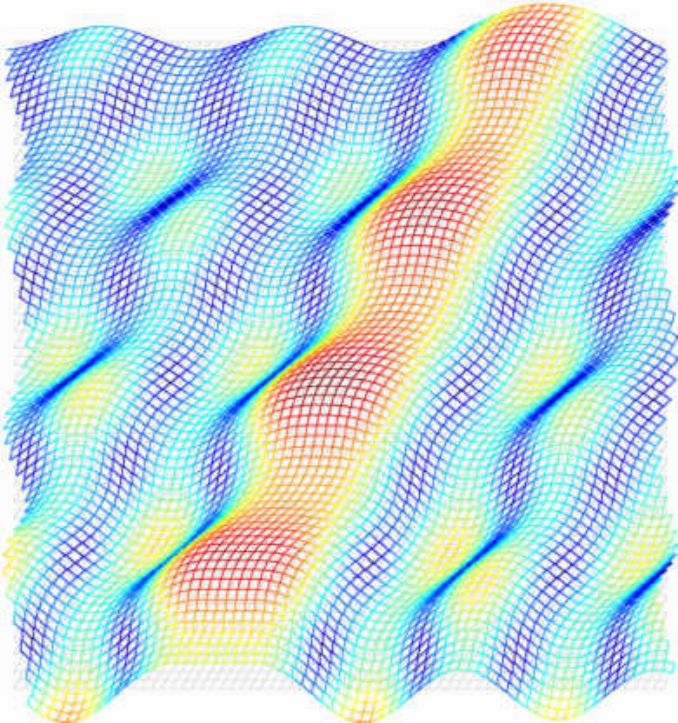}
\end{subfigure}
\begin{subfigure}{0.24\textwidth}
\centering
\includegraphics[width=0.98\linewidth]{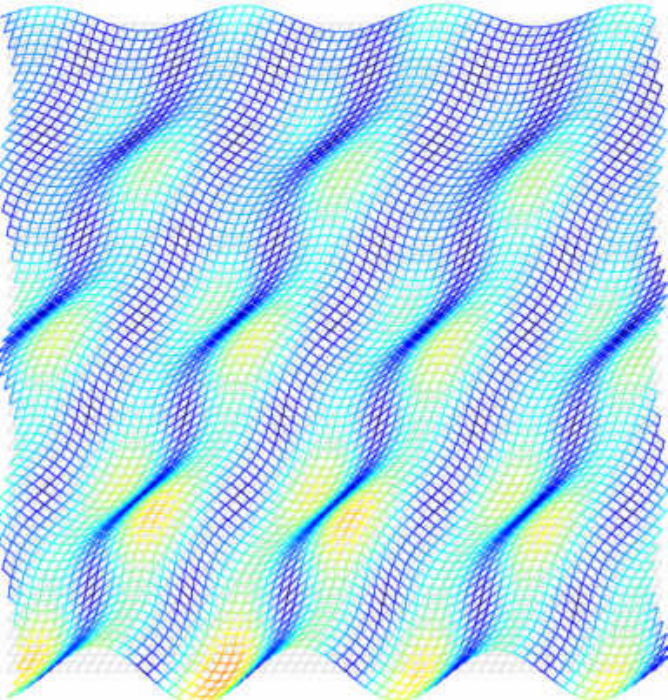}
\end{subfigure}
\caption{\label{fig:rhombus_10_10_deformed}
    Deformed configurations of the rhombic lattice ($\Lambda_1=\Lambda_2=10$) near (the zones are indicated in Fig. \ref{fig:rhombus_10_10_force} \subref{fig:rhombus_10_10_force_h_99} and \subref{fig:rhombus_10_10_force_v_99}) the point of application of a pulsating concentrated force, at a level of prestress close to the elliptic boundary ($\bp=0.99\,\bp_{\text{E}}$).
    The pattern on the first (the second) row shows the motion of the localization induced by the pulsating horizontal (pulsating vertical) load.
    Note that the effect of the load is the generation of bands  almost parallel to each other and possessing almost constant amplitude, except for a modulation of the inclined localization (shown in the lower row).
    In both cases the deformation mode is mostly of shear-type even though an `expansion' component is also present, as indicated by the vectors $\bg_{\text{E}}$,  Fig.~\ref{fig:eigenvalue_rhombus_10_10}.
}
\end{figure}

The localization modes are analyzed in Fig.~\ref{fig:rhombus_10_10_deformed} by inspecting the lattice deformation computed via f.e.m. at different temporal instants through snapshots taken in the neighbourhood of the loading point [two zones are considered, which do not include the concentrated force and are shown in Fig. \ref{fig:rhombus_10_10_force} (\subref{fig:rhombus_10_10_force_h_99}) and (\subref{fig:rhombus_10_10_force_v_99})].

A comparison between the localization band induced by the horizontal load (upper row of Fig.~\ref{fig:rhombus_10_10_deformed}) and that generated by the vertical force (second row in Fig.~\ref{fig:rhombus_10_10_deformed}) shows that the (almost) horizontal band is characterized by an almost perfectly straight wavefront, while the inclined band displays a \textit{periodic modulation along the front}.
This modulation is due to the superposition of the two localization patterns that are activated by the vertical force, where the inclined band prevails on the almost horizontal one, as can be seen in Figs.~\ref{fig:rhombus_10_10_force_v_99} and~\ref{fig:rhombus_10_10_force_v_99_gf}.
\begin{figure}[htb!]
\centering
\begin{subfigure}{0.4\textwidth}
\centering
\caption{\label{fig:rhombus_10_10_force_h_FFT}Horizontal loading}
\includegraphics[width=0.98\linewidth]{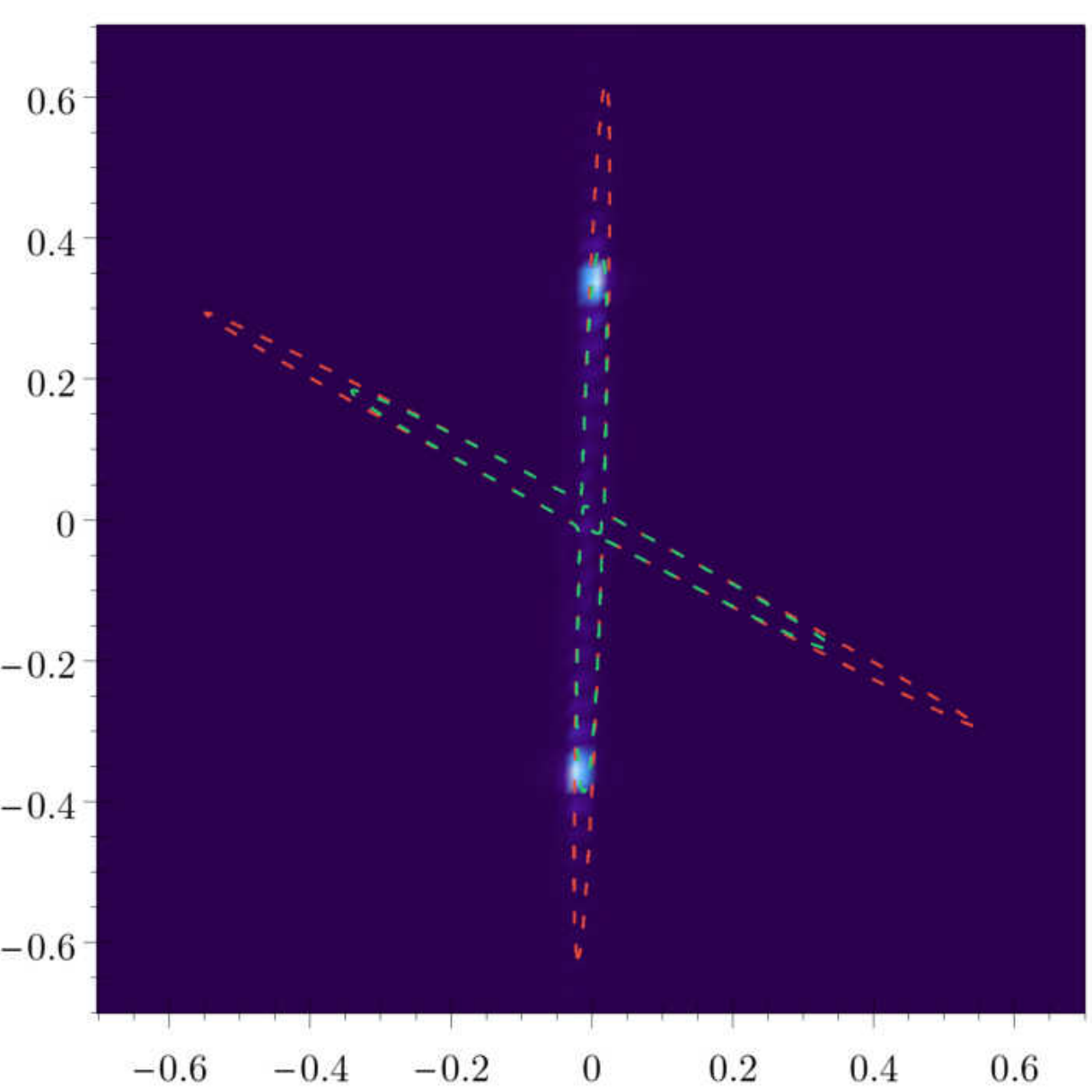}
\end{subfigure} \hspace{2mm}
\begin{subfigure}{0.4\textwidth}
\centering
\caption{\label{fig:rhombus_10_10_force_v_FFT}Vertical loading}
\includegraphics[width=0.98\linewidth]{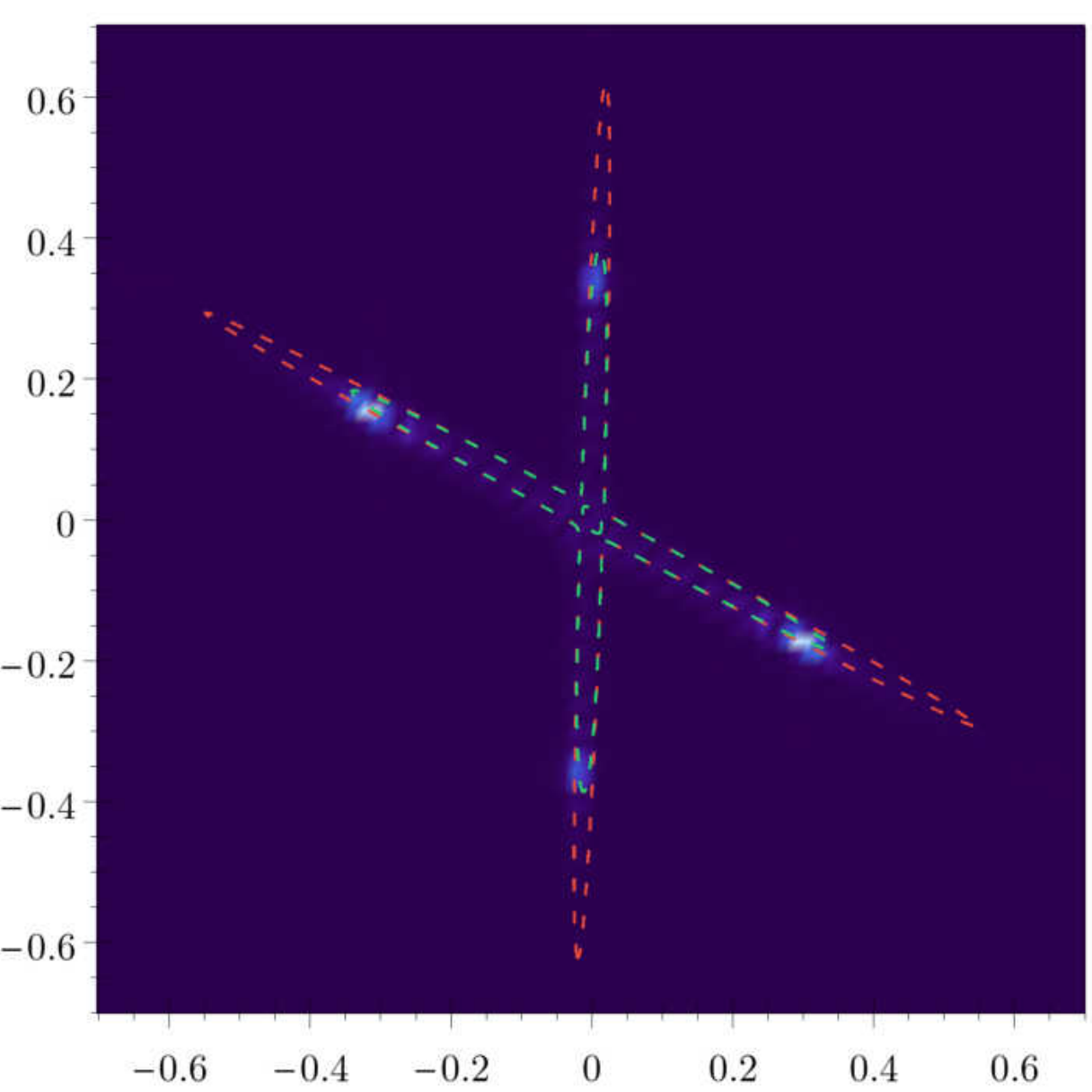}
\end{subfigure}
\caption{\label{fig:rhombus_10_10_force_FFT}
    Fourier transform of the complex displacement fields of the rhombic lattice (reported in Fig. \ref{fig:rhombus_10_10_force}  \subref{fig:rhombus_10_10_force_h_99} and \subref{fig:rhombus_10_10_force_v_99}, with $\Lambda_1=\Lambda_2=10$) subject to an horizontal~(\subref{fig:rhombus_10_10_force_h_FFT}) and vertical~(\subref{fig:rhombus_10_10_force_v_FFT}) pulsating force applied at a prestress close to the elliptic boundary, $\bp=0.99\,\bp_{\text{E}}$. 
    The slowness contours of the lattice (dashed green) and of the effective continuum (dashed red) are superimposed to highlight the Bloch spectrum of waves excited by the forcing source.
    The sharp peaks in the contours are aligned to the directions of ellipticity loss as predicted in Fig.~\ref{fig:eigenvalue_rhombus_10_10}.
    In~(\subref{fig:rhombus_10_10_force_h_FFT}) the horizontal concentrated force only activates waves propagating at $\theta_{\text{cr}}=88.2^\circ$, while in~(\subref{fig:rhombus_10_10_force_v_FFT}) the vertical concetrated force generates four peaks along the directions $\theta_{\text{cr}}=88.2^\circ, 151.8^\circ$.
    This is in excellent agreement with the responses reported in Figs.~\ref{fig:rhombus_10_10_force_h_99} and~\ref{fig:rhombus_10_10_force_v_99}.
}
\end{figure}

The relative contribution of the two localizations can be further investigated through a Fourier transform of the lattice response, to be compared with the Bloch spectrum generated by the forcing source.
Fig.~\ref{fig:rhombus_10_10_force_FFT} shows the Fourier transform of the field generated in the rhombic grid when the material is close to the elliptic boundary ($\bp=0.99\,\bp_{\text{E}}$).
Fig.~\ref{fig:rhombus_10_10_force_h_FFT} and~\ref{fig:rhombus_10_10_force_v_FFT} correspond, respectively, to the Fourier transform of Fig.~\ref{fig:rhombus_10_10_force_h_99} and Fig.~\ref{fig:rhombus_10_10_force_v_99}.
The two sharp peaks of Fig.~\ref{fig:rhombus_10_10_force_h_FFT} clearly show that the source is emanating pure plane waves propagating almost vertically ($\theta_{\text{cr}}=88.2^\circ$) (the slight tilt exactly matches the sub-horizontal wavefronts of the response).
Instead, the four peaks of Fig.~\ref{fig:rhombus_10_10_force_v_FFT} demonstrate that two families of plane waves are activated: the prevailing ones propagate along the inclined direction ($\theta_{\text{cr}}=151.8^\circ$) while the vertically-propagating waves result dimmer (in agreement with the response of Fig.~\ref{fig:rhombus_10_10_force_v_99}).

\begin{figure}[htb!]
\centering
\begin{subfigure}{0.24\textwidth}
\centering
\phantomsubcaption{\label{fig:rhombus_7_15_force_h_0}}
\includegraphics[width=0.98\linewidth]{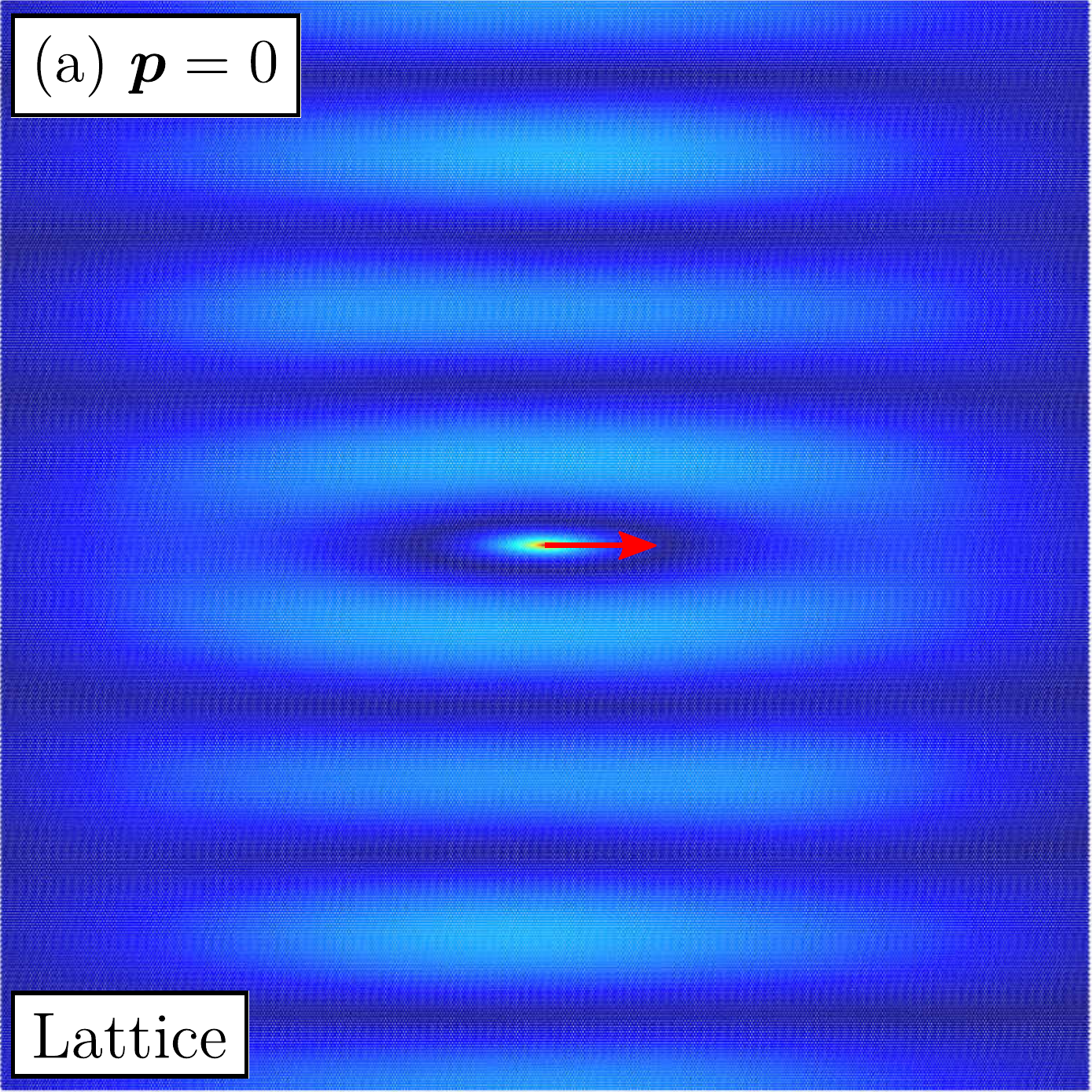}
\end{subfigure}
\begin{subfigure}{0.24\textwidth}
\centering
\phantomsubcaption{\label{fig:rhombus_7_15_force_h_80}}
\includegraphics[width=0.98\linewidth]{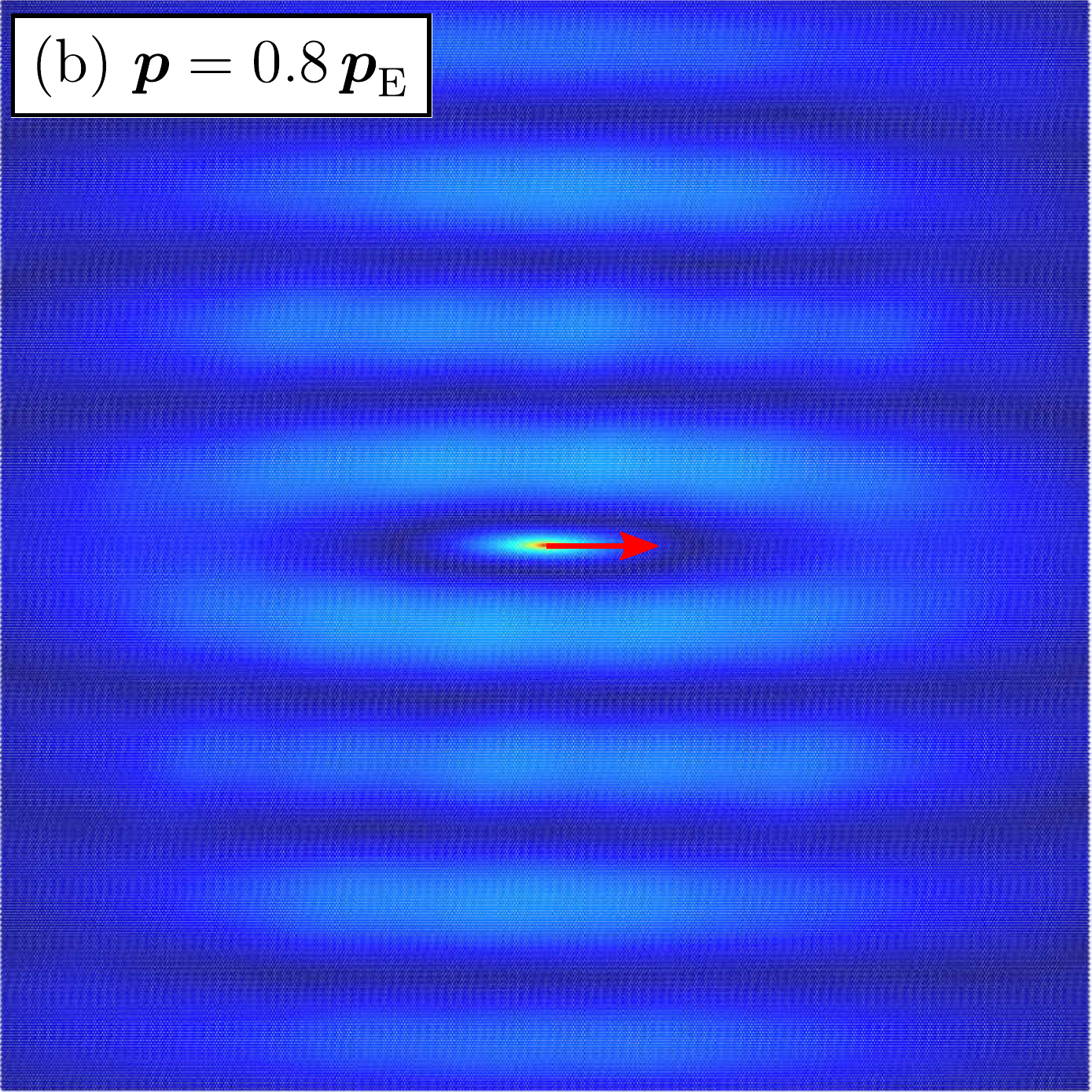}
\end{subfigure}
\begin{subfigure}{0.24\textwidth}
\centering
\phantomsubcaption{\label{fig:rhombus_7_15_force_h_90}}
\includegraphics[width=0.98\linewidth]{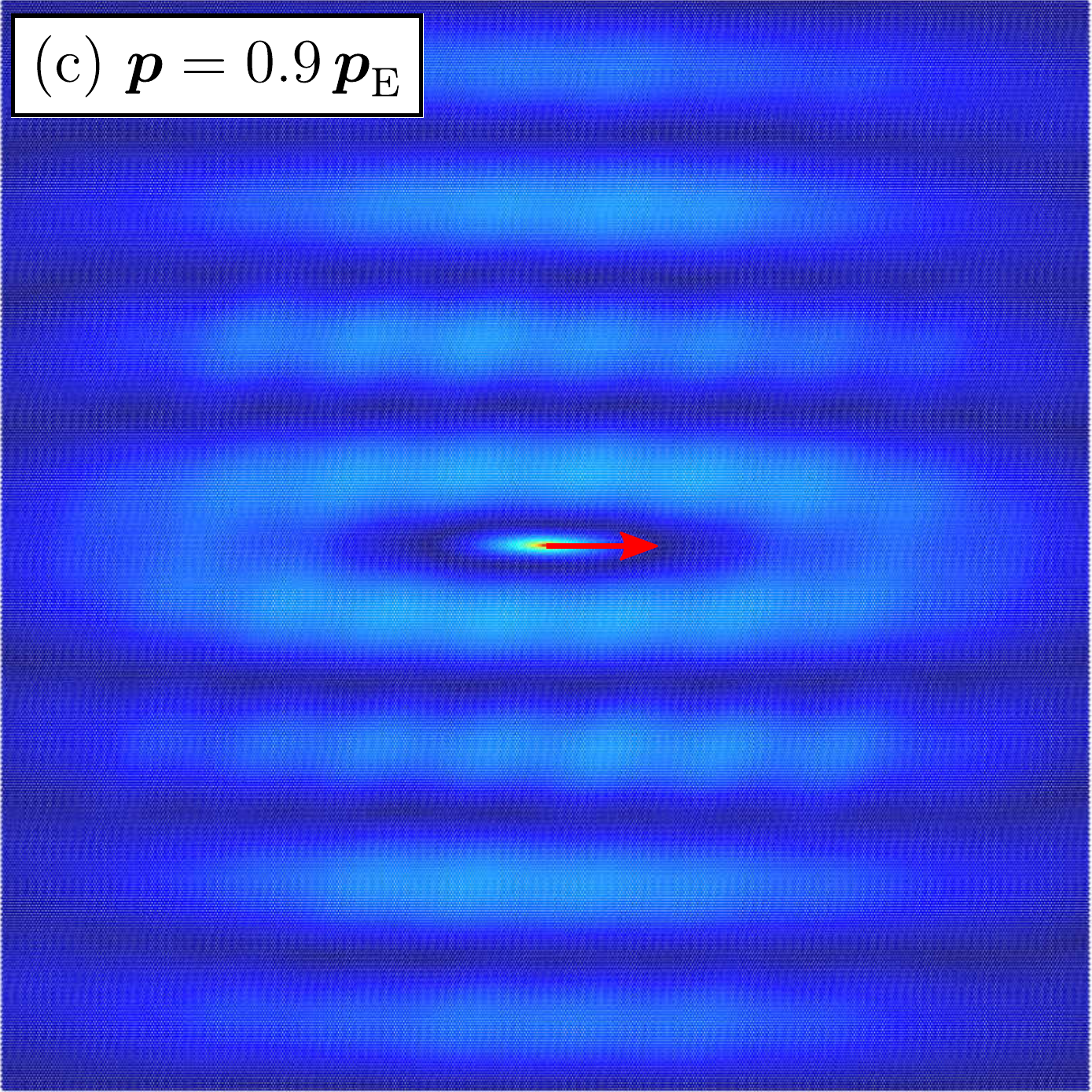}
\end{subfigure}
\begin{subfigure}{0.24\textwidth}
\centering
\phantomsubcaption{\label{fig:rhombus_7_15_force_h_99}}
\includegraphics[width=0.98\linewidth]{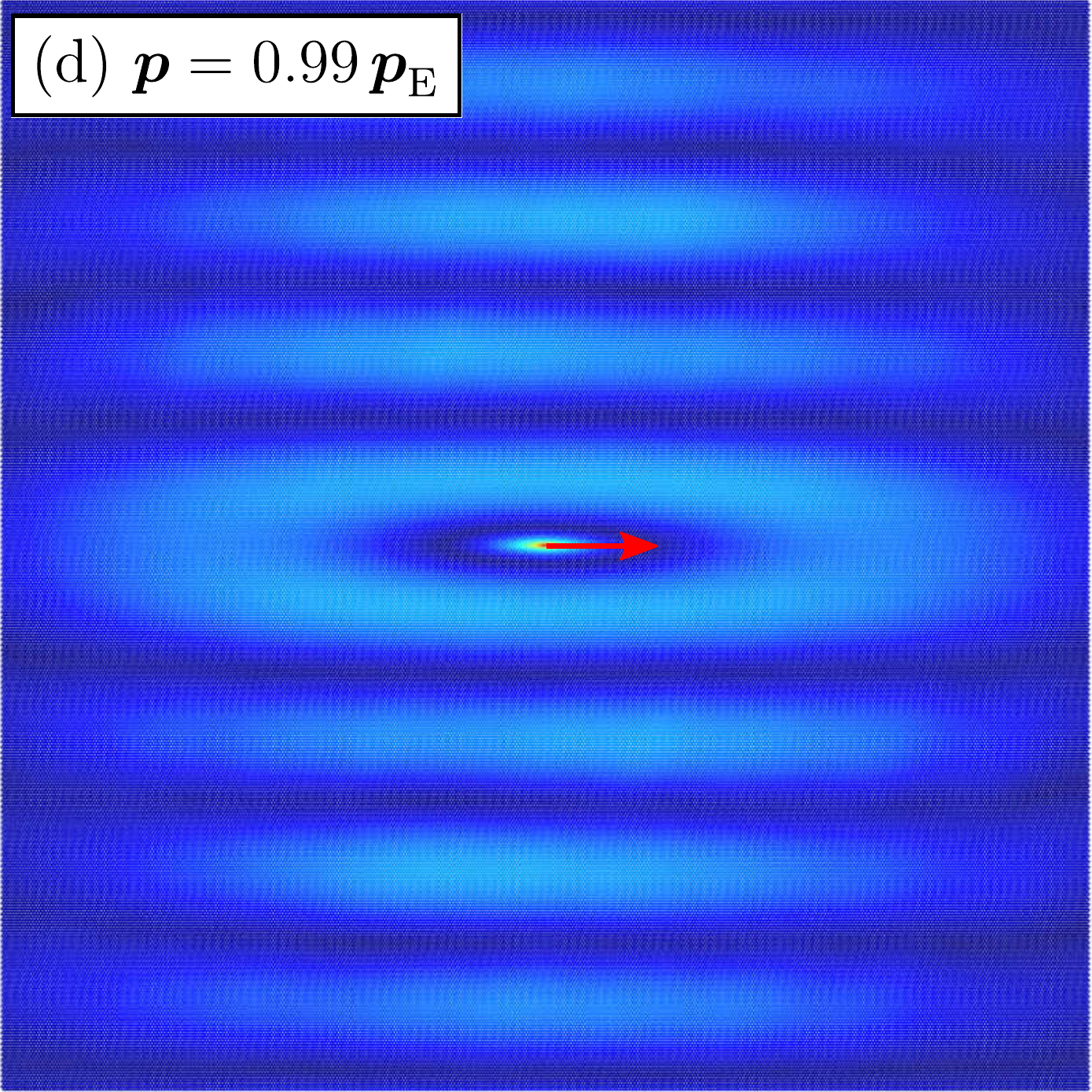}
\end{subfigure}\\
\vspace{0.01\linewidth}
\begin{subfigure}{0.24\textwidth}
\centering
\phantomsubcaption{\label{fig:rhombus_7_15_force_h_0_gf}}
\includegraphics[width=0.98\linewidth]{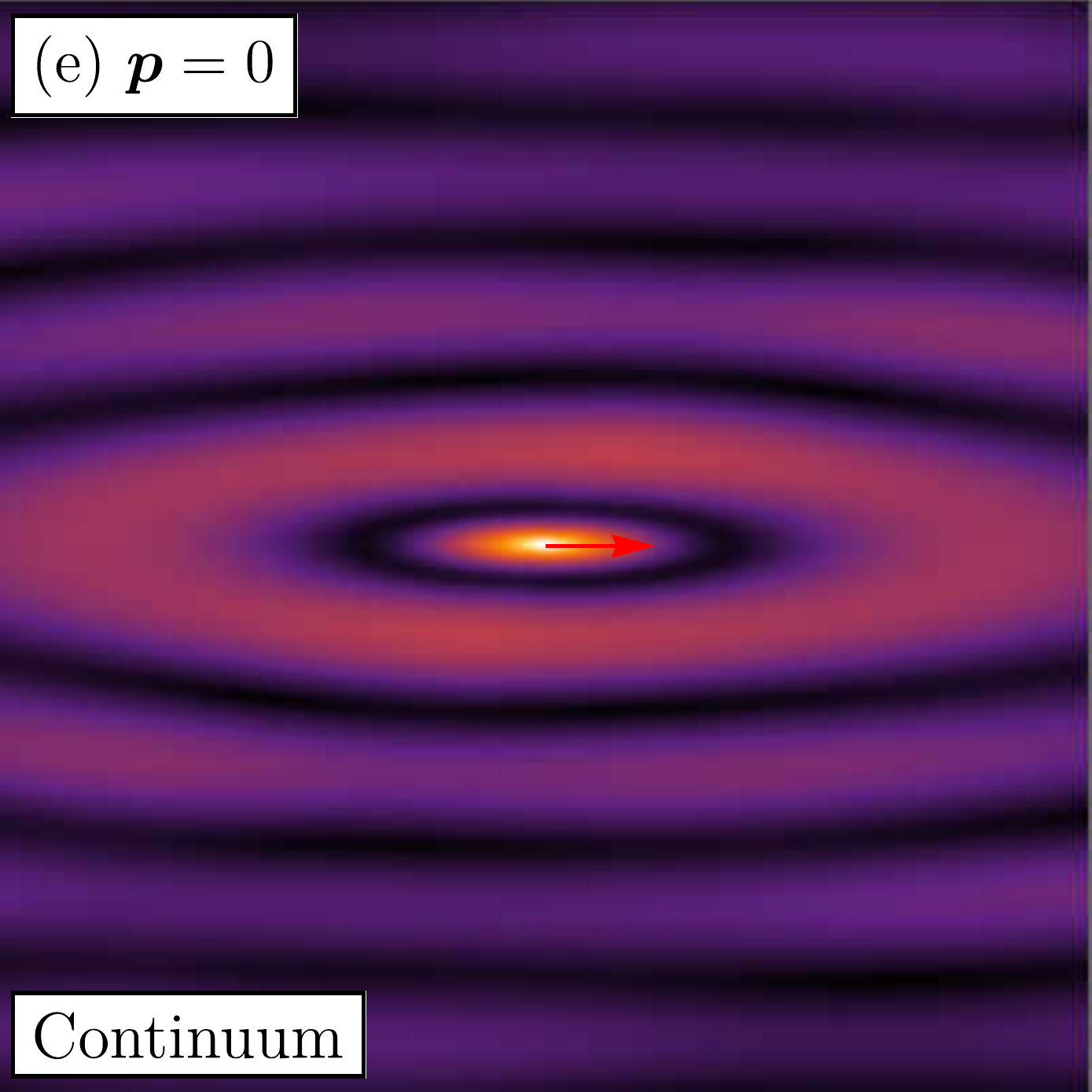}
\end{subfigure}
\begin{subfigure}{0.24\textwidth}
\centering
\phantomsubcaption{\label{fig:rhombus_7_15_force_h_80_gf}}
\includegraphics[width=0.98\linewidth]{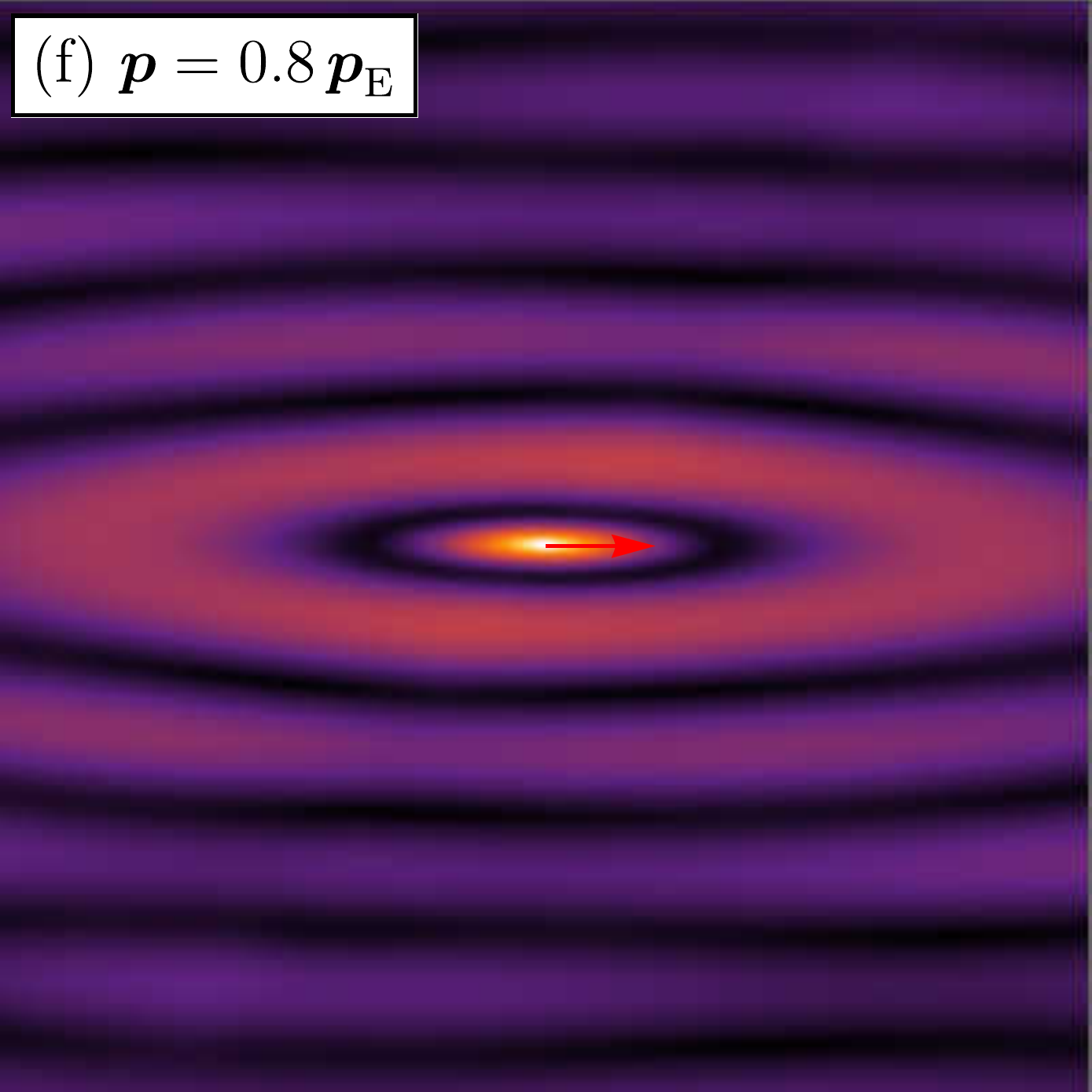}
\end{subfigure}
\begin{subfigure}{0.24\textwidth}
\centering
\phantomsubcaption{\label{fig:rhombus_7_15_force_h_90_gf}}
\includegraphics[width=0.98\linewidth]{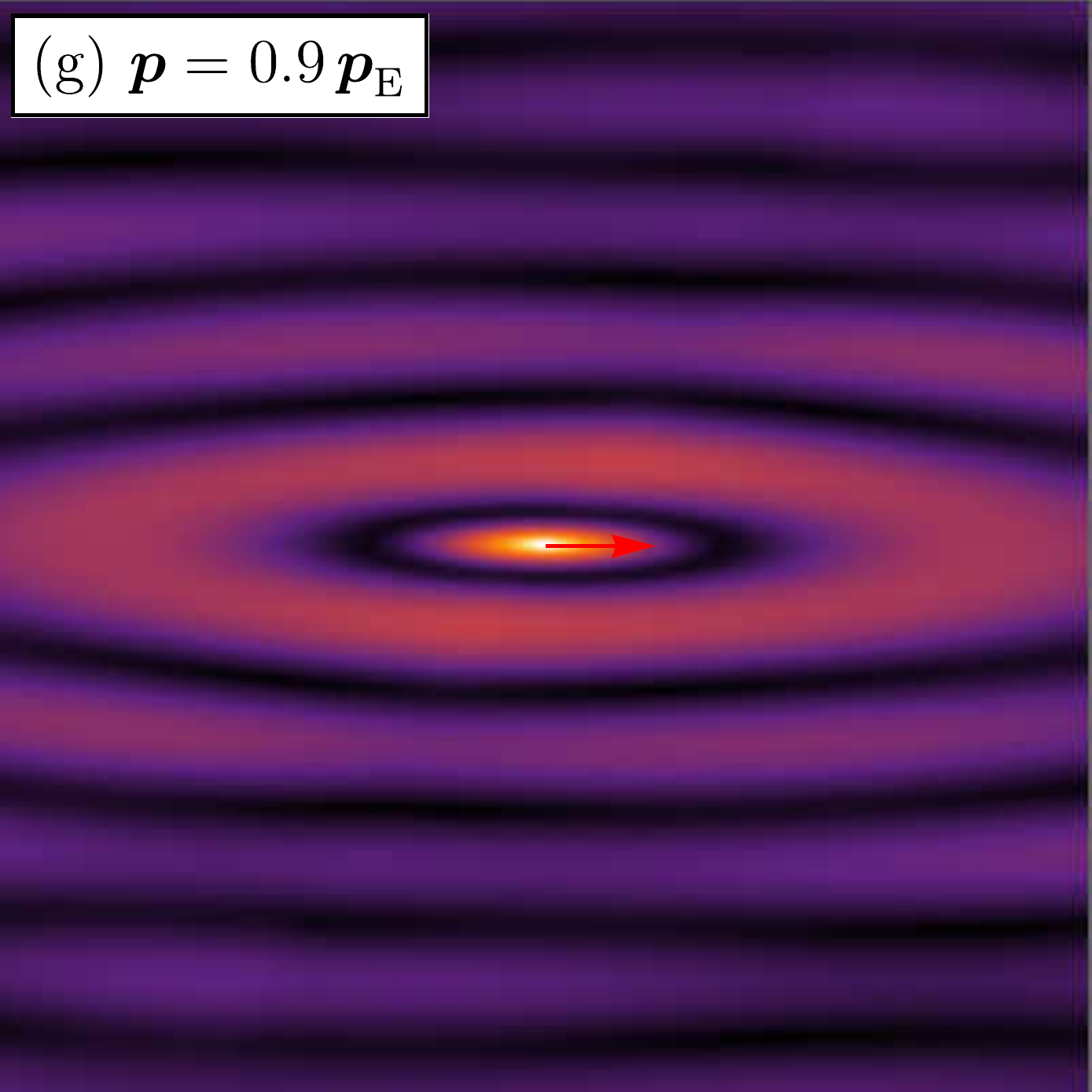}
\end{subfigure}
\vspace{0.015\linewidth}
\begin{subfigure}{0.24\textwidth}
\centering
\phantomsubcaption{\label{fig:rhombus_7_15_force_h_99_gf}}
\includegraphics[width=0.98\linewidth]{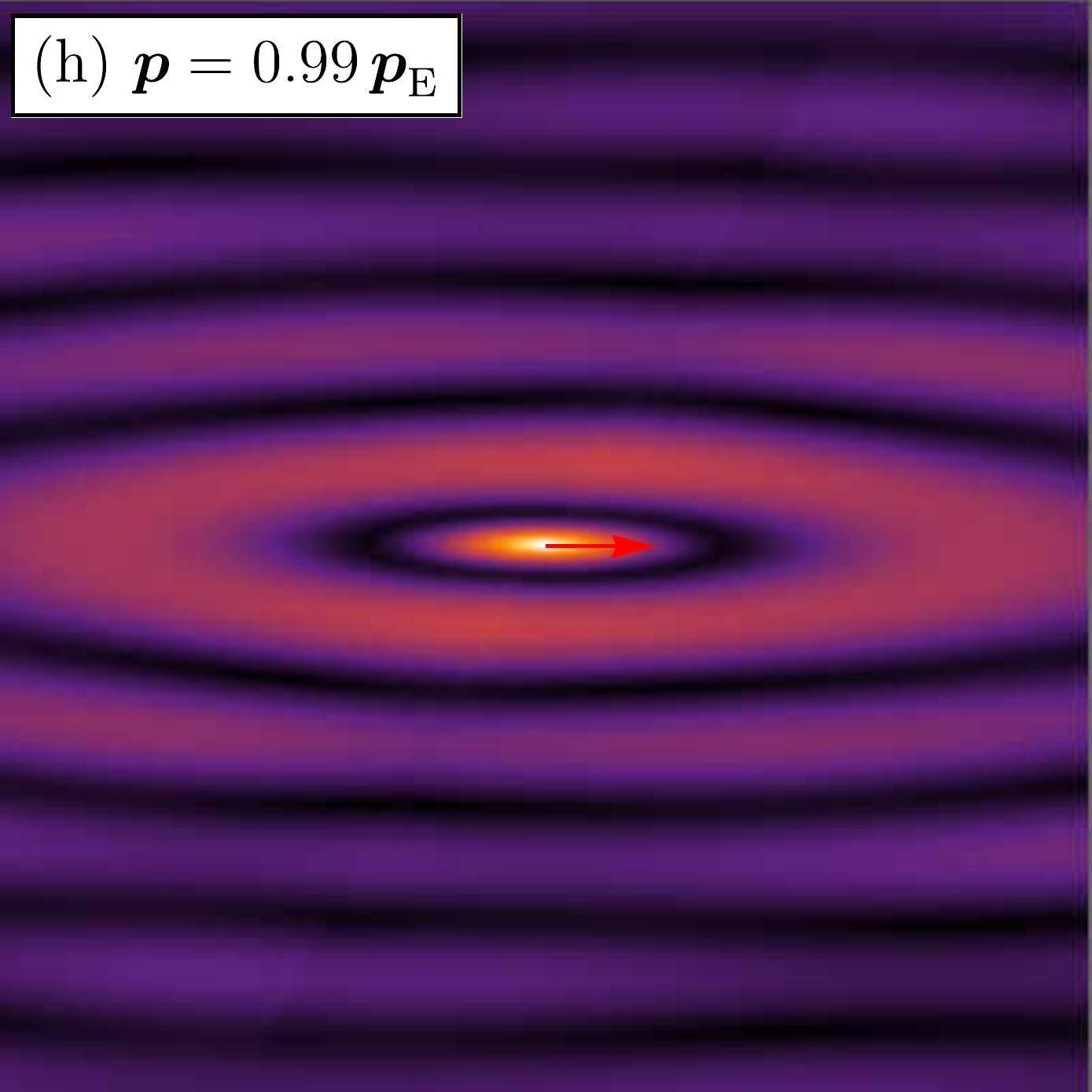}
\end{subfigure}\\
\begin{subfigure}{0.24\textwidth}
\centering
\phantomsubcaption{\label{fig:rhombus_7_15_force_v_0}}
\includegraphics[width=0.98\linewidth]{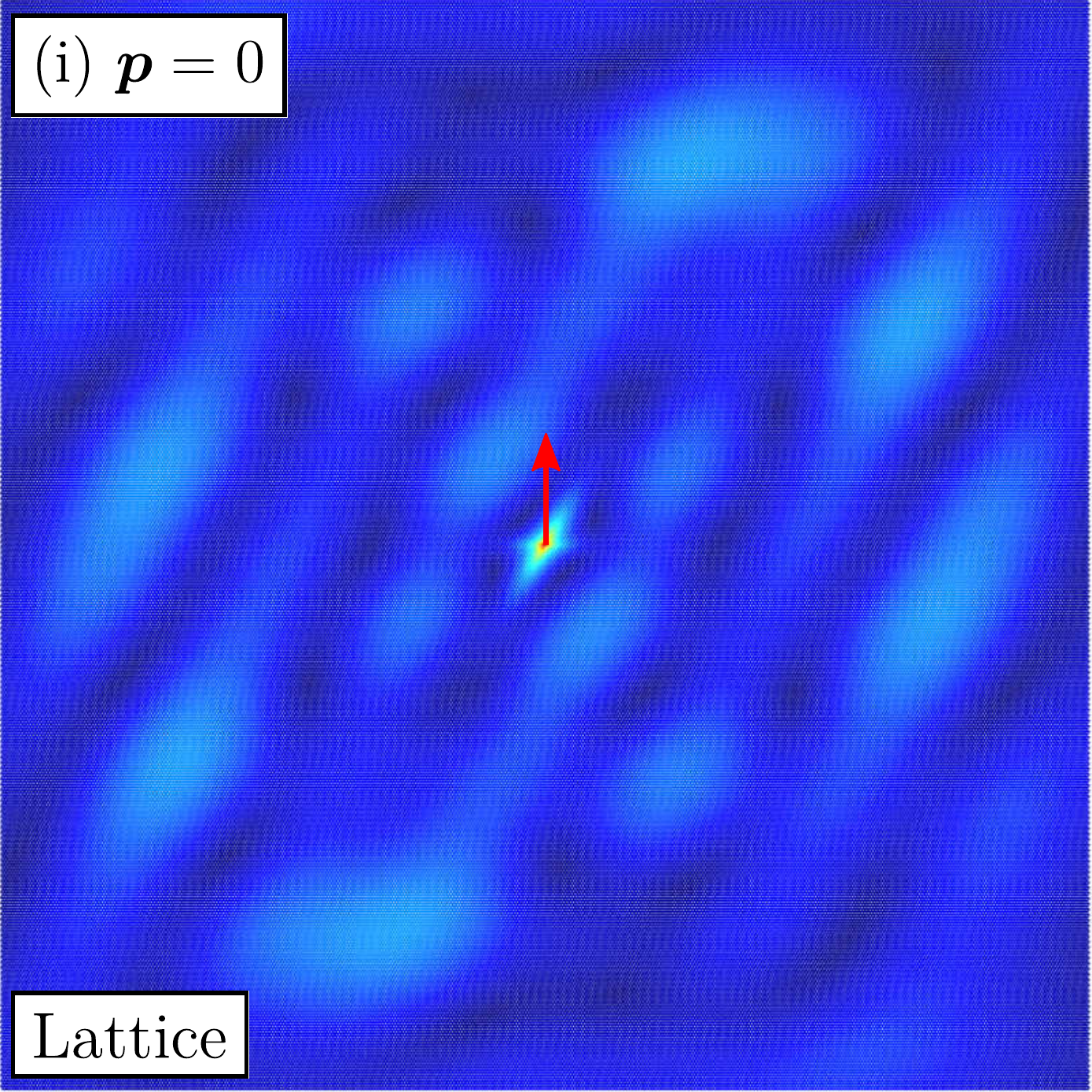}
\end{subfigure}
\begin{subfigure}{0.24\textwidth}
\centering
\phantomsubcaption{\label{fig:rhombus_7_15_force_v_80}}
\includegraphics[width=0.98\linewidth]{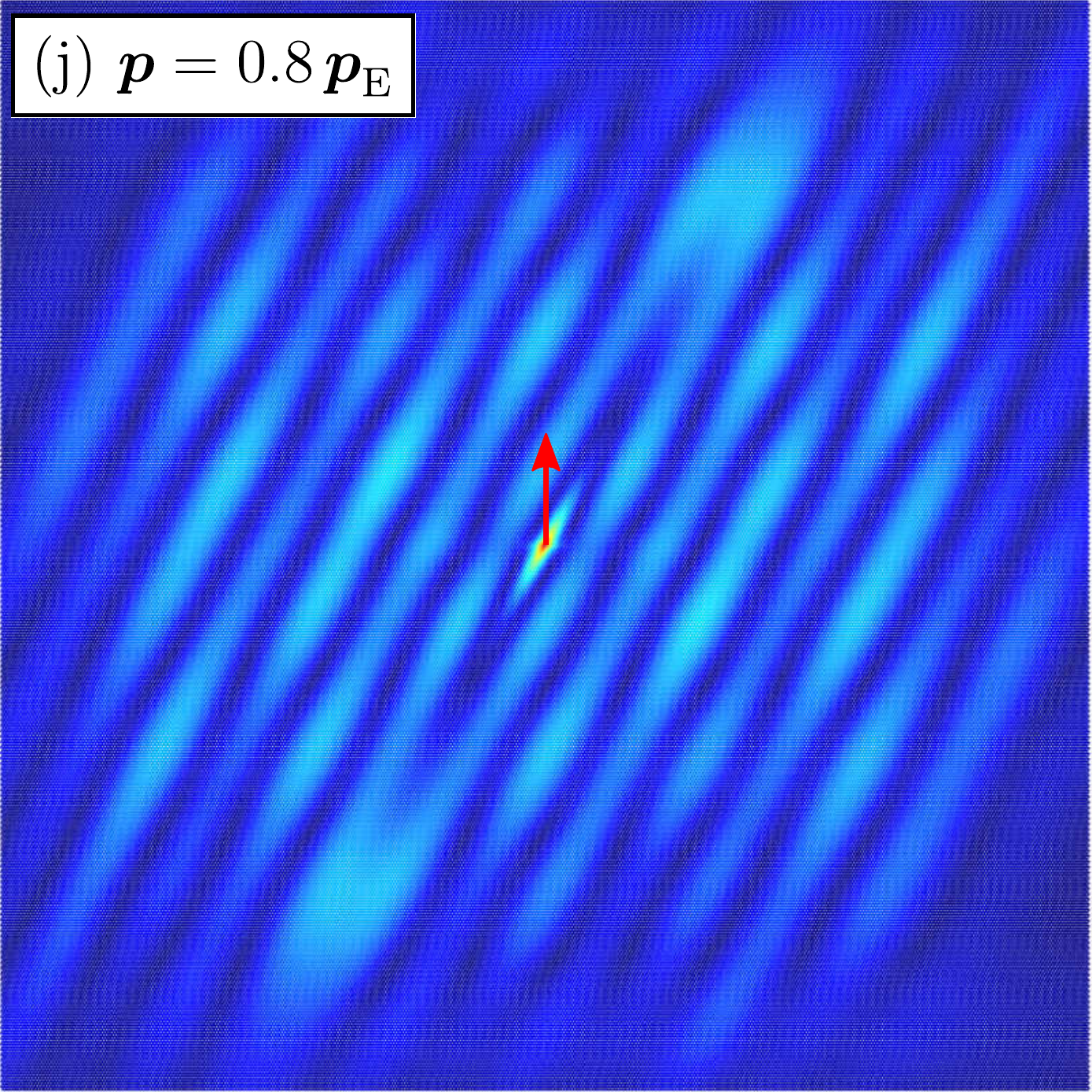}
\end{subfigure}
\begin{subfigure}{0.24\textwidth}
\centering
\phantomsubcaption{\label{fig:rhombus_7_15_force_v_90}}
\includegraphics[width=0.98\linewidth]{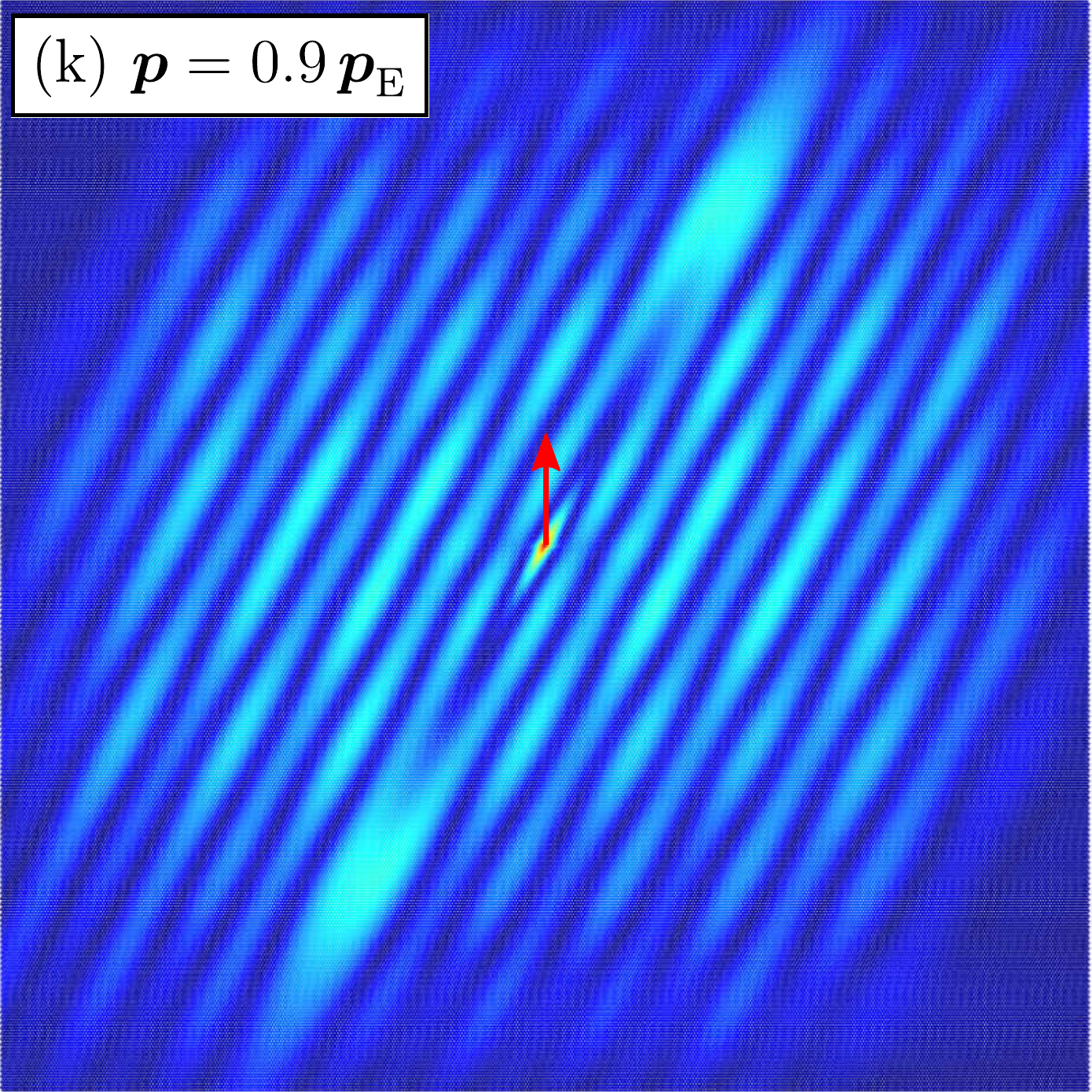}
\end{subfigure}
\begin{subfigure}{0.24\textwidth}
\centering
\phantomsubcaption{\label{fig:rhombus_7_15_force_v_99}}
\includegraphics[width=0.98\linewidth]{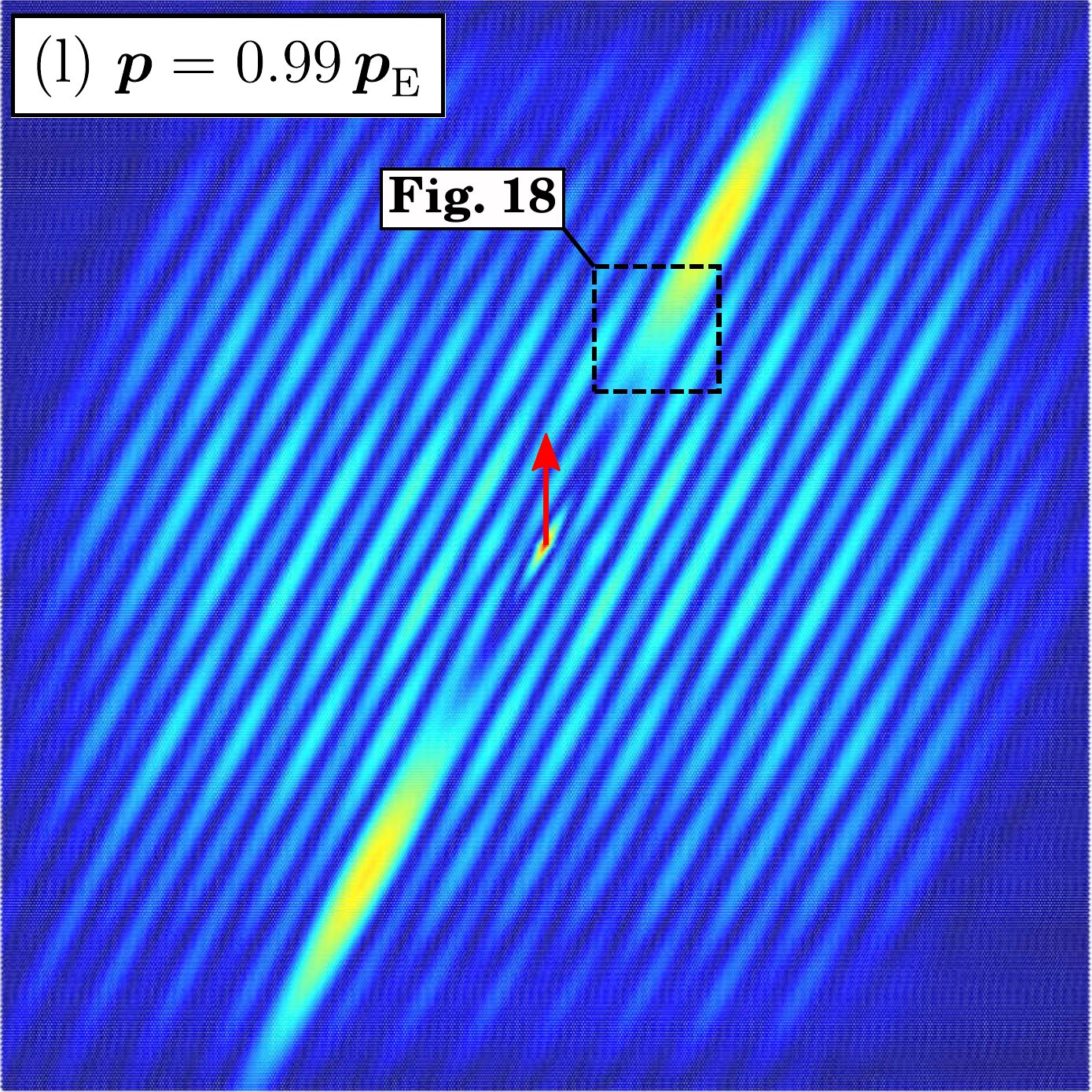}
\end{subfigure}\\
\vspace{0.01\linewidth}
\begin{subfigure}{0.24\textwidth}
\centering
\phantomsubcaption{\label{fig:rhombus_7_15_force_v_0_gf}}
\includegraphics[width=0.98\linewidth]{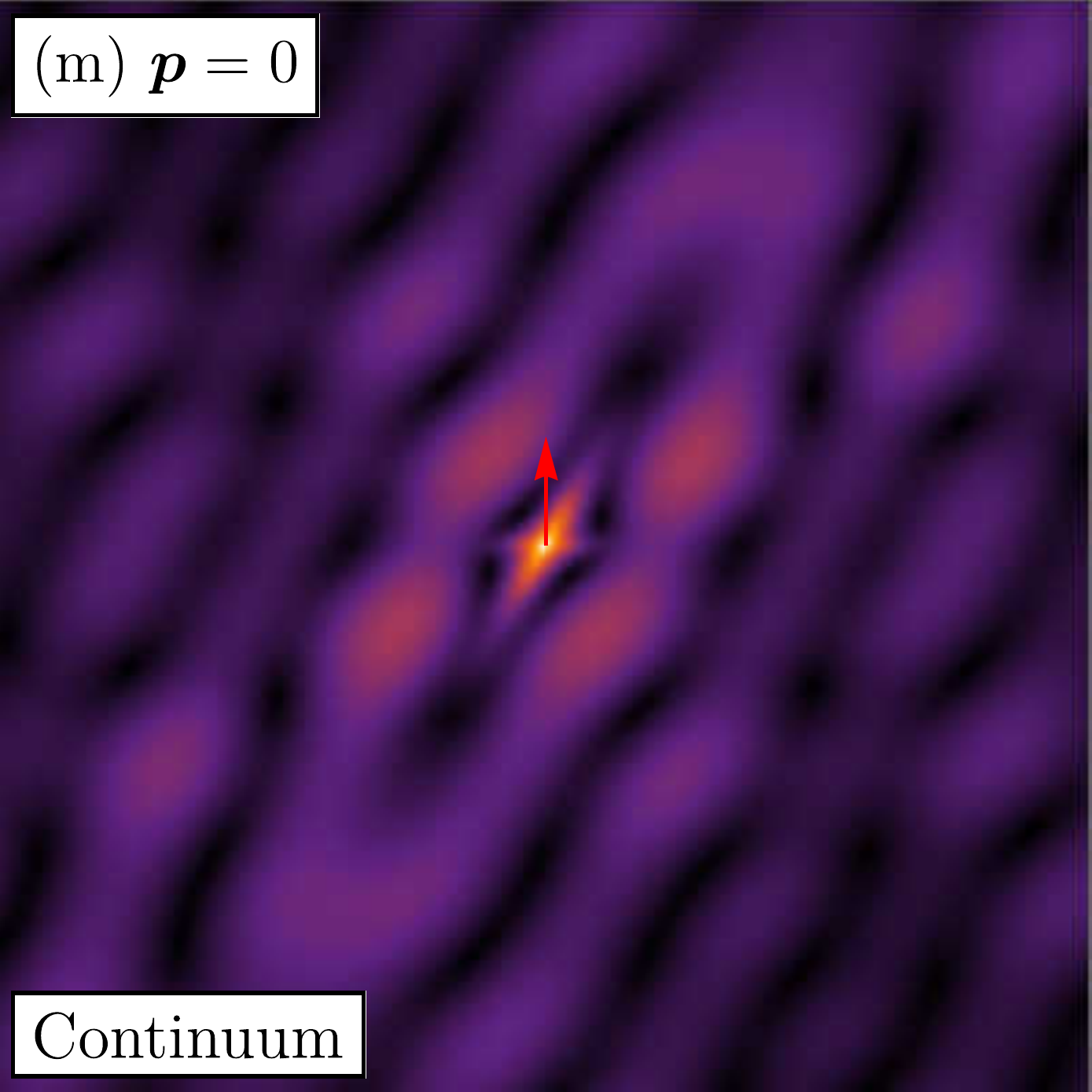}
\end{subfigure}
\begin{subfigure}{0.24\textwidth}
\centering
\phantomsubcaption{\label{fig:rhombus_7_15_force_v_80_gf}}
\includegraphics[width=0.98\linewidth]{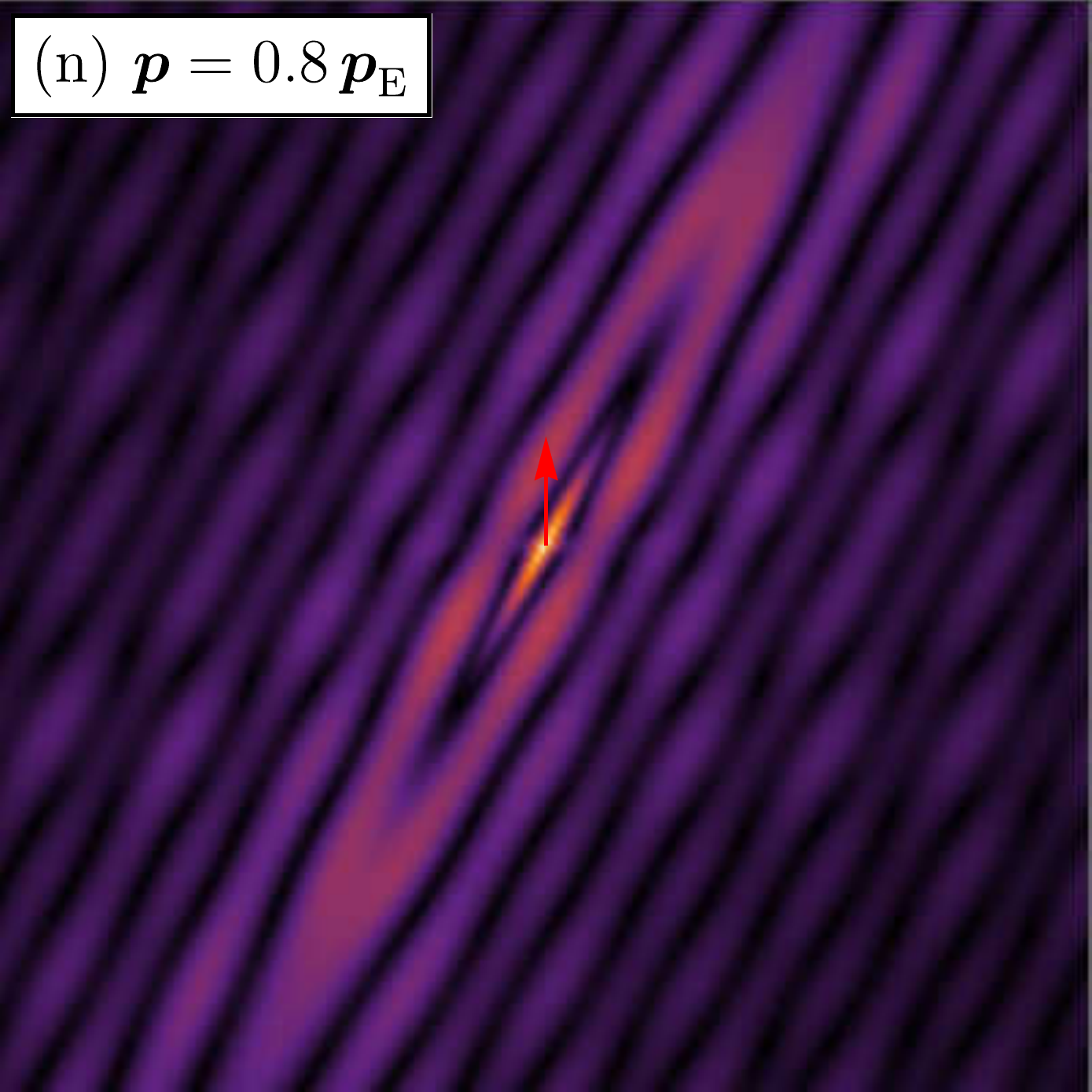}
\end{subfigure}
\begin{subfigure}{0.24\textwidth}
\centering
\phantomsubcaption{\label{fig:rhombus_7_15_force_v_90_gf}}
\includegraphics[width=0.98\linewidth]{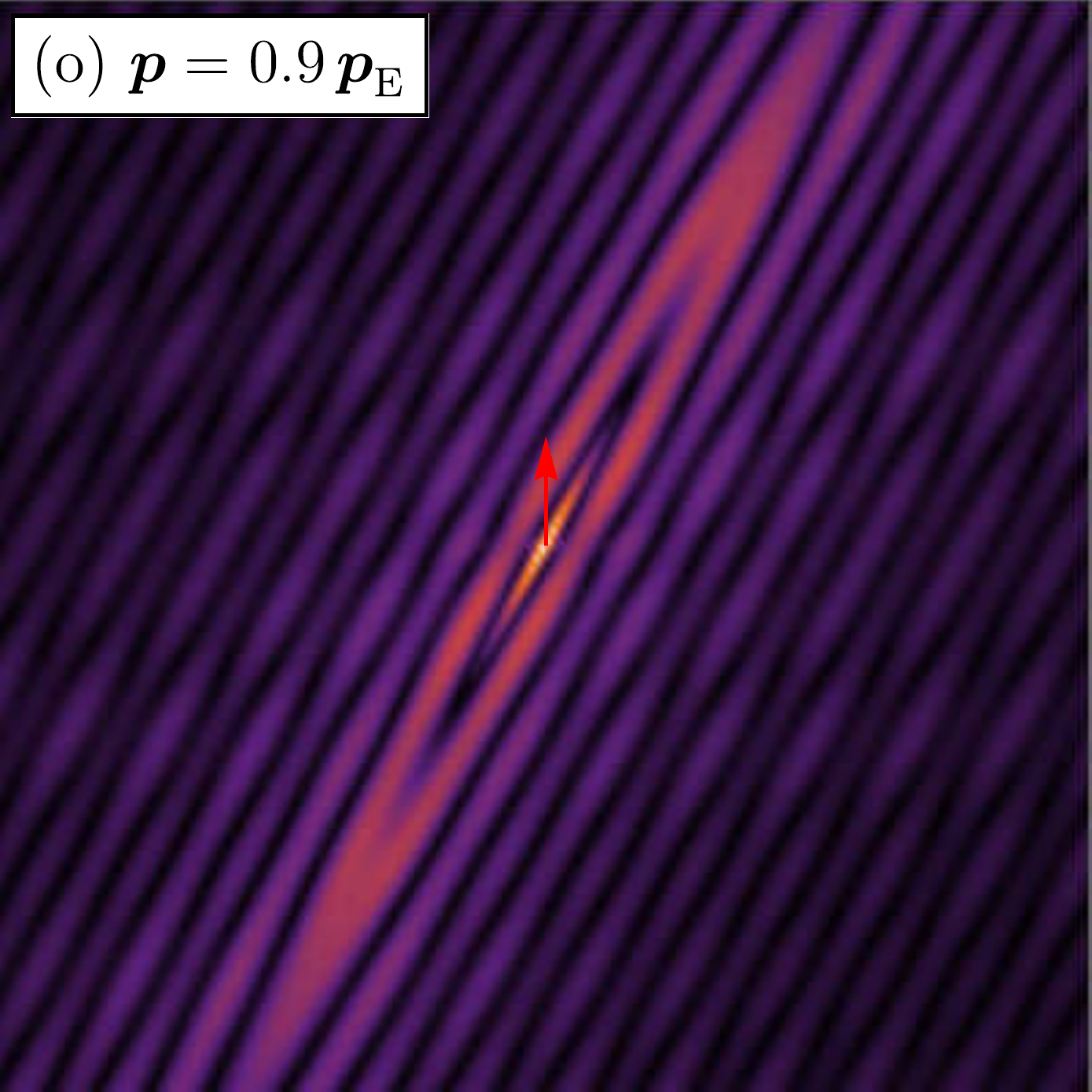}
\end{subfigure}
\begin{subfigure}{0.24\textwidth}
\centering
\phantomsubcaption{\label{fig:rhombus_7_15_force_v_99_gf}}
\includegraphics[width=0.98\linewidth]{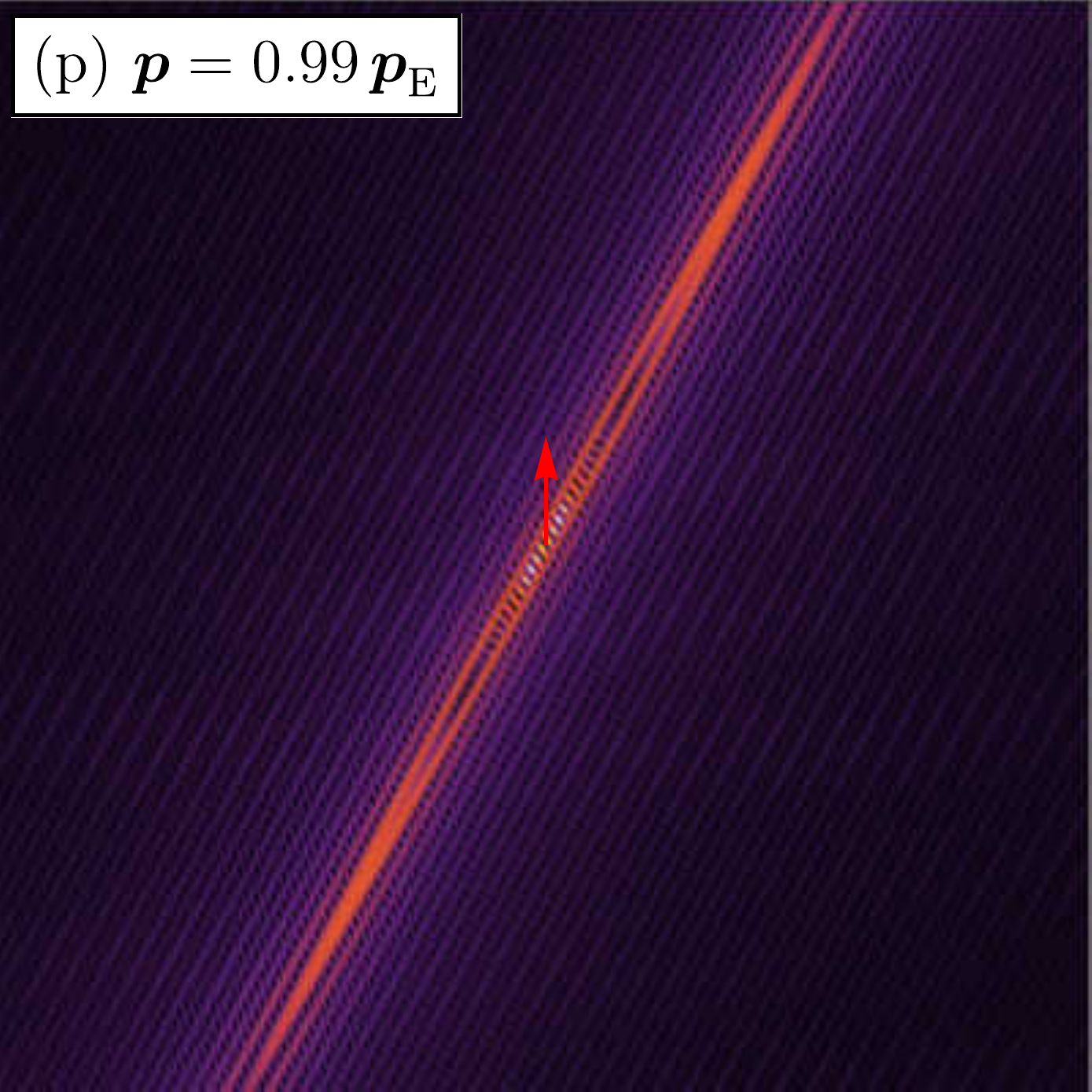}
\end{subfigure}
\caption{\label{fig:rhombus_7_15_force}
    As for Fig. \ref{fig:rhombus_10_10_force}, except that an \textit{anisotropic} rhombic lattice ($\Lambda_1=7,\,\Lambda_2=15$) is considered, for four different levels of preload (the path is shown in Fig.~\ref{fig:ellipticity_domains_7_15}). 
    Note that, even though ellipticity is lost along the direction predicted in Fig.~\ref{fig:eigenvalue_rhombus_7_15} (with normal angle $\theta_{\text{cr}}=151.4^\circ$), the activation of the strain localization depends on the orientation of the applied pulsating force, so that the band is activated by the vertical force while localization is inhibited when the grid is loaded horizontally.
}
\end{figure}

Results obtained with the same setting used to generate Fig.~\ref{fig:rhombus_10_10_force} (relative to an orthotropic rhombic lattice) are reported in Fig.~\ref{fig:rhombus_7_15_force}, which now refers to an \textit{anisotropic} rhombic lattice ($\Lambda_1=7,\,\Lambda_2=15$),  
for four values of prestress, $p_1=p_2=0,-1.634,-1.839,-2.023$.

For a completely anisotropic material only a single localization is expected to occur and in the case of the anisotropic grid considered the localization direction has been predicted in Fig.~\ref{fig:eigenvalue_rhombus_7_15} to occur at an inclination angle $\theta_{\text{cr}}=151.4^\circ$ of the band normal.
However, similarly to the case of the orthotropic grid, the activation of the localization depends on the orientation of the perturbing force.
This can be observed by comparing the lattice response generated by a horizontal and a vertical pulsating force, both reported in Fig.~\ref{fig:rhombus_7_15_force} and showing that strain localization is absent when a horizontal force is applied, regardless of the prestress level (see Figs.~\ref{fig:rhombus_7_15_force_h_0}--\ref{fig:rhombus_7_15_force_h_99_gf}). On the other hand, the vertical concentrated force triggers an inclined localization when the material is brought close to ellipticity loss (see Figs.~\ref{fig:rhombus_7_15_force_v_0}--\ref{fig:rhombus_7_15_force_v_99_gf}).

\begin{figure}[htb!]
\centering
\begin{subfigure}{0.24\textwidth}
\centering
\caption*{$t=0$}
\includegraphics[width=0.98\linewidth]{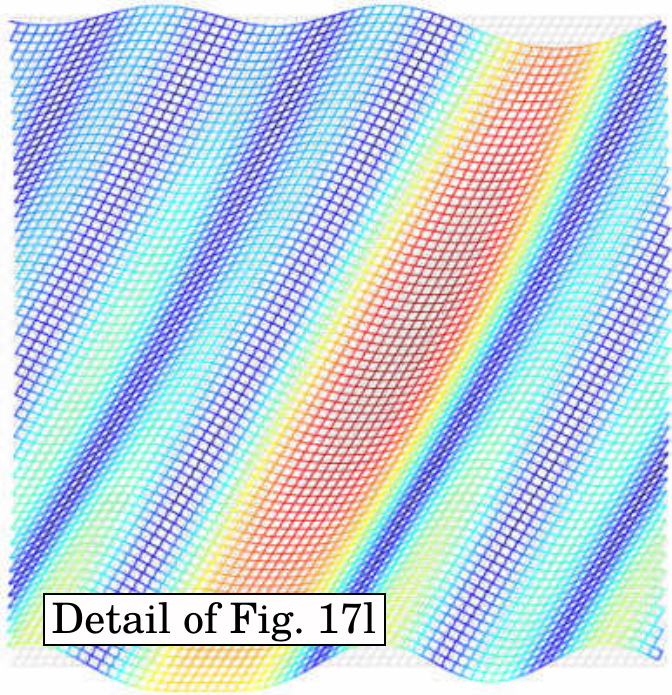}
\end{subfigure}
\begin{subfigure}{0.24\textwidth}
\centering
\caption*{$t=\frac{\pi}{2\omega}$}
\includegraphics[width=0.98\linewidth]{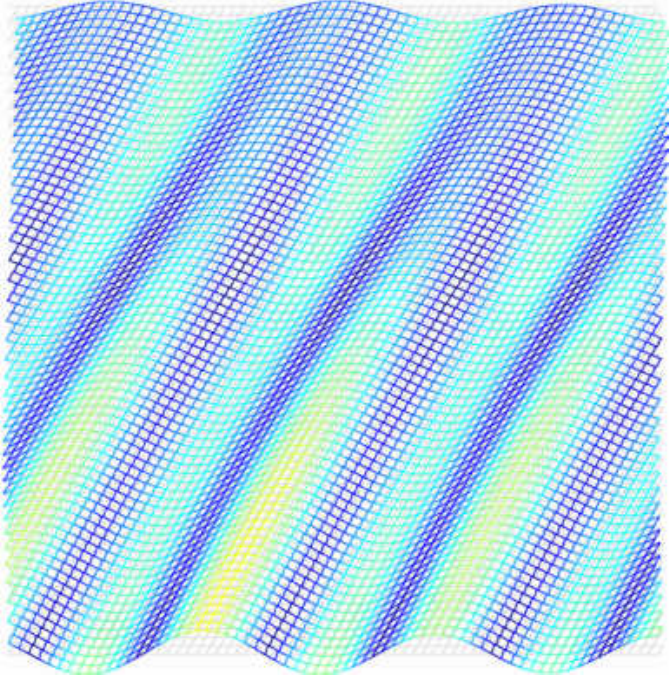}
\end{subfigure}
\begin{subfigure}{0.24\textwidth}
\centering
\caption*{$t=\frac{\pi}{\omega}$}
\includegraphics[width=0.98\linewidth]{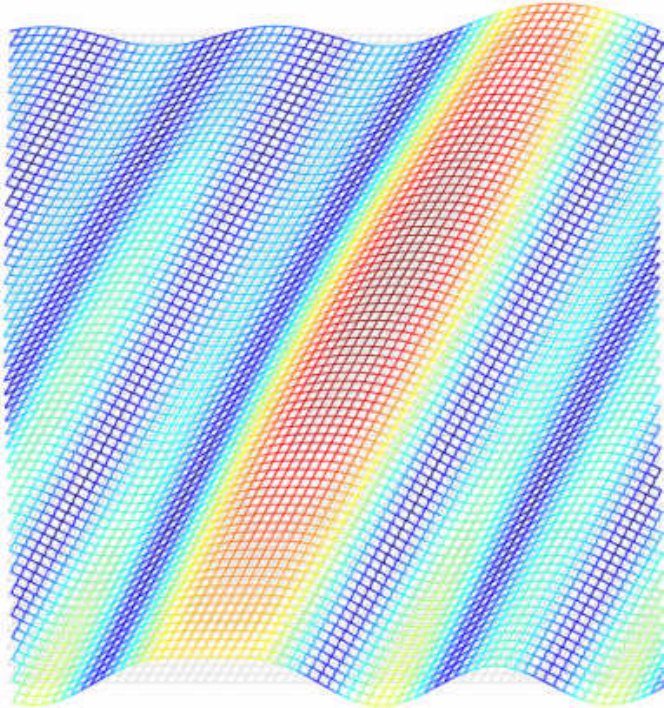}
\end{subfigure}
\begin{subfigure}{0.24\textwidth}
\centering
\caption*{$t=\frac{3\pi}{2\omega}$}
\includegraphics[width=0.98\linewidth]{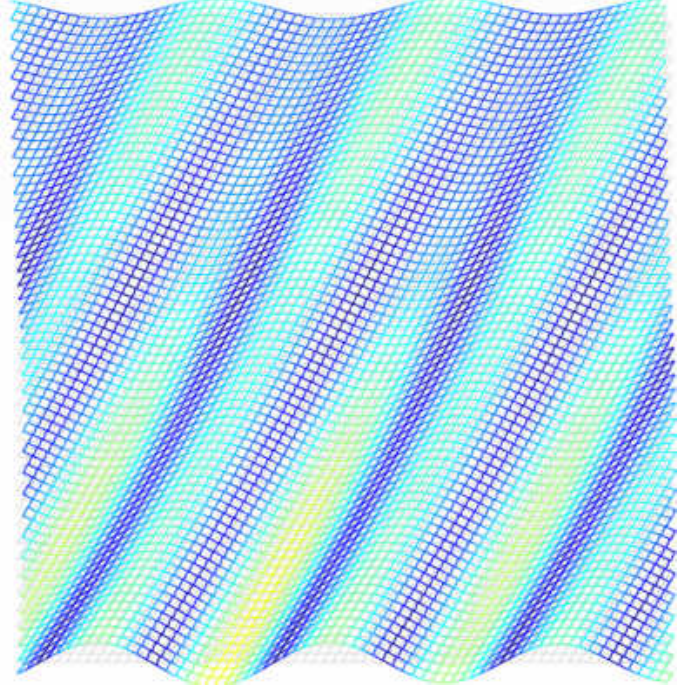}
\end{subfigure}
\caption{\label{fig:rhombus_7_15_deformed}
	As for Fig. \ref{fig:rhombus_10_10_deformed}, except that an \textit{anisotropic} rhombic lattice ($\Lambda_1=7,\,\Lambda_2=15$) is considered, for four different levels of preload (the path is shown in Fig.~\ref{fig:ellipticity_domains_7_15}). 
	The zone reported in the figure is shown in Fig. \ref{fig:rhombus_7_15_force_v_99}.
	The pattern shows the motion of the localization induced by the pulsating vertical load, which generates waves emanating outwards from the localization band, parallel to each other, and possessing an almost constant amplitude. 
    The deformation mode is mostly of shear-type even though an `expansion' component is also present, as indicated by the vector $\bg_{\text{E}}$ reported in  Fig.~\ref{fig:eigenvalue_rhombus_7_15}.
}
\end{figure}

\begin{figure}[htb!]
\centering
\begin{subfigure}{0.4\textwidth}
\centering
\caption{\label{fig:rhombus_7_15_force_h_FFT}Horizontal loading}
\includegraphics[width=0.98\linewidth]{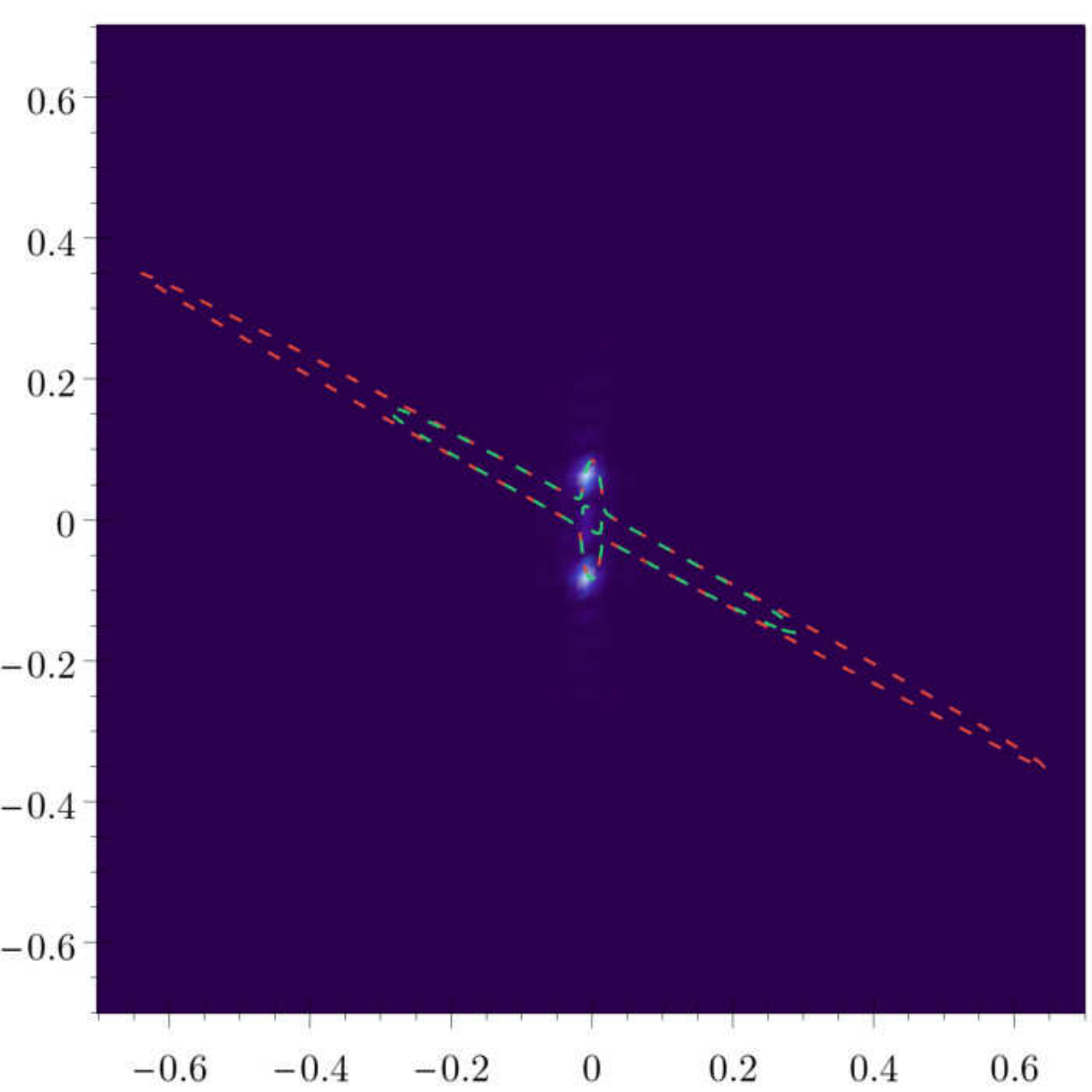}
\end{subfigure} \hspace{2mm}
\begin{subfigure}{0.4\textwidth}
\centering
\caption{\label{fig:rhombus_7_15_force_v_FFT}Vertical loading}
\includegraphics[width=0.98\linewidth]{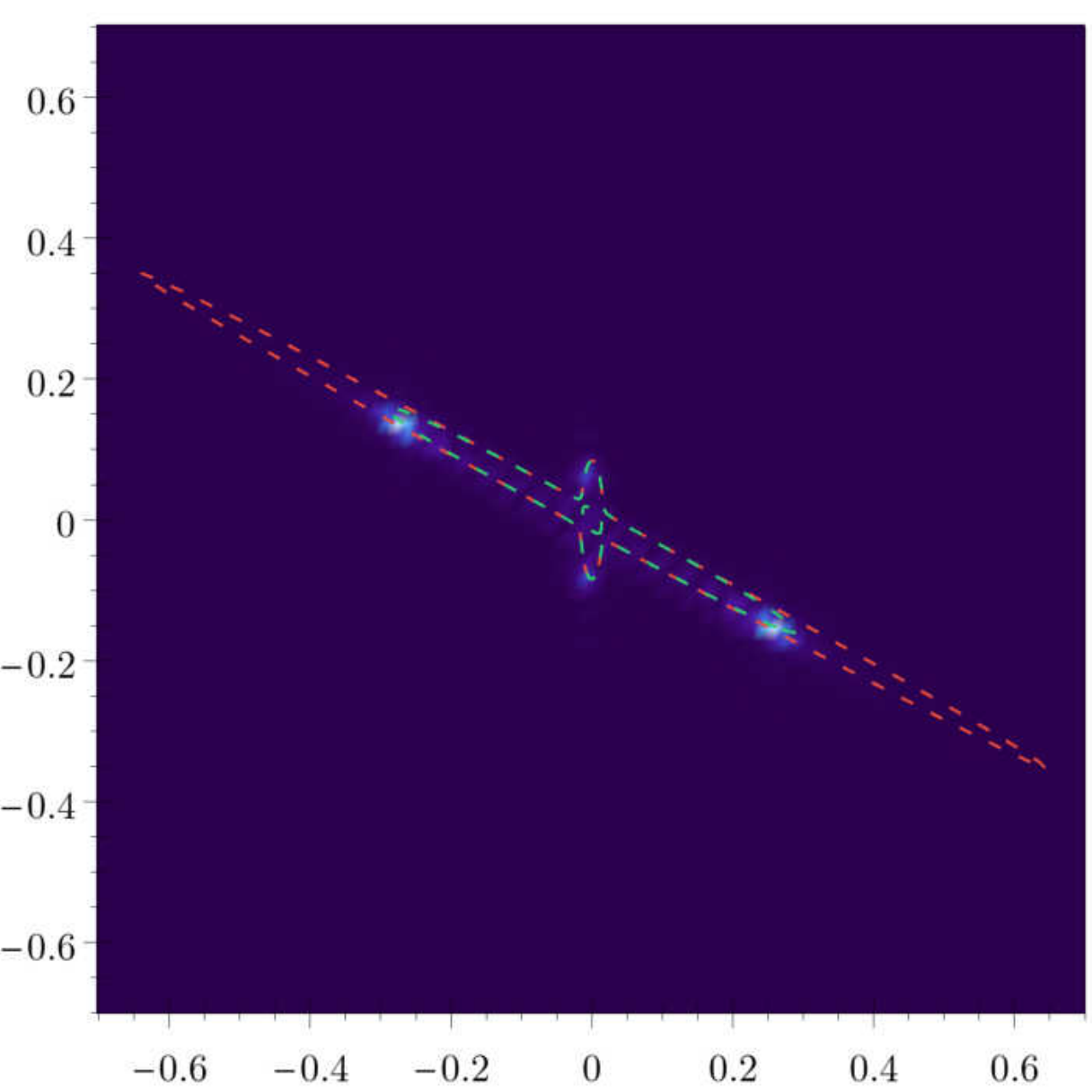}
\end{subfigure}
\caption{\label{fig:rhombus_7_15_force_FFT}
    As for Fig. \ref{fig:rhombus_10_10_force_FFT}, except that an \textit{anisotropic} rhombic lattice ($\Lambda_1=7,\,\Lambda_2=15$) is considered and the Fourier transform refers to the displacement fields reported in Figs. \ref{fig:rhombus_7_15_force_h_99} and \ref{fig:rhombus_7_15_force_v_99}. 
    Localization is absent when the pulsating force is horizontal,~(\subref{fig:rhombus_7_15_force_h_FFT}), as the waves corresponding to the elongated tips of the contours remain inactive. 
    The vertical force triggers two peaks along the direction $\theta_{\text{cr}}=151.4^\circ$, which is in fact the direction normal to the localization band,~(\subref{fig:rhombus_10_10_force_v_FFT}). 
    The Fourier transform is in perfect agreement with the responses shown in Figs.~\ref{fig:rhombus_7_15_force_h_99} and ~\ref{fig:rhombus_7_15_force_v_99}.
}
\end{figure}

Results reported in Figs.~\ref{fig:rhombus_7_15_deformed} and~\ref{fig:rhombus_7_15_force_FFT}, referred to the \textit{anisotropic} rhombic lattice have been obtained with the same setting of Figs.~\ref{fig:rhombus_10_10_deformed} and~\ref{fig:rhombus_10_10_force_FFT}, referring to the orthoptropic case. 

Fig. \ref{fig:rhombus_7_15_deformed} shows that, as only one localization band is present, the deformation pattern is characterized by the generation of essentially straight wavefronts propagating outwards from the localization band. 
The generation of these parallel waves is perfectly captured by the sharp peaks in the Fourier transform of the lattice response, reported in Fig.~\ref{fig:rhombus_7_15_force_FFT}.
Fig.~\ref{fig:rhombus_7_15_force_h_FFT} and~\ref{fig:rhombus_7_15_force_v_FFT} correspond, respectively, to the Fourier transform of Fig.~\ref{fig:rhombus_7_15_force_h_99} and Fig.~\ref{fig:rhombus_7_15_force_v_99}.
The two light spots in Fig.~\ref{fig:rhombus_7_15_force_v_FFT}, superimposed to two tips of the contour aligned parallel to the direction $\theta_{\text{cr}}=151.4^\circ$, denote the peaks of the Fourier transform. 
These clearly shows that the response induced by the vertical force involves pure plane waves propagating with fronts inclined at $\theta_{\text{cr}}=151.4^\circ-90^\circ=61.4^\circ$ with respect to the horizontal axis.

It is also important to note that waves do not propagate vertically when the load is vertical while these become the only propagation mode when the load is horizontal (see the peaks on the short tips of the contour in Fig.~\ref{fig:rhombus_7_15_force_h_FFT}).
This is in agreement with the fact that localization is not generated by the horizontal force and the `slow waves' leading the homogenized continuum to failure of ellipticity remain inactive.

\section{Conclusions}
\label{sec:conclusions}
An analytic formulation has been developed for the time-harmonic dynamics of a grillage of elastic rods (equipped with distributed mass density and rotational inertia), subject to axial forces of arbitrary amount and incrementally loaded in the plane.
Increments are unprescribed, so that incremental axial and shear forces and bending moments are involved. 
The formulation leads, through an asymptotic expansion of Floquet-Bloch waves, to a low frequency approximation for an equivalent prestressed elastic material, which turns out to be coincident with that determined on the basis of an energy match (see Part I of this study). 

The developed technique can be employed to systematically analyze arbitrary lattice geometries and preloaded configurations, therefore predicting both local and global material instabilities, in other words, micro-buckling and strain localizations.
Loss of ellipticity has been analyzed for a skewed grid, to 
\begin{enumerate*}[label=(\roman*)]
    \item explore cubic, orthotropic and fully anisotropic homogenized material responses,
    \item compute the elliptic domain for the homogenized continuum as a function of lattice parameters,
    \item analyse the structure of the acoustic branches close to ellipticity loss, and
    \item investigate forced vibrations (both in physical and Fourier spaces) revealing low-frequency wave localizations.
\end{enumerate*}

The advantage of a time-harmonic formulation with respect to a quasi-static approach lies in the fact that several aspects related to strain localization occurring under dynamic loading can be highlighted and investigated.
In particular, the following conclusions have been reached. 
\begin{itemize}
\item The rotational inertia of the elastic rods does not contribute to the definition of the prestressed elastic solid equivalent to the grillage.
\item The homogenization scheme based on time-harmonic dynamics proves that the long-wavelength asymptotics for waves propagating in the lattice is
governed by the acoustic tensor of the effective medium. This aspect definitely clarifies that a macro-bifurcation in the lattice has to be equivalent to failure of ellipticity in its equivalent continuum.
\item  Short-wavelength (or micro-) bifurcations are provided by the analysis of the dispersion relation of the lattice, interpreted now as a function of the
axial preload state in the rods. Buckling in the lattice becomes the `propagation' of a Bloch wave at vanishing frequency.
Macro-bifurcation in the grid occurs with the progressive lowering, and
eventually vanishing, of the slope of the acoustic branches at the origin, while the dispersion surface attains
non-null frequency for every other wave vector. On the contrary, micro-bifurcation is characterized by a non-vanishing slope of the acoustic branches at the origin, but
the preload-induced lowering of the dispersion surface causes the generation of a zero-frequency wave
with non-null wave vector (corresponding to a finite wavelength buckling).
\item For square grid geometries, the shear wave responsible for the ellipticity loss is the one propagating along the direction of the stiffer 
elastic link (with the lowest slenderness). This conclusion is counterintuitive, as a shear mechanism  would be expected to be generated in the direction of the soft elastic links.
\item Shear bands have been investigated through a Fourier transform of the lattice response, to be compared with the Bloch spectrum generated by the forcing source. 
Sharp peaks in the Fourier transform evidence that the pulsating force is emanating pure plane waves, which is the `signature' of strain localization in a dynamic context. 
\item The asymptotic homogenization scheme is performed near the vanishing frequency. It is therefore believed to be more closely representative of the lattice when the frequency of the pulsating force is sufficiently low. 
When the elliptic boundary is approached, the mismatch in the acoustic properties, between the lattice and its effective continuum approximation, has been found to
become wider for those waves which propagate parallel to the direction of ellipticity loss.
\end{itemize}

Finally, summarizing results from both Part I and Part II, an important conclusion is that, excluding the cases of micro-instabilities, the homogenization is proven to yield a superb approximation to the behaviour of the rod's lattice.
However, micro-instabilities may become explosive, thus extending from a point to an entire medium. 
Strain localization may occur in many different forms, involving one or more bands, which can be of pure shear strain or may contain other strain components. 
The grillages analyzed in the present article represent ideal candidates for the design of structured materials exhibiting various forms of instabilities within the elastic range.

\section*{Acknowledgements}
Financial support is acknowledged from: 
the ERC Advanced Grant `Instabilities and nonlocal multiscale modelling of materials' ERC-2013-ADG-340561-INSTABILITIES (G.B. and L.C.),
PRIN 2015 2015LYYXA8-006 and ARS01-01384-PROSCAN (D.B. and A.P.).
The authors also acknowledge support from the Italian Ministry of Education, University and Research (MIUR) in the frame of the `Departments of Excellence' grant L. 232/2016.

\printbibliography

\appendix
\numberwithin{equation}{section}            %
\numberwithin{figure}{section}              %

\section{Linearized equations of motion for an axially pre-stretched elastica}
\label{sec:linearized_elastica}
A model for an axially stretchable Rayleigh elastic rod can be obtained through a linearization (around a stretched equilibrium configuration) of the equations governing the dynamics of large deflections and flexure of an elastic rod. 
A local axial coordinate $x_0$ is introduced to single out points of the straight, undeformed configuration of an elastic rod, for the moment unstressed. 
This configuration is assumed as reference, so that the potential and kinetic energies are defined as 
\begin{subequations}
\label{eq:potential_kinetic_energy_elastica}
\begin{align}
    \mV &= \int_0^{l_0} \left( \psi_\lambda(\lambda) + \psi_\chi(\chi) - P\,u'(x_0,t) \right) dx_0 \,, \\
    \mT &= \frac{1}{2} \int_0^{l_0} \left(\gamma_0 \left(\dot{u}(x_0,t)^2 + \dot{v}(x_0,t)^2\right) + \gamma_{r,0}\,\dot{\theta}(x_0,t)^2 \right) dx_0 \,,
\end{align}
\end{subequations}
where $l_0$, $\gamma_0$, and $\gamma_{r,0}$ are the initial length, linear mass density, and rotational inertia, while $\psi_\lambda$ and $\psi_\chi$ are strain-energy functions for, respectively, axial  and flexural deformations.
The axial stretch $\lambda$ and the curvature $\chi$ are defined by the kinematics of an extensible, but unshearable, elastica as
\begin{subequations}
\label{eq:strain_measures_elastica}
\begin{align}
    \label{eq:axial_strain}
    \lambda &= (1+u'(x_0,t))\cos{\theta(x_0,t)} + v'(x_0,t)\sin{\theta(x_0,t)} \,, \\
    \label{eq:curvature}
    \chi &= \theta'(x_0,t) = \deriv{}{x_0} \left(\arctan\left(\frac{v'(x_0,t)}{1 + u'(x_0,t)}\right)\right) \,,
\end{align}
\end{subequations}
where in~\eqref{eq:curvature} the unshearability constraint $\theta = \arctan\left(\frac{v'}{1+u'}\right)$ has been explicitly introduced.

A second-order expansion of the functionals~\eqref{eq:potential_kinetic_energy_elastica}, with respect to the independent displacement fields $\{u, v\}$, around the deformed configuration $\{u_0,v_0\} =\{(\lambda_0-1) x_0, 0\}$, yields the linearized response of the rod. The linearization is around a \textit{straight, but axially stretched, configuration}. 
A substitution of Eq.~\eqref{eq:strain_measures_elastica} into Eq.~\eqref{eq:potential_kinetic_energy_elastica} and neglection of an arbitrary constant, leads to the following expansion 
\begin{subequations}
\label{eq:potential_kinetic_energy_elastica_second_order}
\begin{align}
\label{eq:potential_energy_elastica_second_order}
&\begin{aligned}
    \mV(u_0+\delta u, v_0 + \delta v) \sim
    & \int_0^{l_0} \left(\psi_\lambda'(\lambda_0)-P\right)\delta u'(x_0,t) dx_0 +\\
    & +\frac{1}{2} \int_0^{l_0} \psi_\lambda''(\lambda_0)\delta u'(x_0,t)^2 dx_0 +\\
    & +\frac{1}{2} \int_0^{l_0}\left( \frac{\psi_\lambda'(\lambda_0)}{\lambda_0}\delta v'(x_0,t)^2 
    + \frac{\psi_\chi''(0)}{\lambda_0^2}\delta v''(x_0,t)^2 \right) dx_0 \,,
\end{aligned}\\
&\begin{aligned}
    \mT(u_0+\delta u, v_0 + \delta v) \sim
    \frac{1}{2} \int_0^{l_0} \left( \gamma_0 \left(\delta\dot{u}(x_0,t)^2 + \delta\dot{v}(x_0,t)^2\right) + \frac{\gamma_{r,0}}{\lambda_0^2}\delta\dot{v}'(x_0,t)^2 \right)dx_0 \,,
\end{aligned}
\end{align}
\end{subequations}
where it has been assumed that $\psi_\chi'(0) = 0$ in the unloaded configuration.
The vanishing of the first-order term in Eq.~\eqref{eq:potential_energy_elastica_second_order}, occurring when the configuration $\{u_0,v_0\} =\{(\lambda_0 -1) x_0, 0\}$ satisfies equilibrium, implies that the prestretch $\lambda_0$ is the solution of the condition $\psi_\lambda'(\lambda_0)-P=0$. 
This indicates that the axial load $P$ is equal to the axial pre-load.
Moreover, the second-order term in Eq.~\eqref{eq:potential_energy_elastica_second_order} involves the strain energy functions only in terms of second derivatives, $\psi_\lambda''(\lambda_0)$ and $\psi_\chi''(0)$, evaluated on the straight stretched configuration.

An update of the reference configuration from the stress-free configuration to the stretched configuration implies that the second-order functionals~\eqref{eq:potential_kinetic_energy_elastica_second_order} can be used to determine the incremental response of the rod.
To this purpose, the variable of integration is changed from $x_0$ to the current stretched coordinate $s = \lambda_0 x_0$, so that the fields $\{u, v\}$ become functions of $s$.
The second-order terms in Eqs.~\eqref{eq:potential_kinetic_energy_elastica_second_order} can now be written as 
\begin{subequations}
\label{eq:potential_kinetic_energy_elastica_second_order_updated}
\begin{align}
&\begin{aligned}
    \mV(u_0+\delta u, v_0 + \delta v) \sim
    & \frac{1}{2} \int_0^{l} \psi_\lambda''(\lambda_0)\lambda_0\, \delta u'(s,t)^2 ds +\\
    & +\frac{1}{2} \int_0^{l}\left( P\delta v'(s,t)^2 +
    \psi_\chi''(0)\lambda_0\, \delta v''(s,t)^2 \right) ds \,,
\end{aligned}\\
&\begin{aligned}
    \mT(u_0+\delta u, v_0 + \delta v) \sim
    \frac{1}{2} \int_0^{l} \left( \frac{\gamma_0}{\lambda_0} \left(\delta\dot{u}(s,t)^2 + \delta\dot{v}(s,t)^2\right) + \frac{\gamma_{r,0}}{\lambda_0}\delta\dot{v}'(s,t)^2 \right) ds\,,
\end{aligned}
\end{align}
\end{subequations}
where $l=\lambda_0 l_0$ denotes the current length of the rod and the symbol $'$ indicates differentiation with respect to $s$. Moreover, 
$\psi_\lambda''(\lambda_0)\lambda_0 = A(\lambda_0)$ and and $\psi_\chi''(0)\lambda_0 = B(\lambda_0)$  are respectively the current value of axial stiffness and of bending stiffness, both functions of the axial stretch $\lambda_0$.

The second-order functionals, Eqs. (\ref{eq:potential_kinetic_energy_elastica_second_order_updated}),  describe the incremental response of the axially pre-stretched and pre-loaded rod.
Therefore, the equations of motion governing the incremental displacements can be derived via the following functionals
\begin{subequations}
\label{eq:potential_kinetic_energy_elastica_incremental}
\begin{align}
\label{eq:potential_energy_elastica_incremental}
    &\mV(u, v) =
    \frac{1}{2} \int_0^{l} A(\lambda_0)\, u'(s,t)^2 ds
    +\frac{1}{2} \int_0^{l}\left( P v'(s,t)^2 +
    B(\lambda_0)\, v''(s,t)^2 \right) ds \,, \\
\label{eq:kinetic_energy_elastica_incremental}
    &\mT(u, v) =
    \frac{1}{2} \int_0^{l} \left( \frac{\gamma_0}{\lambda_0} \left(\dot{u}(s,t)^2 + \dot{v}(s,t)^2\right) + \frac{\gamma_{r,0}}{\lambda_0}\dot{v}'(s,t)^2 \right) ds \,,
\end{align}
\end{subequations}
where the fields $\{u(s,t),v(s,t)\}$ are current incremental fields.
Note that the linear mass density is divided by the prestretch, indicating that the \textit{current} density governs the incremental inertia of the rod. In fact, mass conservation requires $\gamma_0/\lambda_0=\gamma(\lambda_0)$, where $\gamma$ is the current linear mass density of the stretched rod.
Similarly, the \textit{current} rotational inertia is denoted as $\gamma_{r,0}/\lambda_0=\gamma_r(\lambda_0)$.

Equations~\eqref{eq:governing_beam_EB} are directly obtained through the application of Hamilton's principle on the Lagrangian $\mL=\mT-\mV$ constructed using the second-order functionals~\eqref{eq:potential_kinetic_energy_elastica_incremental}.

\section{Full expression of the effective acoustic tensor}
\label{sec:homogenized_acoustic_tensor}
The complete analytic expression for the effective acoustic tensor of the lattice analyzed in Section~\ref{sec:ellipticity_loss_grid} and~\ref{sec:forced_response} is here reported.
The result is provided in terms of the coefficients $h_{i,j,k,l}^{\text{G}}(p_1,p_2,\Lambda_1,\Lambda_2,\alpha)$, appearing in Eq.~\eqref{eq:acoustic_tensor_grid_contributions}, functions of the non-dimensional parameters defined in Section~\ref{sec:ellipticity_loss_grid} (components that have to be equal by symmetry are not reported and the substitution $\Lambda_2=\Lambda_1/\sqrt{\phi}$, with $\phi=B_2/B_1$, is introduced in order to shorten the result) 
\begin{dgroup*}[style={\footnotesize},breakdepth={20}]
    \begin{dmath*}
        h_{1111}^{\text{G}} = 
        \frac{1}{2 d \sin\alpha} \left(\sinh \left(\frac{\sqrt{p_2}}{2}\right) \left(\sqrt{p_1} p_2 \phi  \cosh \left(\frac{\sqrt{p_1}}{2}\right) \left(4 \Lambda_1^2 \cos (2 \alpha )+11 \Lambda_1^2+\cos (4 \alpha ) \left(\Lambda_1^2-p_2 \phi \right)+p_2 \phi \right)-2 \sinh \left(\frac{\sqrt{p_1}}{2}\right) \left(11 \Lambda_1^2 p_1+4 \Lambda_1^2 \cos (2 \alpha ) (p_1+p_2 \phi )+\cos (4 \alpha ) \left(\Lambda_1^2 p_1+p_2 \phi  \left(\Lambda_1^2-p_2 \phi \right)\right)+p_2 \phi  \left(11 \Lambda_1^2+p_2 \phi \right)\right)\right)+p_1 \sqrt{p_2} \sinh \left(\frac{\sqrt{p_1}}{2}\right) \cosh \left(\frac{\sqrt{p_2}}{2}\right) \left(4 \Lambda_1^2 \cos (2 \alpha )+11 \Lambda_1^2+\cos (4 \alpha ) \left(\Lambda_1^2-p_2 \phi \right)+p_2 \phi \right)\right) ,
    \end{dmath*}
    \begin{dmath*}
        h_{1112}^{\text{G}} = 
        \frac{4 \cos\alpha}{d} \left(\sinh \left(\frac{\sqrt{p_2}}{2}\right) \left(\sqrt{p_1} p_2 \phi  \cosh \left(\frac{\sqrt{p_1}}{2}\right) \left(\Lambda_1^2+\cos (2 \alpha ) \left(\Lambda_1^2-p_2 \phi \right)+p_2 \phi \right)-2 \sinh \left(\frac{\sqrt{p_1}}{2}\right) \left(\Lambda_1^2 p_1+\cos (2 \alpha) \left(\Lambda_1^2 p_1+p_2 \phi  \left(\Lambda_1^2-p_2 \phi \right)\right)+p_2 \phi  \left(\Lambda_1^2+p_2 \phi \right)\right)\right)+p_1 \sqrt{p_2} \sinh \left(\frac{\sqrt{p_1}}{2}\right) \cosh \left(\frac{\sqrt{p_2}}{2}\right) \left(\Lambda_1^2+\cos (2 \alpha ) \left(\Lambda_1^2-p_2 \phi \right)+p_2 \phi \right)\right) ,
    \end{dmath*}
    \begin{dmath*}
        h_{1122}^{\text{G}} = 
        \frac{2 \sin\alpha}{d} \left(\sinh \left(\frac{\sqrt{p_2}}{2}\right) \left(\sqrt{p_1} p_2 \phi  \cosh \left(\frac{\sqrt{p_1}}{2}\right) \left(\Lambda_1^2+\cos (2 \alpha ) \left(\Lambda_1^2-p_2 \phi \right)+p_2 \phi \right)-2 \sinh \left(\frac{\sqrt{p_1}}{2}\right) \left(\Lambda_1^2 p_1+\cos (2 \alpha ) \left(\Lambda_1^2 p_1+p_2 \phi  \left(\Lambda_1^2-p_2 \phi \right)\right)+p_2 \phi  \left(\Lambda_1^2+p_2 \phi \right)\right)\right)+p_1 \sqrt{p_2} \sinh \left(\frac{\sqrt{p_1}}{2}\right) \cosh \left(\frac{\sqrt{p_2}}{2}\right) \left(\Lambda_1^2+\cos (2 \alpha ) \left(\Lambda_1^2-p_2 \phi \right)+p_2 \phi \right)\right) ,
    \end{dmath*}
    \begin{dmath*}
        h_{1211}^{\text{G}} = 
        \frac{4 \cos\alpha}{d} \left(p_1 \sqrt{p_2} \cos^2\alpha \sinh \left(\frac{\sqrt{p_1}}{2}\right) \cosh \left(\frac{\sqrt{p_2}}{2}\right) \left(\Lambda_1^2-p_2 \phi \right)- \sinh \left(\frac{\sqrt{p_2}}{2}\right) \left(\sqrt{p_1} p_2 \phi  \cos^2\alpha \cosh \left(\frac{\sqrt{p_1}}{2}\right) \left(p_2 \phi -\Lambda_1^2\right)+\sinh \left(\frac{\sqrt{p_1}}{2}\right) \left(\cos (2 \alpha ) \left(\Lambda_1^2 p_1+p_2 \phi  \left(\Lambda_1^2-p_2 \phi \right)\right)+p_1 \left(\Lambda_1^2-2 p_2 \phi \right)+p_2 \phi  \left(\Lambda_1^2-p_2 \phi \right)\right)\right)\right) ,
    \end{dmath*}
    \begin{dmath*}
        h_{1212}^{\text{G}} = 
        \frac{4 \sin\alpha}{d} \left(2 p_1 \sqrt{p_2} \cos^2\alpha \sinh \left(\frac{\sqrt{p_1}}{2}\right) \cosh \left(\frac{\sqrt{p_2}}{2}\right) \left(\Lambda_1^2-p_2 \phi \right)+\sinh \left(\frac{\sqrt{p_2}}{2}\right) \left(2 \sqrt{p_1} p_2 \phi  \cos^2\alpha \cosh \left(\frac{\sqrt{p_1}}{2}\right) \left(\Lambda_1^2-p_2 \phi \right)+2 \sinh \left(\frac{\sqrt{p_1}}{2}\right) \left((p_1+p_2 \phi ) \left(-\left(\Lambda_1^2-p_2 \phi \right)\right)-\cos (2 \alpha ) \left(\Lambda_1^2 p_1+p_2 \phi  \left(\Lambda_1^2-p_2 \phi \right)\right)\right)\right)\right) ,
    \end{dmath*}
    \begin{dmath*}
        h_{1222}^{\text{G}} = 
        \frac{-4 \sin^2\alpha \cos\alpha}{d} \left(\sinh \left(\frac{\sqrt{p_2}}{2}\right) \left(2 \sinh \left(\frac{\sqrt{p_1}}{2}\right) \left(\Lambda_1^2 p_1+p_2 \phi  \left(\Lambda_1^2-p_2 \phi \right)\right)+\sqrt{p_1} p_2 \phi  \cosh \left(\frac{\sqrt{p_1}}{2}\right) \left(p_2 \phi -\Lambda_1^2\right)\right)+p_1 \sqrt{p_2} \sinh \left(\frac{\sqrt{p_1}}{2}\right) \cosh \left(\frac{\sqrt{p_2}}{2}\right) \left(p_2 \phi -\Lambda_1^2\right)\right) ,
    \end{dmath*}
    \begin{dmath*}
        h_{2211}^{\text{G}} = 
        \frac{1}{2 d \sin\alpha} \left(\sinh \left(\frac{\sqrt{p_2}}{2}\right) \left(\sqrt{p_1} p_2 \phi  \cosh \left(\frac{\sqrt{p_1}}{2}\right) \left(\Lambda_1^2+8 p_1+\cos (4 \alpha ) \left(p_2 \phi -\Lambda_1^2\right)+4 p_2 \phi  \cos (2 \alpha )+3 p_2 \phi \right)-2 \sinh \left(\frac{\sqrt{p_1}}{2}\right) \left(8 p_1^2-\cos (4 \alpha ) \left(\Lambda_1^2 p_1+p_2 \phi  \left(\Lambda_1^2-p_2 \phi \right)\right)
        +4 p_2 \phi  \cos (2 \alpha ) (2 p_1+p_2 \phi )
        +p_1 \left(\Lambda_1^2+8 p_2 \phi \right)
        +p_2 \phi  \left(\Lambda_1^2+3 p_2 \phi \right)\right)\right)
        +p_1 \sqrt{p_2} \sinh \left(\frac{\sqrt{p_1}}{2}\right) \cosh \left(\frac{\sqrt{p_2}}{2}\right) \left(\Lambda_1^2+8 p_1+\cos (4 \alpha ) \left(p_2 \phi -\Lambda_1^2\right)+4 p_2 \phi  \cos (2 \alpha )+3 p_2 \phi \right)\right) ,
    \end{dmath*}
    \begin{dmath*}
        h_{2212}^{\text{G}} = 
        \frac{-4 \cos\alpha}{d} \left(\sinh \left(\frac{\sqrt{p_2}}{2}\right) \left(2 \sinh \left(\frac{\sqrt{p_1}}{2}\right) \left(-\cos (2 \alpha ) \left(\Lambda_1^2 p_1+p_2 \phi  \left(\Lambda_1^2-p_2 \phi \right)\right)
        +p_1 \left(\Lambda_1^2+2 p_2 \phi \right)+p_2 \phi  \left(\Lambda_1^2+p_2 \phi \right)\right)-\sqrt{p_1} p_2 \phi  \cosh \left(\frac{\sqrt{p_1}}{2}\right) \left(\Lambda_1^2+\cos (2 \alpha ) \left(p_2 \phi -\Lambda_1^2\right)+p_2 \phi \right)\right)-p_1 \sqrt{p_2} \sinh \left(\frac{\sqrt{p_1}}{2}\right) \cosh \left(\frac{\sqrt{p_2}}{2}\right) \left(\Lambda_1^2+\cos (2 \alpha ) \left(p_2 \phi -\Lambda_1^2\right)+p_2 \phi \right)\right) ,
    \end{dmath*}
    \begin{dmath*}
        h_{2222}^{\text{G}} = 
        \frac{-2 \sin\alpha}{d} \left(\sinh \left(\frac{\sqrt{p_2}}{2}\right) \left(2 \sinh \left(\frac{\sqrt{p_1}}{2}\right) \left(\Lambda_1^2 p_1-\cos (2 \alpha ) \left(\Lambda_1^2 p_1+p_2 \phi  \left(\Lambda_1^2-p_2 \phi \right)\right)+p_2 \phi\left(\Lambda_1^2+p_2 \phi \right)\right)-\sqrt{p_1} p_2 \phi  \cosh \left(\frac{\sqrt{p_1}}{2}\right) \left(\Lambda_1^2+\cos (2 \alpha ) \left(p_2 \phi -\Lambda_1^2\right)+p_2 \phi \right)\right)-p_1 \sqrt{p_2} \sinh \left(\frac{\sqrt{p_1}}{2}\right) \cosh \left(\frac{\sqrt{p_2}}{2}\right) \left(\Lambda_1^2+\cos(2\alpha) \left(p_2 \phi -\Lambda_1^2\right)+p_2 \phi \right)\right) ,
    \end{dmath*}
\end{dgroup*}
where
\begin{dmath*}[style={\footnotesize}]
    d =
    e^{-\frac{1}{2} \left(\sqrt{p_1}+\sqrt{p_2}\right)}
    \Lambda_1^2 \left(\left(e^{\sqrt{p_1}} \left(\sqrt{p_1}-2\right)+\sqrt{p_1}+2\right) \left(e^{\sqrt{p_2}}-1\right) p_2   \phi-2 \left(e^{\sqrt{p_1}}-1\right) p_1 \left(e^{\sqrt{p_2}}-1\right)+\left(e^{\sqrt{p_1}}-1\right) p_1 \left(e^{\sqrt{p_2}}+1\right) \sqrt{p_2}\right)
    .
\end{dmath*}

\end{document}